\documentclass[openany,11pt]{book}

\usepackage{pxps}
\newcommand{\OMIT}[1]{}

\newcommand{\PX}{{\slshape Project~X}}

\def\ifmath#1{\relax\ifmmode#1\else$#1$\fi}

\newcommand{\vs}{\emph{vs.}}

\renewcommand{\Im}{\ensuremath{\mathop{\mathrm{Im}}}}

\newcommand{\gsim}{\gtrsim}
\newcommand{\lsim}{\lesssim}

\newcommand{\euvec}[1]{\ensuremath{\bm{#1}}}       % euclidian vectors

\newcommand{\ecm}{\ensuremath{e\,\text{cm}}}

\newcommand{\nnb}{\ensuremath{n\hbox{-}\bar{n}}}
\newcommand{\nnbx}{NNbarX}

\newcommand{\BR}{\ensuremath{\text{BR}}}

\newcommand{\CP}{\ensuremath{CP}}
\newcommand{\CPT}{\ensuremath{CPT}}

\newcommand{\Kplus}{\ensuremath{K^+\rightarrow \pi^+ \nu \bar{\nu}}}
\newcommand{\Kzero}{\ensuremath{K^0_L\rightarrow \pi^0 \nu \bar{\nu}}}

\newcommand{\eff}{\ensuremath{\mathrm{eff}}}

\begin{document}

\pagestyle{empty}
\begin{titlepage}
\setcounter{page}{401} 
\vskip 2in
\begin{center}
    \Huge\bfseries\sffamily
    \textsl{Project~X}: Physics Opportunities
\end{center}
\begin{center}
Andreas~S.~Kronfeld,
Robert~S.~Tschirhart \\
(Editors)
\end{center}
\begin{center}
Usama~Al-Binni,
Wolfgang~Altmannshofer,
Charles~Ankenbrandt,
Kaladi~Babu,
Sunanda~Banerjee,
Matthew~Bass,
Brian~Batell,
David~V.~Baxter,
Zurab~Berezhiani,
Marc~Bergevin,
Robert~Bernstein,
Sudeb~Bhattacharya,
Mary~Bishai,
Thomas~Blum,
S.~Alex~Bogacz,
Stephen~J.~Brice,
Joachim~Brod,
Alan~Bross,
Michael~Buchoff,
Thomas~W.~Burgess,
Marcela~Carena,
Luis~A.~Castellanos,
Subhasis~Chattopadhyay,
Mu-Chun~Chen,
Daniel~Cherdack,
Norman~H.~Christ,
Tim~Chupp,
Vincenzo~Cirigliano,
Pilar~Coloma,
Christopher~E.~Coppola,
Ramanath~Cowsik,
J.~Allen~Crabtree,
Andr\'e~de~Gouv\^ea,
Jean-Pierre~Delahaye,
Dmitri~Denisov,
Patrick~deNiverville,
Ranjan~Dharmapalan,
Markus~Diefenthaler,
Alexander~Dolgov,
Georgi~Dvali,
Estia~Eichten,
J\"urgen~Engelfried,
Phillip~D.~Ferguson,
Tony~Gabriel,
Avraham~Gal,
Franz~Gallmeier,
Kenneth~S.~Ganezer,
Susan~Gardner,
Douglas~Glenzinski,
Stephen~Godfrey,
Elena~S.~Golubeva,
Stefania~Gori,
Van~B.~Graves,
Geoffrey~Greene,
Cory~L.~Griffard,
Ulrich~Haisch,
Thomas~Handler,
Brandon~Hartfiel,
Athanasios~Hatzikoutelis,
Ayman~Hawari,
Lawrence~Heilbronn,
James~E.~Hill,
Patrick~Huber,
David~E.~Jaffe,
Xiaodong~Jiang,
Christian~Johnson,
Yuri~Kamyshkov,
Daniel~M.~Kaplan, % IIT
Boris~Kerbikov,
Brendan~Kiburg,
Harold~G.~Kirk,
Andreas~Klein,
Kyle~Knoepfel,
Boris~Kopeliovich,
Vladimir~Kopeliovich,
Joachim~Kopp,
Wolfgang~Korsch,
Graham~Kribs,
Ronald~Lipton,
Chen-Yu~Liu,
Wolfgang~Lorenzon,
Zheng-Tian~Lu,
Naomi~C.~R.~Makins,
David~McKeen,
Geoffrey~Mills,
Michael~Mocko,
Rabindra~Mohapatra,
Nikolai~V.~Mokhov,
Guenter~Muhrer,
Pieter~Mumm,
David~Neuffer,
Lev~Okun,
Mark~A.~Palmer,
Robert~Palmer,
Robert~W.~Pattie~Jr.,
David~G.~Phillips~II,
Kevin~Pitts,
Maxim~Pospelov,
Vitaly~S.~Pronskikh,
Chris~Quigg,
Erik~Ramberg,
Amlan~Ray,
Paul~E.~Reimer,
David~G.~Richards,
Adam~Ritz,
Amit~Roy,
Arthur~Ruggles,
Robert~Ryne,
Utpal~Sarkar,
Andy~Saunders,
Yannis~K.~Semertzidis,
Anatoly~Serebrov,
Hirohiko~Shimizu,
Robert~Shrock,
Arindam~K.~Sikdar,
Pavel~V.~Snopok,
William~M.~Snow,
Aria~Soha,
Stefan~Spanier,
Sergei~Striganov,
Zhaowen~Tang,
Lawrence~Townsend,
Jon~Urheim,
Arkady~Vainshtein,
Richard~Van~de~Water,
Ruth~S.~Van~de~Water,
Richard~J.~Van~Kooten,
Bernard~Wehring,
William~C.~Wester~III,
Lisa~Whitehead,
Robert~J.~Wilson,
Elizabeth~Worcester,
Albert~R.~Young,
and
Geralyn~Zeller
\end{center}

\begin{center}
\sl
Argonne National Laboratory, Argonne, Illinois \\
University of Alabama, Tuscaloosa, Alabama \\
Brookhaven National Laboratory, Upton, New York \\
University of California, Davis, California \\
University of California, Irvine, California \\
% University of California, Riverside, California \\
% University of California, San Diego, California \\
California State University Dominguez Hills, Carson, California\\
Carleton University, Ottawa, Ontario, Canada \\
Columbia University, New York, New York \\
% CERN, Geneva, Switzerland \\
University of Chicago, Chicago, Illinois \\
University of Cincinnati, Cincinnati, Ohio \\
Colorado State University, Fort Collins, Colorado \\
University of Connecticut, Storrs, Connecticut \\
% Deutsches Elektronen-Synchrotron (DESY), Hamburg, Germany \\
Fermi National Accelerator Laboratory, Batavia, Illinois \\  % file under Fermi
% University of Glasgow, Glasgow, United Kingdom \\
% Harvard University, Cambridge, Massachusetts \\
Hebrew University, Jerusalem, Israel \\
University of Houston, Houston, Texas \\
Illinois Institute of Technology, Chicago, Illinois \\
University of Illinois at Urbana-Champaign, Urbana, Illinois \\
% Imperial College, London, United Kingdom \\
Indiana University, Bloomington, Indiana \\
% INFN, Genoa, Italy \\
% INFN, Milan, Italy \\
INFN, University of Ferrara, Ferrara, Italy \\
INFN, Gran Sasso National Laboratory, Assergi, Italy \\
% Institute for High Energy Physics, Protvino, Russia \\
Institute for Nuclear Research, Moscow, Russia \\
Institute for Nuclear Research, Troitsk, Russia \\
Institute of Theoretical and Experimental Physics, Moscow, Russia \\
Inter University Accelerator Centre, New Delhi, India \\
% Iowa State University, Ames, Iowa \\
Jefferson Laboratory, Newport News, Virginia \\ % file under Jefferson
% The Johns Hopkins University, Baltimore, Maryland \\
University of Kentucky, Lexington, Kentucky \\
University of L'Aquila, L'Aquila, Italy \\
Lawrence Berkeley National Laboratory, Berkeley, California \\
Lawrence Livermore National Laboratory, Livermore, California \\
Los Alamos National Laboratory, Los Alamos, New Mexico \\
Ludwig-Maximillians Universit\"at, Munich, Germany \\
University of Maryland, College Park, Maryland \\
% Massachusetts Institute of Technology, Cambridge, Massachusetts \\
% Universit\'a degli Studi di Milano, Milan, Italy \\
University of Michigan, Ann Arbor, Michigan \\
University of Minnesota, Minneapolis, Minnesota \\
% NIKHEF, Amsterdam, The Netherlands \\
% Northern Illinois University, DeKalb, Illinois \\
Novosibirsk State University, Novosibirsk, Russia \\
Muons Inc., Batavia, Illinois \\
Nagoya University, Nagoya, Japan \\
National Institute of Standards and Technology, Gaithersburg, Maryland \\
New York University, New York, New York \\
North Carolina State University, Raleigh, North Carolina \\
Northwestern University, Evanston, Illinois \\
% Notre Dame University, Notre Dame, Indiana \\
Oak Ridge National Laboratory, Oak Ridge, Tennessee \\
% The Ohio State University, Columbus, Ohio \\
Oklahoma State University, Stillwater, Oklahoma \\
University of Oxford, Oxford, United Kingdom \\
% University of Pennsylvania, Philadelphia, Pennsylvania \\
Physical Research Laboratory, Ahmedabad, India \\
Max-Planck-Institut f\"ur Kernphysik, Heidelberg, Germany \\ % file under Planck
% INFN, Universit\'a e Scuola Normale Superiore di Pisa, Pisa, Italy \\
Universidad Aut\'onoma de San Luis Potos\'i, San Luis Potos\'i, Mexico \\
Saha Institute of Nuclear Physics, Kolkata, India \\
SLAC National Accelerator Laboratory, Stanford, California \\
St. Petersburg Nuclear Physics Institute, Russia \\
State University of New York, Stony Brook, New York \\
% Syracuse University, Syracuse, New York \\
% Technion--Israel Institute of Technology, Tel Aviv, Israel \\
Universidad T\'ecnica Federico Santa Mar\'ia, Chile \\
University of Tennessee, Knoxville, Tennessee \\
% University of Toronto, Toronto, Canada \\
% University of Tsukuba, Tsukuba, Japan  \\
Variable Energy Cyclotron Centre, Kolkata, India \\
% Vanderbilt University, Nashville, Tennessee \\
University of Victoria, Victoria, British Columbia, Canada  \\
Virginia Polytechnic Institute and State University, Blacksburg, Virginia \\
Washington University, St. Louis, Missouri \\
% Weizmann Institute of Science, Rehovot, Israel \\
% University of Wisconsin, Madison, Wisconsin \\
% Yale University, New Haven, Connecticut \\
% York University, Toronto, Canada
\end{center}

\setcounter{page}{0} 
\end{titlepage}

\frontmatter

\pagestyle{fancyplain}

{\sffamily\tableofcontents
\vfill
\begin{minipage}[b]{0.375\textwidth}
FERMILAB-TM-2557 \\
ANL/PHY-13/2 \\
BNL-101116-2013-BC/81834 \\
JLAB-ACP-13-1725 \\
PNNL-22523 \\
SLAC-R-1029 \\
UASLP-IF-13-001
\end{minipage}\hfill
\begin{minipage}[b]{0.125\textwidth}
    June~2013
\end{minipage}

\listoffigures
\addcontentsline{toc}{chapter}{List of Figures}
\listoftables
\addcontentsline{toc}{chapter}{List of Tables}
}
% \include{preface}

% \clearpage{\pagestyle{empty}\cleardoublepage}

\mainmatter

%%%%%%%%%%%%%%%%%%%%%%%%%%%%%%%%%%%%%%%%%%%%%%%%%%%%%%%%%%%%
\chapter{Particle Physics with \PX}
\label{chapt:intro}
%%%%%%%%%%%%%%%%%%%%%%%%%%%%%%%%%%%%%%%%%%%%%%%%%%%%%%%%%%%%

\authors{Andreas S. Kronfeld and Robert S. Tschirhart}

This part of the book presents many aspects of the physics opportunities that become available with the \PX\
superconducting linac.
As discussed in detail in \href{http://arxiv.org/abs/1306.5022}{Part~I}, the key features for physics are the
high intensity, the flexible time structure, and the potential for mounting many experiments simultaneously.
Many components of the \PX\ physics program complement each other: for example, neutrino experiments and
searches for permanent electric dipole moments both aim to find new sources of \CP\ violation, in two
different sectors of particle physics.
Some components of the physics program---from neutrino physics to hadron structure---are stalwarts of
Fermilab fixed-target experiments.
Others---searches for electric dipole moments and for baryon number violation via neutron-antineutron
oscillations---address familiar themes, but the specific research would be new to Fermilab.
The ability and flexibility to simultaneously run such a broad and rich experimental program is what makes
the \PX\ accelerator such an attractive idea.

\section{Themes}
\label{intro:sec:intro}

% {\sl better for Preface}
%
% Major advances in particle physics are driven by facilities of ever greater capacity and capability.
% \PX\ is a proposed new facility that is the evolution of the best assets of the Fermilab accelerator
% campus with the recent revolution in super-conducting RF technology.
% This marriage of existing high quality accelerator infrastructure and state-of-the-art accelerator
% technology will enable a suite of particle beams of unprecedented power and flexibility that can drive
% a rich particle physics experimental research program.

Particle physics aims to understand the nature of matter, space, and time in their most fundamental guise.
Some of the questions that propel our research are as follows:
\begin{itemize}
    \item Are there new forces in nature?
    \label{intro:item:forces}
    \item Do any new properties of matter help explain the basic features of the natural world?
    \label{intro:item:matter}
    \item Are there any new (normal, or fermionic) dimensions to spacetime?
    \label{intro:item:dimensions}
\end{itemize}
In pursuit of these themes, the mainstays of laboratory physics are high-energy colliding-beam 
experiments on the one hand, and intense beams on fixed targets on the other.
Although one usually thinks of the first as the place to discover new particles, and the second as the place 
to tease out rare and unusual interactions, history provides several examples of precise measurements at 
high-energy colliders (for example, the mass of the $W$ boson and the $B_s$ oscillation frequency) and 
unexpected discoveries at high-intensity experiments (for example, flavor mixing in quarks and in neutrinos).

The \PX\ research program discussed in the following chapters addresses these deep questions in several ways:
\begin{itemize}
    \item \emph{New forces}: 
        Experiments have established flavor-violating processes in quarks and neutrinos, so it seems 
        conceivable that charged leptons violate flavor too.
        With \PX, one can search for these phenomena via muon-to-electron conversion and related processes.
        Many of the theoretical ideas unifying forces and flavor violation anticipate baryon-number 
        violation, and \PX\ can extend the limits on neutron-antineutron oscillations by orders of 
        magnitude.
        These same ideas posit measurable flavor-changing neutral currents, thereby mediating rare decays 
        such as (charged and neutral) $K\to\pi\nu\bar{\nu}$.
    \item \emph{New properties of matter}:
        According to the Sakharov conditions, the baryon asymmetry of the universe requires \CP-violating
        interactions, but their strength in the Standard Model is insufficient to account for the 
        observed excess.
        It is not known whether the missing \CP\ violation takes place in the neutrino sector or the quark 
        sector.
        \PX\ will aid both searches, by increasing the reach of neutrino oscillation experiments and by 
        enabling a new suite of searches for nonzero electric dipole moments (EDMs).
        The latter program is broad, looking for an EDM of the neutron, proton, muon directly, and the 
        electron, exploiting amplification in atoms such as $^{225}$Ra, $^{223}$Rn, and $^{211}$Fr.
    \item \emph{New dimensions}:
        Many extensions of the Standard Model introduce extra dimensions: in the case of 
        supersymmetry, the dimensions are fermionic.
        The space of non-Standard interactions opens up possibilities for the interactions mentioned above:\
        quark and neutrino \CP~violation and quark-flavor-changing neutral currents with supersymmetry, and
        flavor-changing neutral currents from a warped fifth spatial dimension.
        Rare kaon decays, EDMs, and neutron-antineutron oscillations are closely tied to these possibilities.
\end{itemize}
In addition to probing these fundamental questions, the \PX\ research program includes experiments that
test and enrich our understanding of quantum chromodynamics and the electroweak theory.
The following chapters spell out in detail the physics motivation and experimental
techniques of this broad program.

% The Reference Design Report volume of the \PX\ book describes in detail an accelerator complex that can
% drive this broad experimental program.
% The \PX\ accelerator architecture presents opportunities to stage construction of the complete facility
% if necessary, with a robust research program at the first stage and each successive stage growing both the
% breadth of beams available and the intensity of those beams.
% A key and unique characteristic of the \PX\ complex is that the multiplicity of beams required by the
% research program can largely be provided {\it simultaneously} enabling multiple experiments to proceed in
% parallel.
% Staging opportunities for the \PX\ accelerator complex are illustrated in Table~\ref{intro:tab:staging}.
% A detailed discussion of the accelerator configuration of each stage can be found in Appendix~I of the
% Reference Design Report volume.
% The research program enabled by each stage is outlined here in the following sections.

The key to \PX\ is that it provides a platform for many experiments requiring high intensity.
Not all of them are documented below, because, once the accelerator and experimental halls have been built, 
creative minds will generate new ideas that we cannot anticipate.
Moreover, \PX\ can, in the farther future, lead to one or more of a neutrino factory, a muon ($\mu^+\mu^-$) 
collider, or a very high-energy proton collider with energy well beyond that of the LHC.

%%%%%%%%%%%%%%%%%%%%%%%%%%%%%%%%%%%%%%%%%%%%%%%%%%%%%%%%%%%%
\section{\PX\ Physics Stage by Stage}
\label{intro:sec:staging}
%%%%%%%%%%%%%%%%%%%%%%%%%%%%%%%%%%%%%%%%%%%%%%%%%%%%%%%%%%%%

The \PX\ linac falls naturally into three stages.
The first accelerates protons (technically H$^-$ ions) to 1~GeV.
It transports a continuous-wave beam, which means that many different time structures can be packed into the
linacs.
The Stage~1 linac, thus, not only drives the existing Booster and Main Injector at higher intensity, but also
can distribute beam to other experiments with no interference to the Booster and Main Injector program.
Interesting new experiments with a spallation target could be mounted, and muon-to-electron conversion could
be studied without antiproton background.
The second stage accelerates the beam (still H$^-$) to 3~GeV.
The Booster and the Main Injector again become more powerful than before, and the 3-GeV linac itself
increases the yield of muons (for flavor-violation experiments) and kaons (for ultrarare kaon decays).
At this intensity, neutrino experiments driven by a 60-GeV Main Injector primary beam attain sufficiently
high event rate to elucidate \CP\ violation (see Sec.~\ref{nu:sec:lbl}) and the possibility of nonstandard
sterile neutrinos (see Sec.~\ref{nu:subsec:nustorm}).
The third stage is a pulsed linac that replaces the forty-year-old Booster, with a further power boost to 
the Main Injector, and no interruption to 1-GeV and 3-GeV operations.

\begin{sidewaystable}
    \centering
    \caption[Physics opportunities for \PX\ by Stage]{Physics opportunities for \PX\ by Stage.
        The accelerator Reference Design (RDR) is described in \href{http://arxiv.org/abs/1306.5022}{Part~I} 
        of this book and comprises Stages~1, 2, and~3.  
        In all Stages, \PX\ beam drives the Main Injector (MI)---in Stages~1 and~2 via the original 
        8-GeV Booster.
        During Stage~2, the Booster cycles at a higher rate, allowing the MI to operate over a wider energy 
        range, 60--120~GeV (instead of 80--120~GeV).
        Examples of 8-GeV muon experiments include Mu2e and muon $g-2$; an example of a 1--3-GeV muon 
        experiment is an extension of Mu2e with optimized time structure and no antiproton background. 
        Muon spin rotation ($\mu$SR) and nuclear irradiation are broader impacts of \PX\ technology, 
        discussed in \href{http://arxiv.org/abs/1306.5024}{Part~III}.}
    \label{intro:tab:staging}
    \begin{tabular}{lc@{\quad\quad}c@{\quad}c@{\quad}c@{\quad\quad}c}
    \hline\hline
                     & Present             &         \multicolumn{3}{l}{\PX\ Accelerator Reference Design}          & Beyond RDR \\
    Program          & NO$\nu$A operations & Stage~1               &        Stage~2        &         Stage~3        & Stage~4 \\
%                      &                     & 1~GeV CW linac        &  $1\to3$~GeV CW linac &  $3\to8$~GeV linac      & power upgrade \\
    \hline
    MI neutrino      & 470--700~kW\footnote{Operating point in range depends on the MI proton beam energy 
    for neutrino production.}$^\textit{,b}$
                                           & 515--1200~kW$^\textit{a,b}$  &   1200~kW             
                                           &   2450~kW             & 2450--4000~kW
        \\
    8 GeV neutrino   & 15--65~kW$^\textit{a,}$\footnote{Operating point in range depends on the MI 
    slow-spill duty factor for kaon and hadron-structure experiments.}
                                           & 0--130~kW$^\textit{a}$& 0--130~kW$^\textit{a}$& 0--172~kW$^\textit{a}$              & 3000~kW
        \\
    8 GeV muon       & 20~kW               & 0--20~kW$^\textit{a}$ & 0--20~kW$^\textit{a}$ & 0--172~kW$^\textit{a}$ & 1000~kW
        \\
    % e.g. $(g-2)$, Mu2e & & & & & 
    % \\
    1--3 GeV muon    & --- & 80~kW & 1000~kW & 1000~kW & 1000~kW 
        \\
    % program, e.g. Mu2e-X & & & & & 
    % \\
    Rare kaon decays & 0--30~kW$^\textit{b,}$\footnote{With less than 30\% duty factor from Main Injector.} 
                                            & 0--75~kW$^\textit{b,}$\footnote{With less than 45\% duty factor from Main Injector.} 
                                                                   & 1100~kW                & 1870~kW                & 1870~kW 
        \\
    Atomic EDMs      & ---  & 0--900~kW & 0--900~kW & 0--1000~kW & 0--1000~kW
        \\
    % Rare isotope & & & & &
    % \\
    Cold neutrons    & --- & 0--900~kW & 0--900~kW & 0--1000~kW & 0--1000~kW
        \\ 
    $\mu$SR facility & --- & 0--900~kW & 0--900~kW & 0--1000~kW & 0--1000~kW
        \\ 
    Irradiation facility & --- & 0--900~kW & 0--900~kW & 0--1000~kW & 0--1000~kW
        \\ 
    \hline
    Number of programs & 4 & 8 & 8 & 8 & 8
        \\
%     Total power      & 735~kW & 2222~kW & 4284~kW & 6492~kW & 11870~kW \\
    Total power      & 740~kW & 2200~kW & 4300~kW & 6500~kW & 12,000~kW 
        \\
    \hline\hline
    \end{tabular}
\end{sidewaystable}

The details of the accelerator staging are shown in Table~\ref{intro:tab:staging}, which includes also the 
capability of the Fermilab accelerator complex following the 2013 shutdown (second column from left).
\nopagebreak
In the rest of this section, we survey the highlights of each Stage of \PX, using this table as a guide.

\subsection{Stage~1}
\label{intro:sec:stage1}

As shown in the third column of Table~\ref{intro:tab:staging}, Stage~1 of \PX\ will increase the Main
Injector beam power for long-baseline neutrino experiments from 700~kW to 1200~kW.
Simultaneously, it will provide substantial power in the 8-GeV Booster for short-baseline neutrino
experiments.
The extra power in the Main Injector would make it easier for ORKA, a proposal to accumulate 1000 events of 
the rare decay $K^+\to\pi^+\nu\bar\nu$, to reach its goals.
In addition to the beam train to feed the Booster and Main Injector, the continuous-wave nature of Stage 1 
means that the beam can be configured to support experiments based on a 1-GeV primary beam itself.
A second beam train can be brought to the Mu2e experiment, increasing the available power from 8~kW to 80~kW.
The lower energy is a further benefit to this experiment, because it produces no antiproton background.
A~third beam train, with aggregated power up to 900~kW, will strike spallation targets optimized for 
particle physics and the programs discussed in \href{http://arxiv.org/abs/1306.5024}{Part~III}.
This facility will provide intense sources of cold neutrons for neutron-antineutron oscillations, ultracold
neutrons for a next-generation neutron-EDM measurement, and isotopes such as $^{225}$Ra,
$^{223}$Rn, and $^{211}$Fr, which are well-suited for electron EDM measurements.
A~straightforward modification of the 1-GeV linac could create and accelerate polarized protons to a momentum
of 0.7~GeV/$c$, which is precisely that needed for a proton EDM experiment in an electrostatic storage ring.
Note that the Standard-Model strong-\CP\ contribution to the EDM changes sign from neutron to proton, whereas 
non-Standard contributions need not be of opposite sign.
Thus, putting commensurate limits on both nucleon EDMs helps to constrain both kinds of \CP\ 
violation.

\subsection{Stage~2}
\label{intro:sec:stage2}

Stage~2 of \PX, as shown in the fourth column of Table~\ref{intro:tab:staging}, would support up to 1200~kW
of power for long-baseline neutrino experiments over a wide range of Main Injector energy, down to 60~GeV
from the usual 120~GeV.
The lower initial energy allows the design of a neutrino beam whose energy spectrum is peaked at somewhat
lower energies.
With the high \PX\ intensity, the flux remains sufficient to study neutrino mixing.
In fact, this setup enhances the sensitivity to neutrino mixing parameters, particularly the \CP-violating
phase of the mixing matrix that affects oscillations.
The high power at 3~GeV can also serve to drive an early phase of a neutrino factory.

Stage~2 is the gateway for very high power for next generation muon and kaon experiments, up to 1000~kW per
experiment.
The energy, 3~GeV, has been chosen because it lies in the optimal ranges for muon and kaon yields.
A third phase of Mu2e and related experiments (e.g., $\mu\to eee$ and oscillations between $\mu^+e^-$ and
$\mu^-e^+$ exotic atoms) will be mounted at the 3-GeV campus.
The 3~GeV is also well suited to a long-recognized goal in kaon physics, the collection of 1000 events (at
the Standard-Model rate) of the decay $K_L\to\pi^0\nu\bar\nu$.
Like its charged partner, it is a discovery mode.
If new particles are found at the LHC, these measurements---on their own and in concert with other
constraints from $K$, $D$, and $B$ physics---lead to excellent discrimination among models of new physics.
Note that these experiments run in parallel with the EDM and $n$-$\bar{n}$ experiments described in Stage~1.
They all use different parts of the continuous-wave beam.

\subsection{Stage~3}
\label{intro:sec:stage3}

Stage~3, summarized in the fifth column of Table~\ref{intro:tab:staging}, fully realizes the Reference
Design.
The total beam power of the Fermilab campus will now exceed 6000~kW, nearly ten times that available today.
The beam power from the Main Injector alone will be 2450~kW, a three-fold increase.
As in Stage~2, the Main Injector can be operated over a wide range of beam energy, 60--120~GeV, depending on
physics needs.
For long-baseline neutrino experiments, the benefit of high power is enormous: increasing the power by a
factor of three increases the reach of an experiment just as much as tripling the detector mass.
Short-baseline experiments at 8-GeV (primary) energy will at this stage have 180~kW of beam power available,
which, again, ten times the current 8-GeV Booster.
Once again, the higher power of Stage~3 at 8~GeV and at 60--120~GeV is a new resource.
The experiments made possible by Stages~1 and~2 continue as before without interruption or penalty.

\subsection{Stage~4: The Longer Term}
\label{intro:sec:stage4}

These three Stages complete the \PX\ Reference Design, but the central idea of physics opportunities enabled
by high beam power need not stop there.
Appendix~II of the Reference Design describes, and the right-most column of Table~\ref{intro:tab:staging}
summarizes, a further upgrade to the entire Fermilab accelerator complex, known as Stage~4.
The key additional capability of Stage~4 is much higher power, 3000--4000~kW at 8~GeV, for example, to drive
more advanced accelerator concepts.
In neutrino physics, these ideas include superbeams (e.g., simultaneous low and high energy neutrino beams
illuminating the same large detector) and neutrino factories with beams produced in muon storage rings.
Furthermore, Stage~4 lays the groundwork for future energy-frontier colliders, such as a multi-TeV muon
collider or a very high energy hadron collider, which would need Stage-4 intensity at the front end.

%%%%%%%%%%%%%%%%%%%%%%%%%%%%%%%%%%%%%%%%%%%%%%%%%%%%%%%%%%%%
\section{Organization of the Physics Chapters}
\label{intro:sec:summary}
%%%%%%%%%%%%%%%%%%%%%%%%%%%%%%%%%%%%%%%%%%%%%%%%%%%%%%%%%%%%

In the following, Chapters~\ref{chapt:nu}--\ref{chapt:nlwcp} flesh out the details of a broad attack on
physics beyond the Standard Model, outlined above.
Participants in the \PX\ Physics Study \cite{pxps:2012} explain, in turn, how the intense, flexible beam of
the \PX~accelerator can be used for neutrino physics, kaon physics, muon physics, electric dipole moments,
neutron-antineutron oscillations, and experiments searching for new, light, weakly-coupled particles.
The research program also has substantial components exploring hadronic structure and spectroscopy, which
are described in Chapters~\ref{chapt:hadron-dy} and~\ref{chapt:hadron-s}.
Chapter~\ref{chapt:lqcd} describes enabling and supportive developments in lattice quantum chromodynamics
that are important to both producing and interpreting measurements and associated scientific insights of the
\PX\ research program.

%%%%%%%%%%%%%%%%%%%%%%%%%%%%%%%%%%%%%%%%%%%%%%%%%%%%%%%%%%%%

%References~\cite{Cabibbo:1963yz,Kobayashi:1973fv} are highly    %cited, and
%Refs.~\cite{Pontecorvo:1957cp,Maki:1962mu} are catching up.

\bibliographystyle{apsrev4-1}
\bibliography{intro/refs}
 % Bob T & Andreas

%%%%%%%%%%%%%%%%%%%%%%%%%%%%%%%%%%%%%%%%%%%%%%%%%%%%%%%%%%%%
\chapter{Neutrino Experiments with \PX}
\label{chapt:nu}
%%%%%%%%%%%%%%%%%%%%%%%%%%%%%%%%%%%%%%%%%%%%%%%%%%%%%%%%%%%%

\authors{Andr\'e de Gouv\^ea, Patrick Huber, Geoffrey Mills, \\
Charles~Ankenbrandt, % Muons Inc.
Matthew~Bass,
Mary~Bishai,
S.~Alex~Bogacz,
Stephen~J.~Brice,
Alan~Bross,
Daniel~Cherdack,
Pilar~Coloma,
Jean-Pierre~Delahaye, % SLAC
Dmitri~Denisov,
Estia~Eichten,
Daniel~M.~Kaplan, % IIT
Harold~G.~Kirk,
Joachim Kopp, 
Ronald~Lipton,
David~Neuffer,
Mark~A.~Palmer,
Robert~Palmer,
Robert~Ryne,  % LBNL
Pavel~V.~Snopok,
Jon~Urheim,
Lisa~Whitehead,
Robert~J.~Wilson,
Elizabeth~Worcester,
and Geralyn~Zeller}

\section{Introduction}

Neutrino oscillations are irrefutable evidence for physics beyond the
Standard Model of particle physics. The observed properties of the
neutrino---the large flavor mixing and the tiny mass---could be
consequences of phenomena which occur at energies never seen since the
Big Bang, and they also could be triggered at energy scales as low as a
few keV. Determining the energy scale of the physics responsible for
neutrino mass is one of the primary tasks at the Intensity Frontier,
which will ultimately require high-precision measurements. High
precision is required because the telltale effects from either a low or
high energy scale responsible for neutrino masses and mixing will be
very small, either because couplings are very small, as in low-energy
models, or the energy scales are very high and thus its effects are
strongly suppressed. 

The three flavor oscillation framework is quite successful in
accounting for many results obtained in very different
contexts: the transformation of $\nu_e$ into $\nu_{\mu,\tau}$ from the
Sun~\cite{Aharmim:2011vm}; the disappearance of $\nu_\mu$ and
$\bar\nu_\mu$ from neutrinos produced by cosmic ray interactions in
the atmosphere; the disappearance of $\nu_\mu$ and
$\bar\nu_\mu$~\cite{Wendell:2010md,Abe:2011ph} from neutrino beams
over distances from
200--740~km~\cite{Ahn:2006zza,Adamson:2012rm,Abe:2012gx}; the
disappearance of $\bar\nu_e$ from nuclear reactors over a distance of
about 160~km~\cite{Abe:2008aa}; the disappearance of $\bar\nu_e$
from nuclear reactors over a distance of about
2~km~\cite{Abe:2012tg,Ahn:2012nd,An:2012eh}; and at somewhat lower
significance also the appearance of
$\nu_e$~\cite{Adamson:2011qu,Abe:2011sj} and, at even lower
significance, the appearance of $\nu_\tau$~\cite{Agafonova:2010dc} has
been observed in experiments using man-made neutrino beams over
200--740~km distance. All these experimental results can be succinctly
and accurately described by the oscillation of three active neutrinos
governed by the following parameters of the Pontecorvo-Maki-Nakagawa-Sakata 
matrix~\cite{Pontecorvo:1957cp,Maki:1962mu}, including their $1\sigma$ ranges~\cite{Fogli:2012ua}
\begin{equation}
    \begin{array}{rclrcrclr}
        \sin^2\theta_{12} & = & 3.07_{-0.16}^{+0.18}\times10^{-1}      &  (16\%);&\quad&
        \Delta m^2 & = & 2.43_{+0.1}^{-0.06}\times10^{-3}~\text{eV}^2  & (3.3\%);\\
        \sin^2\theta_{23} & = & 3.86_{-0.21}^{+0.24}\times10^{-1}      &  (21\%);&\quad&
        \delta m^2 & = & 7.54^{+0.26}_{-0.22}\times10^{-5}~\text{eV}^2 & (3.2\%);\\
        \sin^2\theta_{13} & = & 2.41\pm0.25\times10^{-1}               &  (10\%);&\quad&
        \delta & = & 1.08_{-0.31}^{+0.28}~\text{rad}                   &  (27\%);        
    \end{array}
\label{eq:nu1:parameters}
\end{equation}
where for all parameters whose value depends on the mass hierarchy, we
have chosen the values for the normal mass ordering. The choice of
parametrization is guided by the observation that for those parameters
the $\chi^2$ in the global fit is approximately Gaussian. The
percentages given in parenthesis indicate the relative error on each
parameter. For the mass splitting we reach errors of a few percent,
however, for all of the mixing angles and the \CP\ phase the errors are
in the 10--30\% range. Therefore, while three flavor oscillation is
able to describe a wide variety of experiments, it would seem
premature to claim that we have entered the era of precision neutrino
physics or that we have established the three flavor paradigm at a
high level of accuracy. This is also borne out by the fact that there
are significant hints at short baselines for a fourth
neutrino~\cite{Abazajian:2012ys}. Also, more general, so-called
non-standard interactions are not well constrained by neutrino data;
for a recent review on the topic see Ref.~\cite{Ohlsson:2012kf}. The
issue of what may exist beyond three flavor oscillations, in particular 
the issue of sterile neutrinos, is
discussed below in Sec.~\ref{nu:sec:sbl}.

Once one realizes that the current error bars are uncomfortably
large, the next question is how well one wants to determine the
various mixing parameters. The answer can be given on two, distinct
levels.  One is a purely technical one---if one wants to know $X$ to a
precision of $x$, one must know $Y$ with a precision of $y$; an
example is given by $Y$ taking to be $\theta_{13}$ and $X$  the
mass hierarchy. The other level is driven by theory
expectations of the size of possible phenomenological deviations from
the three flavor framework. In order to address the technical
part of the question, one first has to define the target precision
from a physics point of view. Looking at other fields of high-energy
physics it is clear that the target precision evolves.
For instance, predictions for the top quark mass, in hindsight, seem
to have been always ahead by only a few GeV of the experimental
capabilities, while at the time, there always was a valid physics
argument for why the top quark is just around the corner. A similar
evolution can be observed in $B$ physics. Thus, any argument based on
model-building inspired target precisions is always of a preliminary
nature, as our understanding of models improves. With this
caveat in mind, one argument for a target precision can be based on a
comparison to the quark sector. Based on a theoretical preference for
Grand Unification, one would expect that the answer to the flavor
question should find an answer for leptons and quarks at same time (or
energy scale) and therefore, a test of such a models should be most
sensitive if the precision in the lepton and quark sector were
nearly the same. For instance, the CKM angle $\gamma$, which is the exact
analog of $\delta$ in the neutrino sector, is determined to
$(70.4^{+4.3}_{-4.4})^\circ$~\cite{Lenz:2012az}.
Thus, a precision target for $\delta$ of roughly $5^\circ$ therefore follows.

Another argument for a similar level of precision can be made, based
on the concept of so-called neutrino sum-rules~\cite{King:2005bj}.
Neutrino sum-rules arise in models where the neutrino mixing matrix
has a certain simple form or texture at a high energy scale and the
actual low-energy mixing parameters are modified by a non-diagonal
charged lepton mass matrix. The simplicity of the neutrino mixing
matrix is typically a result of a flavor symmetry, where the overall
Lagrangian possesses an overall flavor symmetry $G$, which can be
separated into two sub-groups $G_\nu$ and $G_l$ for the neutrinos and
charged leptons; it is the mismatch between $G_\nu$ and $G_l$ that
yields the observed mixing pattern, see, e.g., 
Ref.~\cite{Altarelli:2010gt}. Typical candidates for $G$ are given
by discrete subgroups of SU(3) which have a three-dimensional
representation, e.g., $A_4$. In a model-building sense, these
symmetries can be implemented using so-called flavon fields, which
undergo spontaneous symmetry breaking.  This symmetry breaking 
picks the specific realization of $G$; for a recent review
see Ref.~\cite{King:2013eh}. The idea of flavor symmetries is in stark
contrast to the idea that neutrino mixing parameters are anarchic,
i.e., random numbers with no underlying dynamics; for the most
recent version of this argument, see Ref.~\cite{deGouvea:2012ac}. To
find out whether neutrino mixing corresponds to a symmetry or not
should be one of the prime tasks of neutrino physics and furthermore,
finding out which symmetry, should be attempted, as well.

\begin{figure}
    \centering
    \includegraphics[width=0.8\textwidth]{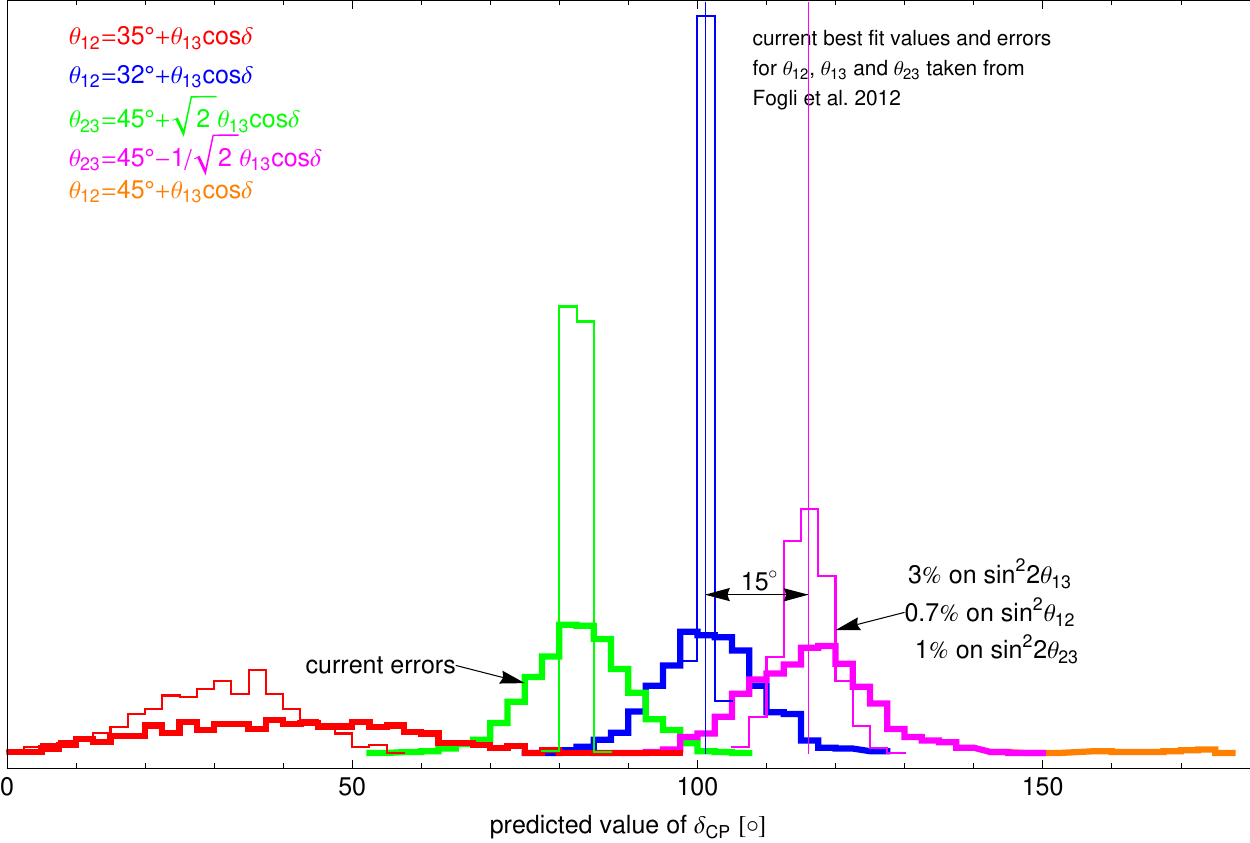}
    \caption[Predicted values of the \CP\ phase $\delta_{\CP}$ from neutrino sum rules]{Distributions of 
        predicted values from $\delta_{\CP}$ from various neutrino sum rules, as denoted in the legend and 
        explained in the text.}
    \label{fig:nu1:sumrules}
\end{figure}

In practice, flavor symmetries lead to relations between
measurable parameters, whereas anarchy does not. For example, if the
neutrino mixing matrix is of tri-bi-maximal form it predicts
$|U_{e3}|=0$, which is clearly in contradiction to observations. In
this case, a non-diagonal charged lepton mass matrix can be used to
generate the right value of $|U_{e3}|$, leading to a sum-rule
\begin{equation}
\label{nu1:eq:sumrule}
\theta_{12}-\theta_{13}\cos\delta=\arcsin \frac{1}{\sqrt{3}}
\end{equation}
that can be tested if sufficiently precise measured values for the
three parameters $\theta_{12},\theta_{13},\delta$ are available.
Depending on the underlying symmetry of the neutrino mixing matrix
different sum-rules arise. In Fig.~\ref{fig:nu1:sumrules}, several
examples are shown and for each case the values of $\theta_{13}$ and
$\theta_{12}$ or $\theta_{23}$ are drawn many times from a Gaussian
distribution where the mean values and ranges are taken from
Eq.~(\ref{eq:nu1:parameters}). The resulting predictions of the value of
the \CP\ phase $\delta$ are histogrammed and shown as colored lines.  The
width of the distribution for each sum-rule arises from the finite
experimental errors on $\theta_{12}$ or $\theta_{23}$ and $\theta_{13}$.
Two observations arise from this simple comparison, first the distance
of the means of the distributions is as small as $15^\circ$ and
secondly the width of the distributions is significant compared to
their separation and a reduction of input errors is mandated.  The
thin lines show the results if the errors are reduced to the value
given in the plot which would be achieved by Daya Bay for
$\sin^22\theta_{13}$, by Daya Bay~II for $\sin^2\theta_{12}$ and by
NO$\nu$A for $\sin^2\theta_{23}$. Assuming that the errors on
$\theta_{12}$, $\theta_{23}$ and $\theta_{13}$ are reduced to this
level, the limiting factor is the natural spread between models, which
is about $15^\circ$, which for a $3\sigma$ distinction between
models translates into a target precision for $\delta$ of $5^\circ$. A
measurement at this precision would allow to obtain valuable
information on whether indeed there is an underlying symmetry behind
neutrino mixing. Moreover, it is likely that is also allows to provide
hints which specific class of symmetries is realized.  This would
constitute a major breakthrough in our understanding of flavor.

In Sec.~\ref{nu:sec:lbl} we discuss long-baseline physics with
subsections on LBNE and muon-based facilities covering the gamut of
measurements from atmospheric parameters over \CP\ violation to
non-standard interactions. The focus in Sec.~\ref{nu:sec:lbl} is
largely on the three flavor oscillation framework, how to test it and
how to discover deviations from it. In Sec.~\ref{nu:sec:sbl} the
physics potential of experiments at a short baseline, i.e., less
than a few kilometers, is highlighted. One of the major physics
motivation for these experiments derives from existing experimental
hints for a eV-scale sterile neutrino.

\section{Long-baseline physics}
\label{nu:sec:lbl}

With the discovery of a large value for $\theta_{13}$, the physics
case for the next generation of long-baseline oscillation experiments
has grown considerably stronger and one of the major uncertainties on
the expected performance has been removed. The remaining questions
are: the value of the leptonic \CP\ phase and the quest for \CP\
violation; the mass hierarchy; whether $\theta_{23}$ is maximal and if
not, whether it is larger or smaller than $\pi/4$; and of course, the
search for new physics beyond the the three active neutrinos paradigm.
Based on our current, incomplete understanding of the origin of
neutrino mass and the observed flavor structure in general it is very
hard to rank these question in their relative importance, but with the
large value of $\theta_{13}$ it is feasible to design and build a
long-baseline facility which can address all three questions with high
precision and significance. Therefore, the question of relative
importance can be avoided.

The error on $\theta_{13}$ will keep decreasing as the reactor
measurements are refined and Daya Bay is expected to yield a precision
which only would be surpassed by a neutrino factory. It is an
important test of the three flavor oscillation model to see whether
the value extracted from disappearance at reactors matches that from
appearance in beams.

A combination of the existing experiments, T2K, NO$\nu$A and reactor
data, allows to obtain a first glimpse on the mass hierarchy and with
extended running and for favorable \CP\ phases a $5\sigma$
determination is possible. Also, new atmospheric neutrino experiments
like PINGU, ICAL at INO and Hyper-K have, in principle, some
sensitivity to the mass hierarchy and the actual level of significance
strongly depends on the obtainable angular and energy resolution for
the incoming neutrino. There are also plans for a dedicated
experiment, called Daya~Bay~2, which would not rely on matter effects
but aims at measuring the interference of the two mass squared
differences at a distance of about 60~km from a nuclear reactor. It
seems likely that global fits will be able to provide a 3--5$\sigma$
determination of the mass hierarchy before the end of the next decade.
It should be noted, that nonetheless a direct and precise method to
test matter effects and to determine the mass hierarchy from a single
measurement would be valuable even in this case.

One of the most commonly used frameworks to discuss physics beyond
oscillations are so-called non-standard interactions (NSI). They can
arise in many different models and their phenomenology is easy to
capture in a model-independent way. For the measurement of NSI, the
fact that $\theta_{13}$ is large means that interference of standard
oscillation amplitudes proportional to $\sin 2 \theta_{13}$ with NSI
effects can enhance sensitivity substantially. If NSI are present, the
extraction of the mass hierarchy from global fits is not likely to
yield the correct result. Note, NSI are a straightforward
mechanism to induce a difference between the reactor and beam
measurements of $\theta_{13}$. Longer baselines generally have more
sensitivity to NSI and also allow a better separation of standard
oscillation and NSI.

Given the likely status of the mass hierarchy measurement by the time
\PX\ becomes active, the other very central physics goal is a
measurement of the leptonic \CP\ phase and potentially the discovery of
\CP\ violation in the lepton sector. It is important to distinguish
these two goals---with large $\theta_{13}$ a measurement of the \CP\
phase at a predetermined level of precision can be virtually
guaranteed, whereas \CP\ violation may or may not be present in the
lepton sector. Therefore, we focus on the measurement of the \CP\
phase and regard the sensitivity towards \CP\ violation as
secondary.\footnote{This is an operational statement, which does not
  imply that \CP\ violation is less interesting. Rather, in practice
  one will have to measure the phase and then one knows whether \CP\ is
  violated or not.} A determination of the \CP\ phase requires to
measure any two out of the following four transitions:
$\nu_e\rightarrow\nu_\mu$, $\bar\nu_e\rightarrow\bar\nu_\mu$,
$\nu_\mu\rightarrow\nu_e$, $\bar\nu_\mu\rightarrow\bar\nu_e$. However,
due to the long baselines, there always will be also matter effects
which yield a contribution to the \CP\ asymmetries as well; it is
necessary to separate this contribution from the genuine \CP\ violation
in the mixing matrix. This separation is greatly facilitated by
exploiting $L/E$ information, ideally spanning a wide enough $L/E$
interval so that more than one node of the oscillation can be
resolved. This requirement, in combination with limitations of
neutrino sources and detectors translates into the need for baselines
longer than 1,000~km~\cite{Diwan:2006qf,Barger:2007jq,Barger:2007yw}.
This is also clearly borne out in the discussion of the LBNE
reconfiguration---shorter baselines like those available in the
existing NuMI beamline, require generally a larger exposure to reach
the same parametric \CP\ sensitivity, in absence of external
information.

For superbeam experiments, the control of systematic errors will be a
major issue, since neither the detection cross sections nor beam
fluxes are known within the required precision. Near detectors,
together with hadron production data, will play an important role.
However, this alone will not be sufficient to obtain per cent level
systematics, since the beam at the near detector is composed mostly of
$\nu_\mu$ and hence a measurement of the $\nu_e$ cross section is not
possible, but in the far detector the signal are $\nu_e$, see e.g., 
Ref.~\cite{Huber:2007em}. Unfortunately, there are no strong theory
constraints on the ratio of muon-to-electron neutrino cross sections
either~\cite{Day:2012gb}. Here, a facility like $\nu$STORM maybe
helpful, which is described in detail in Sec.~\ref{nu:subsec:nustorm}.
Also, better theory calculations of neutrino-nucleon interactions will
certainly be required.  As described in Chapter~\ref{chapt:lqcd}, such 
calculations are possible with lattice QCD and will be carried out over
the next several years.  In this context, these calculations will help
disentangle hadronic from nuclear effects in neutrino-nucleus scattering.

\begin{figure}
    \centering
    \includegraphics[width=0.7\textwidth]{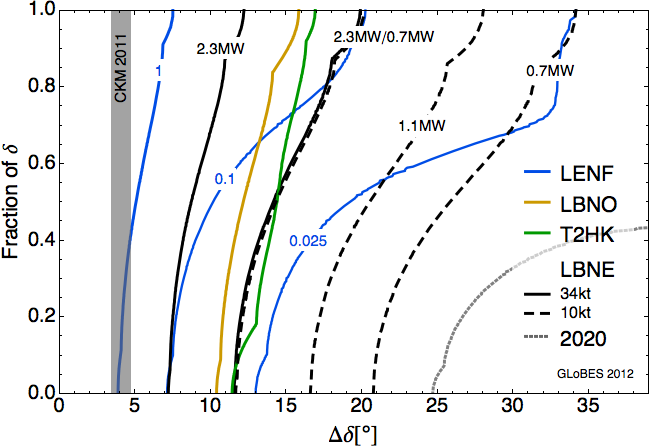}
    \caption[Values of the \CP\ phase $\delta_{\CP}$ for which a given $1\sigma$ precision can 
        be achieved]{Fraction of values of the \CP\ phase $\delta_{\CP}$ for which a given $1\sigma$ 
        precision $\Delta\delta$ can be achieved.
        The various lines are for different setups as indicated in the legend.
        The vertical gray shaded area, labeled ``CKM 2011'', indicates the current errors on the \CP\ phase 
        in the CKM matrix.
        This calculation includes near detectors and assumes consistent flux and cross section uncertainties 
        across different setups.
        The setups are: LENF---a 10-GeV neutrino factory with $1.4\times 10^{22}$ useful muon decays, which
        corresponds to 4-MW proton beam power for $10^8$~s, 2,000~km baseline and a 100~kt magnetized 
        iron detector; LBNO---uses a 100~kt LAr detector at a baseline of 2,300~km and $10^{22}$~POT at 
        50~GeV, which translates into about 800~kW of beam power for $10^8$~s; T2HK---a 560~kt water 
        Cherenkov detector at 295~km using a 1.66~MW beam for $5 \times 10^7$~s, which is equivalent to 
        $1.2\times10^8$~s at 700~kW; LBNE---using LAr detectors of either 10~kt or 34~kt at a distance 
        of 1,300~km with different beam powers as indicated in the legend for $2\times10^8$~s; 
        2020---results obtain from a combined fit to nominal runs of T2K, NO$\nu$A and Daya Bay.
        All detector masses are fiducial.
        Plot courtesy P.~Coloma~\cite{Coloma:2012}.}
    \label{nu:fig:cp}
\end{figure}
In Fig.~\ref{nu:fig:cp}, a comparison of the \CP\ precision for various
facilities, as explained in the legend, is shown. Clearly, the
neutrino factory (blue line, labeled LENF) is the only facility which
approaches the CKM precision, and it has the potential to go even
further. For the superbeams, 2020, LBNE, LBNO, and T2HK, we note that they
span a very wide range of precision, which demonstrates the crucial
importance of achieving sufficient statistics. The
number of events is determined by the product of beam power, detector
mass and running time and each of these ingredients can vary easily
within an order of magnitude.  LBNO has recently submitted an
expression of interest~\cite{LBNO} to CERN which outlines a much
smaller detector and lower beam power which would put its \CP\ precision
somewhere close to any of the reconfigured LBNE options. Obtaining a
sufficient number of events is crucial and clearly, here \PX\ can
help with increasing the beam power at 60~GeV. The sensitivity of
these results to the assumptions made about systematics is not shown
in this plot---but a clear difference does exist, and T2HK exhibits a
very strong sensitivity to the assumed level of
systematics~\cite{Coloma:2012} and thus is significantly more at risk
of running into a systematics limitation.  Both LBNE and LBNO due to
their long baselines and resultant wide $L/E$ coverage are quite safe
from systematics~\cite{Coloma:2012}. Note, at the current stage all
these experiments have to rely on assumptions about their systematics.
In any comparison as presented in Fig.~\ref{nu:fig:cp} the relative
performance can vary greatly depending on these assumptions. In the
end, \emph{both} sufficient statistics combined with small systematics
will be required to perform a precise measurement of the \CP~phase.

%%%%%%%%%%%%%%%%%%%%%%%%%%%%%%%%%%%%%%%%%%%%%%%%%%%%%%%%%%%%%%%%%%
\subsection{Long-Baseline Neutrino Experiment}

The Long-Baseline Neutrino Experiment (LBNE)~\cite{lbnewebsite} plans
a comprehensive program that will fully characterize neutrino
oscillation phenomenology using a high-intensity accelerator
muon-neutrino beam and a massive liquid-argon time-projection chamber (LAr TPC) as a far detector
sited for a 1300-km baseline.  The goals for this program are the
determination of leptonic \CP\ violation, the neutrino mass hierarchy,
precision measurements of neutrino mixing and interactions, as well as
underground physics, including the exploration of proton decay and
supernova neutrino bursts.  The LBNE program assumes a 700~kW Main
Injector (MI) proton beam power, however the beam line and target
station are designed to be able to exploit \PX\ beam power up to 2.3~MW.

For the program of testing and constraining the three-flavor mixing
paradigm underlying neutrino oscillation phenomenology, the key
observables for conventional horn-focused long-baseline neutrino beam
experiments are the survival probabilities of the $\nu_\mu$ and
$\overline{\nu}_\mu$ beam components (in operation with the respective
horn-current polarities), and the corresponding appearance
probabilities for $\nu_e$ (and $\overline{\nu}_e$).  In its simplest
form, the measurements can be reduced to four numbers.  However, as an
on-axis experiment, the LBNE detectors will be exposed to a broad
neutrino energy spectrum, with flux at both the first and second
oscillation maxima.  The interplay of matter effects and both the
\CP-conserving and \CP-violating contributions associated with the phase
$\delta$ present within the standard three-flavor mixing picture, lead
to complex energy dependencies of the $\nu_e$ and $\overline{\nu}_e$
appearance probabilities.  Detailed analysis of these energy
dependencies will enable untangling of overlapping effects, for
example, ambiguities presented by the unknown octant and the currently
limited precision on $\theta_{23}$.  The 1300~km baseline is nearly
optimal for resolving the picture of neutrino mixing: by virtue of the
very long baseline, matter effects are enhanced to the point that
ambiguities between leptonic \CP\ violating effects and the \CP-asymmetry
induced by interactions with electrons as the neutrinos propagate are
well separated.

The significant effort to construct an experiment like LBNE with a
massive, highly sensitive detector and very long baseline is aimed at
minimizing systematic uncertainties to the extent possible.  As a
result, many LBNE measurements are expected to be statistics limited.
To fully capitalize on the LBNE physics potential, it is essential
that investment also be made in the delivery of a neutrino beam with
the highest intensity possible.

\PX\ can provide a significant enhancement to the LBNE neutrino
program.  A staged increase in the MI proton beam power will increase
the neutrino flux proportionately, thus reducing the time required to
reach the science goals and may reduce certain systematic
uncertainties. Stages~2 and~3 of \PX\ would also support further
optimization of the LBNE neutrino energy spectrum while maintaining
high beam power. In the following we provide a few specific examples
of how the science reach of LBNE is substantially accelerated by the
available of different stages of \PX.

%%%%%%%%%%%%%%%%%%%%%%%%%%%%%%%%%%%%%%%%%%%%%%%%%%%%%%%%%%%%%%%%%%%%%%
\subsubsection{Assumptions, Scope and Organization of this Discussion}

In the following discussion, the reach of LBNE toward its neutrino
oscillation physics goals is cast in a context that enables
visualization of the impact of \PX.  LBNE has recently received
DOE CD-1 approval as a phased program, with a far detector fiducial
mass of at least 10 kt in the initial phase.  For the full LBNE
program, a far detector complex with fiducial mass of at least 34 kt
would be deployed.  The actual evolution of the far detector complex
will depend on domestic funding scenarios as well as contributions
from international partners.  For this reason, sensitivities are
plotted as a function of exposure in kt-years.  Thus, a 20-kt far
detector, operating for 5 years in neutrino mode and 5 years in
antineutrino mode with a 700~kW beam would have an exposure of 200
kt-years.  Operating at 2.1~MW beam power, as would be possible with
Stage~3 of \PX, for that same duration would then correspond to
an exposure of 600 kt-years at 700~kW.  Or as indicated above, it
would decrease by a factor of three the time needed to reach a
given physics goal relative to that indicated in these plots.  For a
number of the plots we explicitly show a scenario in which beam power
is increased at specific intervals from 700~kW to 1.1~MW to 2.3~MW, as
the different \PX\ stages begin operation.

Additionally, optimization of the beam line configuration, including
length of the decay volume and energy (nominally 120~GeV) of the
primary MI beam extracted onto the hadron production target, is still
under development.  Consequently, for illustration purposes, a number
of the plots presented here show sensitivity ranges that correspond to
different beam line configurations, ranging from that documented in
the 2012 LBNE Conceptual Design Report~\cite{lbnecdr} to more
optimized configurations including MI operation at 80~GeV and a longer
decay volume ($250~$m, instead of the nominal $200~$m length).

In this discussion, we focus on the $\nu_e$ appearance and $\nu_\mu$
disappearance measurements that are the mainstay of the LBNE
program.  First, while there is a good chance that determination of
the neutrino mass hierarchy will not require \PX, there are
scenarios where the combination of LBNE and \PX\ will be needed,
and this is illustrated briefly in Sec.~\ref{nu:sec:lbne-mh}.  On the
other hand, LBNE sensitivity to \CP\ violation and the
value of the \CP\ phase in the mixing matrix $\delta_{\CP}$ depends
strongly on the beam power and neutrino energy spectrum, and is where
\PX\ is most critical.  This is demonstrated in
Sec.~\ref{nu:sec:lbne-cpv}.  With regard to $\nu_\mu$ disappearance, we
first report in Sec.~\ref{nu:sec:numu_disappearance} the dependence of the
sensitivity to $\theta_{23}$, and specifically its possible departure
from $\pi/4$.  We then describe in Sec.~\ref{nu:sec:nsi} the sensitivity
of LBNE to the presence of non-standard interactions that would modify
the energy-dependence of the $\nu_\mu$ survival probability.  Finally,
comments on the potential impact on precision physics with a highly
capable near detector complex are given in
Sec.~\ref{nu:sec:lbne-precision}.

%%%%%%%%%%%%%%%%%%%%%%%%%%%%%%%%%%%%%%%%%%%%%%%%%%%%%%%%%%%%%%%%%%%%%%
\subsection{LBNE Mass Hierarchy Reach for Unfavorable Scenarios}
\label{nu:sec:lbne-mh}

Unambiguous determination of whether the mass hierarchy (MH) is normal or inverted is one of the most
important questions to be addressed by the current and next generation of neutrino experiments, including the
initial phase of LBNE.
Yet, it is conceivable that neutrino-mixing parameter values will be unfavorable, and additional sensitivity
that could be provided by \PX\ will be needed.
\begin{figure}
    \centering
    \includegraphics[width=0.65\textwidth]{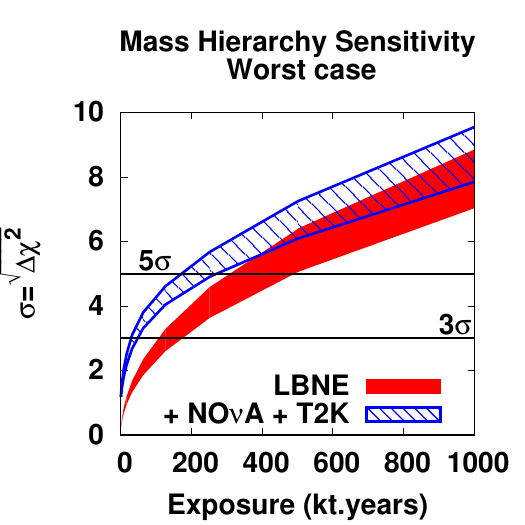}
\caption[LBNE mass hierarchy sensitivity for the worst-case value of $\delta_{\CP}=+90^\circ$]{LBNE mass
hierarchy sensitivity for the worst-case value of $\delta_{\CP}=+90^\circ$.
The bands represent ranges delimiting the two 700~kW beam configurations described in the text.
For $\delta_{\CP}$ less than zero there is only a small sensitivity difference between LBNE alone and when
combined with T2K and NO$\nu$A; greater than $5\sigma$ determination is achieved with an exposure of 100
kt-years.
For higher beam power correspondingly less time is needed to reach the same overall exposure.}
    \label{nu:fig:lbne-mh}
\end{figure}
Figure~\ref{nu:fig:lbne-mh} shows the MH determination significance as function of exposure (the product of
far detector fiducial mass and beam time) for a 700~kW proton beam for the worst case scenario where the
unknown phase in the mixing matrix $\delta_{\CP}$ is $+90^\circ$.
The bands represent the range for two proton beam configurations, as described earlier: The lower edge of the
band is for the nominal 120~GeV proton beam described in the 2012 Conceptual Design Report~\cite{lbnecdr};
the upper edge is for an enhanced beam with an 80~GeV MI beam energy of the same power.
The higher beam power of \PX\ effectively compresses the exposure scale so, for example, a $5\sigma$
measurement that would take 5 years with the 80~GeV/700~kW beam, would be reduced to a little over three
years with a 1.1~MW beam.
Earlier knowledge of the correct mass hierarchy would allow better optimization of the run strategy for other
oscillation parameter measurements.

%%%%%%%%%%%%%%%%%%%%%%%%%%%%%%%%%%%%%%%%%%%%%%%%%%%%%%%%%%%%%%%%%%%%%
\subsubsection{LBNE Reach in \CP\ Violation}
\label{nu:sec:lbne-cpv}

A primary goal of LBNE is observation of \CP\ violation in the neutrino
sector.  Through measurement of the energy-dependent probabilities for
electron-neutrino (antineutrino) appearance in a muon-neutrino
(antineutrino) beam with its source at a distance of 1300 km, LBNE
will be sensitive to terms involving the \CP\ phase $\delta_{\CP}$ that
appears in the standard form of the three-flavor mixing matrix.  If
$\delta_{\CP}$ is zero or $\pi$ radians, there is no \CP\ violating term
in the matrix, and hence deviations from these values would constitute
evidence for \CP\ violation.

\begin{figure}
	\centering
    \includegraphics[width=0.45\textwidth]{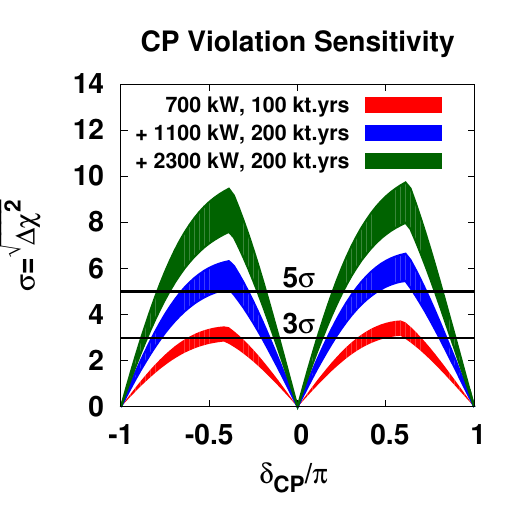}\hfill
    \includegraphics[width=0.45\textwidth]{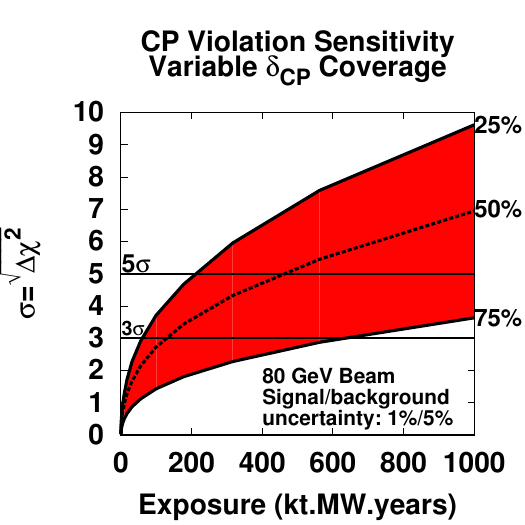}
    \caption[\CP\ violation sensitivity as a function of exposure]{\CP\ violation sensitivity as a function 
        of exposure (far detector mass times beam power times run time) at the indicated proton beam power 
        (corresponding to a straw man \PX\ development scenario).
        Significance of non-zero value of $\delta_{\CP}$ over the full $\delta_{\CP}$ range (left);
        \CP\ violation sensitivity for 50\% coverage (central doted line) of the full $\delta_{\CP}$ range 
        (right).
        The red shaded region in the right hand panel indicates \CP\ fractions from 25\% - 75\%.}
    \label{nu:fig:lbne-cpv}
\end{figure}

The two plots in Fig.~\ref{nu:fig:lbne-cpv} illustrate the significance
of a non-zero (or $\pi$) measurement for different exposures scenarios
with increasing beam power successively from the nominal LBNE 700~kW,
through 1.1~MW (\PX\ Stage 1), to 2.3~MW (\PX\ Stage 2).
Here, significance is defined as the square-root of the difference in $\chi^2$ between the electron-neutrino
spectrum predicted for some the value of $\delta_{\CP}$ and that for a value of 0 or $\pi$ radians.
The left plot shows the significance as a function of $\delta_{\CP}$ itself.
(If the MH were unknown and not measured in the same experiment, as it is for LBNE, ambiguities would make
this distribution asymmetric.) 
The plot on the right shows the exposure, with \PX\ beam power epochs indicated, for which the \CP\ violation
significance is that value or higher for 50\% of the full $\delta_{\CP}$ range.
For example, a 100 kt-year exposure with a 700~kW beam, followed by a 44 kt-year exposure with a 1.1~MW beam
would yield a $3\sigma$ or better CPV significance for half of the $\delta_{\CP}$ range.
For a 35-kt LBNE this corresponds to a little over 4 years.

\begin{figure}
	\centering
    \includegraphics[width=0.45\textwidth]{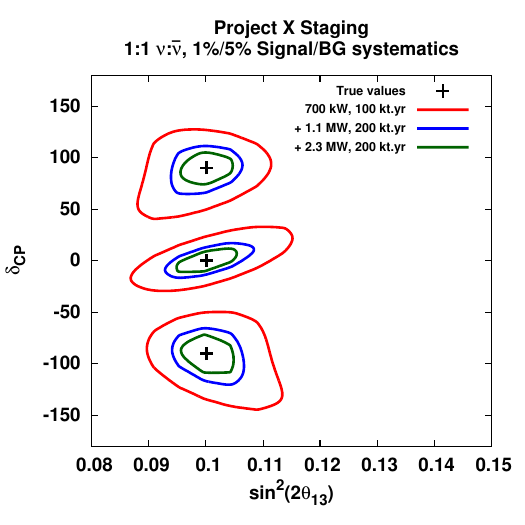}\hfill
    \includegraphics[width=0.45\textwidth]{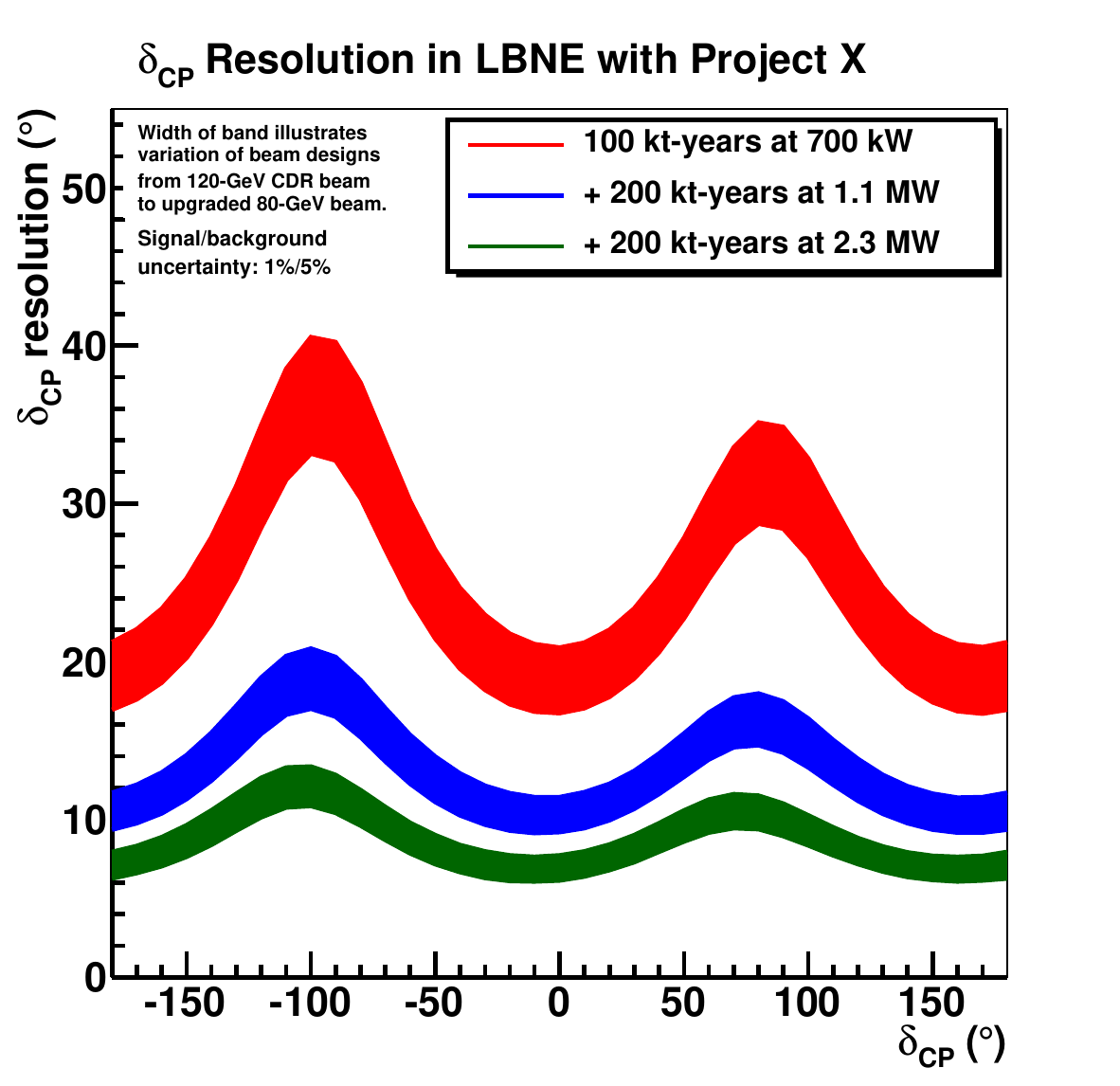}
    \caption[Measurement resolution of the \CP\ phase $\delta_{\CP}$ with \PX\ beam power]{Measurement 
        resolution of the \CP\ phase $\delta_{\CP}$, for a program scenario that includes an evolution of 
        the proton beam as upgraded via \PX. 
        Each time period comprises equal exposure for neutrinos and antineutrinos.
        Normalization uncertainties of 1\% for signal and 5\% for background are assumed.
        At left, $1\sigma$ $\delta_{\CP}$ resolution contours are plotted for three values of $\delta_{\CP}$ 
        ($-90^\circ$, $0^{\circ}$, $+90^\circ$), each with $\sin^2{\theta_{13}} = 0.1$.
        At right, the $1\sigma$ resolution on $\delta_{\CP}$ is plotted as a function of~$\delta_{\CP}$.}
    \label{nu:fig:lbne-delta}
\end{figure}

Figure~\ref{nu:fig:lbne-delta} shows the accuracy in the determination of $\delta_{\CP}$ and $\theta_{13}$.
In the left plot, the bold crosses indicate three different true
values of $\delta_{\CP}$ with the same true value of
$\sin^2{2\theta_{13}} = 0.1$.  The colored solid lines show how the
$1\sigma$ contours shrink by the end of the three successive beam
power phases. The right-hand plot shows the $1\sigma$ resolution on
the \CP\ phase as a function of its true value. The width of the band
illustrates the variation due to beam design alternatives, as in
Fig.~\ref{nu:fig:lbne-mh}.

%%%%%%%%%%%%%%%%%%%%%%%%%%%%%%%%%%%%%%%%%%%%%%%%%%%%%%%%%%%%%%%%%%%%%%%%%%%%%%%%%%%%%
\subsubsection{LBNE Reach with Muon Neutrino Disappearance}
\label{nu:sec:numu_disappearance}

\begin{figure}
	\centering
    \includegraphics[width=0.4\textwidth]{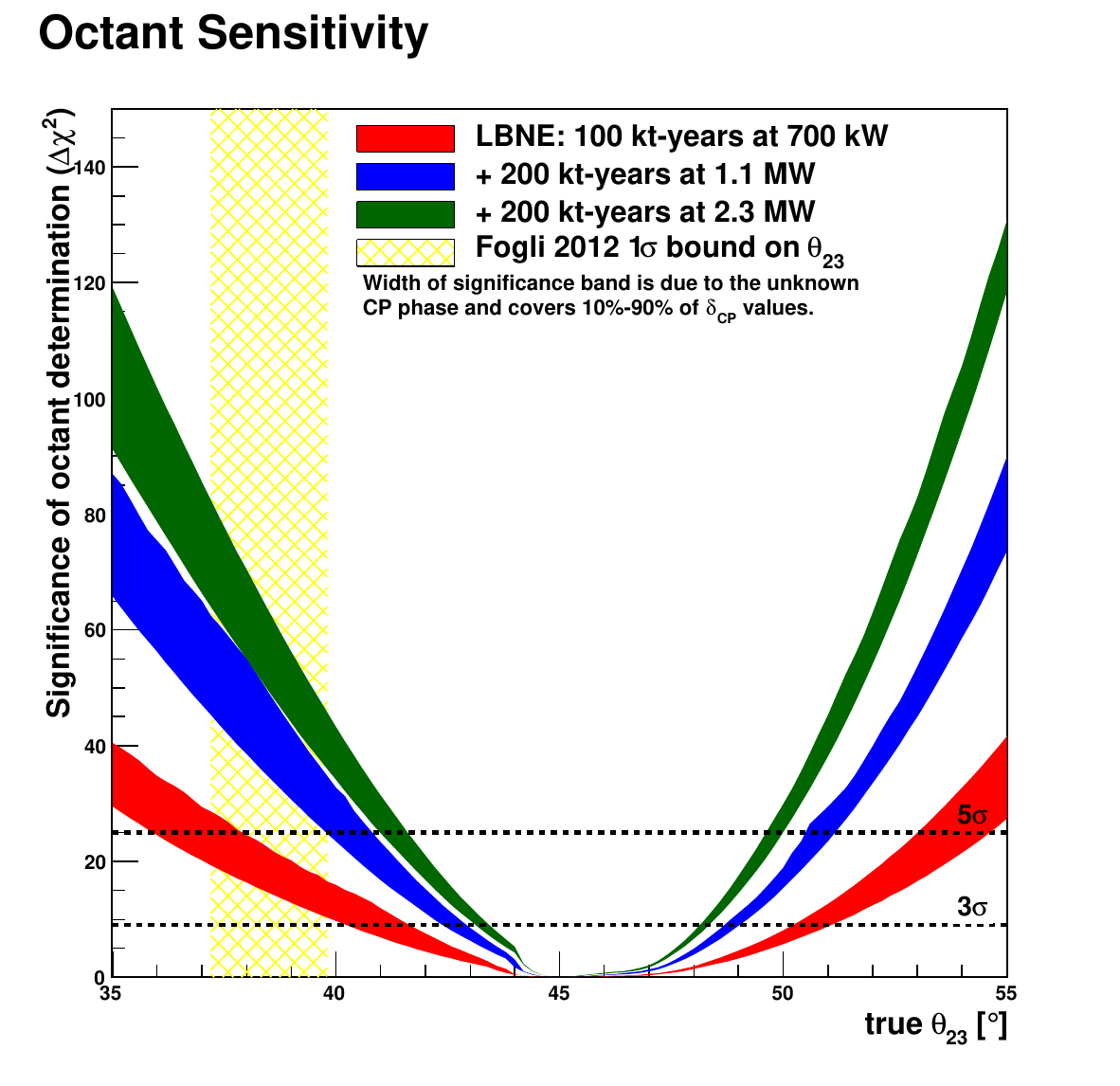} \hfill
    \includegraphics[width=0.575\textwidth]{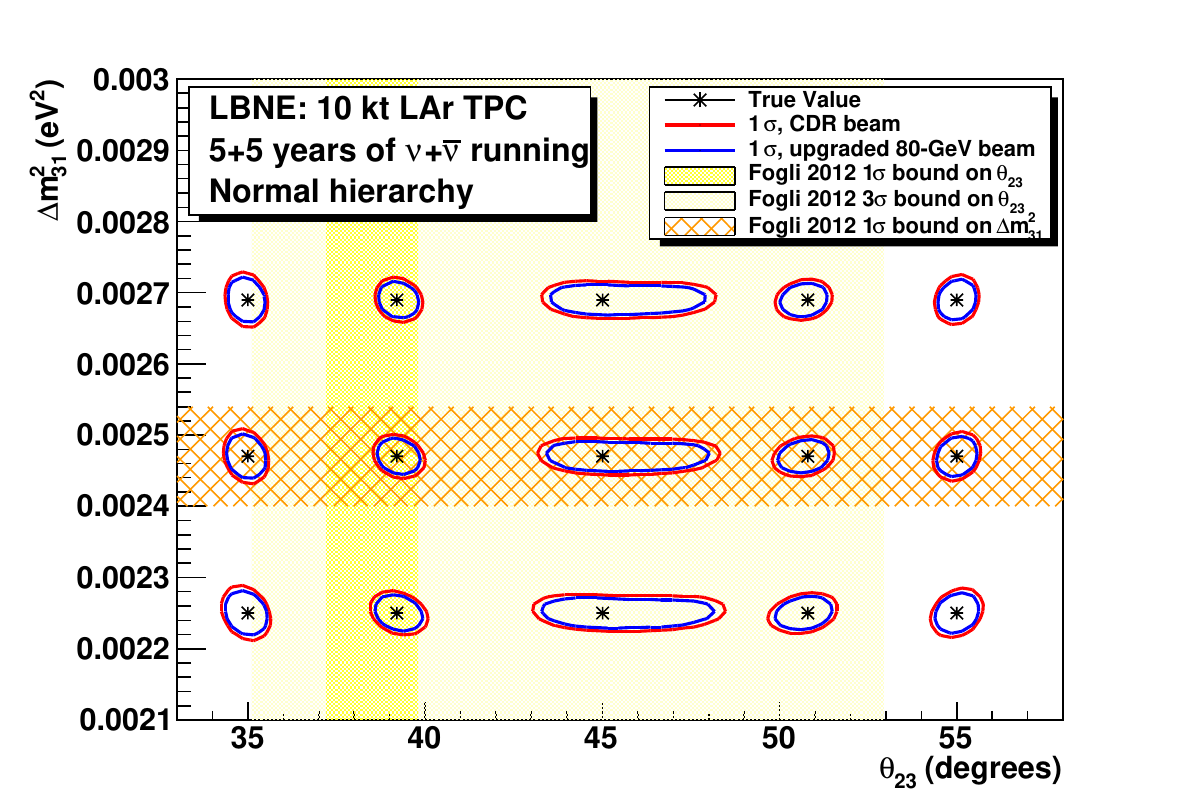}\hspace*{-0.3em}
    \caption[Expected sensitivity to $\theta_{23}$]{Left: Sensitivity of determination of the octant of 
        mixing angle $\theta_{23}$ from $\nu_\mu$ disappearance and $\nu_e$ appearance signals, with the 
        same scenario for beam power evolution and detector exposure assumed in the $\nu_e$ appearance 
        analyses above. 
        Right: Projected precision on the atmospheric $\Delta m^2$ and $\sin^2{2\theta_{23}}$ for the 
        case of a 10-kt far detector operating for 10 years at 700~kW.}
  \label{nu:fig:lbne-octant}
\end{figure}

LBNE capabilities for $\nu_\mu$ disappearance measurements will enable
precision measurement of the mixing angle $\theta_{23}$.  To obtain
maximal sensitivity to both the deviation of $\sin^2{2\theta_{23}}$
from unity and the $\theta_{23}$ octant it is necessary to
simultaneously analyze the $\nu_\mu$ disappearance and $\nu_e$
appearance signals~\cite{Huber:2010dx}.  In Fig.~\ref{nu:fig:lbne-octant}
we show the significance (plotted here for $\Delta\chi^2$, rather than
$\sqrt{\Delta \chi^2}$ used earlier) to determine the octant of
$\theta_{23}$ as a function of its true value.
The range of $\theta_{23}$ values hinted (at the $1\sigma$ level) by
the analysis of existing data by Fogli {\sl et al.}~\cite{Fogli:2012ua}
is indicated by the hatched vertical band for illustrative purposes.
It is important to note, however, that experimental precision on
$\theta_{23}$ itself has a strongly non-linear dependence on the
actual value as one approaches maximal mixing ($45^\circ$).  This
non-linearity is further illustrated in the plot on the right in
Fig.~\ref{nu:fig:lbne-octant} for the case of a 10-kt Far Detector, with
pre-Project-X beams.  Nevertheless, over a considerable range of
plausible $\theta_{23}$ values, the addition of capability from beam
upgrades associated with \PX\ stages can be transformative for
distinguishing $\theta_{23}$ from $45^\circ$.

%%%%%%%%%%%%%%%%%%%%%%%%%%%%%%%%%%%%%%%%%%%%%%%%%%%%%%%%%%%%%%%%%%%%%%%%%
\subsubsection{Sensitivity to Matter Effects from Nonstandard Interactions}
\label{nu:sec:nsi}

Flavor-dependent non-standard interactions (NSI) of neutrinos as they
propagate through matter have been proposed as a way of altering the
pattern of neutrino oscillations without requiring the introduction of
additional neutrino species.  In general, charged-current (CC) and
neutral-current (NC) interactions are possible, and these could either
be flavor-changing or flavor-conserving.  Long-baseline experiments
have especially strong sensitivity to NC NSI-induced effects, since it
is the forward scattering of neutrinos (including $\nu_\mu$ and
$\nu_\tau$) that would give rise to MSW-like distortions of the
survival probability for beam $\nu_\mu$'s as a function of energy.  By
virtue of the 1300~km baseline, LBNE has a unique advantage in this
area compared to other long-baseline experiments, except
atmospheric-neutrino experiments, which may, however, be limited by
systematic effects.

Following Ref.~~\cite{Huber:2010dx}, NC NSI can be parameterized as new contributions
to the MSW matrix in the neutrino-propagation Hamiltonian:
\begin{equation}
  H = U \left( \begin{array}{ccc}
           0 &                    & \\
             & \Delta m_{21}^2/2E & \\
             &                    & \Delta m_{31}^2/2E
         \end{array} \right) U^\dag + \tilde{V}_{\rm MSW} ,
\end{equation}
with
\begin{equation}
  \tilde{V}_{\rm MSW} = \sqrt{2} G_F N_e
\left(
  \begin{array}{ccc}
    1 + \epsilon^m_{ee}       & \epsilon^m_{e\mu}       & \epsilon^m_{e\tau}  \\
        \epsilon^{m*}_{e\mu}  & \epsilon^m_{\mu\mu}     & \epsilon^m_{\mu\tau} \\
        \epsilon^{m*}_{e\tau} & \epsilon^{m*}_{\mu\tau} & \epsilon^m_{\tau\tau}
  \end{array} 
\right)
\end{equation}
Here, $U$ is the leptonic mixing matrix, and the $\epsilon$-parameters
give the magnitude of the NSI relative to standard weak interactions.
For new physics scales of ${\rm few} \times 100$~GeV, $|\epsilon|
\lesssim 0.01$ is expected.

\begin{figure}[p]
    \centering
    \includegraphics[width=0.875\textwidth]{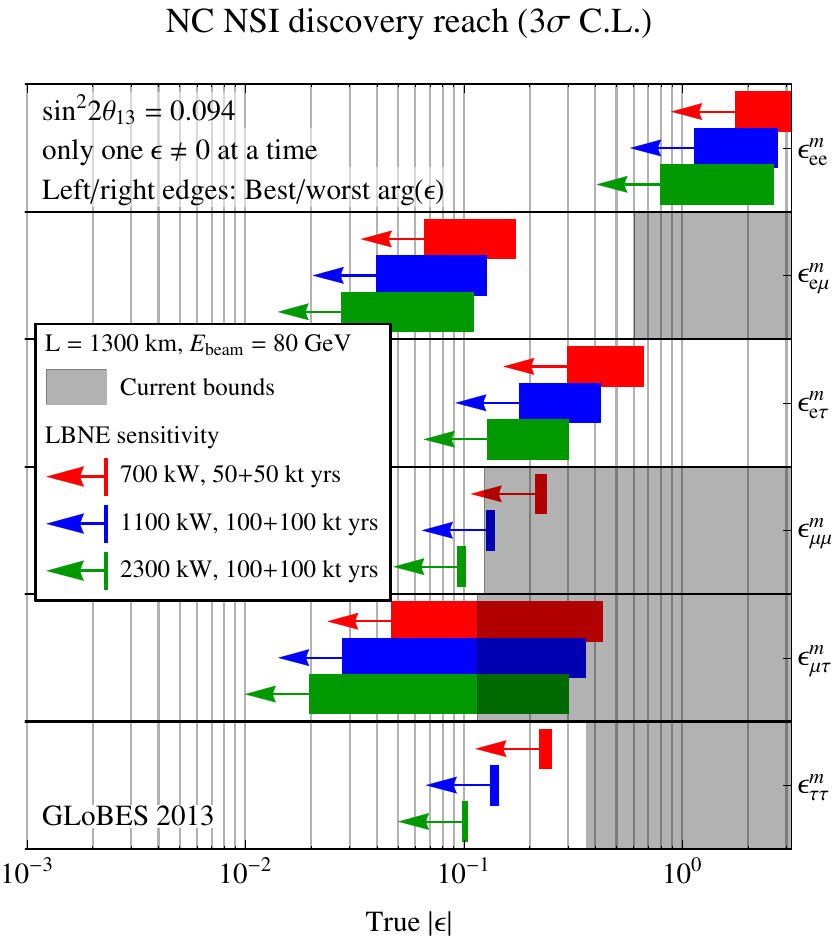}
    \caption[Non-standard interaction discovery reach in LBNE]{Non-standard interaction discovery reach in 
        LBNE at various phases in the evolution of detector mass and beam power.
        The left and right edges of the error bars correspond to the most favorable and the most unfavorable 
        values for the complex phase of the respective NSI parameters.
        The gray shaded regions indicate the current model-independent limits on the different parameters at 
        $3\sigma$~\cite{Davidson:2003ha,GonzalezGarcia:2007ib,Biggio:2009nt}.
        This study takes $\sin^2 2\theta_{13}=0.094$.
        From J.~Kopp.}
\label{nu:fig:LAr-NSI}
\end{figure}

To assess the sensitivity of LBNE to NC NSI, the NSI discovery reach
is defined in the following way: After simulating the expected event
spectra, assuming given ``true'' values for the NSI parameters, one
attempts a fit assuming no NSI. If the fit is incompatible with the
simulated data at a given confidence level, one would say that the
chosen ``true'' values of the NSI parameters are within the
experimental discovery reach. Figure~\ref{nu:fig:LAr-NSI} shows the
NSI discovery reach of LBNE for the case where only one of the
$\epsilon^m_{\alpha\beta}$ parameters at a time is
non-negligible~\cite{Huber:2010dx}.  Even with a 10~kt detector and
700~kW beam power, LBNE can explore substantial new regions of
parameter space.  Enhancing the program with a combination of detector
mass and beam power would extend the discovery reach correspondingly.

%%%%%%%%%%%%%%%%%%%%%%%%%%%%%%%%%%%%%%%%%%%%%%%%%%%%%%%%%%%%%%%%%%%%%%%%%%%
\subsubsection{LBNE Reach in Precision Neutrino Physics}
\label{nu:sec:lbne-precision}

A highly capable neutrino detector to measure the unoscillated
neutrino fluxes and their interactions at the near site will
significantly enhance the core scientific capability of LBNE.  It
would enable a very rich short-baseline physics program with more than
a hundred unique physics and engineering Ph.~D.\ topics.  Among the broad
physics goals of this program~\cite{Akiri:2011dv,lbphys} are to: 
(1)~measure the absolute and relative flux of all four neutrino species
($\nu_mu$, $\nu_e$ and corresponding antineutrinos), including the
energy scales of neutrinos and antineutrinos, as required to normalize
the oscillation signals at the Far Detector; (2)~measure the cross
section of neutrino- and antineutrino-induced inclusive and exclusive
processes in nuclear targets across a large energy range (0.5--50~GeV)
to 3\% precision, to aid in the interpretation of the oscillation
signals in the Far Detector; (3)~measure the yield of particles
produced in neutrino interactions such as neutral and charged
pions/kaons, which are the dominant backgrounds to oscillation
signals; and (4)~measure precisely the fundamental electroweak and
strong interaction parameters that are accessible to neutrino physics;
and (5)~perform sensitive searches for new physics, such as sterile
neutrinos.  While these physics goals will also surely be enhanced
with increased fluxes afforded by \PX, detailed studies of
sensitivities are ongoing at this point.

\afterpage{\clearpage}

%%%%%%%%%%%%%%%%%%%%%%%%%%%%%%%%%%%%%%%%%%%%%%%%%%%%%%%%%%%%%%%%%%%%%%%%%%%
\subsection{Muon-based Neutrino Physics}

The questions of leptonic \CP\ violation and the completeness of
the three-flavor picture, can only by addressed by very high precision
measurements of neutrino and antineutrino oscillation probabilities,
specifically including channels where the initial and final flavor of
neutrino are different. Several neutrino sources have been conceived
to reach high sensitivity and to allow the range of measurements
necessary to remove all ambiguities in the determination of
oscillation parameters. The sensitivity of these facilities is well
beyond that of the presently approved neutrino oscillation program.
Studies so far have shown that, even for the measured large value of
$\theta_{13}$, the neutrino factory, an intense high-energy neutrino
source based on a stored muon beam, gives the best performance for \CP\
measurements over the entire parameter space.  Its time-scale and
cost, however, remain important question marks. Second-generation
super-beam experiments using megawatt proton drivers may be an
attractive option in certain scenarios, but eventually the issue of
systematics control may limit this technology. It should be noted that
once detailed plans are considered, the fiscal and time scales of true
super-beams are very large as well.

\begin{figure}
	\centering
    \includegraphics[width=\textwidth]{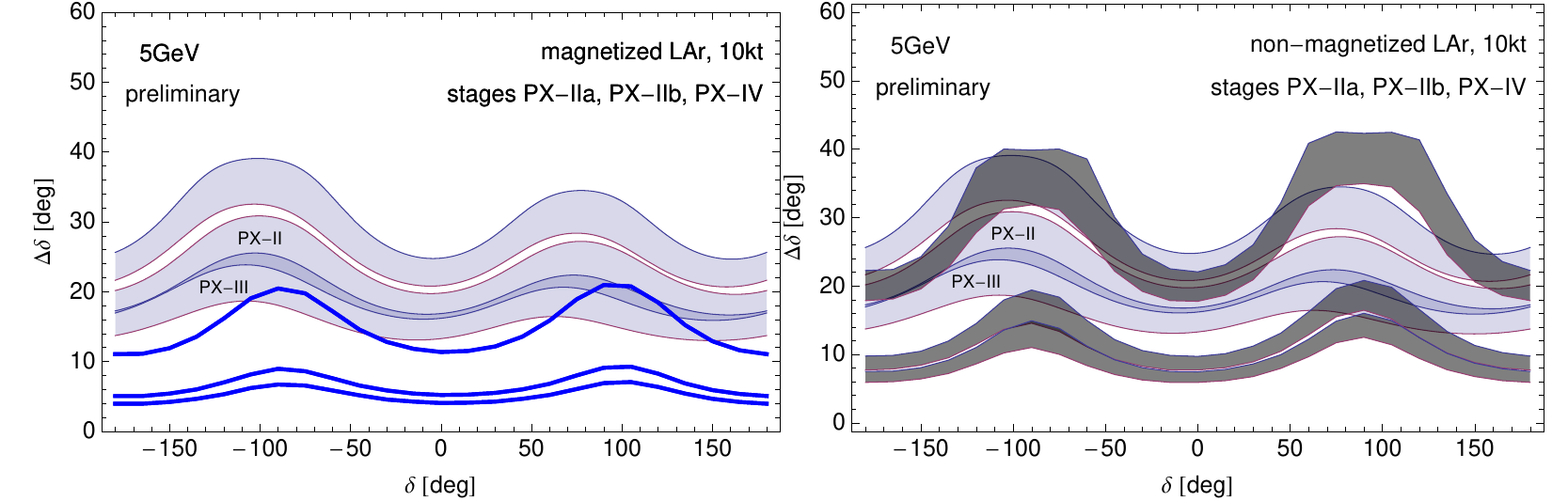}
    \caption[Accuracy on the \CP\ phase as a function of the true value of the \CP\ phase]{Accuracy on the 
        \CP\ phase $\delta$ as a function of the true value of the \CP\ phase at $1\sigma$  
        confidence level.
        The light blue bands depict the accuracy as expected from LBNE using the various beams \PX\ can 
        deliver.
        In the left hand panel, the thick blue lines represent what a neutrino factory beam can do using a 
        magnetized LAr detector.
        In the right hand panel, the gray bands illustrate the accuracy of a neutrino factory using a 
        non-magnetized detector.
        The neutrino factory beam intensities can be found in Table~\ref{nu:tab:nfstages}.
        Adapted from Ref.~\cite{Christensen:2013va}.}
    \label{nu:fig:nfstages}
\end{figure}
In response to the measurement of large $\theta_{13}$, the neutrino factory design has been reoptimized to a
stored muon energy of 10~GeV and a single baseline of 2000~km using a 100~kt magnetized iron detector.
It is possible to further reduce the energy to around 5~GeV and concomitantly the baseline to 1300~km without
an overall loss in performance if one changes the detector technology to improve efficiency around 1--2~GeV;
possible choices could be a magnetized liquid argon detector or a magnetized fully active plastic
scintillator detector.
If one of these technology choices can be shown to be feasible, there currently appears to be no strong
physics performance reason to favor the 10~GeV over the 5~GeV option, or vice versa.
The low-energy option seems attractive due to its synergies with planned super-beams like LBNE and because
the detector technology would allow for a comprehensive physics program in atmospheric neutrinos, proton
decay and supernova detection.
Within the low-energy option detailed studies of luminosity staging have been carried out, which indicate
that at even at 1/20th of the full-scale beam intensity and starting with a 10~kt detector significant
physics gains beyond the initial phases of a pion-decay based beam experiment, like LBNE, can be
realized~\cite{Christensen:2013va}.
At full beam luminosity and with a detector mass in the range of 10--30~kt, a 5~GeV neutrino factory offers
the best performance of any conceived neutrino oscillation experiment, which is shown in
Fig.~\ref{nu:fig:nfstages}.
The gray bands in the right hand panel illustrate the performance using a LAr detector without a magnetic
field, where the charge identification is performed statistically and not on an event-by-event basis, as
explained in detail in Ref.~\cite{Huber:2008yx}.

Muon accelerators offer unique potential for the U.S. high-energy
physics community.  In 2008, and subsequently in 2010, the U.S.
Particle Physics Project Prioritization Panel
(P5)~\cite{nu:p5report,nu:p5report2} recommended that a world-class
program of Intensity Frontier science be pursued at Fermilab as the
Energy Frontier program based on the Tevatron reached its conclusion.
Accordingly, Fermilab has embarked on the development of a next
generation neutrino detector with LBNE and a next generation proton
source with \PX.  However, looking towards the fruition of these
efforts, we must also consider how to provide the next generation of
capabilities that would enable the continuation of a preeminent
Intensity Frontier research program.  Building on the foundation of
\PX, muon accelerators can provide that next step with a high
intensity and precise source of neutrinos to support a world-leading
research program in neutrino physics.  Furthermore, the infrastructure
developed to support such an Intensity Frontier research program can
also enable the return of the U.S. high energy physics program to the
Energy Frontier.  This capability would be provided in a subsequent
stage of the facility that would support one or more muon colliders,
which could operate at center-of-mass energies from the Higgs
resonance at 125~GeV up to the multi-TeV scale.  Thus, muon
accelerators offer the unique potential, among the accelerator
concepts being discussed for the 2013 Community Summer Study process, to
provide world-leading experimental support spanning physics at both
the Intensity and Energy Frontiers.

The U.S. Muon Accelerator Program (MAP) has the task of assessing the
feasibility of muon accelerators for neutrino factory (NF) and Muon
Collider (MC) applications.  Critical path R\&D items, which are
important for the performance of one or more of these facilities,
include
\begin{itemize}
\item Development of a high power target station capable of handling
  4~MW of power.  Liquid metal jet technology has been shown to be
  capable of handling this amount of power. However, a complete
  engineering design of a multi-MW target station with a high field
  capture solenoid (nominal 20~T hybrid normal and superconducting magnet
  with $\sim3~$GJ stored energy) requires considerable further work. While
  challenging, target stations with similar specifications are
  required for other planned facilities (e.g., spallation sources),
  and our expectation is that many of the critical engineering issues
  will be addressed by others over the next several years.

\item Muon cooling is required in order to achieve the beam parameters
  for a high performance NF and for all MC designs under
  consideration.  An ionization cooling channel requires the operation
  of RF cavities in Tesla-scale magnetic fields.  Promising recent
  results from the MuCool Test Area (MTA) at Fermilab point towards
  solutions to the breakdown problems of RF cavities operating in this
  environment~\cite{Jana:2012zza,Jana:2012zz,Freemire:2012zz,Bowring:2012zz,Li:2012zzn}.
\item High-intensity and low-energy beams ($\sim$200~MeV, optimal for
  muon ionization cooling) are susceptible to a range of potential
  collective effects.  Evaluating the likely impact of these effects
  on the muon beams required for NF and MC applications, through
  simulation and experiment, is an important deliverable of the MAP
  feasibility assessment.  These results will be crucial for an
  informed community decision on muon accelerator facilities.
  Furthermore, the proposed staging plan enables confirming R\&D to be
  performed at each stage for the next stage in the plan, thus
  enabling a well-informed decision process moving forward. 

\item For the MC, a new class of backgrounds from muon decays impacts
  both the magnet/shielding design for the collider itself and the
  backgrounds in the detector.  It has been found that the detector
  backgrounds can be managed by means of pixelated detectors with good
  time resolution~\cite{Conway:2013lca,nu:mazzacane}.  Thus, this
  issue appears to present no impediment to moving forward with full
  detector studies and machine design efforts.
\end{itemize}
In the context of the proposed staging plan, baseline parameter
specifications have been developed for a series of facilities, each
capable of providing cutting edge physics output, and at each of which
the performance of systems required for the next stage can be reliably
evaluated. The plan thus provides clear decision points before
embarking upon each subsequent stage.  The staging plan builds on, and
takes advantage of, existing or proposed facilities, specifically:
\PX\ at Fermilab as the MW-class proton driver for muon
generation; Homestake as developed for the LBNE detector, which could
then house the detector for a long baseline neutrino factory.  The
performance characteristics of each stage provide unique physics
reach

\begin{itemize}
\item $\nu$STORM: a short baseline neutrino factory enabling a
  definitive search for sterile neutrinos, see
  Sec.~\ref{nu:subsec:nustorm}, as well as neutrino cross-section
  measurements that will ultimately be required for precision
  measurements at any long baseline experiment.
\item L3NF: an initial long baseline neutrino factory, optimized for a
  detector at Homestake, affording a precise and well-characterized
  neutrino source that exceeds the capabilities of conventional
  superbeam technology.
\item NF: a full intensity neutrino factory, upgraded from L3NF, as
  the ultimate source to enable precision \CP\ violation measurements in
  the neutrino sector.
\item Higgs Factory: a collider whose baseline configurations are
  capable of providing between 5,000 and 40,000 Higgs events per year
  with exquisite energy resolution.
\item Multi-TeV Collider: if warranted by LHC results, a multi-TeV
  Muon Collider likely offers the best performance and least cost for
  any lepton collider operating in the multi-TeV regime.
\end{itemize}

Nominal parameters for a short baseline NF, $\nu$STORM~\cite{Kyberd:2012iz} and two stages of a long baseline
NF optimized for a detector located at Homestake are provided in Table~\ref{nu:tab:nfstages}.
\begin{table}
	\centering
    \caption[Neutrino-factory parameters]{Muon Accelerator Program baseline neutrino-factory parameters for 
        $\nu$STORM and two phases of a neutrino factory located on the Fermilab site and pointed towards a 
        detector at Homestake. 
        For comparison, the parameters of the IDS-NF are also shown.}
    \includegraphics[angle=-90,width=\textwidth]{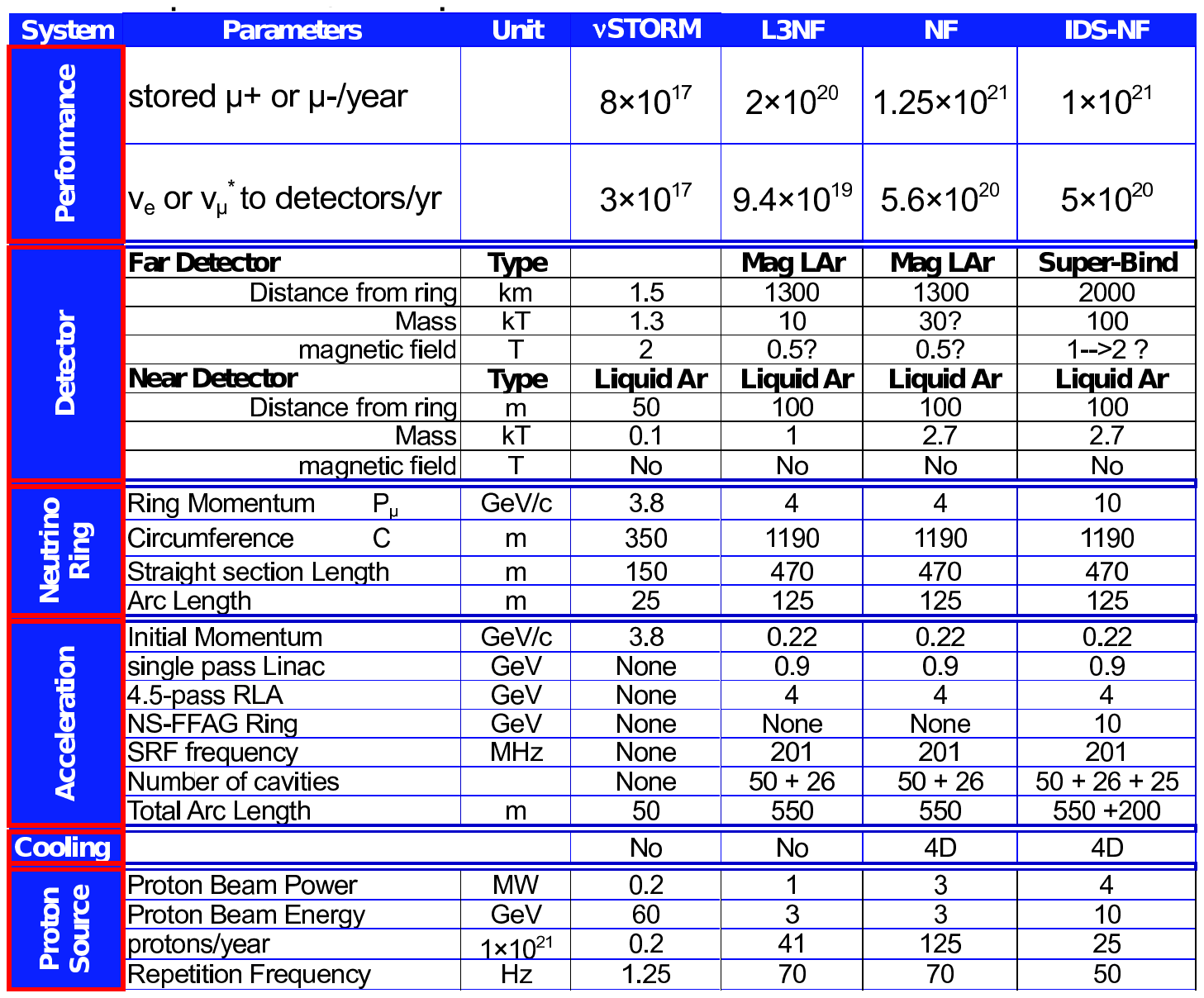}
    \label{nu:tab:nfstages}
\end{table}
MC parameters for two stages of a Higgs Factory as well as 1.5~TeV and 3.0~TeV colliders are provided in
Table~\ref{nu:tab:collider}.
All of these machines would fit readily within the footprint of the Fermilab site.
The ability to deploy these facilities in a staged fashion offers major benefits:
\begin{enumerate} 
    \item the strong synergies among the critical elements of the accelerator complex maximize the size of 
    the experimental community that can be supported by the overall facility; 
    \item the staging plan reduces the investment required for individual steps between stages to levels 
    that will hopefully fit within the future budget profile of the U.S. high energy physics program.
\end{enumerate}

$\nu$STORM's capabilities could be deployed now.
The NF options and initial Higgs Factory could be based on the 3~GeV proton source of \PX\ Stage~2 operating
with 1~MW and, eventually, 3~MW proton beams.
This opens the possibility of launching the initial NF, which requires no cooling of the muon beams, within
the next decade.
Similarly, the R\&D required for a decision on a collider could be completed by the middle of the next decade.

\begin{table}
	\centering
    \caption[Muon-collider parameters]{Muon Accelerator Program baseline muon-collider parameters for both 
        Higgs factory and multi-TeV energy-frontier colliders.
        An important feature of the staging plan is that collider activity could begin with \PX\ Stage~2 
        beam capabilities at Fermilab.}
    \includegraphics[angle=-90,width=\textwidth]{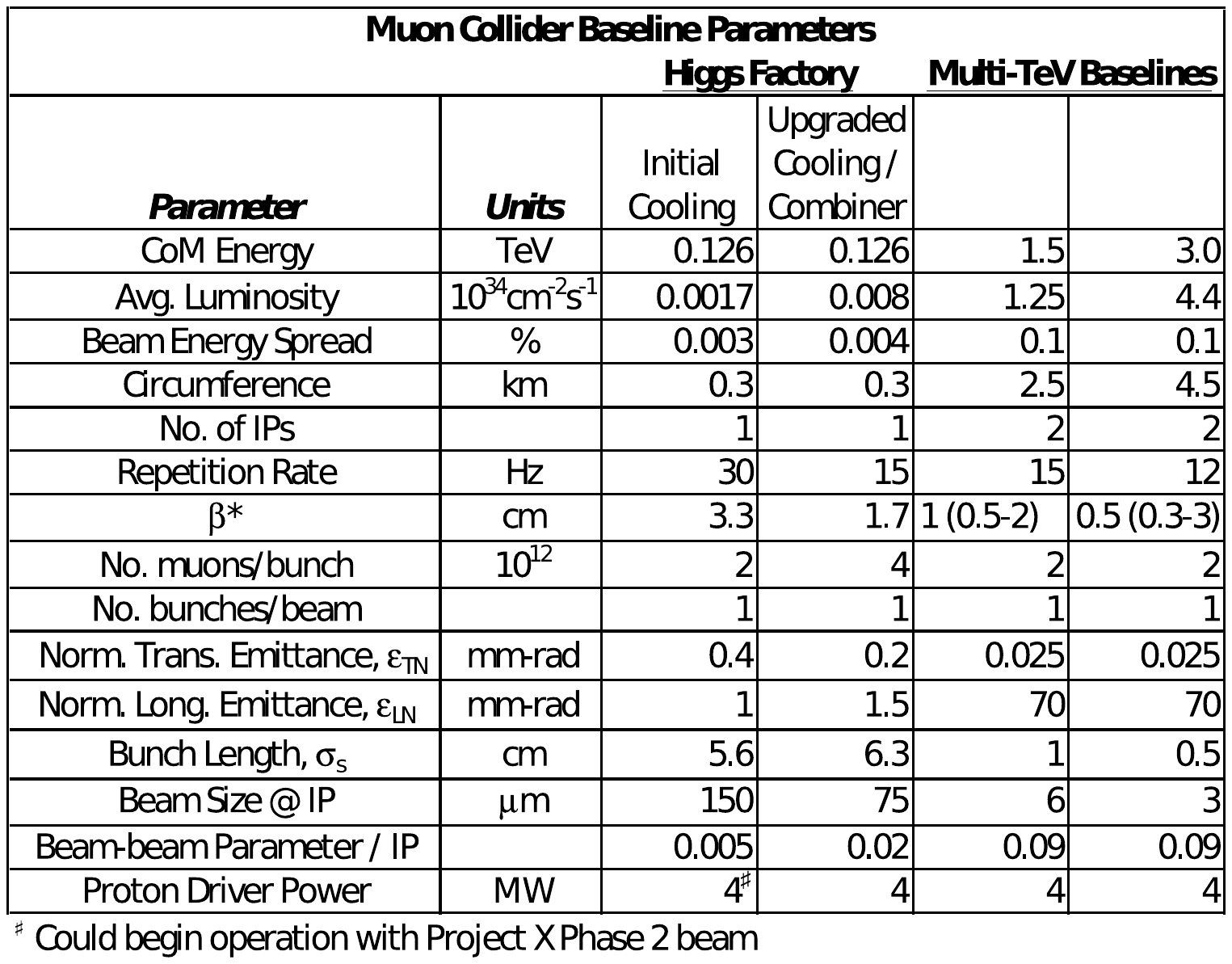}
    \label{nu:tab:collider}
\end{table}

\begin{figure}
	\centering
    \includegraphics[width=0.8\textwidth]{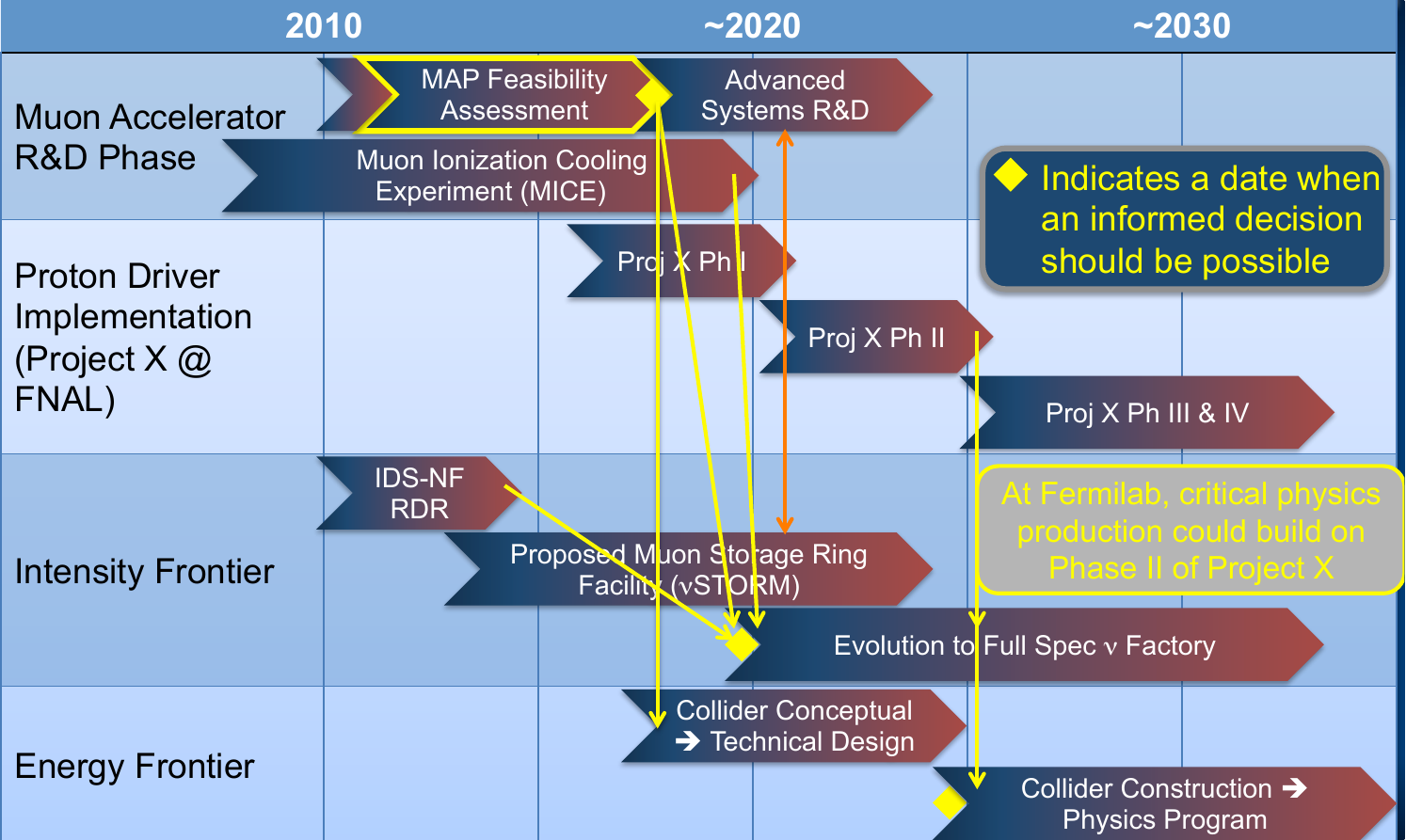}
    \caption[Muon accelerator timeline including the MAP feasibility-assessment period]{Muon accelerator 
        timeline including the MAP feasibility-assessment period. 
        It is anticipated that decision points for moving forward with a neutrino factory project supporting
        intensity-frontier physics efforts could be reached by the end of this decade, and a decision point 
        for moving forward with a muon collider physics effort supporting a return to the energy frontier 
        with a U.S. facility could be reached by the middle of the next decade.
        These efforts are able to build on \PX\ Stage~2 capabilities as soon as they are available.
        It should also be noted that the development of a short baseline neutrino facility, i.e., 
        $\nu$STORM, would significantly enhance MAP research capabilities by supporting a program of 
        advanced systems R\&D.} 
    \label{nu:fig:timeline}
\end{figure}

\begin{figure}
	\centering
    \includegraphics[width=0.75\textwidth]{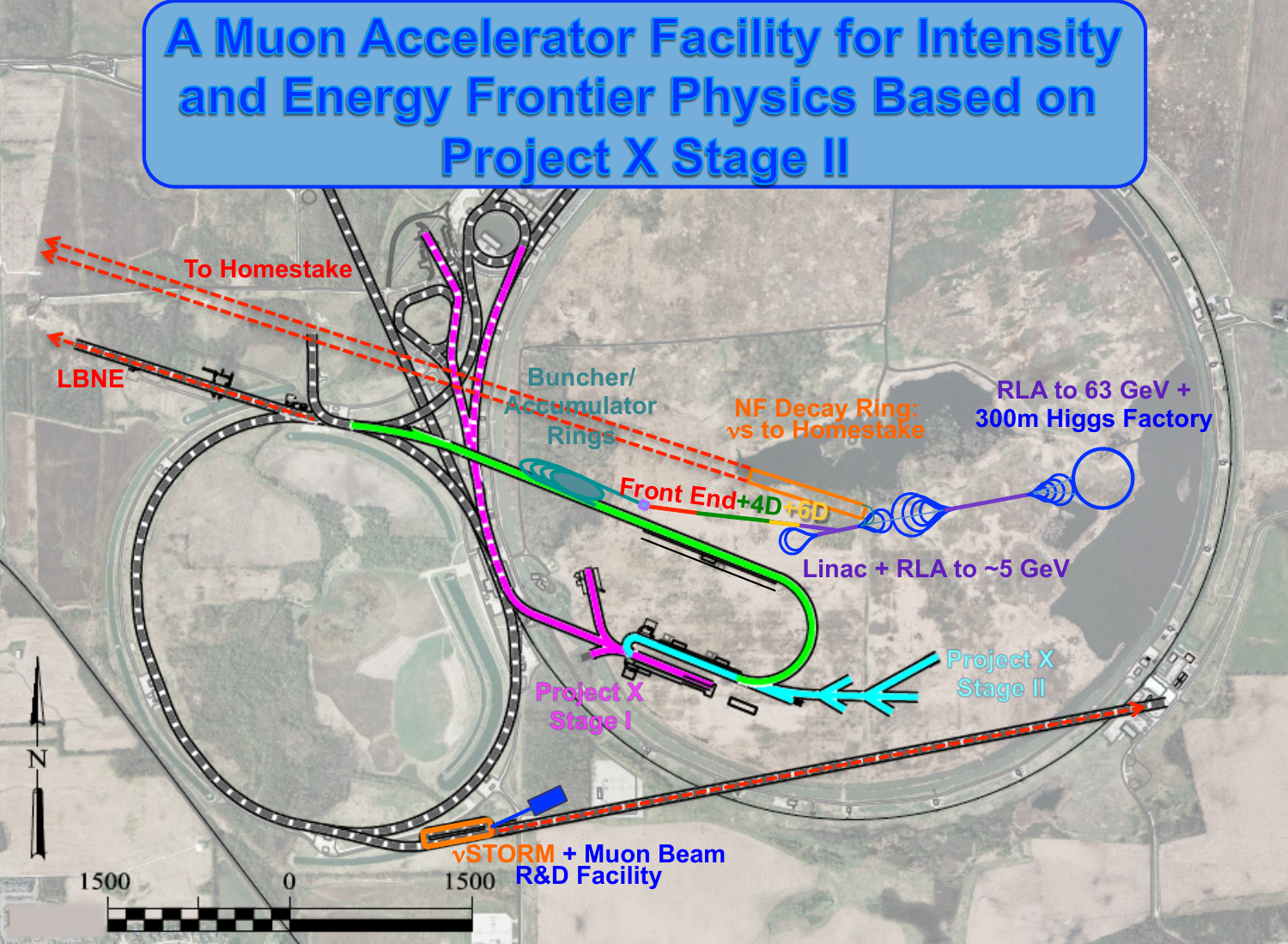}
    \caption[Footprint of neutrino-factory and muon-collider facilities on the Fermilab site.]{Footprint of 
        neutrino-factory and muon-collider facilities, including an initial muon collider Higgs factory, 
        on the Fermilab site.}
\label{nu:fig:layout}
\end{figure}

This timeline is summarized in Fig.~\ref{nu:fig:timeline}, which projects an informed decision on proceeding
with an NF by the end of this decade, and a similar decision point on the first muon collider by the middle
of the next decade.
An MC in the multi-TeV range would offer exceptional performance due to the absence of synchrotron radiation
effects, no beamstrahlung issues at the interaction point, and anticipated wall power requirements at the
200~MW scale, well below the widely accepted 300~MW maximum affordable power requirement for a future high
energy physics facility.
Figure~\ref{nu:fig:layout} shows the potential footprint of a sequence of facilities beginning with
$\nu$STORM and followed by a neutrino factory and Higgs Factory at Fermilab, which could be based on the \PX\
Stage~2 configuration.

To summarize, muon accelerators can enable a broad and world-leading
high energy physics program which can be based on the infrastructure
of the single remaining U.S high energy physics laboratory, Fermilab.
While any decision to move forward with muon accelerator based
technologies rests on the evolving physics requirements of the field,
as well as the successful conclusion of the MAP feasibility assessment
later this decade, the ability of muon accelerators to address crucial
questions on both the Intensity and Energy Frontiers, as well as to
provide a broad foundation for a vibrant U.S. HEP program, argues for
a robust development program to continue. This will enable a set of
informed decisions by the U.S. community starting near the end of this
decade.

% \clearpage

%%%%%%%%%%%%%%%%%%%%%%%%%%%%%%%%%%%%%%%%%%%%%%%%%%%%%%%%%%%%
\section{Short-baseline physics}
\label{nu:sec:sbl}
%%%%%%%%%%%%%%%%%%%%%%%%%%%%%%%%%%%%%%%%%%%%%%%%%%%%%%%%%%%%

Short-baseline oscillation physics in the context of \PX\ deals
with flavor conversion and disappearance phenomena which take place at $L/E$
values which are considerably smaller than those associated
with the mass-squared splittings of atmospheric and solar neutrino
oscillations. This area has seen an increased scientific interest, 
stimulating several workshops and documents,
notably the sterile neutrino white paper \cite{Abazajian:2012ys} and
the report of the short baseline focus group at
Fermilab~\cite{Brice:2012zz}.  The LSND~\cite{Aguilar:2001ty}
and now MiniBooNE~\cite{Aguilar-Arevalo:2012eua} results indicate
a possible flavor conversion of $\bar\nu_\mu$ to $\bar\nu_e$ at the
level of about 0.003. At the same time, MiniBooNE has seen a low energy
excess of events which may or may not be related to their primary
signal and LSND.  The results from calibrations of low energy
radio-chemical solar neutrino experiments using the reaction 
$\text{Ga}+\nu_e\to\text{Ge}+e^-$ based on artificial, mono-energetic neutrino
sources ($^{51}$Cr and $^{37}$Ar) \cite{Giunti:2010zu} seem to show a
deficit in count rate of about 25\% with an error bar of about 10\%.
The so-called reactor anomaly \cite{Mention:2011rk} indicates a 6\%
deficit of $\nu_e$ emitted from nuclear reactors at baselines less
than 100 m. Interestingly, this is entirely based on the re-analysis
of existing data; the deficit is caused by three independent effects
which all tend to increase the expected neutrino event rate. There
have been two re-evaluations of reactor antineutrino
fluxes\cite{Huber:2011wv,Mueller:2011nm} both see an increase of flux
by about 3\%. The neutron lifetime decreased from 887--899~s to 885.7~s
\cite{Beringer:1900zz} and thus the inverse $\beta$-decay cross
section increased by a corresponding amount.  The contribution from
long-lived isotopes to the neutrino spectrum was previously neglected
and enhances the neutrino flux at low energies.

All these hints have a statistical significance around 3$\sigma$ and
may be caused by one or more sterile neutrinos with masses of roughly
0.5 eV to 4 eV. The results of the PLANCK\cite{Ade:2013zuv} satellite
mission data when compared with the measured value of the Hubble
constant hint at new light degrees of freedom in the universe,
possibly sterile neutrinos.

Resolving those anomalies will require a new series of experiments.
More specifically, the short baseline focus group\cite{Brice:2012zz}
recommends that Fermilab pursue accelerator-based experiments which
can definitively address these anomalies on a short timescale. In
conjunction with the global efforts on sterile neutrinos, many of
which do not rely on a large accelerator infrastructure, it seems
plausible and highly likely that, by the time \PX\ starts its
physics program, there will have been either a discovery of sterile
neutrinos, or more generally new physics at short baselines, or
stringent new limits which significantly contradict the current
indications. In the latter case, there will be no short-baseline
program at FNAL in the \PX\ era. In the case of an unambiguous
discovery, the task of \PX\ would be to deliver high intensities
at energies around 8~GeV to allow detailed studies of the newly
discovered sterile neutrino(s), or whatever new physics effect is
behind the short-baseline anomalies.

Several proposals exist, both a Fermilab
(MiniBooNE~II\cite{BooNEProposal} and LAr1~\cite{Chen:2012nv}) and at CERN \linebreak
(ICARUS/NESSIE\cite{Antonello:2012qx}), to use, as MiniBooNE did, pion decay-in-flight
beams.
The crucial difference to MiniBooNE would be
the use of a near detector, and potentially the use of LAr TPCs
instead of Cherenkov detectors. While these new proposals would
constitute a significant step beyond what MiniBooNE has done,
especially in terms of systematics control, it remains to be proven
that a beam which has a 1\% level contamination of $\nu_e$ can be used
to perform a high precision study of a sub-percent appearance effect.
Therefore, not all proposals are able to take full advantage of the
beam intensities \PX\ will deliver.

One proposal to resolve the LSND puzzle is OscSNS
\cite{Garvey:2005pn,OscSNS:2013hua}, which aims to repeat the LSND
measurement while avoiding the shortcomings of LSND.  The idea is to
build a liquid scintillator detector at a powerful 1--3~GeV proton
source to exploit kaon, pion, and muon decay-at-rest. A high beam
power of more than 1~MW and a short duty cycle of less than $10^{-5}$ are key
to improve on LSND's performance. OscSNS is the most
direct test of LSND conceivable and, thus, is entirely model independent
and could be central to resolving the short baseline anomalies.

Another proposed technology is to use a stored muon beam, called
$\nu$STORM . Here, the neutrinos are produced by the purely leptonic,
and therefore well understood, decay of muons.  Thus, the neutrino
flux can be known with sub-percent precision.  The signals
are wrong-sign muons which can be identified quite easily in a
magnetized iron detector. The precise knowledge of the neutrino flux
and the expected very low backgrounds for the wrong-sign muon search
allow one to reduce systematic effects to a negligible level, hence
permitting precise measurements that would shed light on the new physics that may be
behind the short-baseline anomalies.

\subsection{BooNE-X}
\label{nu:subsec:BooNE-X}

\begin{figure}
    \centering
    \vspace*{-12pt}
    \includegraphics[width=0.6\textwidth]{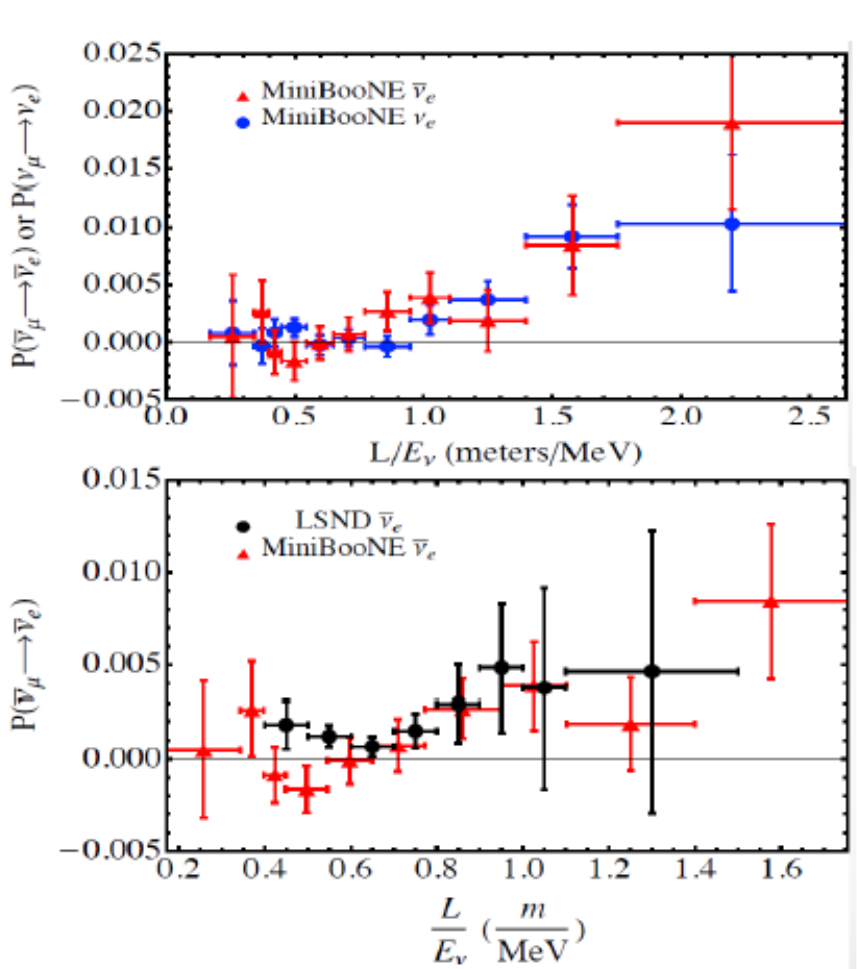}
    \caption[LSND and MiniBooNE anomalies]{LSND and MiniBooNE anomalies as an oscillation probability 
        \vs~$L/E$. 
        The interpretation of the anomalies as an oscillation effect is consistent with the data.}
    \label{nu:fig:SBLAnomaly}
\end{figure}

MiniBooNE has enjoyed 10 years of smooth operation, during which an astounding $6.46\times 10^{20}$ protons
on target (POT) have been delivered in neutrino mode, and an even more astounding $1.14\times 10^{21}$~POT
have been delivered in antineutrino mode.
The results of those data are compared to the LSND data in Fig.\ref{nu:fig:SBLAnomaly} in the context of an
oscillation phenomena.
The neutrino mode data has yielded a excess of $162.0 \pm 28.1_{\text{stat}} \pm 38.7_{\text{syst}}$ events 
at reconstructed neutrino energies below 475~MeV.
That excess is not described well by a simple two-neutrino model, but can be accommodated by an extended 3
active + 2 sterile neutrino model, fit to the world's relevant neutrino data.
While the statistical significance of the excess is ~ $6\sigma$, the overall significance is limited to
$3.4\sigma$ by the systematic error in the estimation of the background.
That systematic error is related to the error in the detector acceptance or efficiency for $\pi^0$ background
events, and to a lessor extent, the flux of neutrinos, and the neutrino-nucleus cross sections.
Similarly, an excess of is observed in antineutrino mode of $78.4 \pm 20.0_{\text{stat}}\pm
20.3_{\text{syst}}$ events, consistent with the neutrino-mode data.

Given the success of the MiniBooNE program, we believe that
constructing a MiniBooNE detector (reusing the mineral oil,
electronics, and PMTs from MiniBooNE if necessary) at a new location
$\sim$ 200 meters from the Booster Neutrino Beam proton target, will
be the most expedient way understand whether or not the excess events
observed by MiniBooNE are caused by an oscillation process. The
primary motivation for a near detector, rather than a detector further
away, is that the neutrino interaction rate will be over seven times
larger, and the measurement will precisely determine the
neutrino-related backgrounds within six months of running.  The
combination of the present MiniBooNE neutrino-mode data, plus a
4-month ($1\times 10^{20}$~POT or $\sim 700,000$ neutrino events)
neutrino-mode run with the BooNE\cite{BooNEProposal} detector, would
result in a $5\sigma$ sensitivity to whether or not the excess is an
oscillation effect.

In the \PX\ era, the linac will enable a much brighter Booster
Neutrino Beam (BNB), and if oscillation phenomena are indeed verified,
a detailed exploration of oscillations would be possible with the
addition of a third detector at a distance of ~1--2 km from the BNB
proton target. The BooNE-X detector with a mass of 2--5~kT would be
suitable with the higher neutrino flux available. The MiniBooNE
technology costs scale with $(\text{mass})^{2/3}$, which is favorable
compared with liquid argon costs, which scale with $(\text{mass})^1$.
Measurements \nopagebreak of a precision of $6\sigma$ would be possible with such
a three-detector system.

\subsection{LarLAr: A One-kiloton Liquid Argon Short Baseline Experiment}
\label{nu:subsec:larlar}

An interesting and powerful way to probe the MiniBooNE/LSND anomalies would be to combined the MicroBooNE
detector with another, larger, liquid-argon time projection chamber (LAr~TPC) in a near/far configuration.
A near/far configuration, dubbed LAr1, would considerably reduce the systematic errors, while the size of the
second detector would increase statistics significantly, which are expected to be the limiting factor for a
MicroBooNE-only search.
With a two detector system, a definite statement regarding oscillations could be made.

\begin{figure}
    \centering
    \includegraphics[width=0.6\textwidth]{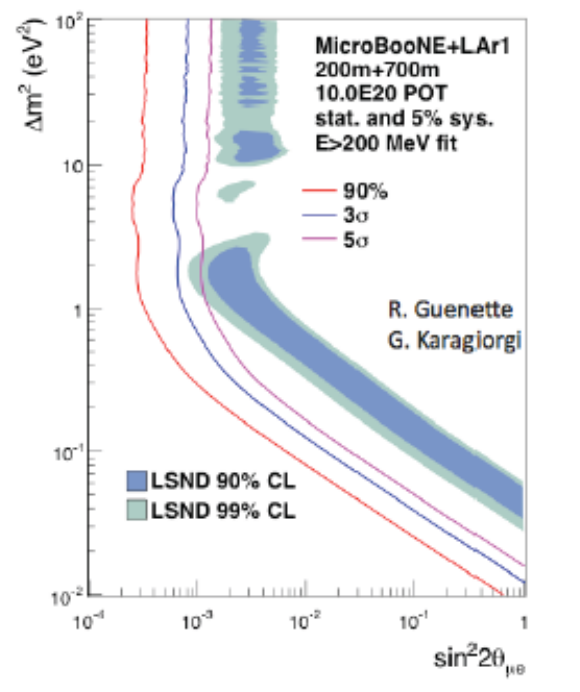}
    \caption[LarLAr sensitivity to the LSND anomaly in neutrino mode]{LarLAr sensitivity to the LSND anomaly 
        in neutrino mode.} 
    \label{nu:fig:larlar}
\end{figure}

The LBNE collaboration is currently designing a 1~kt LArTPC as an
engineering prototype.  It has been pointed out that this detector
could be instrumented and placed in the BNB at Fermilab to study
short-baseline oscillations.  Several configurations have been
considered for this experiment. The MicroBooNE detector, used as the
near detector, could be located either at 200~m or 470~m from the BNB.
The far detector, LarLAr, could be placed either at 470~m or 700~m. Note
that no further optimization has been done on the chosen detector
locations, which leaves room for improvement.  In the sensitivity
studies presented here, the fiducial volumes assumed for MicroBooNE
and LarLAr are 61.4~t and 347.5~t respectively. A flat 80\% efficiency
was assumed. All results shown below are for statistical errors only,
which are assumed to be the dominant source of uncertainty.  
Fig.~\ref{nu:fig:larlar} shows sensitivity curves to a 3+1 neutrino model,
for different configurations with both MicroBooNE and LarLAr detectors
combined in neutrino and antineutrino modes, for a total of
$6.6\times10^{20}$~POT in each mode. Such a sample is achievable in
two years under an improved-linac \PX\ scenario.  It is clear
from these studies that combining two LAr detectors is a very powerful
way to probe short-baseline oscillations. If systematic uncertainties
can be reasonably mitigated, this two LAr-detector experiment would
offer definitive measurements (at the 5 $\sigma$ level) of the Mini-
BooNE/LSND anomalies in both neutrino and antineutrino modes. Note
that in the antineutrino case, more than $6.6\times10^{20}$~POTs would
be required to reach the 5 $\sigma$ level for the whole allowed
parameter space.

\subsection{$\nu$Storm: Neutrinos from Stored Muons}
\label{nu:subsec:nustorm}

\begin{figure}[p]
\centering
    \includegraphics[width=0.75\textwidth]{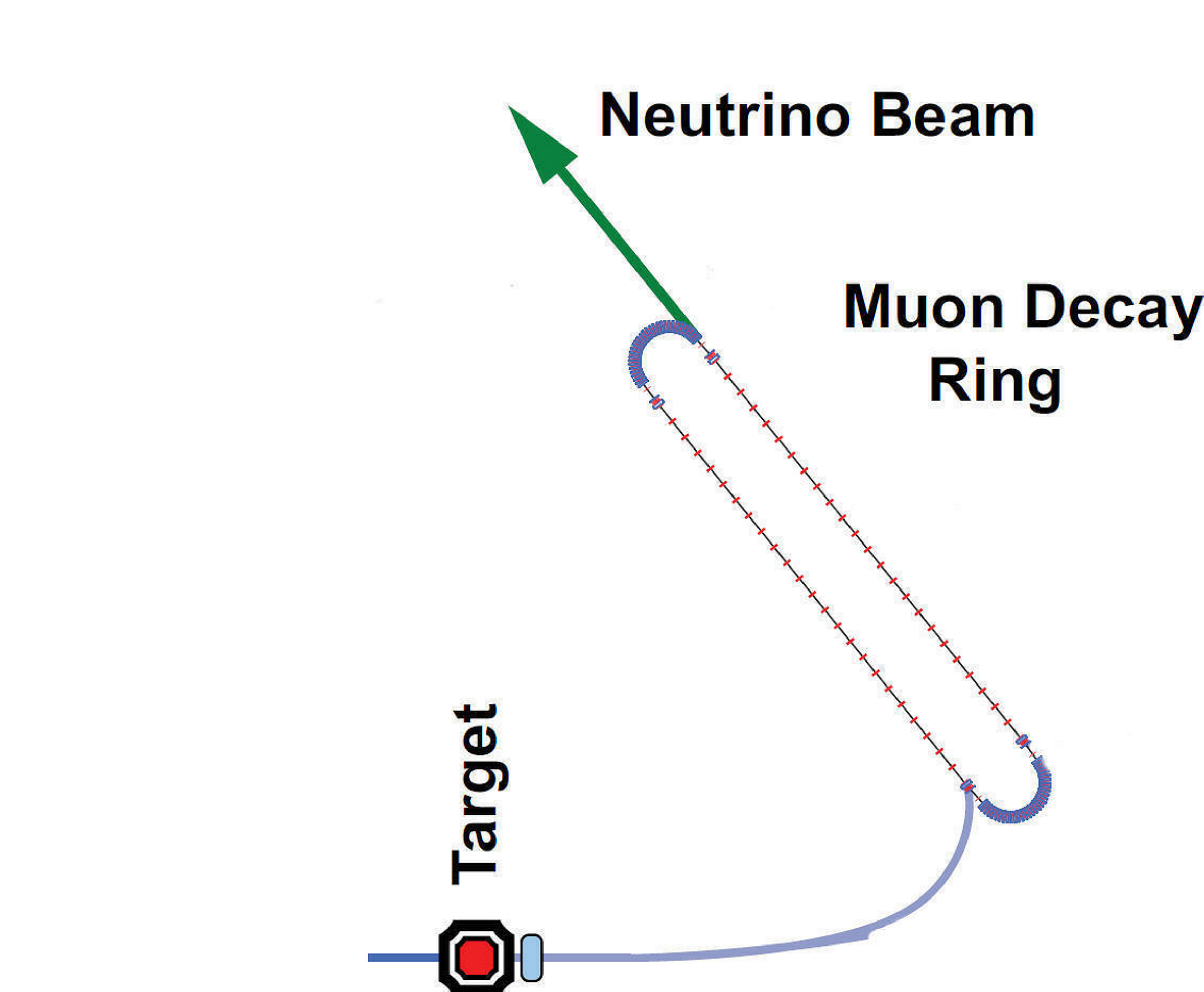}
    \caption[Schematic of the $\nu$Storm facility]{Schematic of the $\nu$Storm facility.}
    \label{nu:fig:STORM}
\end{figure}

The idea of using a muon storage ring to produce a high-energy ($\simeq$ 50~GeV) neutrino beam for
experiments was first discussed in 1974 by Koshkarev \cite{Koshkarev:1974xx}.
A detailed description of a muon storage ring for neutrino oscillation experiments was first produced in 1980
by Neuffer \cite{NeufferTelmark}.
In his paper, Neuffer studied muon decay rings with $E_\mu$ of 8, 4.5 and 1.5~GeV.
With his 4.5~GeV ring design, he achieved a figure of merit of $\simeq 6\times10^9$ useful neutrinos per
$3\times10^{13}$ protons on target.
The facility we describe here---$\nu$STORM\cite{Kyberd:2012iz}---is essentially the same facility proposed 
in 1980 and would use a 3--4~GeV/$c$ muon storage ring to study eV-scale oscillation physics and, in addition,
could add significantly to our understanding of $\nu_e$ and $\nu_\mu$ cross sections.
In particular the facility can
\begin{enumerate}
    \item address the large $\Delta$m$^2$ oscillation regime and make a major contribution to the study of 
        sterile neutrinos; 
    \item make precision $\nu_e$ and $\bar{\nu}_e$ cross-section measurements;
    \item provide a technology ($\mu$ decay ring) test demonstration and $\mu$ beam diagnostics test bed;
    \item provide a precisely understood $\nu$ beam for detector studies.
\end{enumerate}
See Fig.~\ref{nu:fig:STORM} for a schematic of the facility.

\begin{figure}[p]
	\centering
    \includegraphics[angle=-90,width=0.8\textwidth]{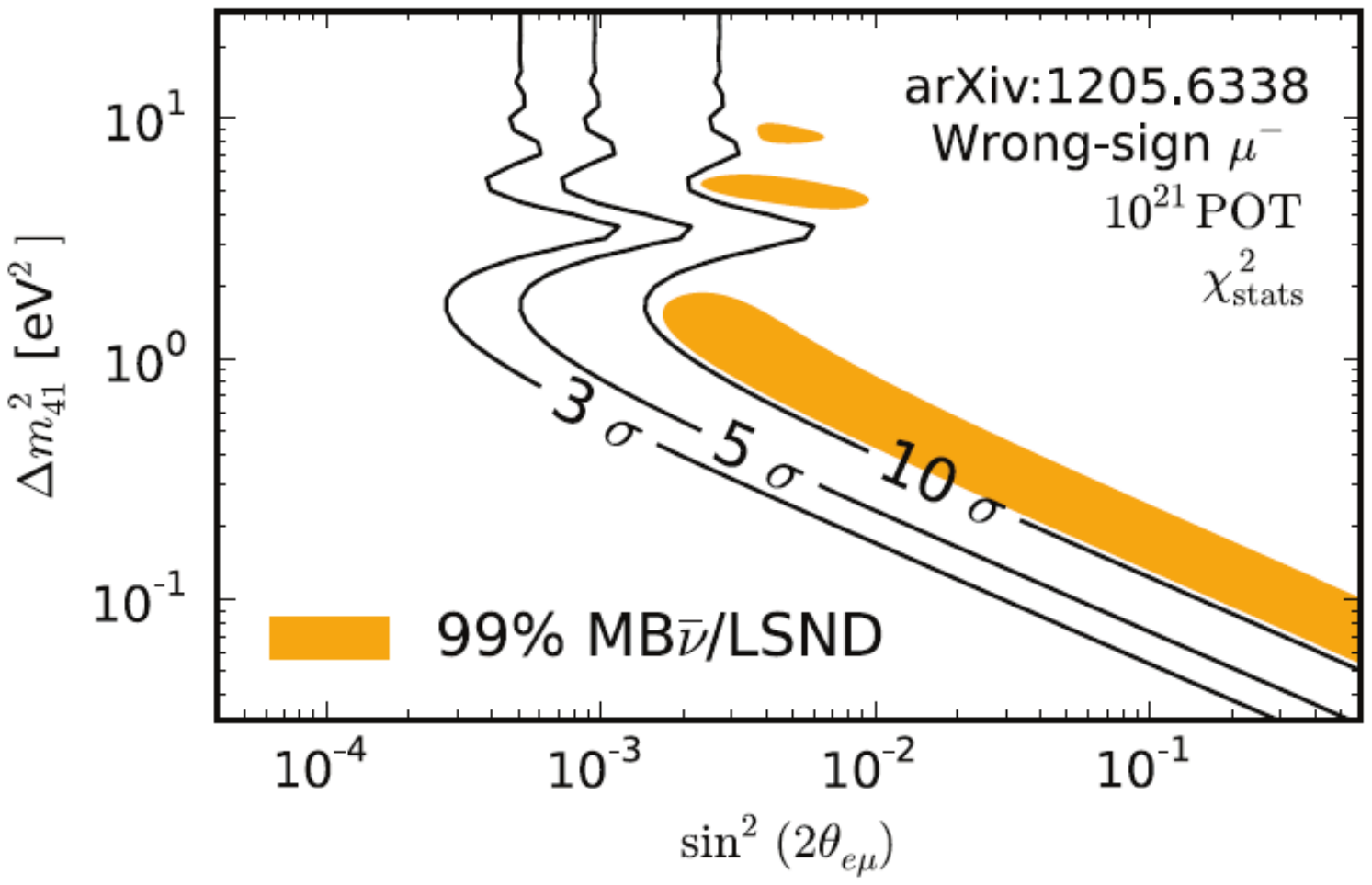}
    \caption[Exclusion limits from a five year run of $\nu$STORM]{Exclusion limits (statistical uncertainties
        only) from a five year run of~$\nu$STORM. 
        The orange-shaded areas show the combined 99\%~CL allowed region from MiniBooNE and 
        LSND.}
    \label{nu:fig:nustorm}
\end{figure}
\afterpage{\clearpage}

The facility is the simplest implementation of the neutrino-factory concept~\cite{Geer:1997iz}.
In our case, 60~GeV/$c$ protons are used to produce pions off a conventional solid target.
The pions are collected with a focusing device (horn or lithium lens) and are then transported to, and
injected into, a storage ring.
The pions that decay in the first straight of the ring can yield a muon that is captured in the ring.
The circulating muons then subsequently decay into electrons and neutrinos.
We are starting with a storage ring design that is optimized for 3.8~GeV/$c$ muon momentum.
This momentum was selected to maximize the physics reach for both oscillation and the cross section physics.

It would also be possible to create a $\pi\to\mu$ decay channel and inject the muons into the decay ring with
a kicker magnet.
This scheme would have the advantage that the transport channel could be longer than the straight in the
decay ring and thus allow for more $\pi$ decays to result in a useful $\mu$.
This does complicate the facility design, however, due to the need for the kicker magnet and the desire to
use single-turn extraction from the Main Injector.

Muon decay yields a neutrino beam of precisely known flavor content and energy.
For example for positive muons: $\mu^+ \rightarrow e^+$ + $\bar{\nu}_\mu$ + $\nu_e$.
In addition, if the circulating muon flux in the ring is measured accurately (with beam-current transformers,
for example), then the neutrino beam flux is also accurately known.
Near and far detectors are placed along the line of one of the straight sections of the racetrack decay ring.
The near detector can be placed at 20--50 meters from the end of the straight.
A near detector for disappearance measurements will be identical to the far detector, but only about one
tenth the fiducial mass.
It will require a $\mu$ catcher, however.
Additional purpose-specific near detectors can also be located in the near hall and will measure
neutrino-nucleon cross sections.
$\nu$STORM can provide the first precision measurements of $\nu_e$ and $\bar{\nu}_e$ cross sections which are
important for future long-baseline experiments.
A far detector at $\simeq$ 2000 m would study neutrino oscillation physics and would be capable of performing
searches in both appearance and disappearance channels.
The experiment will take advantage of the ``golden channel'' of oscillation appearance $\nu_e \rightarrow
\nu_\mu$, where the resulting final state has a muon of the wrong-sign from interactions of the
$\bar{\nu}_\mu$ in the beam.
In the case of $\mu^+$s stored in the ring, this would mean the observation of an event with a $\mu^-$.
This detector would need to be magnetized for the wrong-sign muon appearance channel, as is the case for the
current baseline neutrino factory detector \cite{Choubey:2011xz}.
A number of possibilities for the far detector exist.
However, a magnetized iron detector similar to that used in MINOS\cite{Ables:1995wq} is likely to be the most
straight forward approach for the far detector design.
We believe that it will meet the performance requirements needed to reach our physics goals.
For the purposes of the $\nu$STORM oscillation physics, a detector inspired by MINOS, but with thinner plates
and much larger excitation current (larger B field) is assumed.

\subsection{Neutrinos from Stopped Kaons, Pions, and Muons}
\label{nu:subsec:nustop}

The \PX\ facility provides a unique opportunity for US science to
perform a definitive search for sterile neutrinos. The MW beam power
of \PX\ is a prodigious source of neutrinos from the decay of
$K^+$, $\pi^+$ and $\mu^+$ at rest. These decays produce a well
specified flux of neutrinos via $K^+ \rightarrow \mu^+ \nu_\mu$,
$\tau_K$ = $1.2\times 10^{-8}$ s, $\pi^+ \rightarrow \mu^+ \nu_\mu$,
$\tau_\pi$ = $2.7\times 10^{-8}$ s, and $\mu^+ \rightarrow e^+ \nu_e
\bar\nu_\mu$, $\tau_\mu$ = $2.2\times 10^{-6}$ s.  With the \PX\
RCS option, the low duty factor is more than 1000 times less than
LAMPF, and this smaller duty factor reduces cosmic backgrounds and
allows the induced events from $\pi^+$ decay to be separated from the
$\nu_e$ and $\bar\nu_\mu$ induced events from $\mu^+$ decay.
 
The detector would be based on the LSND and MiniBooNE detector
technologies, similar to the OscSNS proposal\cite{OscSNS:2013hua} and
would consist of an $\sim$ 1~kT tank of mineral oil, covered by
approximately 3500 8-inch phototubes, and located about 60~m from the
production target.  The $K^+$ decays provide a mono-energetic
$\nu_\mu$ which can be seen via charged current reactions. A direct
measurement of oscillations can be made by measuring their rate as a
function of flight path.  The experiment will use the mono-energetic
29.8~MeV to investigate the existence of light sterile neutrinos via
the neutral-current reaction $\nu_\mu {\,}^{12}\text{C} \rightarrow {\,}^{12}\text{C}^*
(15.11~\text{MeV})$, which has the same cross section for all active
neutrinos but is zero for sterile neutrinos. An oscillation of this
reaction, with a known neutrino energy, is direct evidence for sterile
neutrinos. The experiment can also carry out a unique and decisive
test of the LSND appearance $\bar\nu_e$ signal. In addition, a
sensitive search for $\nu_e$ disappearance can be made by searching
for oscillations in the detector of the reaction $\nu_e {\,}^{12}\text{C}
\rightarrow e^+{\,}^{12}\text{N}(gs)$, where the $\text{N}(gs)$ is identified by its
beta decay.  It is important to note that all of the cross sections
involved are known to a few percent or better.  The existence of light
sterile neutrinos would be the first major extension of the Standard
Model, and sterile neutrino properties are central to dark matter,
cosmology, astrophysics, and future neutrino research. An experiment
at \PX\ would be able to prove whether sterile neutrinos can
explain the existing short-baseline anomalies.

\subsection{Dark Sector Physics at SBL Neutrino Experiments}
\label{nu:subsec:dark}

Finally, short-baseline neutrino oscillation experiments are also ideal tools to search for more exotic
physics\cite{Hewett:2012ns}.
The production of other weakly interacting particles---such as axions, dark gauge bosons, and WIMP
particles---is in many cases expected to be detectable in SBL neutrino experiments.
The "portal" to the dark sector is dark-photon mixing with normal photons, and $\pi^0$ and $\eta$ decays to
photons produce the dark-sector particles: $\pi^0, \eta \rightarrow \gamma V$, $V\rightarrow\bar\chi\chi$.
As an example, the 3$\sigma$ anomaly in the muon $g-2$ could be explained by such a model and account for the
dark matter observed in the universe.
Indeed, MiniBooNE has already proposed\cite{Dharmapalan:2012xp} to test some or those models with a run where
the beam is steered off-target to suppress neutrino production.
For further discussion, see Chapter~\ref{chapt:nlwcp}.

\subsection{Neutrino Scattering Physics Experiments}
\label{nu:subsec:xsec}

In general, the higher intensity beams offered by \PX\ would
enable more precise measurements of Standard-Model processes. They
include leptonic processes span the range from $\nu e \rightarrow \nu
e$ at a stopped $K/\pi/\mu$ neutrino source, all the way up to the
deep-inelastic scattering of neutrinos off of nuclei in the many-GeV
energy range. Those measurements can probe non-standard interactions
and yield important information about nuclear physics.
Neutrino-nucleus scattering data is essential for interpreting
precision long-baseline oscillation experiments.  For example, a
detector in the LBNE neutrino beam, if designed properly, could make
dramatic progress on further understanding of those processes with the
high event rates of a \PX\ beam.
As discussed above, a better understanding of neutrino-nucleon form factors,
obtained via lattice QCD, will also help.

\section{Summary}

For the short-baseline program, \PX\ most likely will play a role
after a discovery has been made and in that case, the goal would be a
precise measurement of the parameters of the newly discovered physics.
If there is no discovery in the short-baseline program prior to
\PX, it is doubtful that this program would be pursued in the
\PX\ era.  The only technology which seems to have a clear
upgrade path to high precision short-baseline physics without running
into systematics issues is $\nu$STORM.  $\nu$STORM would profit
considerably from increased beam power at 120~GeV.

The LBNE experiment is strong motivation for \PX\ in order to fully capitalize on the considerable investment
made on a large underground detector and new beamline.
The currently approved LBNE Project scope includes a new beamline capable of accepting all the beam power
\PX\ can deliver.
However, the initial detector mass is rather small, 10 kt, and on the surface, which may require a further
effective reduction of fiducial mass to cope with cosmogenic backgrounds.
The initial small detector mass and the risk of it becoming effectively smaller with surface operations is
strongly motivating the LBNE collaboration grow in number and resources in order to place the initial
detector underground, and with a mass greater than 10~kT.

A staged muon-based program starting with $\nu$STORM can evolve in various, adjustable steps to a full
neutrino factory, which, eventually, stes the stage for a muon collider.
This pathway seems to be a very attractive option, producing outstanding physics with every step.
At the same time, it crucially requires \PX\ and, thus, could be one of the most compelling motivations for
\PX.
Obviously, going beyond $\nu$STORM requires a vigorous R\&D effort, which in the form of the IDS-NF and MAP
is already ongoing, but would benefit from increased funding.

%%%%%%%%%%%%%%%%%%%%%%%%%%%%%%%%%%%%%%%%%%%%%%%%%%%%%%%%%%%%

\bibliographystyle{apsrev4-1}
\bibliography{nu/refs}
 % Geoff, Patrick & Andre, Sam, Mary, \ldots

%%%%%%%%%%%%%%%%%%%%%%%%%%%%%%%%%%%%%%%%%%%%%%%%%%%%%%%%%%%%
\chapter{Kaon Physics with \PX}
\label{chapt:kaon}
%%%%%%%%%%%%%%%%%%%%%%%%%%%%%%%%%%%%%%%%%%%%%%%%%%%%%%%%%%%%

\authors{Vincenzo Cirigliano, David E. Jaffe, Kevin Pitts, \\
Wolfgang~Altmannshofer, 
Joachim~Brod, 
Stefania~Gori, 
Ulrich~Haisch, and
Robert~S.~Tschirhart}

\section{Introduction}  

Kaon decays have played a key role in the shaping of the Standard Model (SM)
\cite{Glashow:1961tr,Salam:1968rm,Weinberg:1967tq} from the discovery of kaons~\cite{Rochester:1947mi} until
today.
Prominent examples are the introduction of internal flavor quantum numbers (strangeness)
\cite{Pais:1952zz,GellMann:1953zza}, parity violation ($K\to 2\pi, 3\pi$ puzzle)
\cite{Dalitz:1954cq,Lee:1956qn}, quark mixing~\cite{Cabibbo:1963yz,Kobayashi:1973fv}, meson-antimeson
oscillations, the discovery of \CP\ violation~\cite{Christenson:1964fg}, suppression of flavor-changing
neutral currents (FCNC) and the Glashow-Iliopoulos-Maiani (GIM) mechanism~\cite{Glashow:1970gm}.
Kaon properties continue to have a high impact in constraining the flavor sector of possible extensions of
the~SM.
As we explain in this chapter, their influence will extend into the \PX~era.

In the arena of kaon decays, a prominent role is played by the FCNC modes mediated by the quark-level
processes $s\to d (\gamma, \ell^+\ell^-, \nu\bar{\nu})$, and in particular the four theoretically cleanest
modes $K^+\to\pi^+\nu\bar{\nu}$, $K_L\to\pi^0\nu\bar{\nu}$, $K_L\to\pi^0 e^+ e^-$, and
$K_L\to\pi^0\mu^+\mu^-$.
Because of the peculiar suppression of the SM amplitude (loop level proportional to $|V_{us}|^5$) which in
general is not present in SM extensions, kaon FCNC modes offer a unique window on the flavor structure of SM
extensions.
This argument by itself already provides a strong and model-independent motivation to study these modes, even
while the TeV-scale is probed a the LHC: rare $K$ decays can teach us about the flavor structure of SM
extensions at much, much higher energies.
For further discussion of the role of quark and lepton flavor physics in the search for new phenomena,
see the recent review in Ref.~\cite{Buras:2013ooa}.

The discovery potential of rare decays depends on how well we can calculate their rates in the SM, how strong
the constraints from other observables are, and how well we can measure their branching ratios (BRs).
State-of-the-art predictions are summarized in Table~\ref{kaon:fig:tab} and show that we currently know the BRs
$K^+\to\pi^+\nu\bar{\nu}$ at the 10\% level, $K_L\to\pi^0\nu\bar{\nu}$ at the 15\% level, while 
$K_L\to\pi^0e^+e^-$, and $K_L\to\pi^0\mu^+\mu^-$ at the 25--30\% level.
Note that the charged and neutral $K\to\pi\nu\bar{\nu}$ modes are predicted with a precision surpassing
any other FCNC process involving quarks.

Within a general effective field theory (EFT) analysis of new physics effects, $K\to\pi\nu\bar{\nu}$
probe a number of leading dimension-six operators.
A subset of these operators is essentially unconstrained by other observables, and therefore on general
grounds one can expect sizable deviations from the SM in $K\to\pi\nu\bar{\nu}$ (both modes), depending on
the flavor structure of the Beyond the Standard Model (BSM) scenario.
Moreover, an analysis of the correlations among various rare $K$ decay modes allows one to disentangle the
size of different BSM operators, thus enhancing our model-discriminating power and making the case for
building a broad $K$ physics program, that involves all rare FCNC decays.

If one restricts the analysis to the subset of $Z$-penguin BSM operators, which are the dominant in several
explicit models of new physics, a number of constraints on $K \to \pi \nu \bar{\nu}$ emerges.
In fact, $Z$-penguin operators affect a large number of kaon observables ($K \to \pi \ell^+ \ell^-$,
$\epsilon_K$, $\epsilon'/\epsilon$, and in the case of one operator $K \to \pi \ell \nu$ through
$\text{SU}(2)$ gauge invariance).
Figure~\ref{kaon:fig:1} illustrates that currently the strongest constraints on $K \to \pi \nu \bar{\nu}$ arise
from direct \CP\ violation in $K \to \pi \pi$ decays, which excludes order-of-magnitude deviations in $K_L \to
\pi^0 \nu \bar{\nu}$ while still allowing for dramatic effects in $K^+ \to \pi^+ \nu \bar{\nu}$.
While this is true only in models in which the $Z$-penguin dominates contributions to $K \to \pi \nu
\bar{\nu}$, we think this constraint should be used as a target for future ``discovery" searches in $K_L \to
\pi^0 \nu \bar{\nu}$ at \PX.
As discussed in detail later in this chapter, there is strong evidence to support a Day-1 \PX\ \Kzero\
experiment with $\sim1000$ SM event sensitivity, which would retain plenty of discovery potential even in
presence of the constraint from $\epsilon'/\epsilon$.

This chapter is organized as follows.
Section~\ref{kaon:sect:theory} elaborates further on the physics case supporting the search for rare FCNC K decays
well into the next decade.
We begin with a review of the SM predictions (Sec.~\ref{kaon:sect:SM}) and we then discuss the physics reach,
first in a model-independent effective theory framework (Sec.~\ref{kaon:sect:BSM1}), then within supersymmetric
models (Sec.~\ref{kaon:sect:BSM2}), and last within Randall-Sundrum models of warped extra dimensions
(Sec.~\ref{kaon:sect:BSM3}).
We briefly comment on the reach of other decay modes in Sec.~\ref{kaon:sect:beyondrare}.
In Sec.~\ref{kaon:sect:exp} we first summarize the landscape of kaon experiments in this decade
(Sec.~\ref{kaon:sect:explandscape}), and then discuss the opportunity and impact of rare kaon decay measurements
at \PX\ (Sec.~\ref{kaon:sect:projectX}).
We summarize in Section.~\ref{kaon:sect:summary}.

\begin{table}
    \centering
    \caption[SM predictions and experimental limits for the four cleanest rare kaon decays]{Summary of 
        current SM predictions and experimental limits for the four cleanest rare kaon decays.
        In the SM predictions, the first error is parametric, the second denotes the intrinsic theoretical 
        uncertainty.}
    \label{kaon:fig:tab}  
    \begin{tabular}{lcc}
    \hline \hline
    Mode  &  Standard Model & Experiment \\ \hline
    $K^+ \to \pi^+ \nu \bar{\nu}$ & $7.81(75)(29)\times 10^{-11}$ & $(1.73^{+1.15}_{-1.05})\times 10^{-10}$  \ E787/949 \\
    $K_L \to \pi^0 \nu \bar{\nu}$ & $2.43(39)(6)\times 10^{-11}$ & $< 2.6\times 10^{-8}$ \  E391a \\
    $K_L \to \pi^0 e^+e^-$ & $(3.23^{+0.91}_{-0.79})\times 10^{-11}$ & $< 28 \times 10^{-11}$ \ KTEV \\
    $K_L \to \pi^0 \mu^+\mu^-$ & $(1.29^{+0.24}_{-0.23})\times 10^{-11}$ & $< 38 \times 10^{-11}$ \ KTEV \\ 
    \hline \hline
    \end{tabular}
\end{table}

\section{Rare Kaon Decays as Deep Probes of New Physics}
\label{kaon:sect:theory}

Rare kaon decays are severely suppressed in the Standard Model (SM).
Therefore they are highly sensitive to possible new physics (NP) effects.
Given the high precision of the SM predictions---in particular those of the ``golden modes''
$K\to\pi\nu\bar\nu$ (charged and neutral)---as well as the expected future experimental sensitivities, even
deviations from the SM predictions as small as 20--30\% could allow to establish the existence of~NP.
Moreover, visible deviations from the SM predictions are possible within many well motivated NP models like
the Minimal Supersymmetric Standard Model (MSSM) and in Randall-Sundrum (RS) Models.
In the following we give more details on the SM predictions of rare $K$ decays and their sensitivity to the
flavor structure of SM extensions.

\subsection{The baseline: Rare Kaon Decays in the Standard Model}
\label{kaon:sect:SM}

\begin{figure}
    \centering 
    \includegraphics[scale=0.35]{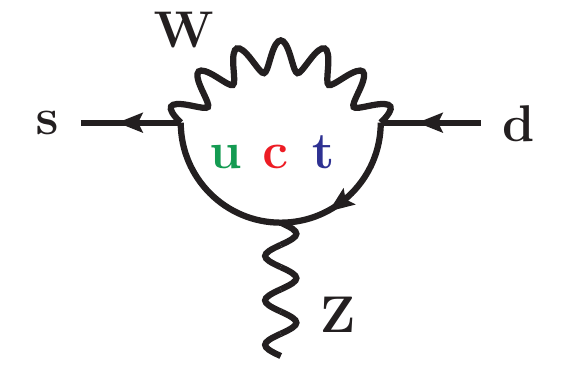}\hspace{2em}
    \includegraphics[scale=0.35]{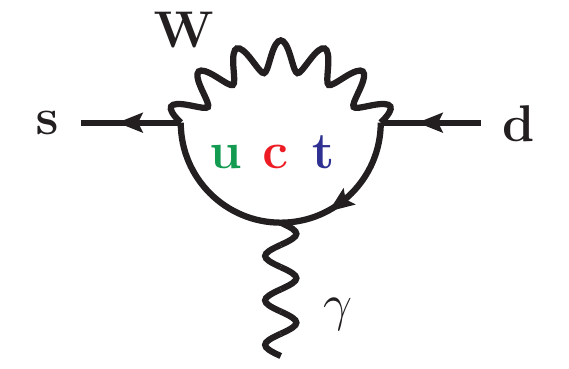}\hspace{2em}
    \includegraphics[scale=0.35]{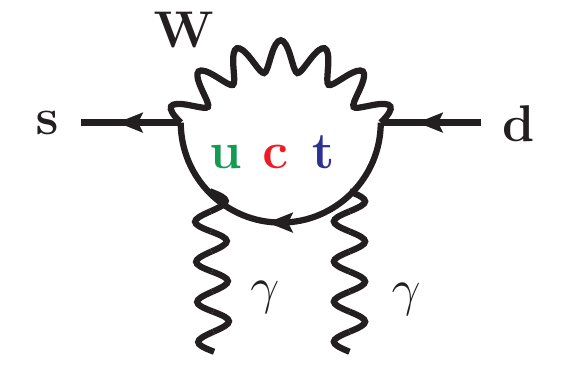}
    \caption[$Z$ penguin, single- and double-photon penguin]{$Z$ penguin, single- and double-photon penguin.}
    \label{kaon:fig:1sm}
\end{figure}

The decays $K^{+}\to\pi^{+}\nu\bar{\nu}$, $K_{L}\to\pi^{0}\nu\bar{\nu}$,
$K_{L}\to\pi^{0}e^{+}e^{-}$ and $K_{L}\to\pi^{0}\mu^{+}\mu^{-}$ proceed dominantly through
heavy-quark induced flavor-changing neutral currents (FCNC).
Within the standard model, the electroweak processes inducing the rare $K$ decays arise first at the one-loop
level and are of three types: $Z$ penguin and $W$ box, single photon penguin, and double photon penguin, each
being a function of the ratios $m_q^2/M_W^2$ (see Fig.~\ref{kaon:fig:1sm}).
Here $m_q$, $q=u,c,t$, are the up-type quark masses, and $M_W$ is the $W$ boson mass.
(The GIM mechanism cancels the constant part of the loop functions when summing over the three up-quark
flavors.)

The relative importance of each type of process contributing to the rare $K$ decays can be neatly understood
in terms of the limit of the loop functions for large or small quark masses.
The $Z$ penguin, as well as the \CP-violating single-photon penguin, are dominated by short-distance physics
(top- and charm-quark), due to the powerlike breaking of the Glashow-Iliopoulos-Maiani (GIM) mechanism.
On the contrary, the \CP-conserving photon penguins are fully dominated by the long-distance up-quark
contribution, arising from the logarithmic behavior of the corresponding loop functions.

Theory predictions for the decay rates are obtained using an effective theory framework, which allows us to
separate the different energy scales involved in the decay processes and to use appropriate methods of
calculation~\cite{Buchalla:2001fh}.
The short-distance part is encoded explicitly into the Wilson coefficients of the weak effective Hamiltonian.
However, computing the hadronic matrix elements of operators involving quark fields is a nontrivial problem,
which can be addressed with lattice QCD (see Chapter~\ref{chapt:lqcd}) or other nonperturbative methods based
on symmetries and dispersion relations.

%The appropriate effective theory to
%describe the long-distance physics is the chiral effecting Lagrangian,
%which is expressed in terms of the physical hadronic (pion, kaon,
%\ldots) fields. However, the dependence on the short distance
%parameters is only implicit, mainly due to our incomplete
%understanding of the nonperturbative QCD dynamics in terms of quarks
%and gluons. 

\boldmath
\subsubsection{$K^+ \to \pi^+ \nu \bar\nu$ and $K_L \to \pi^0 \nu \bar\nu$}
\unboldmath

\begin{figure}
\centering
    \includegraphics[width=0.8\textwidth]{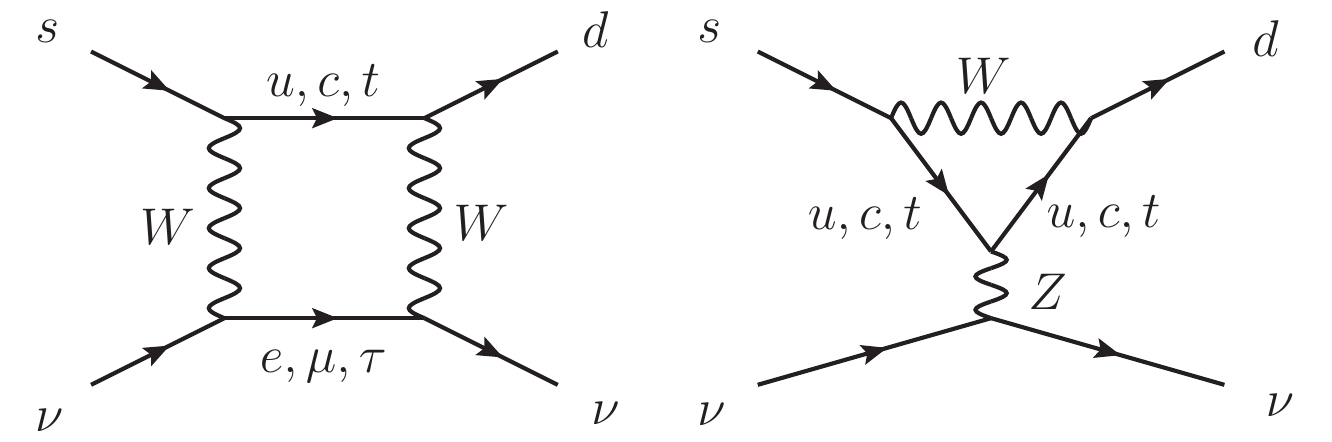}
    \caption[Leading-order diagrams for $K\to\pi\nu\bar\nu$ in the SM]{Leading-order diagrams contributing 
        to the decay amplitude for $K\to\pi\nu\bar\nu$ in the SM.} 
    \label{kaon:fig:kpnnLO}
\end{figure}

\begin{figure}
    \centering
    \includegraphics[width=0.8\textwidth]{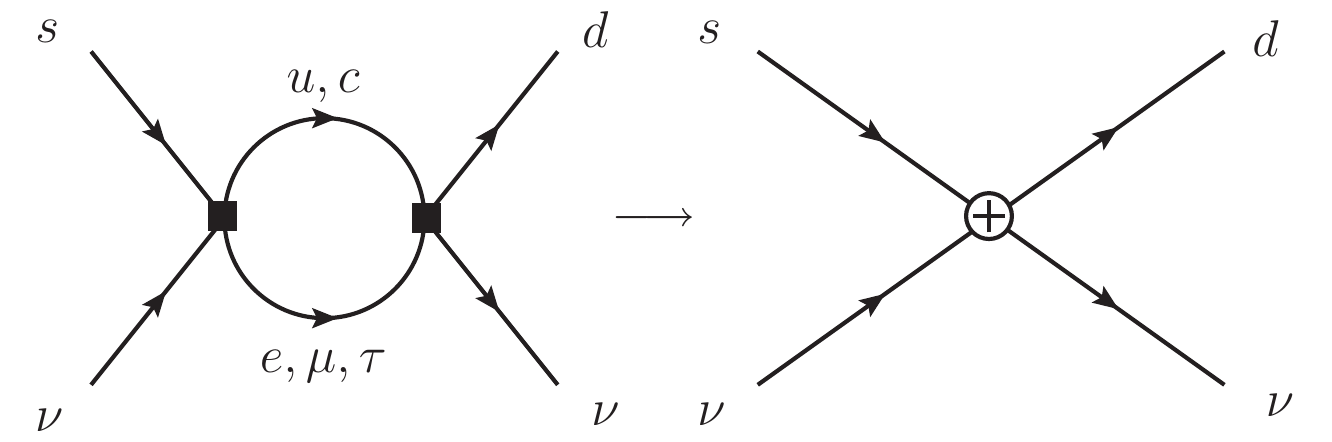}
    \caption[Leading-order mixing of current-current and penguin operators into $Q_\nu$]{Leading-order mixing of 
        current-current and penguin operators into $Q_\nu$.}
    \label{kaon:fig:kpnneffLO}
\end{figure}

For the the decays $K^{+} \to \pi^{+} \nu \bar{\nu}$ and $K_{L} \to \pi^{0} \nu \bar{\nu}$ short-distance
physics dominates because of the absence of photon penguins.
The effective Hamiltonian for the $K^+ \to \pi^+ \nu \bar\nu$ decay involves, to a good approximation, only
the operator $Q_\nu = (\bar s d)_{V-A} (\bar\nu \nu)_{V-A}$.
Its Wilson coefficient, induced at leading order by the SM box and penguin diagrams shown in
Fig.~\ref{kaon:fig:kpnnLO}, contains two terms proportional to $\lambda_t$ and $\lambda_c$, respectively,
where $\lambda_i\equiv V_{id}V_{is}^*$.
We have used the CKM unitarity relation $\lambda_u = - \lambda_c - \lambda_t$ to eliminate $\lambda_u$.
The leading behavior of the top-quark contribution $X_t$, proportional to $\lambda_t$, is given by
$m_t^2/M_W^2$.
The smallness of $\lambda_t$ compensates the effect of the large top-quark mass and makes it comparable in
size to the charm-quark contribution $P_c$, proportional to $\lambda_c$, with the leading behavior
$(m_c^2/M_W^2)\ln(m_c^2/M_W^2)$.
The appearance of the large logarithm is related to the bilocal mixing of current-current and penguin
operators into $Q_\nu$ through charm-quark loops shown in Fig.~\ref{kaon:fig:kpnneffLO}.
This introduces large scale uncertainties, which have been removed by computing the next-to-next-to-leading
order (NNLO) QCD corrections to $P_c$ in renormalization-group (RG) improved perturbation
theory~\cite{Buras:2005gr}.
In addition, the electroweak corrections are known.
They sum the LO and next-to-leading order (NLO) QED logarithms to all orders and fix the renormalization
scheme of the electroweak input parameters in the charm-quark sector, leading to the final prediction $P_c =
0.368(25)$~\cite{Brod:2008ss}.

The top-quark contribution $X_t$ does not contain a large logarithm and can be computed in fixed-order
perturbation theory.
The NLO QCD corrections have been known for a long time~\cite{Buchalla:1998ba,Misiak:1999yg}, and also the
full two-loop electroweak corrections to $X_t$ have been computed recently, fixing the renormalization scheme
of the electroweak input parameters also in the top-quark sector and rendering the remaining scale and scheme
dependence essentially negligible~\cite{Brod:2010hi}.
The final result is $X_t = 1.465(17)$, where the error is largely due to the remaining QCD scale uncertainty.

The branching ratio of the charged mode is given by
\begin{equation}
\label{kaon:eq:brch}
  \text{BR}_\text{ch} = \kappa_+
  (1+\Delta_{\text{EM}})
  \Bigg[\left(\frac{\Im \lambda_t}{\lambda^5} X_t\right)^2 + 
  \left(\frac{\text{Re}\lambda_c}{\lambda} \left(P_c + \delta P_{c,u}
    \right) + \frac{\text{Re}\lambda_t}{\lambda^5} X_t\right)^2
  \Bigg].
\end{equation}
Here, the quantity $\kappa_+ = 0.5173(25)\times10^{-10}$~\cite{Mescia:2007kn} contains the hadronic matrix
element of~$Q_\nu$.
It has been determined from the full set of $K_{\ell 3}$ data using isospin symmetry, including NLO and
partially NNLO corrections in chiral perturbation theory ($\chi$PT) and QED radiative
corrections~\cite{Mescia:2007kn}.
The quantity $\Delta_{\text{EM}} = -0.3\%$~\cite{Mescia:2007kn} accounts for the effects of real soft photon
emission.
Moreover, the CKM parameter $\lambda = |V_{us}| = 0.2255(7)$~\cite{Antonelli:2010yf}.

The effects of soft charm and up quarks as well as of higher-dimensional operators have been estimated in
$\chi$PT and are lumped into $\delta P_{c,u}=0.04(2)$, which enhances the branching ratio by roughly
6\%~\cite{Falk:2000nm}.
The error on $\delta P_{c,u}$ could in principle be reduced by a lattice-QCD
calculation~\cite{Isidori:2005tv}.

Using $m_t(m_t)=163.7(1.1)\text{GeV}$~\cite{Group:2010ab},
$m_c(m_c)=1.279(13)\text{GeV}$~\cite{Chetyrkin:2009fv}, and the remaining input from
Ref.~\cite{Charles:2004jd,Nakamura:2010zzi}, we find the following numerical prediction:
\begin{equation}
  \text{BR}_\text{ch} = (7.81 \pm 0.75 \pm 0.29)
  \times 10^{-11}\, , 
\end{equation}
where the first error is related to the uncertainties of the input parameters, and the second error
quantifies the remaining theoretical uncertainty.
The parametric error is dominated by the uncertainty in the CKM inputs $|V_{cb}|$ (56\%) and $\bar\rho$
(21\%) and could be reduced significantly in the future by better determinations of these parameters.
The main contributions to the theoretical uncertainty are ($\delta P_{c,u}: 46\%$, $X_t: 24\%$, $P_c: 20\%$,
$\kappa_{\nu}^+: 7\%$), respectively.
The branching ratio has been measured with a value $\text{BR}_\text{ch} = (1.73^{+1.15}_{-1.05}) \times
10^{-10}$~\cite{Artamonov:2008qb}, consistent with the SM prediction within the (still large) experimental
error.

The neutral mode $K_L \to \pi^0 \nu \bar\nu$ is purely \CP-violating~\cite{Littenberg:1989ix,Buchalla:1998ux},
so only the top-quark contribution is relevant for the decay rate because of the smallness of $\Im\lambda_c$.
It is given by the same function $X_t$ as for the charged mode.

The branching ratio is given by
\begin{equation}
  \text{BR}_\text{neutr} = \kappa_L\left(
    \frac{\Im \lambda_t}{\lambda^5}X_t \right)^2 \, ,
  \label{kaon:eq:brl}
\end{equation}
where $\kappa_L=2.231(13)\times 10^{-10}$ comprises the hadronic matrix element of $Q_\nu$ has been extracted
from the $K_{\ell 3}$ decays, as for the charged mode~\cite{Mescia:2007kn}.
There are no further long-distance contributions, which is the reason for the exceptional theoretical
cleanness of this mode.

Including also a factor taking into account the small ($\approx -1\%$) effect of indirect \CP\
violation~\cite{Buchalla:1996fp}, we find for the branching ratio
\begin{equation}
\text{BR}_\text{neutr} = (2.43 \pm 0.39 \pm 0.06)
\times 10^{-11}\, ,
\end{equation}
using the same input as for the charged mode.
Again, the first error corresponds to the parametric and the second to the theoretical uncertainty.
Here, the parametric uncertainty is dominated by the error in the CKM parameters $V_{cb}\,(54\%)$ and
$\bar\eta\,(39\%)$ and could again be reduced in the future by better determinations of these parameters.
The main contributions to the second, theoretical uncertainty are ($X_t: 73\%$, $\kappa_{\nu}^L: 18\%$),
respectively.
All errors have been added in quadrature.

The neutral mode has not been observed yet; an upper bound for the branching ratio is given by
$\text{BR}_\text{neutr} < 6.7 \times 10^{-8}\,(90\%\,\text{CL})$~\cite{Ahn:2007cd}.

\subsubsection{\boldmath$K_L \to \pi^0 \ell^+ \ell^-$}
\label{kaon:sect:KLp0ll}

Unlike the neutrino modes, the $K_L \to \pi^0 \ell^+ \ell^-$ modes, $\ell=e$, $\mu$, have sizeable
long-distance contributions.
Although these contributions are difficult to calculate, these processes are relevant because they are
sensitive to helicity-suppressed contributions, which allows the disentangling scalar and pseudoscalar
operators from vector and axial-vector operators~\cite{Mescia:2006jd}.
It can be exploited because of the good theoretical control over the individual contributions to the
branching ratios, which we now consider in turn:
\begin{enumerate}
\item
The direct \CP-violating contribution (DCPV) is contained in two Wilson coefficients $C_{7V}$ and $C_{7A}$
induced by $Z$ and $\gamma$ penguins, which are known at NLO QCD~\cite{Buchalla:1995vs}.
The matrix elements of the corresponding operators $Q_{7V} = (\bar s d)_{V-A} (\ell^+\ell^-)_V$ and $Q_{7A} =
(\bar s d)_{V-A} (\ell^+\ell^-)_A$ can be extracted from $K_{\ell 3}$ decays in analogy to the neutrino
modes~\cite{Mescia:2007kn}.
\item
The indirect \CP-violating contribution (ICPV) is related via $K^0$--$\bar K^0$ mixing to the decay $K_S \to
\pi^0 \ell^+ \ell^-$.
It is dominated by a single $\chi$PT coupling $a_S$~\cite{D'Ambrosio:1998yj}, whose absolute value can be
extracted from the experimental $K_S \to \pi^0 \ell^+ \ell^-$ decay rates to give
$|a_S|=1.2(2)$~\cite{Batley:2003mu}.
\end{enumerate}
Both ICPV and DCPV can produce the final lepton pair in a $1^{--}$ state, leading to interference between the
two amplitudes.
Whether the interference is constructive or destructive is determined by the sign of $a_S$, which is unknown
at the moment (see also~\cite{Bruno:1992za,Buchalla:2003sj,Friot:2004yr}).
It can be determined by measuring the $K_L \to \pi^0 \mu^+ \mu^-$ forward-backward
asymmetry~\cite{Mescia:2006jd}.
In addition, lattice QCD calculations of these modes will be able to determine the relative sign of the two
amplitudes in the next few years; see discussion in Chapter~\ref{chapt:lqcd}.

The purely long-distance \CP-conserving contribution is induced by a two-photon intermediate state $K_L \to
\pi^0 \gamma^* \gamma^* \to \pi^0 \mu^+ \mu^-$ and produces the lepton pair either in a phase-space
suppressed $2^{++}$ or in a helicity suppressed $0^{++}$ state.
The former is found to be negligible~\cite{Buchalla:2003sj}, while the latter is only relevant for the muon
mode because of helicity suppression.
It can be extracted within $\chi$PT from experimental information on the $K_L \to \pi^0 \gamma \gamma$
decay~\cite{Isidori:2004rb}.

The prediction for the branching ratio is~\cite{Mertens:2011ts}
\begin{equation}\begin{split}
    \text{BR}_{e^+e^-}     = 3.23^{+0.91}_{-0.79} \times 10^{-11} \quad
        (1.37^{+0.55}_{-0.43} \times 10^{-11}), \\
    \text{BR}_{\mu^+\mu^-} = 1.29^{+0.24}_{-0.23} \times 10^{-11} \quad
        (0.86^{+0.18}_{-0.17} \times 10^{-11}),
\end{split}\end{equation}
for constructive (destructive) interference.
The error of the prediction is completely dominated by the uncertainty in $a_S$ and could be reduced by
better measurements of the $K_S \to \pi^0 \ell^+ \ell^-$ modes~\cite{Mescia:2006jd}.
Experimental upper limits for the two decays~\cite{AlaviHarati:2003mr,AlaviHarati:2000hs} are
\begin{equation}\begin{split}
    \text{BR}_{e^+e^-}     < 28 \times 10^{-11}~ (90\%~\text{CL}), \\
    \text{BR}_{\mu^+\mu^-} < 38 \times 10^{-11}~ (90\%~\text{CL}),
\end{split}\end{equation}
which lie still one order of magnitude above the SM predictions.

\subsubsection{Future Improvements}

Over the next decade, we expect improvements in lattice-QCD calculations combined with progress in $B$ meson
measurements (LHCb and Super-Belle) to allow one to reduce the parametric uncertainty on both
$K\to\pi\nu\bar{\nu}$ to the 5\% level.
Substantial improvements in $K _L\to \pi^0 \ell^+ \ell^-$ will have to rely on lattice QCD computations,
requiring the evaluation of bi-local operators.
Exploratory steps exist in this direction, but these involve new techniques and it is hard to forecast the
level of uncertainty that can be achieved, even in a ten-year timescale.
Therefore, from a theory perspective, the golden modes remain both $K \to \pi \nu \bar{\nu}$ decays, because
they suffer from small long-distance contamination, indeed negligible in the \CP\ violating $K_L$ case.

% \subsection{Rare K decays as probes of new physics: model-independent considerations} 

\subsection{Beyond the Standard Model: Model-independent Considerations} 
\label{kaon:sect:BSM1}

New-physics searches in rare kaon decays can be approached using a top-down or a bottom-up approach.
In the former case one starts with a concrete NP model and predicts the observables and their correlations,
while in the latter case one maps classes of models onto an effective theory with the goal to get insights
free from personal taste and prejudices.
In this subsection we give a concise review of the bottom-up approach to kaon physics.

The starting point to obtain an effective description is to make the reasonable assumption that the scale
$\Lambda$ associated to the new dynamics is sufficiently above the weak scale $v$, which in turn allows for a
systematic expansion in powers of $v/\Lambda \ll 1$.
If one furthermore assumes that SM particles are weakly coupled to the NP sector (a technical
assumption which could be relaxed), then one may classify the new interactions in terms of 
$\text{SU}(2)_L\times\text{U}(1)_Y$ invariant operators of increasing dimension.

Our discussion here parallels the one given in Ref.~\cite{sebastian}. 
To leading order in $v/\Lambda$,  six operators can affect the $K \to \pi \nu \bar{\nu}$ decays. 
Three of  these are four-fermion operators and affect the  $K \to \pi \ell^+ \ell^-$ decays  as well 
(one of these operators  contributes to $K \to \pi \ell \nu$  by $\text{SU}(2)$  gauge invariance). 
The coefficients of these operators are largely unconstrained by other observables, and therefore one can expect  sizable 
deviations from the SM in  $K \to \pi \nu \bar{\nu}$  (both modes)  and $K \to \pi \ell^+ \ell^-$. 

The other leading operators contributing to $K \to \pi \nu \bar{\nu}$ involve the Higgs doublet $\phi$ and
reduce, after electroweak symmetry breaking, to effective flavor-changing $Z$-boson interactions, with both
left-handed (LH) and right-handed (RH) couplings to quarks.
These $Z$-penguin operators (both LH and RH) are the leading effect in many SM extensions, and affect a large
number of kaon observables ($K \to \pi \ell^+ \ell^-$, $\epsilon_K$, $\epsilon'/\epsilon$, and in the case of
one operator $K \to \pi \ell \nu$ through $\text{SU}(2)$ gauge invariance), so we discuss them in some detail.
The set of dimension-6 operators with a $\phi$ field include 
\begin{equation} \label{kaon:eq:1}
Q =  \big (\phi^\dagger \! \stackrel{\leftrightarrow}{D}_\mu \! \phi \big) \left ( \bar D_L \gamma^\mu S_L \right ) \,, \qquad 
\widetilde Q = \big (\phi^\dagger \! \stackrel{\leftrightarrow}{D}_\mu \! \phi \big ) \left ( \bar d_R \gamma^\mu s_R \right ) \,,
\end{equation}
where $D_L,S_L$ ($d_R,d_R$) are $\text{SU}(2)_L$ quark doublets (singlets) and
$\stackrel{\leftrightarrow}{D}_\mu=D_\mu-\stackrel{\leftarrow}{D}_\mu$ with $D_\mu$ denoting the electroweak
covariant derivative.
After electroweak symmetry breaking, one has
\begin{equation} \label{kaon:eq:2}
Q \, \to \, \bar d_L \gamma_\mu s_L Z^\mu +  \bar u_L \gamma_\mu c_L Z^\mu + \ldots \,, \qquad 
\widetilde Q \, \to \, \bar d_R \gamma_\mu s_R Z^\mu  + \ldots  \,,
\end{equation}
where the ellipses represent additional terms that are irrelevant for the further discussion.
We see that $Q$ induces the left-handed (LH) $Z$-penguin well-known from the minimal supersymmetric SM
(MSSM), Randall-Sundrum (RS) models, {\it etc.}, while $\widetilde Q$ leads to a right-handed~(RH)
$Z$-penguin, which is highly suppressed in the SM by small quark masses.
The results (\ref{kaon:eq:2}) hence imply that flavor-changing $Z$-boson interactions relevant for kaon physics
can be parameterized to leading order in $v/\Lambda$ by
\begin{equation} \label{kaon:eq:3}
{\cal L}_\text{eff} \propto \left(\lambda_tC_\text{SM} + C_\text{NP} \right )  
    \bar d_L \gamma_\mu s_L Z^\mu  + \widetilde C_\text{NP} \, d_R \gamma_\mu s_R Z^\mu,
\end{equation}
where $\lambda_q = V_{qs}^\ast V_{qd}$ with $V_{qp}$ denoting the elements of the quark mixing matrix and
$C_\text{SM} \approx 0.8$ encodes the SM contribution to the LH $Z$-penguin.

\begin{figure}
    \centering
    \includegraphics[width=0.6\textwidth]{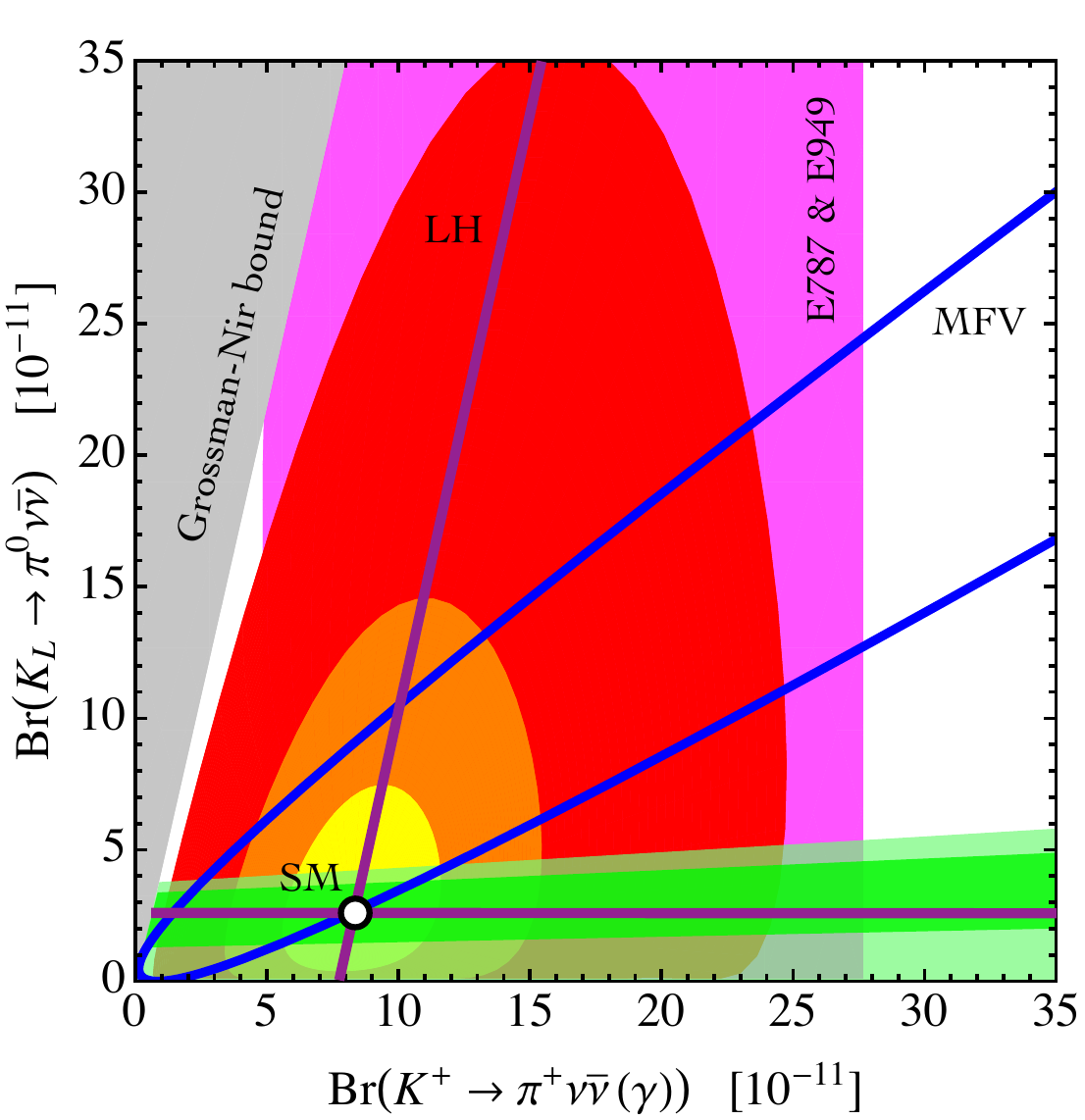} 
    \caption[Predictions for the $K\to \pi \nu \bar \nu$ branching ratios with $Z$-penguin 
        dominance]{Predictions for the $K \to \pi \nu \bar \nu$ branching ratios assuming dominance of the 
        $Z$-penguin operators, for different choices of the effective couplings 
        $C_\text{NP},\tilde{C}_\text{NP}$.
        The SM point is indicated by a white dot with black border.
        The yellow, orange, and red shaded contours correspond to 
        $|C_\text{NP},\tilde{C}_\text{NP}|\leq\left\{0.5,1,2\right\}|\lambda_tC_\text{SM}|$;
        the magenta band indicates the 68\% confidence level~(CL) constraint on 
        $\BR(K^+\to\pi^+\nu\bar{\nu}(\gamma))$ from experiment~\cite{Adler:2008zza};
        and the gray area is theoretically inaccessible.
        The blue parabola represents the subspace accessible to MFV models.
        The purple straight lines represent the subspace accessible in models that have only LH currents, 
        due to the constraint from $\epsilon_K$~\cite{Blanke:2009pq}.
        The green band represents the region accessible after taking into account the correlation of 
        $K_L\to\pi^0\nu\bar{\nu}$ with $\epsilon^\prime_K/\epsilon_K$: the (light) dark band corresponds to 
        predictions of $\epsilon^\prime_K/\epsilon_K$ within a factor of (5) 2 of the experimental value, 
        using central values for the hadronic matrix elements.
        See text for additional details.}
\label{kaon:fig:1}
\end{figure}

In terms of the effective NP couplings $C_\text{NP}$ and $\widetilde C_\text{NP}$, the branching ratios of
the $K \to \pi \nu \bar \nu$ decays take a simple form, namely
\begin{equation}\begin{split}
    \text{BR}(K_L \to \pi^0 \nu \bar\nu) \propto \left(\Im  X \right)^2, \\
    \text{BR}(K^+ \to \pi^+ \nu \bar \nu(\gamma)) \propto |X|^2 ,
\end{split}
\label{kaon:eq:4}
\end{equation}
with
\begin{equation} \label{kaon:eq:5}
    X = X_\text{SM} + \frac{1}{\lambda^5} \left( C_\text{NP} + \widetilde C_\text{NP} \right), 
\end{equation} 
where $X_\text{SM}\approx1.2e^{2.9i}$ represents the SM contribution and $\lambda\approx0.23$ denotes the
Cabibbo angle.
Treating the magnitude and phase of $C_\text{NP}$ $(\widetilde C_\text{NP})$ as free parameters one can then
determine the possible deviations in the $K \to \pi \nu \bar \nu$ branching ratios in a model-independent
fashion.
The outcome of such an exercise is shown in Fig.~\ref{kaon:fig:1}.
Here the yellow, orange, and red shaded contours correspond to
$|C_\text{NP}|\leq\left\{0.5,1,2\right\}|\lambda_tC_\text{SM}|$, and the magenta band indicates the 68\%
confidence level~(CL) limit on $\text{BR} (K^+ \to \pi^+ \nu \bar \nu (\gamma))$ from the
combination of the E787 and E949 results~\cite{Adler:2008zza}.
The gray area is inaccessible because, cf.\ Eqs.~(\ref{kaon:eq:4}), $|X|^2\ge(\Im X)^2$ for any $X$, a constraint
known as the Grossman-Nir bound~\cite{Grossman:1997sk}.
It is evident from the figure that $\text{O}(1000\%)$ enhancements of $\text{BR} (K_L \to \pi^0\nu \bar \nu)$
are in principle possible without violating the experimental constraint on $\text{BR} (K^+ \to \pi^+ \nu \bar
\nu)$, if NP were to generate large \CP-violating effects in the LH $Z$-penguin.
Since (\ref{kaon:eq:5}) is symmetric under the exchange of $C_\text{NP}$ and $\widetilde C_\text{NP}$, the same
conclusions hold in the case of the RH $Z$-penguin.

The situation changes dramatically if one restricts oneself to scenarios of minimal-flavor violation~(MFV),
where the effective couplings satisfy $C_\text{NP}\propto\lambda_tC_\text{SM}$ and $\widetilde C_\text{NP}
\approx 0$ by definition.
The subspace accessible to MFV models is indicated by the blue parabola in Fig.~\ref{kaon:fig:1}.
As one can see the pattern of deviations is very restricted in this class of models, which implies that
precision measurements of both $K \to \pi \nu \bar \nu$ modes provide a unique way to test and to possibly
refute the MFV hypothesis.
Still one has to bear in mind that explicit MFV realizations such as the MSSM predict effects that do not
exceed $\text{O} (10\%)$~\cite{Buras:2000qz}, which sets the benchmark for the precision that upcoming kaon
experiments should aim for.

\begin{figure}
    \centering
    \includegraphics[width=0.6\textwidth]{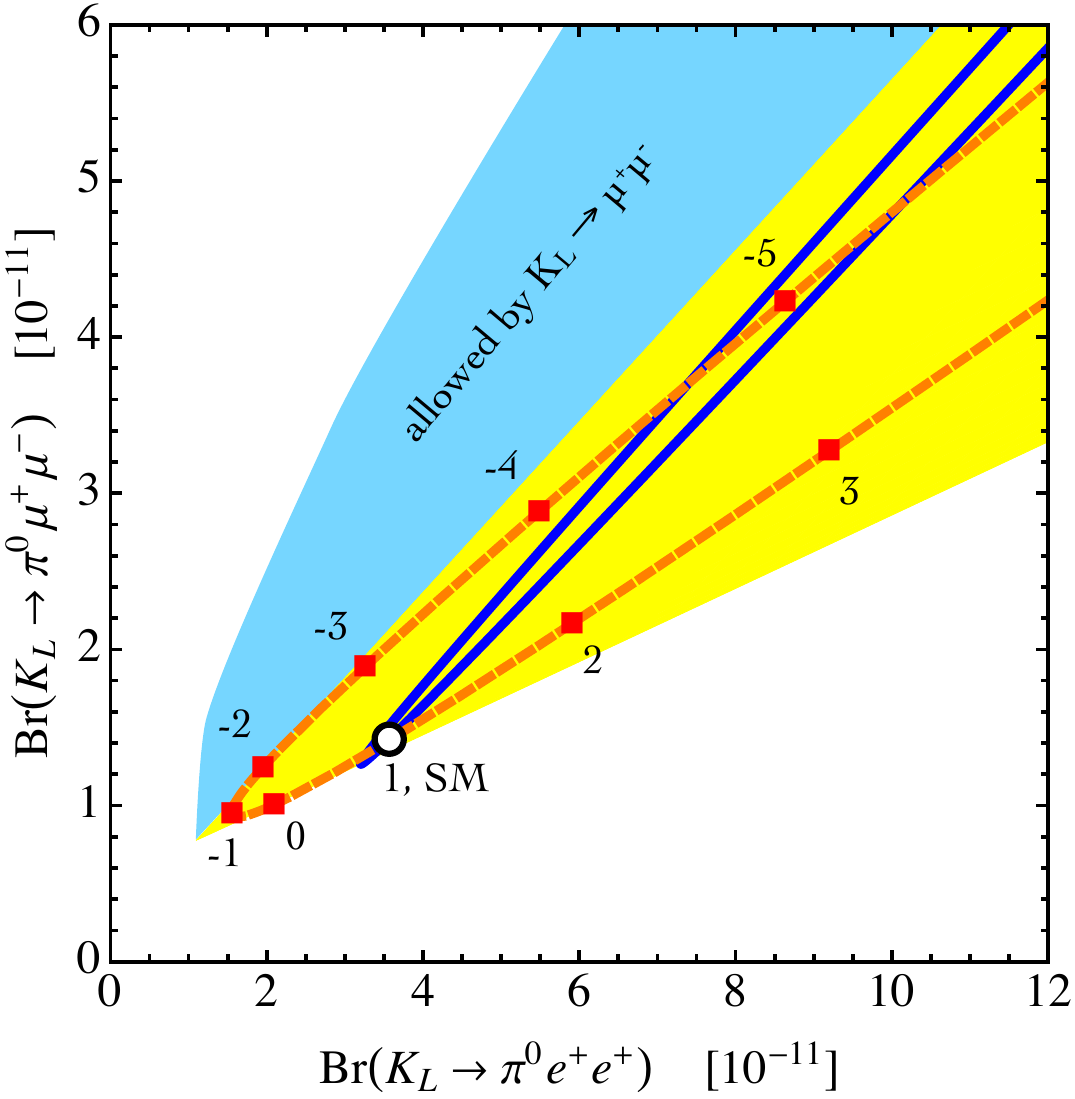}
    \caption[Predictions for the $K_L\to\pi^0\ell^+\ell^-$ branching ratios with new physics]{Predictions 
        for the $K_L\to\pi^0\ell^+\ell^-$ branching ratios assuming different types of NP contributions.
        The SM point is indicated by a white dot with black border.
        The blue parabola represents the region accessible by allowing only for $C_\text{NP}$ with arbitrary 
        modulus and phase.
        The subspace accessible when $C_{V,A} \neq 0$ is represented by the dashed orange parabola (common 
        rescaling of $C_{A,V}$) and the yellow shaded region (arbitrary values of $C_{A,V}$).
        The subspace accessible when $C_{S,P} \neq 0$ (compatibly with $K_L \to \mu^+ \mu^-$) is represented 
        by the light blue shaded region.}
\label{kaon:fig:1bis}
\end{figure}

The number of operators that can leave an imprint in the $K_L \to \pi^0 \ell^+ \ell^-$ ($\ell = e,\mu)$
decays is larger than the one in the case of $K \to \pi \nu \bar \nu$.
Besides (axial-)vector operators resulting from $Z$- and photon-penguin diagrams also (pseudo-)scalar
operators associated to Higgs exchange can play a role~\cite{Mescia:2006jd}.
In a model-independent framework, one hence should consider
\begin{equation} \label{kaon:eq:6}
Q_A = (\bar d \gamma^\mu s )(\bar \ell \gamma_\mu \gamma_5 \ell), \quad
Q_V = (\bar d \gamma^\mu s ) (\bar \ell \gamma_\mu \ell), \quad 
Q_P = (\bar d s ) (\bar \ell \gamma_5 \ell), \quad
Q_S = (\bar d s ) (\bar \ell \ell) .
\end{equation} 
In Fig.~\ref{kaon:fig:1bis} we depict the accessible parameter space corresponding to various classes of NP.
The blue parabola illustrates again the predictions obtained by allowing only for a contribution $C_\text{NP}$
with arbitrary modulus and phase.
We see that in models with dominance of the LH $Z$-penguin the deviations in $K_L \to \pi^0 \ell^+ \ell^-$
are strongly correlated.
A large photon-penguin can induce significant corrections in $C_V$, which breaks this correlation and opens
up the parameter space as illustrated by the dashed orange parabola and the yellow shaded region.
The former predictions are obtained by employing a common rescaling of $C_{A,V}$, while in the latter case
the coefficients $C_{A,V}$ are allowed to take arbitrary values.
If besides $Q_{A,V}$ also $Q_{P,S}$ can receive sizable NP corrections a further relative enhancement of
$\text{BR} (K_L \to \pi^0 \mu^+ \mu^-)$ compared to $\text{BR} (K_L \to \pi^0 e^+ e^-)$ is possible.
This feature is exemplified by the light blue shaded region that corresponds to the parameter space that is
compatible with the constraints on $C_{P,S}$ arising from $K_L \to \mu^+ \mu^-$.

In many explicit SM extensions such as RS scenarios~\cite{Blanke:2008yr,Bauer:2009cf}, little Higgs
models~\cite{Blanke:2007wr}, {\it etc.} the pattern of deviations in the $K_L \to \pi^0 \ell^+ \ell^-$
channels is however less spectacular than suggested by Figure~\ref{kaon:fig:1bis}.
In fact, this is a simple consequence of the relations
\begin{equation} \label{kaon:eq:7}
C_A \propto  -\frac{1}{s_w^2} \left ( C_\text{NP} - \widetilde C_\text{NP} \right ) , \quad
C_V \propto  \left (\frac{1}{s_w^2} - 4 \right ) \left ( C_\text{NP} + \widetilde C_\text{NP} \right ) , \quad 
C_{P,S} \propto m_s m_\ell \,,
\end{equation}
where the factors $-1/s_w^2 \approx -4.4$ and $1/s_w^2 - 4 \approx 0.4$ arise from the axial-vector and
vector coupling of the $Z$-boson to charged leptons, while the mass factors $m_{s,\ell}$ reflect the helicity
suppression of pseudoscalar and scalar interactions.
To overcome this suppression requires the presence of an extended gauge and/or Higgs sector.
Explicit NP models that produce such pronounced effects in $K_L \to \pi^0 \ell^+ \ell^-$ as shown in
Fig.~\ref{kaon:fig:1bis} have not been built.

So far we have only considered NP in rare kaon decays.
It is, however, also important to consider how effects in $K \to \pi \nu \bar \nu$ and
$K_L \to \pi^0 \ell^+ \ell^-$ are linked to deviations in well-measured kaon observables like $\epsilon_K$
and $\epsilon^\prime_K/\epsilon_K$.
In fact, \CP\ violation in kaon mixing provides the most stringent constraint on
possible new flavor structures in many non-MFV scenarios.
This is a consequence of the strong chiral and renormalization group enhancement of the left-right operator
$Q_{LR} = (\bar d_R s_L) (\bar d_L s_R)$ relative to the SM contribution $Q_{LL} = (\bar d_L \gamma^\mu s_L)
(\bar d_L \gamma_\mu s_L)$.
For NP scales $\Lambda = \text{O}(1~{\rm TeV})$, one has approximately
\begin{equation} \label{kaon:eq:8}
    \epsilon_K \propto \Im  \left( 97 C_{LR}+ C_{LL} \right ) ,
\end{equation}
with $C_{LR,LL}$ denoting the effective coupling of $Q_{LR,LL}$.
Concerning $\epsilon_K$, SM extensions fall, hence, into two classes: those with currents of only one 
chirality (LH or RH) and those with both (LH and RH).
In the former case it can be shown~\cite{Blanke:2009pq} that under mild assumptions there are stringent
correlation between $\Delta S = 2$ and $\Delta S =1$ observables, while no such link exists in the latter
case.
For $K \to \pi \nu \bar \nu$ this model-independent correlation leads to two branches of solutions, one
parallel to the $\text{BR} (K^+ \to \pi^+ \nu \bar \nu)$ axis and one parallel to the Grossman-Nir bound.
These two branches are indicated in Fig.~\ref{kaon:fig:1} by purple lines.
Certain little Higgs~\cite{Blanke:2007wr} and $Z^\prime$ models~\cite{Promberger:2007py}, in fact, show this
distinctive pattern, while in the generic MSSM~\cite{Buras:2004qb} and the RS framework
\cite{Blanke:2008yr,Bauer:2009cf} the correlation is completely washed out.
The $\epsilon_K$ constraint thus does not restrict the $K \to \pi \nu \bar \nu$ decay rates in general.

A second kaon observable that is known
\cite{Buras:2000qz,Bauer:2009cf,Blanke:2007wr,Buras:1998ed,Buras:1999da} to bound NP in the $K \to \pi \nu
\bar \nu$ modes is the ratio of the direct and indirect \CP\ violation in $K_L \to \pi \pi$ as measured by
$\epsilon^\prime_K/\epsilon_K$.
The reason for this correlation is simple to understand from the approximation
\begin{equation}  \label{kaon:eq:9}
    \frac{\epsilon_K^\prime}{\epsilon_K} \propto 
    - \Im \left [ \lambda_t \left ( -1.4 + 13.8 R_6 - 6.6 R_8 \right ) + 
    \left ( 1.5 + 0.1 R_6 - 13.3 R_8 \right ) \left (C_\text{NP} - \widetilde C_\text{NP} \right) \right],
\end{equation}
where the first (second) term in brackets encodes the SM (NP) contribution.
Typical values of the hadronic matrix elements $R_{6,8}$ quoted in the literature are $R_6 \in [0.8, 2]$ and
$R_8 \in [0.8,1.2]$.
The current status and prospects for lattice-QCD calculations of $K\to\pi\pi$ matrix elements are discussed 
in Chapter~\ref{chapt:lqcd}: a complete three-flavor lattice-QCD calculation of $\epsilon_K'/\epsilon_K$ is expected in a couple
of years, with a total error as small as~$\sim$20\%.

The hierarchy of the numerical coefficients multiplying $R_{6,8}$ in Eq.~(\ref{kaon:eq:9}) is explained by
recalling that while the QCD- ($R_6$) and $Z$-penguins ($R_8$) are both strongly enhanced by
renormalization-group effects, the former correction results mainly from the mixing of the current-current
operators $Q_{1,2}$ into the QCD-penguin, which is essentially free from NP.
In contrast, mixing with $Q_{1,2}$ plays only a minor role in the case of the $Z$-penguins, so that any NP
contribution to the initial conditions in this sector directly feeds through into
$\epsilon^\prime_K/\epsilon_K$.
This implies that a strong cancellation of QCD- and $Z$-penguins is present only within the SM, but not
beyond.
The coefficients $C_\text{NP}$ and $\widetilde C_\text{NP}$ hence have in general a considerable impact on
both $K \to \pi \nu \bar \nu$ and $\epsilon^\prime_K/\epsilon_K$ and this leads to a stringent
model-independent correlation between the observables.
This feature is illustrated by the (light) green band in Fig.~\ref{kaon:fig:1}, which corresponds to
predictions of $\epsilon^\prime_K/\epsilon_K$ within a factor of (5) 2 of the experimental value.
One observes that even under mild theoretical assumptions, the constraint on $\epsilon^\prime_K/\epsilon_K$
disfavors order of magnitude enhancements of the \CP-violating channel $K_L \to \pi^0 \nu \bar \nu$, while it
has little impact on the \CP-conserving $K^+ \to \pi^+ \nu \bar \nu$ mode.
Let us add that $\epsilon^\prime_K/\epsilon_K$ can also receive large contributions from
chromomagnetic-dipole operators~\cite{Bauer:2009cf,Buras:1999da}.
Since these effects are in general not linked to those associated to the $Z$-penguins, the aforementioned
correlation between $\epsilon^\prime_K/\epsilon_K$ and $K_L \to \pi^0 \nu \bar \nu$ is expected to be robust,
in that it can be evaded only by cancellations among different contributions to
$\epsilon^\prime_K/\epsilon_K$.

%\subsection{Beyond the SM: SUSY and warped extra dimensions}

\subsection{The Minimal Supersymmetric Standard Model}
\label{kaon:sect:BSM2}

Supersymmetric extensions of the SM contain several new sources of flavor violation.
In particular, the bilinear and trilinear soft SUSY breaking terms, which lead to the masses of the squarks,
are not necessarily aligned in flavor space with the quark masses.
The result are flavor and \CP\ violating gluino-squark-quark interactions that can induce large NP effects in
various low energy flavor observables.
Indeed, the good agreement of the experimental data on FCNC processes with the SM predictions leads to strong
constraints on the new sources of flavor violation in the MSSM.
Interestingly, rare kaon decays can give important complementary information on the flavor structure of the
MSSM.
In the following we focus on the clean $K \to \pi \nu\bar\nu$ decays.

\subsubsection{Minimal Flavor Violation}

Assuming completely generic flavor mixing among the squarks leads to excessively large contributions to
several well measured FCNC processes, unless the masses of the SUSY particles are well beyond the reach of
the LHC.
One way to avoid the strong experimental flavor constraints, is to assume that the SM Yukawa couplings are
the only sources of flavor violation, the so-called Minimal Flavor Violation (MFV) Ansatz.
In such a framework, FCNCs are suppressed by the same small CKM matrix elements as in the SM, and
experimental bounds can be naturally avoided.

One finds that in the MSSM with MFV, the corrections to the branching ratios of the $K_L \to \pi^0
\nu\bar\nu$ and $K^+ \to \pi^+ \nu\bar\nu$ decays are generically tiny and can only reach
O(10\%)~\cite{Isidori:2006qy}.
Moreover, this is only possible if stops and charginos are extremely light with masses below 200~GeV.
Given the expected experimental and theoretical uncertainties, an observation of one of the $K \to \pi
\nu\bar\nu$ decays, with a branching ratio that differs significantly from the SM prediction, would therefore
not only rule out the SM, but would also be strong evidence for sources of flavor violation beyond MFV.
Note that this statement holds in the context of the MSSM.
In general, the MFV framework does allow for sizable NP contributions to both branching ratios that are
strongly correlated.

\subsubsection{Beyond Minimal Flavor Violation}

Within the MSSM, sizable non-Standard effects in the $K \to \pi \nu\bar\nu$ decays can only be generated if
the soft SUSY breaking terms have a non-MFV structure.
The leading amplitudes that can give rise to large effects are generated by: (i) charged-Higgs--top-quark
loops~\cite{Isidori:2006jh} and (ii) chargino--up-squark loops~\cite{Nir:1997tf,Colangelo:1998pm}.

In the case (i), deviations from the SM can be generated in the large $\tan\beta$ regime by non-MFV terms in
the soft masses of the right-handed down squarks.
The recently improved bounds on the branching ratios of the rare decays $B_s \to \mu^+\mu^-$ and $B_d \to
\mu^+\mu^-$, however, put strong constraints on such flavor structures.

In the case (ii), large effects can be induced if the trilinear couplings of the up squarks contain new
sources of flavor violation beyond MFV.
This provides the exciting opportunity to probe flavor violation in the {\it up sector} with rare kaon decays.
Interestingly enough, the required up squark flavor structures are only mildly constrained by current flavor
data, with the strongest constraints coming from $\epsilon_K$ and
$\epsilon^\prime/\epsilon$~\cite{Buras:1999da}.
However, as these observables are also highly sensitive to other, independent, flavor violating sources, huge
effects in the $K \to \pi \nu\bar\nu$ decays cannot be excluded.
Large flavor violating entries in the up squark trilinear couplings are well motivated.
They are for example required in certain models of radiative flavor violation~\cite{Crivellin:2011sj} and can
also provide a NP explanation of the surprisingly large observed difference in the direct \CP\ asymmetries in
$D \to K^+K^-$ and $D \to \pi^+\pi^-$ decays~\cite{Giudice:2012qq}.
Generically, one can expect uncorrelated O(1) corrections to both $K \to \pi \nu\bar\nu$ branching ratios in
these models, but even enhancements by an order of magnitude cannot be excluded~\cite{Buras:2004qb}.
Note that extreme enhancements, however, require considerably fine tuning to avoid the constraints from
$\epsilon_K$ and $\epsilon^\prime/\epsilon$.
Even neglecting fine tuned scenarios, extremely valuable information on the MSSM flavor sector can be
obtained from $K \to \pi \nu\bar\nu$, thanks to the high precision of the envisioned future experiments.

\subsubsection{Very Light Neutralinos}

The MSSM allows the interesting possibility that the mass of the lightest neutralino $\chi$ is so small that
the $K \to \pi \chi \chi$ decays become possible~\cite{Dreiner:2009er}.
As neither neutrinos nor neutralinos are detected in experiment, the signature is the same: $K \to \pi +
E\!\!\!\!/$.
The decay with neutralinos in the final state can have appreciable rates only if beyond-MFV flavor structures
in the down squark sector are present which allow the $K \to \pi \chi \chi$ decay already at the tree level.
Interestingly, a small finite mass of the neutralinos of O(100~MeV) would lead to a considerable distortion of
the pion momentum spectrum, allowing to disentangle $K \to \pi \nu\bar\nu$ and a possible $K \to \pi \chi
\chi$ contribution.

%%%%%%%%%%%%%%%%%%%%%%%%%%%%%%%%%%%%%%%%%%%%%%%%%%%%%%%%%%%%%%%%%%
\subsection{The Randall-Sundrum Model}
\label{kaon:sect:BSM3}

Randall-Sundrum models, in which all SM fields are allowed to propagate in the bulk, represent a very
exciting alternative to more traditional extensions of the SM, like the MSSM.
The model contains important new sources of flavor violation beyond the MFV framework.
The explanation of the hierarchies of the SM fermion masses and mixings leads to non universal shape
functions of the SM fermions in the bulk and therefore to nonuniversalities in the interactions of the
Kaluza-Klein (KK) and SM gauge bosons with SM fermions.
This implies FCNCs at the tree level mediated by the several gauge bosons and by the Higgs.
However, the tree level flavor violating couplings are proportional to the mass splitting between the two
fermions, hence leading to a suppression of the flavor transitions involving the first two generation
fermions, through the so called RS-GIM mechanism~\cite{Agashe:2004cp}.
Additionally, it has been shown that enlarging the bulk gauge symmetry to
$\text{SU}(3)_c\times\text{SU}(2)_L\times\text{SU}(2)_R\times\text{U}(1)$ guaranties the protection of the Z
boson flavor changing (and flavor conserving) couplings with left handed quarks and a not too large NP
contribution to the T parameter even for low KK scales~\cite{Agashe:2003zs,Csaki:2003zu,Agashe:2006at}.
The latter model is the so called RS model with custodial protection.
In spite of these protection mechanisms, the flavor structure of the RS model is very rich and it generically
leads to too large NP contributions to $\epsilon_K$~\cite{Csaki:2008zd}.
In the following we will focus on the discussion of several kaon rare decays in the subspace of parameter
space that predicts $\epsilon_K$ compatible with the experimental constraints.

\subsubsection{The $K\to\pi\nu\bar\nu$ Decays}

The most important NP contribution to the $K\to\pi\nu\bar\nu$ rare decays arises from tree level electroweak
(EW) penguin diagrams.
In general enhancements of the neutral $K_L\to\pi^0\nu\bar\nu$ decay by almost an order of magnitude are
possible even for a multi-TeV KK scale~\cite{Bauer:2009cf}.
The NP contributions to $K^+\to\pi^+\nu\bar\nu$ are in general uncorrelated with those entering the neutral
decay and can also be sizable: the model can predict enhancements of the branching ratio of
$K^+\to\pi^+\nu\bar\nu$ by a factor 2, sufficient to reach the central value of the present measurement of
the charged kaon decay~\cite{Blanke:2008yr}.

\begin{figure}
    \centering
    \includegraphics[width=0.5\textwidth]{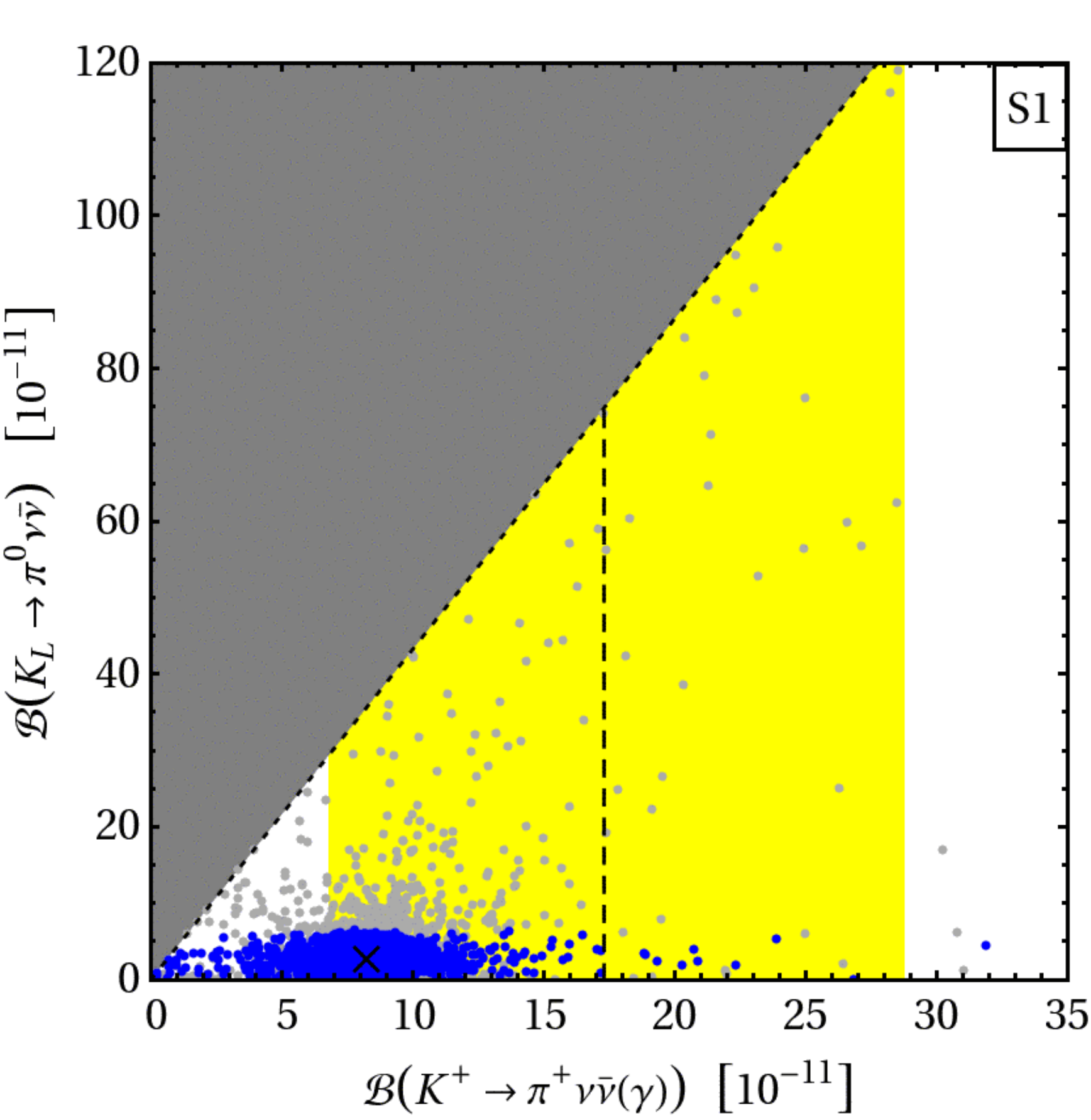} 
    \caption[$\epsilon_K'/\epsilon_K$ and $\BR(K\to\pi\nu\bar{\nu})$ in Randall-Sundrum models.]{Impact of 
        $\epsilon_K'/\epsilon_K$ on $K\to\pi\nu\bar{\nu}$ branching ratios in Randall-Sundrum models.
        The blue (light gray) scatter points are consistent (inconsistent) with the measured value of
        $\epsilon_K'/\epsilon_K$.
        Plot taken from Ref.~\cite{Bauer:2009cf}.}
    \label{kaon:fig:RS1}
\end{figure}

However, EW penguins generically give also the dominant correction to the direct \CP\ violation in
$K\to\pi\pi$, as discussed in Section~\ref{kaon:sect:BSM1}.
This results in a strong anti-correlation between $K_L\to\pi^0\nu\bar\nu$ and the \CP\ violating observable
$\epsilon^\prime$.
Imposing the constraint from $\epsilon^\prime/\epsilon$ disfavors large deviations of the branching ratio of
$K_L\to\pi^0\nu\bar\nu$ from its SM prediction.
Sizable NP effects in $K^+\to\pi^+\nu\bar\nu$ are instead unconstrained by $\epsilon^\prime/\epsilon$, since
in general there is no correlation between the charged \CP\ conserving kaon decay and direct \CP\ violation in
$K\to\pi\pi$.
These points are illustrated in Fig.~\ref{kaon:fig:RS1}.

\subsubsection{The $K_L\to\pi^0\ell^+\ell^-$ and $K_L\to\mu^+\mu^-$ Decays}

The \CP\ violating $K_L\to\pi^0\ell^+\ell^-$ decays are not as clean as the $K\to\pi\nu\bar\nu$ modes.
However they offer the opportunity to constrain additional $\Delta F=1$ effective operators that are not
entering in the neutrino decay modes.
In the RS model the dominant NP effect arises from the tree level exchange of a Z-boson with axial-vector
couplings to the SM fermions.
This results in a direct correlation between the branching ratios BR$(K_L\to\pi^0 e^+ e^-)$ and
BR$(K_L\to\pi^0 \mu^+\mu^-)$ and also between $K_L\to\pi^0\ell^+\ell^-$ and the \CP\ violating process
$K_L\to\pi^0\nu\bar\nu$.
This implies that: (i) too large NP contributions to $K_L\to\pi^0\ell^+\ell^-$ are disfavored by the
constraint from $\epsilon^\prime/\epsilon$ and (ii) a precise measurement of both decays, $K_L\to\pi^0
\mu^+\mu^-$ and $K_L\to\pi^0\nu\bar\nu$, would test the operator structure of the model.

The NP contributions to the leptonic \CP\ conserving $K_L\to\mu^+\mu^-$ decay are encoded by the same effective
Hamiltonian describing the $K_L\to\pi^0\ell^+\ell^-$ decays.
However, contrary to the latter decays, the short distance (SD) contribution to $K_L\to\mu^+\mu^-$ is by far
dominated by the absorptive contribution with two internal photon exchanges.
Consequently, the SD contribution constitutes only a small fraction of the branching ratio.
Nonetheless this decay can lead to interesting constraints in the RS model.
The correlation between $K^+\to\pi^+\nu\bar\nu$ and $K_L\to\mu^+\mu^-$ offers in fact a clear test of the
handedness of the NP flavor violating interactions.
The former decay is sensitive to the vector component of the flavor violating $Z s\bar d$ coupling, while the
latter measures its axial-vector component.
Therefore, since the SM flavor changing Z penguin is purely left-handed, in the original RS model, in which
the NP contributions to these decay are dominated by left-handed Z couplings, the two decay modes show a
direct correlation~\cite{Bauer:2009cf}.
On the contrary, in the RS model with custodial protection where the main NP effect in $K_L\to\mu^+\mu^-$
arises from right-handed Z couplings, the correlation has been found to be an inverse
one~\cite{Blanke:2008yr}.

\subsection{Beyond Rare Decays}
\label{kaon:sect:beyondrare}

While the main focus of this chapter was on the  FCNC probes, 
it is worth mentioning that kaons provide as well  unique probes of 
the charged-current (CC) sector of SM extensions. 
Two prominent examples involve 
precise measurements of the  ratio  $R_K = \Gamma (K \to e \nu) / \Gamma (K \to \mu \nu)$, 
which  test  lepton universality, 
and measurements of the transverse muon polarization $P_\mu^T$  in the 
semileptonic  decay $K^+  \to \pi^0 \mu^+ \nu_\mu$, 
which is sensitive to BSM  sources of \CP\ violation in scalar CC  operators. 
In both cases there is a clean discovery window provided 
by the precise SM  theoretical prediction of $R_K$~\cite{Cirigliano:2007xi} 
and by the fact that in the SM  $P_\mu^T$ is generated only by small and theoretically known final state interactions~\cite{Zhitnitsky:1980he,Efrosinin:2000yv}.

 %%%%%%%%% %%%%%%%%% %%%%%%%%% %%%%%%%%% %%%%%%%%% %%%%%%%%% %%%%%%%%% %%%%%%%%% %%%%%%%%%

\section{Experiments} 
\label{kaon:sect:exp}

\subsection{Experimental Landscape in This Decade}
\label{kaon:sect:explandscape}

\noindent {\bf NA62.} The NA62 experiment~\cite{NA62} at CERN is an in-flight measurement of \Kplus.
The experiment will have a commissioning run with a partial detector later in 2012.
Full commissioning followed by a physics run will begin in 2014.
The NA62 goal is a measurement of the \Kplus\ branching ratio with 10\% precision.
The NA62 experiment anticipates a very robust and diverse kaon physics program beyond the primary measurement.

\medskip 

\noindent {\bf KOTO.} The KOTO experiment~\cite{KOTO} at JPARC is an in-flight measurement of \Kzero.
Significant experience and a better understanding of the backgrounds to this rare decay mode were obtained in
E391a, the predecessor of KOTO.
The anticipated sensitivity of the experiment is a few signal events (assuming the SM branching ratio) in
three years of running with 300~kW of beam.
A commissioning run will occur later in 2012, but the longer term performance of the experiment will depend
upon the beam power evolution of the JPARC accelerator.

\medskip

\noindent {\bf TREK.} The TREK Experiment (E06) at JPARC~\cite{TREK} will search for T violation in charged
kaon decays by measuring the polarization asymmetry in $K^+ \to \pi^0 \mu^+ \nu_m$ decays.
TREK needs at least 100 kW (proposal assumes 270 kW) for this measurement.
While the accelerator is running at lower power, collaborators have proposed P36, which will use much of the
TREK apparatus to perform a search for lepton flavor universality violation through the measurement of
$\Gamma (K \to e \nu)/\Gamma (K \to \mu \nu)$ at the 0.2\% level.
The P36 experiment requires only 30 kW of beam power and will be ready to run in 2015.
The uncertainty of the JPARC beam power profile and potential conflicts for beamline real estate make the
long term future of the TREK experiment unclear.

\medskip

\noindent {\bf ORKA.} The ORKA experiment~\cite{Comfort:2011zz}, is proposed to measure \Kplus\ with 1000
event sensitivity at the Main Injector later this decade.
ORKA is a stopped kaon experiment that builds on the experience of the E787/949 experiments at Brookhaven.
Like NA62, ORKA offers a wide variety of measurements beyond the \Kplus\ mode.

\vspace{.3cm}

Let us look at the experimental landscape at the end of this decade under optimistic assumptions.
The NA62 experiment will have measured the \Kplus\ branching ratio to 10\% precision.
The KOTO will have measured the \Kzero\ mode with standard model sensitivity.
The P36 experiment will have improved precision on lepton flavor universality.
The ORKA and TREK experiments would be in progress.
Even under the optimistic scenario spelled out above, we would be far from exploring the full physics reach
of kaons.
Therefore, there are significant opportunities for important measurements in the kaon sector at \PX.

\subsection{\PX\ Kaon Program}
\label{kaon:sect:projectX}

The flagship measurement of the \PX\ kaon era would be an experiment to measure the \Kzero\ branching ratio
with 5\% precision.
This effort will need to build upon the KOTO experience, benefit from significant detector R\&D and take
advantage of the beam power and flexibility provided by Stage~2 of \PX.
Based upon the \Kzero\ experience at JPARC, it seems likely that an effort to achieve this ultimate
sensitivity will take two generations.
Depending upon the outcome of the TREK experiment at JPARC, a T violation experiment would be an excellent
candidate for \PX, as would a multi-purpose experiment dedicated to rare modes that involve both charged and
neutral particles in the final state.
This experiment might be able to pursue $K_L\to \pi^0 \ell^+ \ell^-$ as well as many other radiative and
leptonic modes.

\subsection{A \Kzero\ Experiment at \PX}

\begin{figure}
    \centering
    \includegraphics[width=0.5\textwidth]{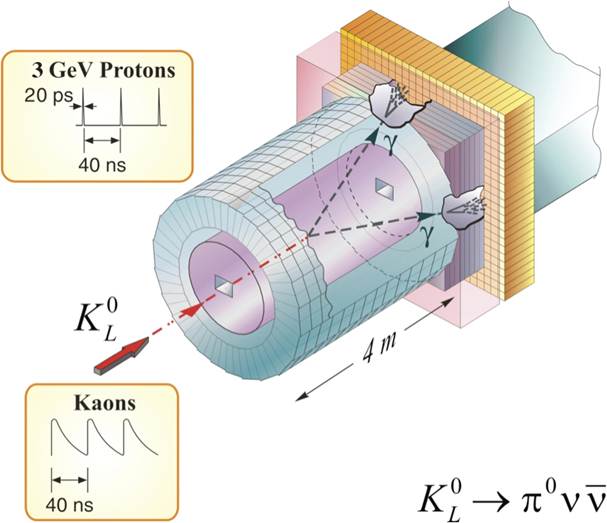} 
    \vspace{0.2cm}
    \caption[Illustration of the KOPIO concept for \PX]{Illustration of the KOPIO concept for \PX.
        Precision measurement of the photon arrival time through time-of-flight techniques is critical.
        Good measurement of the photon energies and space angles in a high rate environment is also critical 
        to controlling backgrounds.}
    \label{kaon:fig:KOPIO_PX}
\end{figure}
Several years ago, the KOPIO initiative~\cite{KOPIO} proposed to measure \Kzero\ with a SM sensitivity of 100
events at the BNL AGS as part of RSVP (Rare Symmetry Violating Processes) project.
The experimental technique and sensitivity were well-developed and extensively reviewed.
KOPIO was designed to use a neutral beam at $42^\circ$ targeting angle produced by 24 GeV protons from the
BNL AGS.
The produced neutral kaons would have an average momentum of 800 MeV$/c$ with a range from 300 to 1200
MeV$/c$.
A low momentum beam was critical for the Time-Of-Flight (TOF) strategy of the experiment.

The TOF technique is well-matched to the kaon momentum that would be produced
by a proton beam of 3 GeV kinetic energy at \PX. Performance of the TOF strategy 
was limited by the design bunch width of 200 ps at the AGS. The \PX\ beam pulse timing,
including target time slewing, is expected to be less than 50 ps and would substantially
improve the momentum resolution and background rejection capability of the \Kzero\ 
experiment driven with \PX\ beam.  
The KOPIO concept for \PX\ is illustrated in Fig.~\ref{kaon:fig:KOPIO_PX}.  

\begin{figure}
    \centering
    \includegraphics[width=0.5\textwidth]{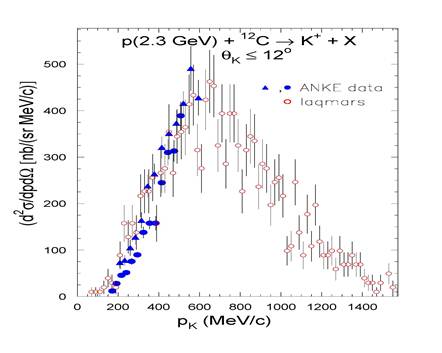} 
    \caption[$K^+$ momentum spectrum produced from 2.3~GeV protons]{LAQGSM/MARS (MARS15) 
        simulation~\cite{Gudima:2009zz} of the $K^+$ momentum spectrum 
        produced from 2.3~GeV protons (kinetic) on a thin carbon target (open circles).
        Absolutely normalized measurements (closed circles and triangles) from the ANKE 
        experiment~\cite{Buescher:2004vn} are overlaid indicating excellent validation of the simulation in 
        this production region.} 
    \label{kaon:fig:MARS_COSY}
\end{figure}

\begin{figure}
\centering
\includegraphics[width=0.5\textwidth]{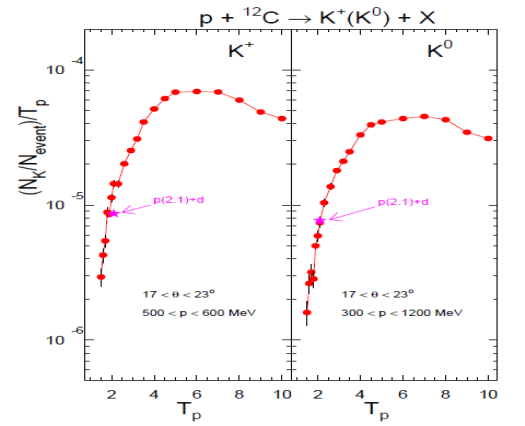} 
\vspace{0.2cm}
    \caption[$K^+$ and $K_L$ production yield at constant beam power]{LAQGSM/MARS (MARS15) 
        simulation~\cite{Gudima:2009zz} of the $K^+$ and $K_L$ production yield at constant beam power 
        (yield/$T_p$) for experimentally optimal angular and energy regions as a function of beam kinetic 
        energy $T_p$ (GeV).}
\label{kaon:fig:MARS_YIELD}
\end{figure}

The Fermilab Accelerator Physics Center has recently developed a comprehensive simulation module in the
LAQGSM/MARS (MARS15) framework~\cite{Gudima:2009zz} for particle production in the challenging $T_p$ region
of 1--4~GeV.
Kaon production in this module is treated as a sum of well measured exclusive channels with little tuning.
The simulations have been benchmarked against the high quality data sets from the COSY/ANKE
experiment~\cite{Buescher:2004vn}.
One such benchmark, shown in Fig.~\ref{kaon:fig:MARS_COSY}, is an absolute prediction of forward $K^+$
production yield on carbon and is in excellent agreement with COSY/ANKE data.
The estimated (LAQGSM/MARS15) kaon yield at constant beam power (yield/$T_p$) is shown in
Fig.~\ref{kaon:fig:MARS_YIELD}.
The yield on carbon saturates at about 5 GeV, and the $T_p=3.0$~GeV yield is about a factor of about two
times less than the peak yield in the experimentally optimal angular region of 17--23 degrees which mitigates
the high forward flux of pions and neutrons.
The 3.0 GeV operational point is a trade-off of yield with accelerator cost.
The enormous beam power of \PX\ more than compensates for operation at an unsaturated yield point.

The comparative $K_L$ production yields from 
thick targets fully simulated with LAQGSM/MARS15 are shown in Table~\ref{kaon:tab:klyield}.
\begin{table}
    \centering
    \caption[Comparison of the $K_L$ production yield]{Comparison of the $K_L$ production yield.
        The BNL AGS kaon and neutron yields are taken from RSVP reviews in 2004 and 2005.
        The \PX\ yields are for a thick target, fully simulated with LAQGSM/MARS15 into the KOPIO beam 
        solid angle and momentum acceptance.}
    \label{kaon:tab:klyield}
    \begin{tabular}{lccccc}
    \hline \hline
       & Beam energy & Target ($\lambda_I$) & $p(K)$ (MeV$/c$) & $K_L/s$ into 500 $\mu$sr & $K_L:n$ ($E_n>10$ MeV)\\
    \hline
    BNL AGS & 24 GeV & 1.1 Pt & 300-1200 & $60\times10^6$ & $\sim\!1:1000$ \\
    \PX\ & 3 GeV & 1.0 C & 300-1200 & $450\times10^6$ & $\sim\!1:2700$ \\
    \hline\hline
    \end{tabular}
\end{table}

The AGS $K_L$ yield per proton is 20~times the \PX\ yield; however, \PX\ compensates with a 0.5 mA proton
flux that is 150 times the RSVP goal of $10^{14}$ protons every 5~seconds.
Hence the neutral kaon flux would be eight times the AGS flux goal into the same beam acceptance.
The nominal five-year \PX\ run is 2.5~times the duration of the KOPIO AGS initiative and hence the reach of a
\PX\ \Kzero\ experiment could be 20~times the reach of the RSVP goals.

A TOF-based \Kzero\ experiment driven by \PX\ would need to be re-optimized
for the \PX\ $K_L$ momentum spectrum, TOF resolution and corresponding
background rejection. It is likely that this optimization would be based on a smaller
neutral beam solid angle which would simply the detector design, increase the
acceptance and relax the requirement to tag photons in the fierce rate environment 
of the neutral beam. Optimizing the performance will probably require a proton pulse train
frequency of 20-50 MHz and an individual proton pulse timing of $\sim\!20$ ps. 
Based on the E391a and KOTO experience, 
a careful design of the target and neutral 
beam channel is required to minimize the neutron halo. 
The optics from a long ($\sim\!39$ cm) carbon target (Table~\ref{kaon:tab:klyield}) 
may be inconsistent with the neutron halo requirements. A shorter and denser target
would have to be engineered to handle the beam power while maintaining the kaon flux.
The  high $K_L$ beam flux, the potential of break-through TOF performance and improvements
in calorimeter detector technology support the plausibility of a Day-1 \Kzero\ experiment
with $\sim\!1000$ SM event sensitivity.

\subsection{$K^+$ Experiments at \PX}

In the case where a significant non-SM result were observed by ORKA~\cite{Comfort:2011zz}, 
the \Kplus\ 
decay mode could be studied with higher statistics with a $K^+$ beam driven by
\PX. The high-purity, low-momentum $K^+$ beam designed for 
ORKA could also 
serve experiments to precisely measuring the polarization asymmetry 
in $K^+ \to \pi^0 \mu^+ \nu_m$ decays and to continue the search for 
lepton flavor universality violation  through  the
measurement of $\Gamma (K \to e \nu)/\Gamma (K \to \mu \nu)$ at high precision.

\section{Summary} 
\label{kaon:sect:summary}

Rare kaon decays are extremely sensitive probes of the flavor and \CP-violating sector of any SM extension.
The $K \to \pi \nu \bar{\nu}$ golden modes have great discovery potential: (i) sizable ($O(1)$) deviations
from the SM are possible; (ii) even small deviations can be detected due to the precise theoretical
predictions.
Next generation searches should aim for a sensitivity level of $10^3$ SM events (few \%) in both $K^+$ and
$K_L$ modes, so as to maximize discovery potential.
We foresee the search for $K_L \to \pi^0 \nu \bar{\nu}$ as the flagship measurement of the kaon program at
\PX, with the potential to uncover novel BSM sources of \CP\ violation.
But we also stress the importance of pursuing the broadest possible set of measurements, so as to enhance the
model discriminating power of \PX.

The \PX\ kaon program will benefit greatly from an ongoing R\&D effort to produce hermetic, highly efficient
low-energy calorimetry; high precision calorimetric timing; particle identification for $\pi/\mu$ and $\pi/K$
separation at low energies; and very low mass tracking with excellent momentum and spatial resolution.
Although R\&D can move forward in the near term, there is a significant concern that domestic expertise in
kaon physics will be completely depleted if there is no near-term kaon program in the U.S.
As a consequence, the ORKA experiment at the Main Injector is an absolutely integral part of the \PX\ 
kaon program.
If ORKA does not run this decade, there will be little hope of carrying out the extremely challenging kaon
program that the science motivates and \PX\ can facilitate.

\bibliographystyle{apsrev4-1}
\bibliography{kaon/refs}

 % Kevin P. & Vincenzo; David J and Bob T

%%%%%%%%%%%%%%%%%%%%%%%%%%%%%%%%%%%%%%%%%%%%%%%%%%%%%%%%%%%%
\chapter{Muon Experiments with \PX}
\label{chapt:muon}
%%%%%%%%%%%%%%%%%%%%%%%%%%%%%%%%%%%%%%%%%%%%%%%%%%%%%%%%%%%%

\authors{Robert Bernstein, Graham Kribs, \\
% Kaladi~Babu,
Vincenzo~Cirigliano,
Andr\'e de~Gouv\^ea,
Douglas~Glenzinski,
Brendan~Kiburg, \\
Kyle~Knoepfel,
Nikolai~V.~Mokhov, 
Vitaly~S.~Pronskikh,
    and
Robert~S.~Tschirhart}

\section{Introduction}
\label{mu:sec:intro}

The fundamental origin of flavor in the Standard Model remains a mystery.
Despite the roughly eighty years since Rabi asked ``Who ordered that?'' upon learning of the discovery of the
muon, we have not understood the reason that there are three generations or, more recently, why the quark and
neutrino mixing matrices are so different.
The solution to the flavor problem would give profound insights into physics beyond the Standard Model (BSM)
and tell us about the couplings and the mass scale at which the next level of insight can be found.
Rare muon decays provide exceptional probes of flavor violation beyond the Standard Model physics.
The observation of charged lepton flavor violation (CLFV) is an unambiguous signal of new physics and muons,
because they can be made into intense beams, are the most powerful probe.
Experiments at \PX\ using charged lepton flavor violation can probe mass scales up to
$\text{O}(10^4)$~TeV/$c^2$.

\PX's unique combination of intensity and flexibility of time structure make it possible to envisage a
range of experiments.
Searches for $\mu\to e\gamma$ and $\mu\to 3e$ are stopped muon experiments that
require as low, constant instantaneous rates as are practical.
Muon-to-electron conversion experiments use captured muons, and current designs benefit more from a pulsed 
beam structure.
The spacing between pulses and the requirements on the width of pulses depends on the $Z$ of the element in
which the capture occurs, and a range of elements is often required to either map out or exclude a given
BSM interaction.
Certainly if a signal is observed before \PX\ a systematic study of different $Z$ materials is required
with pulse separations varying by an order of magnitude from hundreds of nanoseconds to a few microseconds.

The ability to switch the time structure of the beam to fit the needs of an individual experiment is as much
a part of the strength of \PX\ as is the raw intensity: if you can't use the intensity because of the
time structure, you can't do the physics.

%%%%%%%%%%%%%%%%%%%%%%%%%%%%%%%%%%%%%%%%%%%%%%%%%%%%%%%%%%%%
\section{Physics Motivation}
\label{mu:sec:theory}
%%%%%%%%%%%%%%%%%%%%%%%%%%%%%%%%%%%%%%%%%%%%%%%%%%%%%%%%%%%%

As is well known, Yukawa couplings in the quark and lepton sectors break the global flavor symmetries of the
Standard Model to $\text{U}(1)_B \times \text{U}(1)_\ell$ (with Dirac neutrino masses) or just
$\text{U}(1)_B$ (with Majorana neutrino masses).
Parameterizing the flavor mixing as CKM~\cite{Cabibbo:1963yz,Kobayashi:1973fv} and
PMNS~\cite{Pontecorvo:1957cp,Maki:1962mu} mixing for the quark and neutrino sectors very successfully
accommodates all experimental observations to date.

Rare muon decays provide exceptional probes of flavor violation
beyond the Standard Model physics.  This is because the predicted
rates for $\mu \to e$ processes in the Standard Model resulting 
from a neutrino mass mixing insertion are unobservably small 
\cite{Marciano:1977wx,Bilenky:1977du,Cheng:1977nv,Lee:1977qz,Lee:1977tib}
\begin{eqnarray}
    \BR(\mu \to e\gamma) &=& \frac{3\alpha}{32\pi}
        \left| \sum_{i=2,3} U_{\mu i}^* U_{e i} \frac{\Delta m_{i 1}^2}{M_W^2} \right|^2
        < 10^{-54},
\end{eqnarray}
where $U_{\alpha i}$ are elements of the PMNS neutrino mixing matrix and $\Delta m_{ij}^2$ are the neutrino
mass-squared differences.
Hence, the observation of charged lepton flavor violation (CLFV) is an unambiguous signal of new physics.

\subsection{Effective Theory Discussion}
\label{mu:sec:eff}

\begin{figure}
    \centering
    \includegraphics[scale=0.6]{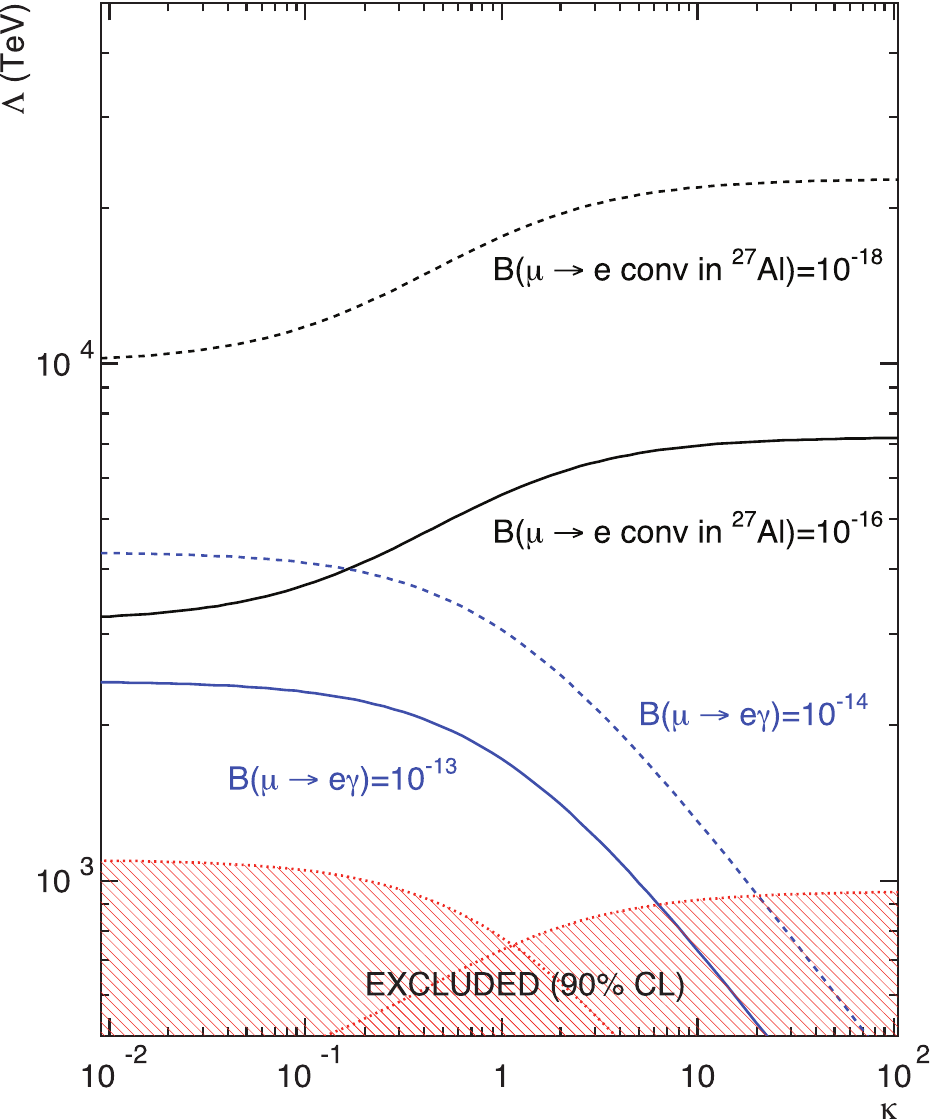}
    \caption[Mass scale $\Lambda$ vs.\ $\kappa$ for selected CLFV experiments]{Mass scale $\Lambda$ 
        \vs~$\kappa$ for selected experiments.  
        The left-hand side of the plot for small $\kappa$ can be probed by experiments such as MEG looking 
        for $\mu\to e\gamma$.
        The right-hand side can be tested by $\mu N\to eN$.  
        Comparing and contrasting measurements and limits pins down or constrains new physics more 
        powerfully than either experiment alone.
        A~similar plot can be made for the $\mu\to3e$ process, but the meaning of $\kappa$ would be 
        different.
        From Ref.~\cite{deGouvea:2013zba}.}
    \label{muon:fig:kappaVsLambda}
\end{figure}
Physics beyond the Standard Model generically can lead to new sources of flavor violation that far exceed the
Standard Model values.
A simple model-independent approach to CLFV is to simply write the effective operators that mediate the lepton flavor-violating processes, 
including \cite{deGouvea:2004dg,deGouvea:2007xp}
\begin{eqnarray}
    \mathcal{L}_{CLFV} &=& \frac{m_\mu}{(1 + \kappa) \Lambda^2} \bar{\mu}_R \sigma_{\mu\nu} e_L F^{\mu\nu}
        + \frac{\kappa}{(1 + \kappa) \Lambda^2} \bar{\mu}_L \gamma_\mu e_L \bar{f}_L \gamma^\mu f_L.
    \label{mu:eq:clfv}
\end{eqnarray}
These are parameterized by $1/\Lambda^2$ with a coefficient $\kappa$ that weights the relative importance of
the ``magnetic moment'' type operator versus the four-fermion interaction.
Specific scenarios of physics beyond the Standard Model will lead to 
different combinations of these and additional operators, such as scalar
and tensor~\cite{Kuno:1999jp,Kitano:2002mt,Cirigliano:2009bz}
(as well as different relative weights for the four-fermion interactions 
as the other fermion $f$ varies).
There are basically two classes of possibilities:
\begin{itemize}
\item models that directly impact electroweak symmetry breaking, 
such as supersymmetry
\item models that directly impact the flavor puzzle, such as ones
that attempt to \emph{explain} the flavor mixings and mass hierarchy
\end{itemize}
Constraints in the $\kappa$-$\Lambda$ plane from current and potential limits on
$\mu\to e\gamma$ and $\mu N\to eN$ are shown in Fig.~\ref{muon:fig:kappaVsLambda}.

In the context of a general effective theory
analysis~\cite{Kuno:1999jp,Kitano:2002mt,Cirigliano:2009bz}, 
it has been shown that information about the relative strength of the different
four-fermion 
operators that mediate CLFV  can be obtained by studying 
$\mu \to e$ conversion on different target nuclei. 
There are three types of effective operators that contribute to
the coherent $\mu \to e$ conversion process: the dipole, the
vector, and the scalar operators. Using the nonrelativistic 
approximation for   the muon wave function, the three operators give 
the same form of overlapping integrals among the wave functions of 
the initial muon and the final electron and the nucleon density in 
the target nuclei. However, as the relativistic and finite nuclear size effects
become 
important for heavy
nuclei~\cite{Shanker:1979ap,Czarnecki:1998iz,Kitano:2002mt},   
the transition amplitudes for the three 
operators  show different dependences on the atomic number $Z$. 
The relative numbers of neutrons and protons also change as $Z$ 
increases: this fact allows one to find out if the 
LFV  operators couples to up-type or down-type  
quarks again by looking at the target atom dependence. 
The theoretical uncertainties  of such an analysis arise 
predominantly from  the nucleon ``sigma~terms''  $\langle N | m_q \bar{q} q |N
\rangle$ ($q=u,d,s$). 
These uncertainties can be largely reduced with input from lattice
QCD (see Sec.~\ref{lqcd:subsec:Mu2e} and Ref.~\cite{Kronfeld:2012uk})
and do not constitute a limiting factor  in 
discriminating models where one or at most two underlying operators 
(dipole, scalar, vector)  provide the dominant source of lepton flavor
violation. 
A realistic discrimination among underlying mechanisms 
requires a measurement of  the ratio of conversion rates 
at the $5 \%$ level (in the case of two light nuclei)  or at the  $20 \%$ level
(in the case of one light and one heavy nucleus)~\cite{Cirigliano:2009bz}.

Operators besides those in Eq.~(\ref{mu:eq:clfv}) can also signal new physics.
One possibility is muonium-antimuonium oscillations, where $\mu^+e^-$ oscillates into $\mu^-e^+$ via 
four-fermion interactions such as
\begin{eqnarray}
    \frac{\overline{\mu}_R \gamma_\mu e_R \overline{\mu}_R \gamma^\mu e_R}{\Lambda^2} + \text{H.c.}
\end{eqnarray}
and other chiralities.
This type of interaction was considered in a recent paper on flavor-violating Higgs couplings to leptons
\cite{Harnik:2012pb}.
Reinterpreting their limit, which used the MACS experimental result \cite{Willmann:1998gd}, one obtains the
bound $\Lambda \gsim 1.6$~TeV\@.
It is also possible to change lepton flavor and charge through scattering off nuclei, %
$\mu^\pm N\to e^\mp N'$, which proceeds through further higher dimensional operators.

In the LHC era, the motivations for the continued search for new sources of flavor violation should be clear:
if new physics is discovered at the LHC, searches for and measurements of charged lepton flavor violation can
have enormous discriminating power in differentiating among models.
On the other hand, if no direct evidence of new physics is found, experiments at \PX\ using charged lepton
flavor violation can probe mass scales up to $\text{O}(10^4)$~TeV, far beyond the reach of any planned
collider.
In this chapter, based largely on the ideas discussed at the meeting of the Project~X Physics
Study~\cite{pxps:2012}, we discuss several examples where large flavor violation arises in the lepton sector.

\subsection{Supersymmetry}
\label{mu:sec:susy}

Weak scale supersymmetry remains an intriguing possibility to understand the stability of the electroweak
scale.
In the minimal supersymmetric standard model (MSSM), the Higgs mass is constrained to be very light, less
than about $130$~GeV (a value computed long ago; see, for example, S.P.~Martin's classic ``Supersymmetry
Primer'' \cite{Martin:1997ns}).
The observation at the LHC of a particle consistent with being a Higgs boson of about $126$~GeV is, thus, a
tantalizing clue that a \emph{weakly} coupled description of electroweak symmetry breaking is a viable
possibility.
The lack of evidence for superpartners at the LHC challenges the supersymmetry paradigm, but a version of
supersymmetry called ``natural supersymmetry,'' with Higgsinos and stops of order the electroweak scale and
gluinos not too heavy remains viable, and is only now being probed~\cite{stopsearchatlas,stopsearchcms}.

The MSSM contains slepton mass matrices that are otherwise undetermined.
Arbitrary slepton mixing would lead to a huge rate for CLFV
\cite{Lee:1984tn,Lee:1984kr,Hisano:1995cp,Hisano:1998fj,Masina:2002mv,Paradisi:2005fk,Ciuchini:2007ha}.
Instead, the nonobservation of $\mu \to e$ processes can be used to constrain the slepton flavor mixing,
often parameterized by $\delta^\ell_{XY} \equiv \delta m^2_{XY}/m^2$ where $\delta m^2_{XY}$ is the
off-diagonal $(12)$-entry appearing in the sfermion mass matrix connecting the $X$-handed slepton to the
$Y$-handed slepton, and $m^2$ is the average slepton mass.
Reference~\cite{Ciuchini:2007ha} found $\delta^\ell_{LR} \lsim 3 \times 10^{-5}$, while
$\delta^\ell_{LL}\lsim6\times 10^{-4}$, over a scan of the mSUGRA parameter space.
Similarly strong bounds on $\delta^\ell_{RR}$ can also be found, though cancellations between diagrams in the
amplitude can in some cases allow for much larger mixing \cite{Masina:2002mv,Paradisi:2005fk,Ciuchini:2007ha}.

One of the interesting developments over the past five years is the possibility that slepton flavor mixing
may \emph{not} lead to such large rates for $\mu \to e$, when the MSSM is extended to include an approximate
$R$-symmetry \cite{Kribs:2007ac}.
Unlike the MSSM, the most important constraint is not necessarily $\mu \to e\gamma$ \cite{Fok:2010vk}.
This is easily seen by inspection of the $R$-symmetric flavor-violating operators: $\mu \to e\gamma$ requires
a chirality-flip via a muon Yukawa coupling, whereas $\mu \to e$ conversion has no such requirement.
We find that $\mu \to e$ conversion rules out maximal mixing throughout the right-handed slepton mixing
parameter space for sub-TeV superpartner masses.
This is complementary to $\mu \to e\gamma$, where we find cancellations between the bino and Higgsino
diagrams, analogous to what was found before in the MSSM \cite{Masina:2002mv,Paradisi:2005fk,Ciuchini:2007ha}.
For left-handed slepton mixing, we find possible cancellations in the amplitudes for $\mu \to e$ conversion,
and instead $\mu \to e\gamma$ provides generally the strongest constraint.
Finally, we find that $\mu \to 3e$ provides the weakest constraint on both left-handed and
right-handed slepton mixing throughout the parameter space considered.

\subsection{Neutrino Flavor Oscillations}
\label{mu:sec:neutrino}

The right--handed neutrino mass matrix that is central to the understanding of small neutrino masses via the
seesaw mechanism can arise either (1) from renormalizable operators or (2) from nonrenormalizable or
super-renormalizable operators, depending on the symmetries and the Higgs content of the theory beyond the
Standard Model.
In Ref.~\cite{Babu:2002tb}, lepton flavor violating (LFV) effects were studied in the first class of seesaw
models wherein the $\nu_R$ Majorana masses arise from renormalizable Yukawa couplings involving a $B-L = 2$
Higgs field.
In this model, detailed predictions for $\tau\to\mu\gamma$ and $\mu\to e\gamma$ branching ratios
were found after taking the-then neutrino oscillation data into account.
In minimal supergravity models, a large range of MSSM parameters (suggested by the relic abundance of
neutralino dark matter and that was consistent with Higgs boson mass and other constraints) have radiative
decays are in the range accessible to planned experiments.
This compares with predictions of lepton flavor violation in the second class of models that arise entirely
from the Dirac Yukawa couplings.
The ratio $r \equiv \BR(\mu \to e\gamma)/\BR(\tau\to\mu\gamma)$ provided crucial insight into the origin of
the seesaw mechanism for neutrino mass generation.

In Ref.~\cite{Dutta:2003ps}, the predictions for $\BR(\mu\to e\gamma)$ and $\BR(\tau\to\ell\gamma)$,
$\ell=\mu,e$, were studied in a class of horizontal $\text{SU}(2)$ models that lead to a $3\times 2$ seesaw
model for neutrino masses.
Two such models were considered that obtained the correct pattern for the PMNS matrix.
In these models, the effective low energy theory below the $\text{SU}(2)_H$ scale is the MSSM\@.
Assuming a supersymmetry breaking pattern as in the minimal SUGRA models (with consistency to $g-2$,
$b\to s\gamma$ and WMAP dark matter constraints on the parameters of the model), the $\BR(\mu\to e\gamma)$
prediction was expected to be accessible to the MEG experiment.
Given that Ref.~\cite{Dutta:2003ps} is nearly ten years old, it remains interesting to update the theoretical
analysis with the latest constraints (including the Higgs mass) and determine the impact of future CLFV
experiments in this class of models.

\subsection{Extra Dimensions}
\label{mu:sec:extrad}

In Ref.~\cite{Csaki:2010aj}, a detailed calculation of the $\mu\to e\gamma$ amplitude in a warped
Randall-Sundrum (RS) model was presented using the mixed position/momentum representation of 5D propagators
and the mass insertion approximation, where the localized Higgs VEV was assumed to be much smaller than the
Kaluza-Klein (KK) masses in the theory.
The calculation reveals potential sensitivity to the specific flavor structure known as ``anarchic Yukawa
matrices.'' 
While generic flavor bounds can be placed on the lepton sector of RS models, one can systematically adjust
the structure of the $Y_e$ and $Y_\nu$ matrices to alleviate the bounds while simultaneously maintaining
anarchy.
In other words, there are regions of parameter space which can improve agreement with experimental
constraints without fine tuning.
Conversely, one may generate anarchic flavor structures which---for a given KK scale---cannot satisfy the
$\mu\to e\gamma$ constraints for \emph{any} value of the anarchic scale $Y_*$.
Over a range of randomly generated anarchic matrices the KK scale may be pushed to $4$~TeV\@.
It is interesting to consider the case where $M_{\text{KK}}=3$~TeV where KK excitations are accessible to the
LHC\@.
The minimal model suffers a $\text{O}(10)$ tension between the tree-level lower bound, $Y_* > 4$ and the
loop-level upper bound $Y_* < 0.5$.
This tension is slightly alleviated in the custodial model, where the tree-level lower bound, $Y_* > 1.25$
and the loop-level upper bound $Y_* < 0.3$.
Thus, even for $M_\text{KK}=3$~TeV, some mild tuning in the relative sizes of the 5D Yukawa matrix is
required.
Now, anarchic models generically lead to small mixing angles.
This feature fits the observed quark mixing angles well, but is in stark contrast with the lepton sector 
where neutrino mixing angles are large, pointing to additional flavor structure in the lepton sector.
For example in~\cite{Csaki:2008qq} a bulk $A_4$ non-Abelian discrete symmetry is imposed on the lepton sector.
This leads to a successful explanation of both the lepton mass hierarchy and the neutrino mixing angles (see
also~\cite{delAguila:2010vg}) while all tree-level lepton number-violating couplings are absent, so the only
bound comes from the $\mu\to e\gamma$ amplitude.

In Ref.~\cite{Chang:2005ag}, LFV processes were studied in 5D gauge 
models that are related to neutrino mass generation.
Two complete models which generate neutrino masses 
radiatively were examined. 
They give rise to different neutrino mass patterns thus, 
it is not surprising that they give different prediction for
LFV rates. The first model, with a low unification scale, 
makes essential use of bileptonic scalars. 
It also contains characteristic doubly-charged gauge bosons. 
In this model, the rare $\tau$ decays are much more enhanced compare 
to their counterpart $\mu$ decays. Among the $\tau\to\ell\gamma$ decays 
the largest mode is the $\mu\gamma$, at a level $<10^{-14}$.
The second model, by contrast, has a high unification scale 
(a 5D orbifold version of the usual GUT).  The important 
ingredient for LFV and neutrino masses is using an symmetric
representation under the GUT [$\mathbf{15}$ under \text{SU}(5)] for the Higgs bosons. 
In this model, $\mu \to e$ conversion in nuclei can be within the 
experimental capability of future experiments. As in the first model, 
$\mu\to e\gamma$ will not be observable. This is very different 
from conventional four-dimensional unification models.
It was also noticed \cite{Chang:2005ag} that the split fermion model 
has the characteristic of $L\to 3l$ and $\mu\to e$ conversion
dominating over $L\to l\gamma$.

\subsection{Connections between CLFV and the Muon Magnetic Moment}
\label{muon:sec:gm2-theory}

In Sec.~\ref{muon:sec:gm2}, the current experimental and theoretical status of the muon anomalous magnetic
moment, along with expectations for the near and intermediate futures, is discussed.
In a nutshell, the world's most precise measurement of the anomalous magnetic moment, $g-2$, of the muon
disagrees with the world's best standard model estimate for this observable at around the $3.6\sigma$ level.
The existence of new, heavy degrees of freedom may be responsible for the observed discrepancy.

It is useful to compare, in a model-independent way, new physics that could mediate CLFV to that which may
have manifested itself in precision measurements of the muon anomalous magnetic moment.
Similar to the discussion in Sec.~\ref{mu:sec:eff}, new, heavy physics contributions to the muon $g-2$ are
captured by the effective Lagrangian
\begin{equation}
{\cal L}_{g-2}\supset
\frac{m_{\mu}}{\Lambda^2}\bar{\mu}_R\sigma_{\mu\nu}\mu_L F^{\mu\nu}+h.c.
\,.
\label{muon:eq:l_g-2}
\end{equation}
Current $g-2$ data point to $\Lambda\sim 8$~TeV.
Equation~(\ref{muon:eq:l_g-2}) is, however, very similar to Eq.~(\ref{mu:eq:clfv}) in the limit $\kappa\ll
1$, keeping in mind that $\Lambda$ in Eq.~(\ref{muon:eq:l_g-2}) need not represent the same quantity as
$\Lambda$ in Eq.~(\ref{mu:eq:clfv}) in the limit $\kappa\ll 1$.

We can further relate the effective $g-2$ effective new physics to that of charged-lepton flavor violating
processes as follows: $(\Lambda_\text{CLFV})^{-2}=\theta_{e\mu}(\Lambda_{g-2})^{-2}$.
Here the parameter $\theta_{e\mu}$ measures how well the new physics conserves flavor.
For example, if $\theta_{e\mu}=0$, the new physics is strictly flavor conserving, while if the new physics is
flavor-indiferent, $\theta_{e\mu}\sim 1$.
In either case, a lot can be inferred regarding whether new physics has manifested itself in the muon $g-2$,
and what properties this new physics ought to have.
If $\theta_{e\mu}\sim 1$, negative searches for $\mu\to e\gamma$ already preclude a new physics
interpretation to the muon $g-2$ results, since, as discussed earlier, these constrain
$\Lambda\gtrsim1000$~TeV.
On the other hand, if the muon $g-2$ discrepancy is real evidence for new physics, current searches for
$\mu\to e\gamma$ already reveal that flavor violation in the new-physics sector is highly
suppressed: \linebreak $\theta_{e\mu}<10^{-4}$.
A~similar analysis can be carried out for $\kappa\gg 1$.
In this case, the relative sensitivity of the most relevant charged-lepton flavor violating processes (either
$\mu\to e$-conversion or $\mu\to eee$) is much higher.

The comparison of data on the anomalous magnetic moment of the muon is, of course, also quite powerful when
it comes to concrete models.
A detailed analysis of quite generic versions of the MSSM allows one to directly related the branching ratio
for $\mu\to e\gamma$ to the supersymmetric contributions to the muon $g-2$ \cite{Hisano:2001qz}.

%%%%%%%%%%%%%%%%%%%%%%%%%%%%%%%%%%%%%%%%%%%%%%%%%%%%%%%%%%%%
\section{Experiments}
\label{mu:sec:expt}
%%%%%%%%%%%%%%%%%%%%%%%%%%%%%%%%%%%%%%%%%%%%%%%%%%%%%%%%%%%%

Searches for CLFV searching for muons changing into electrons have been the most important for several reasons.
First, as soon as the muon was realized to be a heavier version of the electron, there was every reason to
ask why it did not decay into its lighter relative, and the discovery of the muon long predates the discovery
of the tau.
Second, we can make muon beams but not tau beams.
Even today, in the era of flavor factories, the intensity of muon beams makes up for the (usually) smaller
smaller per-particle effect.
The kaon CLFV processes are also not as powerful as muon-based searches.
Therefore, muon-based CLFV experiments have dominated the field.
There is an active program to improve muon-based limits by four orders-of magnitude in key processes (so
roughly an order of magnitude in mass reach) and remain ahead of the competition from other channels.
There are three important muon-based searches: muon-to-electron conversion, $\mu^-N\to e^-N$, $\mu\to
e\gamma$, and $\mu\to3e$.
A fourth process that is ripe for improvement and of increasing interest is $\mu^-N(A,Z)\to e^+N(A,Z-2)$.
Finally, the muonium-antimuonium transition provides a unique window into BSM physics, and it may be possible
to improve the searches by two orders of magnitude.

The experiments and their beam requirements are summarized in Table~\ref{mu:tab:overall}.
\begin{table}
    \centering
    \caption[Beam requirements for muon experiments]{Summary of beam requirements for muon experiments.}
    \label{mu:tab:overall}
    \begin{tabular}{ccccc}
        \hline\hline
        Process & Time Structure & Capture or stop & Accepted muons & Muon KE\\
        \hline
        $\mu \rightarrow 3e$ &continuous& stop& $\text{O}(10^{19})$&surface\\
        $\mu \rightarrow e \gamma$&continuous&stop & $\text{O}(10^{19})$&surface \\
%         $\mu N \rightarrow e N$& pulsed &capture& $ & $\text{O}(10^{19})$  &     $\leq 50 $ MeV \\
        $\mu^- N \rightarrow e^- N$&pulsed &capture & $\text{O}(10^{19})$  &$\leq 50 $ MeV       \\
        $\mu^- N \rightarrow e^+ N(A,Z-2)$&pulsed &   capture & $\text{O}(10^{19})$  &$\leq 50 $ MeV     \\
        $\mu^+ e^- \rightarrow \mu^- e^+$&pulsed&stop&$\text{O}(10^{13})$&surface \\
        \hline\hline
    \end{tabular}
\end{table}

\subsection{$\mu \rightarrow e \gamma$}
\label{mu:subsec:muegamma}

\subsubsection{Current Status}

MEG at the Paul Scherrer Institute (PSI) in Zurich, Switzerland, has just reached a limit of 
$5.7\times10^{-13}$ at 90\% CL with $3.6 \times 10^{14}$ stopped muons~\cite{Adam:2013}.
The experiment is now background limited.
An upgrade proposal to reach a limit of $6 \times 10^{-14}$ has been approved at PSI~\cite{Baldini:2013}.
Here, we provide an equation from Ref.~\cite{Baldini:2013}, explained in Ref.~\cite{rhb:2013},
which gives the relationship among resolutions and the level at which the experiment observes background:
\begin{eqnarray}
    {\cal B} & \propto & \frac{R_{\mu}}{D} \, \Delta t_{e \gamma} \, \frac{\Delta E_e}{m_{\mu}/2} 
        \left(\frac{\Delta E_{\gamma}}{15 m_{\mu}/2} \right)^2
        \left(\frac{\Delta\theta_{e\gamma}}{2} \right)^2,
    \label{mu:eqn:muegback}
\end{eqnarray}
where $\mathcal{B}^{-1}$ is the number of muons collected in order to reach one background event.
The factors are the muon stop rate divided by the beam duty factor,~$R_\mu/D$;
the detector time resolution,~$\Delta t_{e\gamma}$;
the positron energy resolution,~$\Delta E_e$;
the photon energy resolution,~$\Delta E_{\gamma}$; and 
the angular resolution,~$\Delta\theta_{e\gamma}$.
Improving the vertex determination lowers the background quadratically through the last factor.

\subsubsection{Next Steps}

The MEG upgrade proposes to use either a surface muon beam with momentum $\approx29$~MeV/$c$ or subsurface 
muon beam at $\approx25$~MeV/$c$, along with the thinnest possible stopping target.
The use of a subsurface beam is motivated by reducing the range straggling to stop muons.
Hence the thinnest target gives the best constraints on the event vertex in the reconstruction of the
back-to-back $e$ and~$\gamma$ in $\mu \rightarrow e \gamma$.
The straggling in range is given by
\begin{eqnarray}
\Delta R &\propto& P^{3.5} \sqrt{( (0.09)^2 + (3.5 \Delta P/P)^2} ,
\end{eqnarray}
where $P$ is the momentum and $\Delta P$ its spread, so a reduction in beam momentum gives a rapid decrease
in the distance over which the muon stops~\cite{Baldini:2013}.

There are two choices for going beyond the MEG upgrade proposal.
MEG did not convert the photon, and its approved upgrade continues to use this method to achieve a ten-fold
improvement in the limit.
MEG is also considering an active target, although this is still an option rather than part of the baseline
design.
With photon conversion, a thin converter is needed so that multiple scattering and energy loss do not spoil
the resolution.
Then, however, the statistical power suffers, because only a small fraction of the photons can be converted,
although the remaining events have the superior resolution of tracking, relative to calorimetry.
How to resolve the conflict between statistics and resolution requires further study.

\subsubsection{Beam Requirements}

Since the background is effectively a function of the square of the instantaneous intensity, as continuous a
beam as possible, with minimal instantaneous fluctuation, is required.
The beam should either be surface or slightly subsurface as explained above.

\subsection{Muon-to-electron Conversion}
\label{mu:subsec:mu2e}

Muon-to-electron conversion experiments \cite{rhb:2013} bring negative muons to rest by stopping them in
a target.
The muons fall into orbit around an atomic nucleus.
The muons can then (1) decay while in orbit (known as either DIO or MIO in the literature), (2) undergo 
nuclear capture, or (3) convert into electrons.
The first process is a background; the second, the normalization for the signal; the third, the signal
itself: a mono-energetic electron at an energy of the muon mass minus binding and recoil energy.
Typical signal energies for the converted electrons are therefore close to 100~MeV, depending on $Z$.
The nucleus recoils coherently in the process and does not change state.
One might think that muon decays would not be a significant background since the peak and upper limit of the
muon free-decay Michel spectrum is at 52.8 MeV, far from the 100 MeV signal energy.
The spectrum of a muon decaying from an atomic orbital differs from the free decay spectrum because the
outgoing electron can exchange a photon with the nucleus.
The endpoint now becomes the conversion energy.
This is simple to understand: transform to the rest frame of the outgoing neutrinos.
Then neglecting the tiny neutrino mass, the final state is an outgoing electron recoiling against a nucleus,
precisely the same state as the conversion signal.
Modern evaluations of the spectrum can then be combined with realistic resolutions and other effects to
extract an expected amount of background~\cite{Czarnecki:2011}.
Improving the resolution and minimizing energy loss in the apparatus (a stochastic process that increases the
$\delta$-function signal width, increasing all backgrounds) are therefore central to both improving existing
limits and future \PX\ experiments.

The other major background comes from radiative pion capture, in which $\pi^- N \rightarrow \gamma N$ and the
photon either internally or externally converts and produces an electron indistinguishable from signal.
By spacing the beam pulses further part, one can use the pion lifetime to reduce the background.
Pulsed beams, with (for Mu2e) $10^{-10}$ protons between pulses per protons in pulses, are therefore a key
ingredient in the next generation of experiments; this suppression is known as extinction.
A~related source of background is antiproton production.
Fermilab Booster experiments use 8~GeV kinetic energy protons and thus are above the antiproton production
threshold.
Antiprotons do not decay (so far as we know) and move slowly, with kinetic energies of $\sim 5$ MeV;
therefore much of the time information associated with the beam pulse is lost.
Antiprotons can therefore evade the extinction requirements.
If they reach the stopping target, they will then annihilate in the same material used to capture muons and
produce pions that then undergo radiative pion capture.
Experiments must then place absorbers in the beam to annihilate the antiprotons before they reach the stopping
target.
The absorber also stops muons, lowering the flux on the stopping target.
The need to reduce the antiproton rate to an acceptable level without an unacceptable loss of muons is a
limitation of the upcoming generation of experiments.
\PX, with 1--3~GeV proton beams instead of 8~GeV, will produce negligible numbers of antiprotons and 
eliminate this problem.

\subsubsection{Current Status}

The best existing searches for muon-to-electron conversion have been performed at PSI by the SINDRUM-II 
collaboration.
SINDRUM-II used a variety of materials; the best limits were set on Au, with $R_{\mu e} < 7 \times 10^{-13}$
at 90\%~CL.
The SINDRUM-II series had three relevant limitations:
\begin{enumerate}
 \item the time between beam pulses at PSI is just under 20~ns, which leaves the experiments vulnerable to 
     backgrounds from radiative pion capture since the pulse separations approximately the pion lifetime;
 \item only $\text{O}(10^8)$ muons/s were available;
 \item the $\pi e5$ area at PSI required SINDRUM-II to use a degrader and beam vetoes.  
 \end{enumerate}

\subsubsection{Next Steps}
Fermilab's Mu2e experiment at the Booster will reach  be able to set a limit of $6 \times 10^{-17}$ at 90\% CL for conversions on aluminum, a
four order-of-magnitude improvement over SINDRUM-II limits on titanium and gold.
There are two possibilities: (1) Mu2e sees a signal, or (2) it does not.
If it does not, a huge part of SUSY parameter space, and that of other models, will be ruled out up to mass
scales near $10^4$ TeV/$c^2$.
\PX\ can improve the statistical power by as much as two orders of magnitude.
If a signal \emph{is} seen, then Mu2e's aluminum target needs to be changed to other elements in order to
probe the nature of the new physics.
However, as $Z$ increases the lifetime of the $\mu N$ muonic atom decreases until the conversion is obscured
by the beam flash and backgrounds.
The \PX\ flexible time structure and short beam pulses can be used to mitigate the experimental
difficulties.
Other, new technologies, such as FFAGs or helical cooling channels might be used as well.
In either case, to study a signal or to improve a limit, the intensity provided at \PX\ would be required to
advance.
  
% It should be noted that the current beam timeline for Mu2e uses only one~third of the available time because
% of details of the beam structure that are irrelevant for this discussion.
% In principle one could immediately gain a factor of three, but then with no other changes the background
% would be approximately one event or more from cosmic-ray induced events and the decay-in-orbit process.
% The increase in delivered protons may also cause a problem from increased radiation damage to the solenoids
% requiring more frequent warm-ups to anneal the Al-stabilized superconductor.
  
Recent studies have concentrated on a ten-fold improvement using the current Mu2e tripartite solenoid
design~\cite{Glenzinski:2013a}.
The goal is to have an experiment with a single-event sensitivity approximately ten times better than Mu2e with fewer than 
1~background event.
The experiment would use the lower energy proton beams at 1 or 3~GeV from \PX.
Enumerating the assumptions:
\begin{enumerate}
  \item proton pulses with a full-base width of 100~ns;
  \item duty factor of 90\%;
  \item intrinsic extinction from the machine of $\leq 10^{-6}$ followed by an additional $\leq 10^{-6}$ as 
      in Mu2e through the ``extinction dipole'';
  \item protons at 1 or 3~GeV to eliminate antiproton-induced RPC backgrounds;
  \item a beam transport system to the current Mu2e beam line.
\end{enumerate}
These studies have found that either an Al or Ti stopping target seemed workable.
The yield was $1.4 \times 10^{-4}$ stopped muons/proton at 1~GeV and $6.7 \times 10^{-4}$ at 3~GeV, compared
to $1.6 \times 10^{-3}$, an order-of-magnitude higher at the Booster's 8~GeV.
Straightforward calculations then lead to the requirements in Table~\ref{mu:tab:protons}
(for Al; Ti is similar).
\begin{table}
    \centering
    \caption[Protons required to reach a ten-fold improvement in sensitivity in $\mu\to e$ conversion]
    {For three  
        different beam energies, the number of protons required to reach a ten-fold improvement in 
        sensitivity for a next generation Mu2e experiment using \PX\ beams.}
    \label{mu:tab:protons}
    \begin{tabular}{cccc}
        \hline\hline
        Proton energy&Protons on target&Beam power&Protons/pulse\\
        \hline
        8 GeV&$3.6 \times 10^{21}$&80 kW&$1.0 \times 10^{8}$\\
        3 GeV&$8.6 \times 10^{21}$&72 kW&$2.5 \times 10^{8}$\\
        1 GeV& $40 \times 10^{21}$&112 kW&$1.2 \times 10^{9}$\\
        \hline\hline
    \end{tabular}
\end{table}
  
\begin{figure}
    \centering
    \includegraphics[scale=0.6]{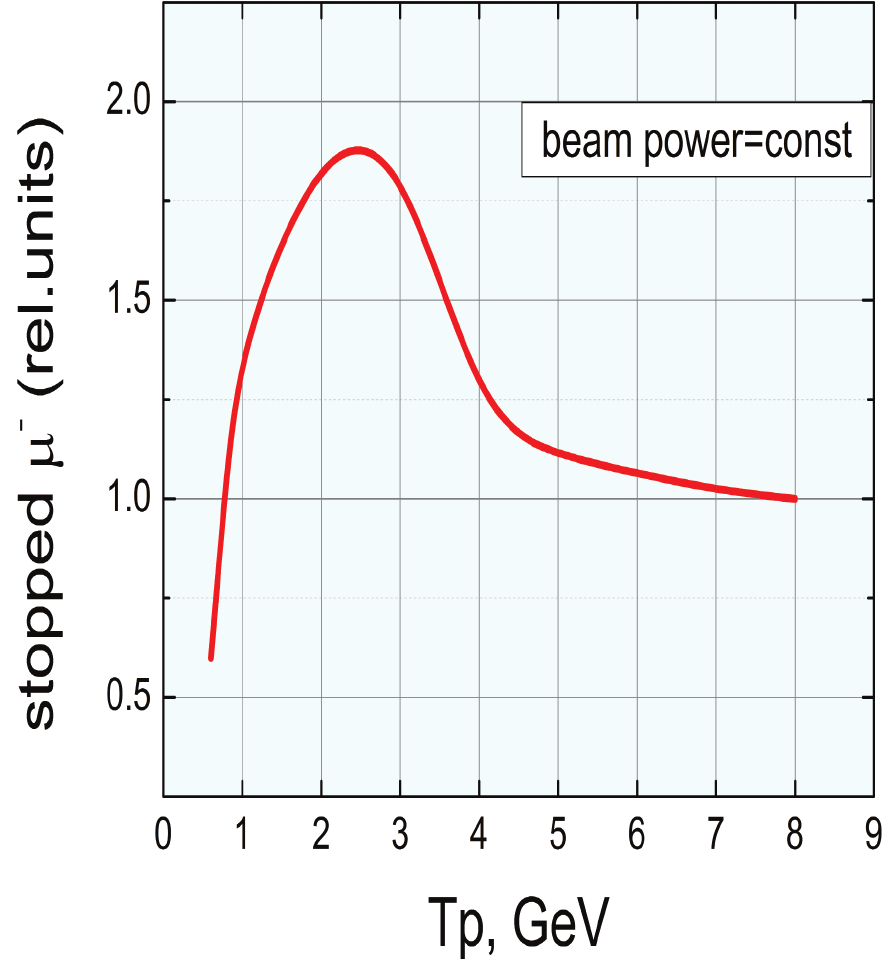}
    \caption[Stopped-muon yield in the current design of the Mu2e apparatus]{Stopped-muon yield in the 
        current design of the Mu2e apparatus as a function of proton kinetic energy~$T_p$, normalized to 
        unity at the Booster-era Mu2e of 8~GeV kinetic energy.
        Note the 1~GeV yield is slightly better than the 8~GeV yield, and the 3~GeV yield is almost twice as 
        high. 
        Beam power is kept constant as the proton energy varies.}
    \label{mu:fig:muonYield}
\end{figure}

The rates are approximately 3--4 times the Mu2e rates, which will likely require upgrades to the detector but
does not seem an unachievable goal.
Mu2e at the Booster is designed for 8 kW on the proton target.
There are then at least two immediate challenges:
\begin{enumerate}
    \item the increased power requires additional shielding to the production solenoid;
        radiation damage to the Al-stabilized superconductor will cause quenching and although Al-stabilized
        superconductor can be annealed the loss of data will be too great;
    \item the number of neutrons will scale along with the protons and beam power, and those neutrons
        can fire the cosmic ray veto, punch through to the detector, or cause other problems.
\end{enumerate}
We show two relevant graphs from Ref.~\cite{Glenzinski:2013}.
The first, Fig.~\ref{mu:fig:muonYield}, shows that in fact the muon yield, with the current Mu2e solenoid
design, not only can be maintained but it can be \emph{higher} at 3~GeV and about equal at 1~GeV, holding
power fixed.
This result is far from obvious, since the collection efficiency of the Mu2e solenoids were optimized for the
Booster-era 8~GeV protons.
 
\begin{figure}
    \centering
    \includegraphics[scale=0.6]{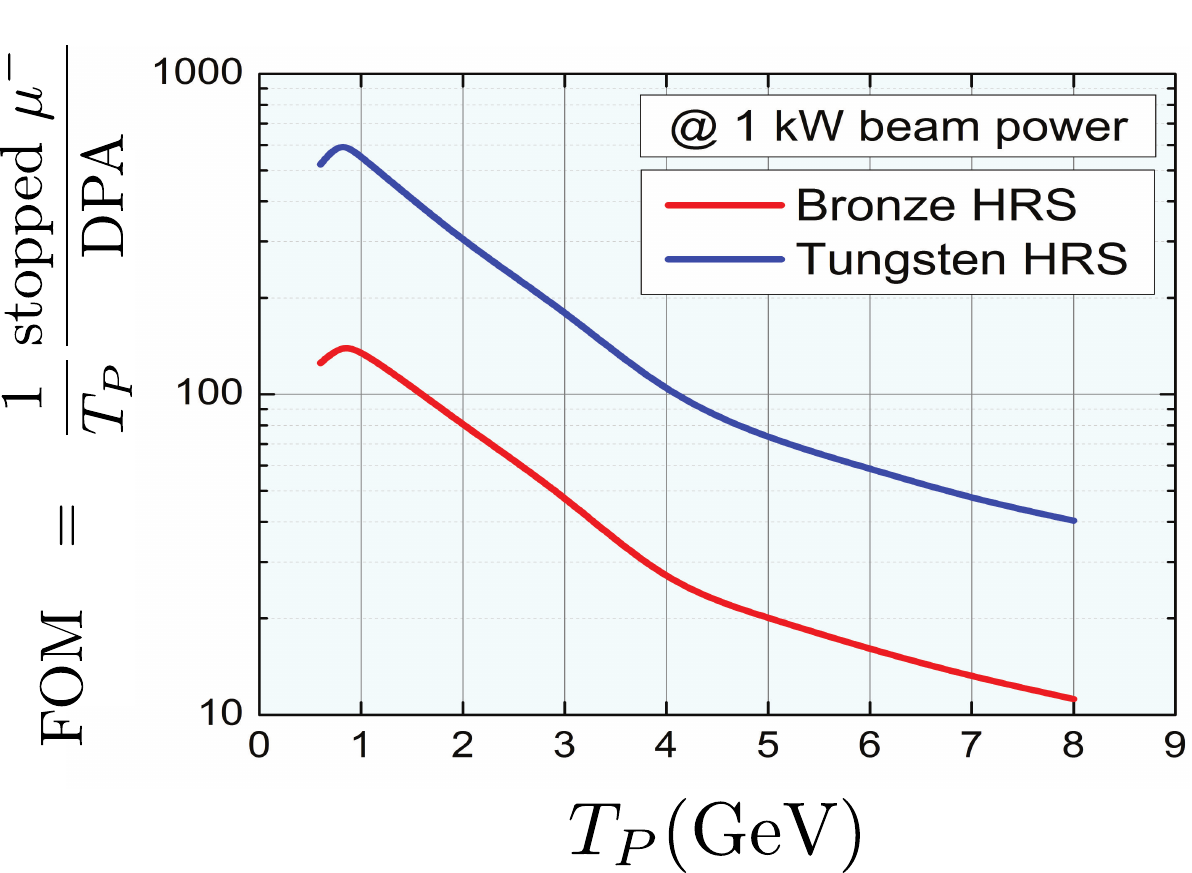}
    \caption[Stopped muons per GeV \vs\ incoming proton kinetic energy]{Figure of merit FOM, defined as 
        stopped muons per GeV per peak DPA for 1 kW beam power in the current Mu2e solenoid, plotted against 
        the kinetic energy of the incoming proton.
        Two heat and radiation shields (HRS) are shown; the bronze HRS is close to the current Mu2e design, 
        and the tungsten shield is a more effective, but more expensive, variant.}
    \label{mu:fig:radDam}
\end{figure}

The second problem is that of radiation damage in the Al stabilizers for the solenoid coils, 
addressed in Fig.~\ref{mu:fig:radDam}.
A figure of merit (FOM) is defined as the number of stopped muons per~GeV (so for 1 kW fixed beam power) per
peak radiation damage.
The radiation damage in metals is measured in displacements per atom (DPA), which is simply the number of times an atom
is displaced from its site in the crystal lattice, per unit fluence
(fluence is the flux per unit area, integrated over time).
It is a standard metric for radiation damage~\cite{Li:2012}.
The plot shows that the FOM peaks at about 1~GeV: the number of DPAs at the peak radiation damage location is
small, driving the FOM upwards.
The peak DPA drops in part because a lower energy proton beam will produce more isotropic secondaries,
driving the peak DPA to smaller values as the energy decreases.
The current Mu2e heat-shield design corresponds to the curve labeled ``bronze HRS.''
The ``tungsten HRS'' would be a better absorber, but tungsten is more expensive.

\subsubsection{Beam Requirements}

There are three general requirements:
\begin{enumerate}
    \item A pulsed beam.  In a Mu2e-style experiment on Al or Ti, pulses no longer than 50~ns would be best.  
    \item A variable time separation between pulses.
        The requirement on the time separation between pulses is governed by the experimental details and the
        lifetime of muonic atoms in the converting nucleus: too short, and the detector may be overwhelmed by the
        beam flash; too long, and the muons will decay away.
        For Au, one would want about 100--200~ns between pulses.
    \item As little beam as possible between pulses.
        The issue here is often referred to as extinction.
        The RPC background is suppressed by waiting for pions to decay after the pulse.
        A proton arriving between pulses restarts the clock and removes the suppression factor.
\end{enumerate}

While an order-of-magnitude improvement seems plausible, if Mu2e sees no signal it will be difficult to take
full advantage of \PX\ intensities for a $\times 100$ improvement.
Backgrounds from cosmic rays and the absolute calibration of the momentum scale (related to separating the
decay-in-orbit background from the signal) would limit the experiment in its current design.
In the case of a signal, it would be imperative to measure the conversion rate on heavy nuclei such as Au.
These backgrounds would likely not preclude the measurement but would likely be significant limitations.
Nor would it be straightforward to improve the limit switching from Al to heavy nuclei, where one might
expect the signal to be larger and a limit therefore better.
The first two beam-related requirements above will be difficult to meet simultaneously in a Mu2e-style
experiment.
With a pion lifetime of 26~ns and a muonic lifetime of 72.6~ns, combined with a beam pulse of order 1~ns,
the radiative pion backgrounds cannot be suppressed even with extinction methods: the muonic lifetime is just
too short.
There are next-generation concepts that form circulating beams in which the pions can decay, effectively
creating a long flight path before forming the final muon beam.\cite{Kuno:2008zz} These reduce the radiative
pion capture background sufficiently.
They also manage the prompt background and beam flash, allowing the experiment to access short-lifetime
capture materials such as Au.

\subsection{$\mu \rightarrow 3e$}
\label{mu:subsec:mu3e}

\subsubsection{Current Status}

A new measurement should strive to set a limit $< ~\text{O}(10^{-16})$ to be competitive with existing limits
and other planned measurements.
The current limit  in SINDRUM is $\BR(\mu \rightarrow 3e) < 1.0 \times 10^{-12}$ at
90\%~CL \cite{Bellgardt:1988}.
Therefore a factor of $10^4$ improvement is required.
With a one-year run, one then requires $10^9$--$10^{10}$ decays/s, before acceptances, etc., are included.
The current $\pi e5$ (MEG) beamline yields about $10^9$ muons/s, barely enough.
A spallation neutron source at PSI (called SINQ, \href{http://www.psi.ch/sinq/}{http://www.psi.ch/sinq/}) 
could provide $5 \times 10^{10}$ muons/s, probably an effective minimum requirement.

Existing experiments have used stopped muons and muon decay-at-rest.
In that case the outgoing electron and positrons can be tracked and the kinematic constraints 
$\sum\bm{p}=\bm{0}$ and $\sum E = m_e$, along with timing, can then be used to identify the rare
decay.

Unfortunately, this mode suffers from many of the same problems as $\mu \rightarrow e \gamma$.
Because it is a decay, unlike muon-to-electron conversion, $\mu \rightarrow 3e$ electrons are in the same
momentum range as ordinary Michel decays.
Therefore there are accidental backgrounds from Michel positrons that coincide with $e^+e^-$ pairs from
$\gamma$ conversions or from other Michel positrons that undergo Bhabha scattering.
(One could cut on the opening angle between the positrons and each of the electrons, since conversions tend
to have a small opening angle, but if the $\mu \rightarrow 3e$ process occurs through processes with a
photon, one then loses acceptance.)

\subsubsection{Next Steps}

A new $\mu \rightarrow 3e$ experiment, using monolithic active pixel sensors, has just received
approval at PSI~\cite{Berger:2011}.
As described in Ref.~\cite{Blondel:2012}, the proponents plan to overcome the difficulties above by making
the tracking material so thin that multiple scattering is small and backgrounds from radiative muon decay
($\mu^+\to e\nu\bar{\nu}\gamma$ with a subsequent photon conversion) are negligible.
The location of the experiment is a matter of logistics, time-sharing with MEG, etc.
A first-round would achieve $10^{-15}$ with eventual improvements in the beam (possibly moving to a
spallation neutron source at PSI) and the detector yielding a potential limit of $10^{-16}$.
Potential limits on the experiment from the decay $\mu^+ \rightarrow e^+e^-e^+ \nu_e \bar{\nu}_{\mu}$ are
discussed in Ref.~\cite{Djilkibaev:2009}; the phase space for accepting such decays and their being
indistinguishable from a $\mu \rightarrow 3e$ signal may be the ultimate limitation of these experiments.

\subsubsection{Beam Requirements}

The beam requirements are quite similar to MEG: a nearly monochromatic beam with a high stopping rate over a
small volume.
The experiment will use a high-intensity surface beam with small emittance and a momentum-bite of $\leq 10$\%.
The initial phase for the PSI run will be $10^7$--$10^8$ muons/s rising closer to $10^8$ for the second
phase.

The second phase of $\mu \rightarrow 3e$ hopes to reach $\text{O} (10^{-16})$.
An unpleased stopping rate of order GHz is then required.
PSI's HiMB project could supply the necessary intensity in the current experimental area, about $2 \times
10^{9}$ stops/s, and a detailed feasibility study of HiMB has just started as of this writing.
Assuming the HiMB area is built and successful, then $10^7$ $\mu$/s (for perfect acceptance) are needed to
reach the sensitivity at which the radiative decay background limits the experiment in a few-year run.

\subsection{Muonium-antimuonium Oscillations}
\label{mu:subsec:muonium}

Hydrogenic bound states of $\mu^+e^-$ (muonium, or ``Mu") can convert through a $\Delta L=2$ process to
$\mu^-e^+$ ($\overline{\text{Mu}}$).
This process is analogous to $K^o\bar{K}^o$ mixing; Pontecorvo~\cite{Pontecorvo:1958} suggested the process
could proceed through an intermediate state of two neutrinos.
Part of the calculation is performed in Ref.~\cite{Willmann:1998}.
One typically states the result of a search as an upper limit on an effective coupling analogous to $G_F$:
$G_{ {\text{Mu}} \overline{ \text{Mu} } }$, where the exchange is mediated by such particles as a doubly
charged Higgs, a dileptonic gauge boson, heavy Majorana neutrinos, or a supersymmetric $\tau$-sneutrino.
The new interaction leads to a splitting of the otherwise degenerate energy levels (recall the coupling is
$V-A$).
Such a new interaction would break the degeneracy by an amount
\begin{eqnarray}
\frac{\delta}{2} &=& \frac {8 \, G_F}{\sqrt{2} n^2 \pi a_o^3} \left (\frac{G_{ {\text{Mu}}  \overline{ 
\text{Mu} } }}{G_F} \right ),
\end{eqnarray}
where $n$ is the principal quantum number and $a_o$ is the Bohr radius of the muonium atom.
For $n=1$,
\begin{eqnarray} 
    \delta &=& 2.16 \times 10^{-12} \,  \frac{G_{ {\text{Mu}}  \overline{ \text{Mu} } }}{G_F} \,\, \text{eV}.
\end{eqnarray}
Assuming an initially pure $\mu^+e^-$ state, the probability of transition is given by:
\begin{eqnarray}
{\cal P}(t) &=& \sin^2 \left ( \frac{\delta t}{2 \hbar} \right ) \,\,\lambda_{\mu}e^{-\lambda_{\mu} t},
\end{eqnarray}
where $\lambda_{\mu}$ is the muon lifetime.
Modulating the oscillation probability against the muon lifetime tells us the maximum probability of decay as
antimuonium occurs at $t_\text{max}= 2 \tau_{\mu}$.
The overall probability of transition is
\begin{eqnarray}
P_\text{total}&=& 2.5 \times 10^{-3} \left (  \frac{G_{ {\text{Mu}}  \overline{ \text{Mu} } }}{G_F} \right).
\label{mu:eq:totalprobmuonium}
\end{eqnarray}

Normally the experiments quote a limit on $G_{ {\text{Mu}} \overline{ \text{Mu} } } $.
Experimentally, of course, no such thing is measured; one measures a probability of transition.
The limit is set assuming an interaction of $(V\pm A)\times(V\pm A)$ although one can also set limits on
masses of, for example, dileptonic gauge bosons.
We follow the practice of quoting a limit on the ratio of coupling constants.

The experimental signature of antimuonium decay is an energetic electron from normal muon decay in
coincidence with an approximately 13.5~eV kinetic energy positron (the Rydberg energy in the $1s$ state).
Because the negative muon can be captured, the signal rate is suppressed by the capture fraction (depending
on $Z$, around a factor of two for (V$\mp$A)$\times$(V$\pm$A) processes).
This measurement suffers rate-dependent backgrounds not dissimilar to those found in $\mu\to e\gamma$ and
$\mu\to3e$, from accidentals and radiative decay processes:
\begin{enumerate}
    \item The rare decay mode $\mu^+ \rightarrow e^+ e^+e^- \nu_{e}\bar{\nu}_{\mu}$ with a branching ratio 
        of $(3.4\pm0.4)\times 10^{-5}$ (from the 2008 PDG).
        If one of the positrons has low kinetic energy and the electron is detected, this channel can fake a 
        signal. 
    \item The system starts as muonium, hence $\mu^+ \rightarrow e^+ \nu_e \bar{\nu}_{\mu}$ yields a 
        positron.
        If the $e^+$ undergoes Bhabha scattering, an energetic electron can be produced.
        Background results from the coincidence of that scattering with a scattered $e^+$.
\end{enumerate}

\subsubsection{Current Status}

Modern experiments rely on the coincident detection of the muon and positron; the most recent experiment is
described in Ref.~\cite{Willmann:1999}.
A subsurface $\mu^+$ at $\approx$ 20 MeV/$c$ is passed into SiO$_2$ powder (the material provides stopping
power with voids for the muonium system to escape).
The apparatus could detect the decay of both muonium and antimuonium.
Decay positrons or electrons were observed in a spectrometer at right angles to the beam and after passing
through a pair of MWPCs were detected in CsI.
Atomic electrons (or positrons) were electrostatically collected, focused, and accelerated to 5.7 keV.
A dipole then charge- and momentum-selected the particles, which were finally detected by an MCP.
The advantages of observing the thermal muonium are obvious: one can verify the experimental method and
calibrate the detectors, study acceptances with reversed polarities, etc.
The most recent experiment set a limit $G_{\text{Mu}\overline{\text{Mu}}}/G_F< 3.0 \times 10^{-3}$ at 
90\%~CL~\cite{Willmann:1999}.

\begin{figure}
    \centering
    \includegraphics[scale=0.6]{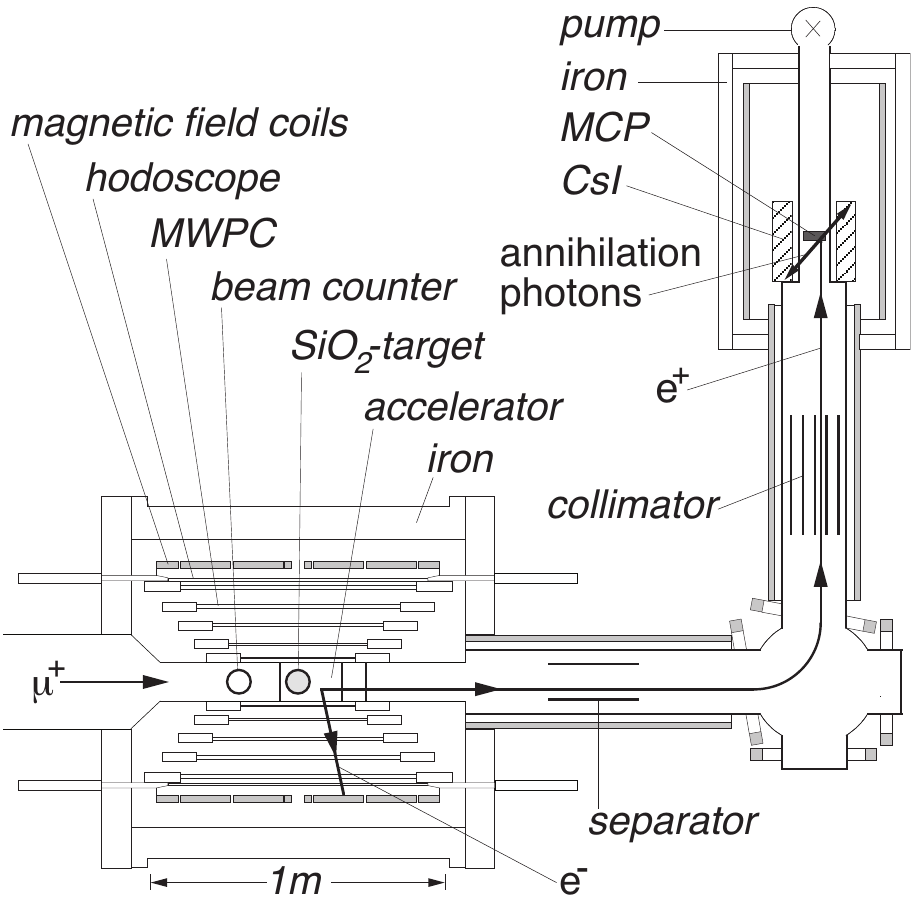}
    \caption[MACS apparatus at PSI]{MACS apparatus at PSI.
        The signature requires the energetic $E^-$ from the $\mu^-$ decay of $\overline{\text{Mu}}$ in a 
        magnetic spectrometer, in coincidence with the atomic shell $e^+$, which is accelerated and 
        magnetically guided onto a microchannel plate; at least one annihilation photon is then detected in 
        a CsI calorimeter.
        From Ref.~\cite{Willmann:1999}.}
    \label{mu:fig:muonium_apparatus}
\end{figure}

\subsubsection{Next Steps}

It is interesting to consider placing the muonium system in a magnetic field, since the muonium energy levels
will be split.
We refer the reader to Refs.~\cite{Kuno:1991, Feinberg:1961} for a fuller discussion of the physics.
Because the spectrometers used to detect and measure electron momenta require a magnetic field, this effect
must be included in the calculation of the transition rate.
In this more general case, $\delta \rightarrow \sqrt{\delta^2 + \Delta^2}$.
The effect is significant even for a weak ($\sim$ 0.1T) field because of the Zeeman splitting of the energy
levels.
The reduction factor for fields of about 0.1 Gauss to 0.1 Tesla is nearly flat at a factor of two, but
\cite{Hou:1995} shows the reduction becomes rapidly more suppressed at higher fields.

The technology of earlier experiments is now obsolete.
One significant limit was the rate capabilities of the available MWPCs.
New technologies could certainly improve on 1998-style methods.
The best existing experiment used CsI; modern scintillating crystals such as LYSO have much better rate
capabilities.

\subsubsection{Beam Requirements}

A pulsed beam seems a necessity to reduce backgrounds from the $\mu^+ \rightarrow e^+ e^+e^-
\nu_{e}\bar{\nu}_{\mu}$ muon decay.
The intensity should be commensurate with a few order-of-magnitude improvement in the limit.
The efficiency for muonium formation is already about 60\%, and the earlier experiments ran for $\approx 210$
hours with $1.4 \times 10^{9}$ decaying muonium atoms~\cite{Willmann:1998}.

%%%%%%%%%%%%%%%%%%%%%%%%%%%%%%%%%%%%%%%%%%%%%%%%%%%%%%%%%%%%

% 
% \clearpage

\subsection{Muon Anomalous Magnetic Moment $g-2$}
\label{muon:sec:gm2}

\subsubsection{Current Status}
%CURRENT STATUS
%E989 STATUS
A concrete plan is in place to probe new physics at the TeV scale by improving the precision on the
measurement of the anomalous magnetic moment of the muon, $a_\mu\equiv(g_\mu-2)/2$.
Within the framework of the Dirac equation, $g_\mu$ is expected to equal~2.
However, quantum loop corrections associated with QED, electroweak, and QCD processes lead to a deviation
from this value.
These SM loop corrections have been calculated with a precision of 420~ppb (part-per-billion)
\cite{PhysRevD.86.010001}:
\begin{equation}
    a_{\mu}^\text{SM} = 116591802(49)\times 10^{-11}.
\end{equation} 
The largest contributor to the theoretical uncertainty stems from two hadronic effects.
The first, with larger error on $a_\mu$, is hadronic vacuum polarization, which can, however, be extracted
from $e^+e^-\to\text{hadrons}$ and from hadronic $\tau$ decays.
The second, with smaller but less solid error, comes from the hadronic light-by-light process.
The prospects for calculating both hadronic contributions with lattice QCD and, thereby
reducing and solidifying the uncertainties, is discussed in Sec.~\ref{lqcd:subsec:gm2}.

The most precise experimental determination of $a_\mu$ was conducted at Brookhaven by the E821 experiment.
Muons with a momentum of 3.094~GeV/$c$ are injected into an $\approx$~7~m-radius magnetic storage ring.
Precise measurements of both the 1.45~T magnetic field and the precession frequency of the muons in that
field allows the anomaly to be determined to 540~ppb\cite{PhysRevD.73.072003}:
\begin{equation}
    a_{\mu}^{\text{E821}} = 116592089(63)\times 10^{-11}.
\end{equation}
The comparison between the standard model prediction and the measurement is $\Delta a_\mu = 287(80)\times
10^{-11}$, which amounts to a $3.6\sigma$ deviation.
To highlight one example, if supersymmetry exists there would be new contributions to $a_\mu$ via
supersymmetric particle loops, analogous to the standard model QED, weak, and QCD loops.
This current discrepancy between the experimental observation and the standard model could be a hint that
such contributions actually exist.

\subsubsection{Next Steps}

The Brookhaven E821 experiment finished statistics limited.
In the near future, the Fermilab E989 Collaboration aims to reduce the experimental uncertainty on $a_\mu$ to
140~ppb\cite{E989prop}.
This relies on a factor 20 increase in the statistics and a variety of improvements in the measurements of
the magnetic field and the decay electrons' energies and times.
The proposal has received Mission Need (CD-0) from the Department of Energy, and the storage ring is being
transported to Fermilab.
Slated to start taking data in 2016, the New Muon $g-2$ Experiment (E989) will accumulate the necessary
statistics using a $\mu^{+}$ beam in about two years.
If the central value remains the same, i.e., $a_\mu^{\text{E989}}$= $a_\mu^{\text{E821}}$, the improved
experimental precision will result in a $5.5\sigma$ deviation with the standard model.
Further expected improvements in theory could increase this significance to $8\sigma$, which would amount to
a discovery of new physics.

At that point, it becomes interesting to consider if the current experimental configuration would be able to accommodate the higher intensity $\mu^{-}$ or $\mu^{+}$ beams of \PX. 
Without significant reductions in the theory uncertainties of $a_{\mu}$, there would be minimal motivation to continue with $\mu^{+}$. 
However, switching to $\mu^{-}$ is a natural followup that would instill confidence that the E989 systematic uncertainties are well-understood. 
The $\mu^{-}$ production rate is about 2.5 times lower than for $\mu^+$ \cite{vanDyck}, so a $\mu^{-}$
measurement with comparable statistics would take a prohibitive amount of time without \PX.

Running with $\mu^{-}$ requires flipping the polarity of the storage ring magnet, which would help
demonstrate that the field-related systematics are under control.
Almost all aspects of the current design would be appropriate for a measurement of $g-2$ of the negative muon.
However, the electric quadrupole plates that provide vertical beam focusing in the muon storage ring would
need to be upgraded for running with $\mu^-$.
This is because the electrons can get trapped in the high field regions near the quadrupole plates, whereas
the positrons produced during $\mu^+$ running disappear at the surfaces.
This necessitates an order of magnitude more stringent vacuum requirements and an improved design of the
plate surfaces.

The biggest challenge for handling the additional primary beam would come from the pion production target
station.
E989 will use the Inconel production target that was used for antiproton production at the end of the
Tevatron Run II.
Four booster batches will be redistributed into 16 bunches during each 1.33~s main injector cycle.
A lithium lens that collects the pions has been developed to handle the increased pulse rate of 12~Hz, up
from 0.45~Hz during Tevatron running.
With an increased repetition rate, the cooling capacity of the lens is a significant issue.
Care must be taken to avoid reaching the melting point of lithium (453.75~K).
The lithium lens has been simulated in ANSYS to understand the thermal constraints.
These simulations were validated with a test stand that pulsed the lens with 12~Hz repetition rates.
Additional preliminary simulations have been performed to model the lens performance at higher repetition
rates.
The studies determined that running with the lens in the current configuration was not feasible at rates 
greater than 21~Hz, because of melting.
Rates of 15~Hz and 18~Hz were not ruled out, but more work is needed to understand how we would safely handle the additional rate heading into the \PX\ era.

When the secondary production target is upgraded to accept additional rate, the systematic errors of E989 could be further reduced by modifying the experiment to use a smaller beam aperture.
In the current configuration, the magnetic field must be highly uniform -- and precisely measured -- over the storage aperture radius of 4.5~cm.
With the higher beam flux, the collimators could be reduced to a few centimeters where the magnetic field gradients are smallest, reducing the systematic contribution from the knowledge of the field.

%At that point, a natural followup would be to try to run with the same configuration and a higher intensity beam at \PX.
Alternatively, we could try to use all of the additional $\mu^{-}$ beam that \PX\ delivers to the storage ring. 
Several experimental components are well-suited to handle additional beam rate.
Each calorimeter is segmented into 54 lead fluoride crystals, which would be critical for resolving pileup at
\PX\ beam rates.
The data acquisition system for E989 is designed to digitize all calorimeter channels continuously during the
beam spills.
This means that the incoming raw data rate is determined by the chosen digitization frequency rather than the
instantaneous decay electron rate.
The data is then processed to identify electrons and read out during the 800~ms period when the Booster
batches are being delivered to the NO$\nu$A experiment.
Assuming some reasonably similar proton economics occurs in the LBNE era, the E989 detector system will be
able to handle the additional rate.

Another possible outcome of the E989 $\mu^+$ running could be that the discrepancy between experimental and
Standard-Model values of $a_\mu$ disappears.
In this event, the path to reusing the storage ring to significantly higher precision is not immediately
clear.
E821 was designed to have sensitivity to the electroweak contribution to the anomalous magnetic moment,
$a_\mu^{\text{EW}}=154(1)\times10^{-11}$ (or 130~ppb).
If the current anomaly disappears then the next contribution within the Standard Model would be the two-loop
Higgs terms, $a_\mu^{\text{H}}\approx4\times 10^{-11}$ \cite{Czar-Krause-Marc}.
This contribution of $\approx 35$~ppb is an additional factor of four beyond the proposed sensitivity of E989
(140~ppb).
Assuming that advances in lattice QCD lead to improvements in the uncertainties of the leading hadronic terms
over the next decade, a next generation muon $g-2$ experiment could continue to probe the standard model.
 
A new approach to measure $a_\mu$ to 100~ppb with an ultracold muon beam has been proposed at \linebreak 
J-PARC \cite{Mibeprop,Iinuma}.
The basic strategy is to bring 3~GeV protons to a production target and collect surface muons ($\mu^+$).
These muons are then stopped in a secondary target where they form muonium and diffuse.
Lasers are used to remove the electrons from the muonium, resulting in ultracold muons with a kinetic energy
of around 25~mV.
These muons would then be accelerated to 300~MeV/$c$ and injected into a 66~cm diameter, 3~Tesla storage ring
via a novel three-dimensional injection spiral scheme.

The success of such a proposal relies on an ultrahigh intensity surface muon beam.
To reach a precision of 100~ppb in about a year of running, $10^6$ ultra cold muons per second must be
produced.
The expected efficiency of converting surface muons to ultracold muons is on the order of $10^{-5}$ to
$10^{-3}$, implying a surface muon beam rate requirement of $10^9$ to $10^{11}$ muons per second.
The lower end of this range is comparable to the current rates produced at~PSI.

\subsubsection{Beam Requirements}

The extension of the muon $g-2$ storage ring experiments require a muon beam with a momentum of 3.094~GeV/$c$.
A high intensity $\mu^-$ beam would be the natural extension to the current experiment.
The beam should be pulsed with bunch spacings no smaller than 10~ms to allow adequate time for muon decay and
data acquisition.
The beam pulses should be no longer than 120~ns so that the leading edge of the pulse does not lap the
trailing edge during injection into the storage ring.
Additional muons per bunch can be utilized by the existing $g-2$ experimental design.
However, the number of booster batches delivered to the pion production target is not easily accommodated by
the lithium collection lens.

The small storage ring experiments would utilize a high intensity surface muon beam with required rates of
$10^{9}$--$10^{11}$ $\mu^+$ per second.
Significant technology advances in laser ionization to produce ultra cold muons is required.
A novel 3-dimensional injection spiral scheme into a tabletop scale 3~T cyclotron would also need to be
developed.

%%%%%%%%%%%%%%%%%%%%%%%%%%%%%%%%%%%%%%%%%%%%%%%%%%%%%%%%%%%%
\section{Summary}
\label{mu:sec:summary}
%%%%%%%%%%%%%%%%%%%%%%%%%%%%%%%%%%%%%%%%%%%%%%%%%%%%%%%%%%%%

Studies of charged lepton flavor violation with muons are of paramount importance.
If the LHC experiments discover new physics, these processes can discriminate and distinguish among models.
If the mass scale of BSM physics is beyond that accessible at the LHC then charged lepton flavor violation
can probe up to $10^4$~TeV/$c^2$.
Since such probes are indirect, regardless of the LHC results, one experiment will not suffice.
\PX\ offers an opportunity to perform the key experiments required in one place in a staged manner.

There are five processes that are essential to this campaign, namely, $\mu^+\to e^+\gamma$, 
$\mu^+\to e^+e^+e^-$, $\mu^-N\to e^-N$, $\mu^-N\to e^+N$, and $\mu^+e^-\leftrightarrow\mu^-e^+$.
The first two ``decay experiments" require a very different time structure from the next two ``capture''
experiments involving conversion in the field of a nucleus, and the last, muonium-antimuonium oscillations,
requires yet another time structure.
All require intense beams at megawatts of power at 1--3 GeV proton kinetic energy.
\PX\ has the flexibility of time structure required and can supply the requisite intensity for a full
set of measurements.
Initial studies of the conversion experiments are promising: it is plausible that an order-of-magnitude
improvement over the Booster-era Mu2e is achievable.
The $\mu \rightarrow e \gamma$ and $\mu \rightarrow 3e$ experiments are being pursued at PSI, with an
approved upgrade to MEG to reach $\BR(\mu \rightarrow e \gamma < 6 \times 10^{-14}$ at 90\% CL and an
approved experiment for $\mu \rightarrow 3e$ at roughly $10^{-16}$.
Progressing past these experiments will require new experimental techniques since they are likely to be
background-limited.
One likely requirement is hard cuts on the data to eliminate backgrounds, and here the \PX\ intensity
can make up for acceptance loss.
In muonium-antimuonium, a two-order of magnitude improvement is likely possible, and both the flexible
time structure and intensity are essential.
Finally, the $g{-}2$ anomaly can be probed as well for the opposite sign of muons, permitting CPT tests and
systematic cross-checks presuming the current anomaly survives the Booster experiment and improvements in the
lattice calculations.
If new physics is seen, further investigation will be required.

\bibliographystyle{apsrev4-1}
\bibliography{muon/refs}
 % Bob B & Graham

%%%%%%%%%%%%%%%%%%%%%%%%%%%%%%%%%%%%%%%%%%%%%%%%%%%%%%%%%%%%
\chapter[Measurements of EDMs with \PX]{Measurements of Electric Dipole Moments with \PX}
\label{chapt:edm}
%%%%%%%%%%%%%%%%%%%%%%%%%%%%%%%%%%%%%%%%%%%%%%%%%%%%%%%%%%%%

\authors{Tim~Chupp, Susan~Gardner, Zheng-Tian~Lu, \\
Wolfgang~Altmannshofer,
Marcela~Carena,
Yannis~K.~Semertzidis}

\section{Introduction}
\label{edm:sec:intro}

%suitable for cribbing into an executive summary

A permanent electric dipole moment (EDM) $\bm{d}$ of a nondegenerate system is 
proportional to its spin $\bm{S}$, and it is nonzero if the energy of the
system shifts in an external electric 
field, in a manner controlled by $\bm{S}\cdot\bm{E}$. 
Such an interaction breaks the discrete symmetries of parity $P$ and 
time reversal $T$. According to the \CPT\ theorem, it reflects the existence of \CP\ violation, 
i.e., of the product of charge conjugation $C$ and parity $P$, as well.  
A nonzero EDM has yet to be established, and the existing experimental limits, 
as we report for a variety of systems in Table \ref{edm:tab:limits}, are extremely sensitive 
probes of new physics, probing the existence of new particles and new sources of 
\CP\ violation beyond the TeV scale. 
While the discovery of a nonzero EDM in any system must be our first and foremost
goal, increasingly sensitive EDM measurements in a variety of systems are also essential
to constraining and ultimately determining the nature of any new sources of \CP\ violation found. 

EDM searches of enhanced experimental sensitivity are a key step in the exploration of
the fundamental nature of our Universe, particularly in regards to the manner in which it came to have such a
markedly large baryon asymmetry of the universe (BAU).
Sakharov tells us that particle physics is capable of a microscopic explanation of the BAU, but baryon
number, $C$, and \CP\ violation are all required in concert with a departure from thermal equilibrium in 
order to realize a nonzero result~\cite{Sakarov:1967dj}.
Interestingly, all the necessary ingredients appear in the Standard Model (SM), but numerical assessments of
the BAU in the SM fall far short of the observed
value~\cite{HEP-PH/9312215,HEP-PH/9406289,Huet:1994jb,PHLTA.B155.36,NUPHA.B287.757,HEP-PH/9305274}.
This motivates the ongoing hunt for new sources of \CP\ violation. 
Currently we know as a result of 
the experiments at the $B$-factories, with key input from the Tevatron, that 
the Cabibbo-Kobayashi-Maskawa (CKM) mechanism 
serves as the dominant source of flavor and \CP\ violation 
in flavor-changing processes~\cite{Charles:2004jd,Isidori:2010kg}. 
Nevertheless, these definite conclusions do not end our search 
because we have not yet understood the origin of the BAU. 
\begin{table}
    \centering
    \caption[Upper limits on EDMs from different experiments]{Upper limits on EDMs ($|d|$) from different
        experiments.
        For the ``Nucleus'' category, the EDM values are of the $^{199}$Hg atom that hosts the nucleus.
        No \emph{direct} limit yet exists on the proton EDM, though such could be realized through a storage 
        ring experiment, possible at \PX\ and elsewhere; see Sec.~\ref{edm:sec:expt}.
        Here we report the best inferred limit in brackets, which is determined by asserting that the 
        $^{199}$Hg limit is saturated by $d_p$ exclusively.}
    \label{edm:tab:limits}
    \begin{tabular}{lccc}
    \hline\hline
     Category & Limit (\ecm)  & Experiment &  Standard Model Value (\ecm)   \\
    \hline
     Electron             & $1.0\times10^{-27}\,(90\%\, {\rm C.L.})$    &   YbF molecules in a beam \cite{Hudson:2011zz}  &   10$^{-38}$       \\
     Muon             &  $1.9\times10^{-19}\,(95\%\, {\rm C.L.})$   &   Muon storage ring \cite{Bennett:2008dy} &   10$^{-35}$       \\
     Neutron             & $2.9\times10^{-26}\,(90\%\, {\rm C.L.})$    &   Ultracold neutrons in a bottle \cite{Baker:2006ts}   &   10$^{-31}$       \\
     Proton             &  $[7.9\times10^{-25}]$   &   Inferred from $^{199}$Hg \cite{Griffith:2009zz} &   10$^{-31}$       \\
     Nucleus              & $3.1\times10^{-29}\,(95\%\, {\rm C.L.})$    &  $^{199}$Hg atoms in a vapor cell  \cite{Griffith:2009zz}   &   10$^{-33}$      \\
    \hline\hline
    \end{tabular}
\end{table}

Searches for EDMs have a particularly high priority in the LHC era. 
If new physics is discovered at the LHC, then EDMs offer a unique window on its nature.  
EDMs act as exquisitely sensitive 
probes of 
the existence of possible new \CP-violating phases beyond those present in the SM. 
In particular, EDMs are uniquely suitable to probing 
additional 
sources of \CP\ violation in the Higgs sector. 
On the other hand, in the absence of any direct new physics signals at the LHC, 
increasingly sensitive searches for EDMs 
provide 
access to energy scales well beyond our direct reach, probing new physics 
at \emph{much} higher scales as long as the new physics 
is assumed to contain sizable sources of \CP\ violation.

In the next section we explore these ideas in greater detail, 
describing the experimental observables, the theoretical frameworks to analyze
them, and the windows opened on TeV scale physics and beyond. 
In subsequent sections, we offer a broad overview of the current and planned
experiments, showing how the program at \PX\ can both complement and 
enhance current plans. Finally we turn to a discussion of the broader opportunities
the \PX\ concept offers for the study of new sources of \CP\ violation and close
with a summary. 

%%%%%%%%%%%%%%%%%%%%%%%%%%%%%%%%%%%%%%%%%%%%%%%%%%%%%%%%%%%%
\section{Physics Motivation}
\label{edm:sec:physics}
%%%%%%%%%%%%%%%%%%%%%%%%%%%%%%%%%%%%%%%%%%%%%%%%%%%%%%%%%%%%

\subsection{Overview}
\label{edm:subsec:overview}

In complex systems, the observation of a violation of a symmetry (or symmetries) of the SM 
constitutes evidence for physics beyond the Standard Model (BSM). 
Searches for permanent EDMs are being developed in a variety of systems, including
nuclei, atoms, molecules, and solids, and
are particularly prominent examples of such tests. Although \CP\ is not a symmetry of the SM, 
EDM searches are 
null tests nevertheless, because observing a nonzero EDM at current levels of sensitivity would attest to the
existence of physics beyond the electroweak~SM.
The SM without neutrino masses nominally has two sources of \CP\ violation: through a single phase $\delta$ in
the Cabibbo-Kobayashi-Maskawa (CKM) matrix, as well as through the $T$-odd, $P$-odd product of the gluon field
strength tensor and its dual, the latter product being effectively characterized in the full SM by the
parameter $\bar\theta$.
The CKM mechanism of \CP\ violation does give rise to nonzero EDMs; however, the first nontrivial contributions
to the quark and charged lepton EDMs come in three- and four-loop order, respectively, so that for the down quark
$|d_d| \sim 10^{-34}$ \ecm~\cite{Khriplovich:1985jr,HEP-PH/9704355}, whereas for the electron $|d_e| \sim
10^{-38}$ \ecm~\cite{Pospelov:1991zt} with massless neutrinos.
In the presence of neutrino mixing, the lepton EDMs can become much larger, though they are still orders of
magnitude beyond experimental reach~\cite{HEP-PH/9510306}.
Turning to the neutron EDM, $d_n$, a plurality of nonperturbative enhancement mechanisms can act.
There is a well-known chiral enhancement, under which the neutron EDM is estimated to be 
$|d_n| \sim 10^{-31}$--$10^{-33}$ \ecm~\cite{Gavela:1981sk,Khriplovich:1981ca,He:1989xj}, 
making it several orders of magnitude below 
current experimental sensitivity nonetheless---and likely experimentally inaccessible 
for decades. A distinct enhancement
arising from the nucleon's intrinsic flavor structure may also operate \cite{ARXIV:1202.6270}. 
The second mechanism, known as strong \CP\ violation, 
appears with an operator of mass dimension four; consequently, it is 
unsuppressed by any mass scale and need not be small, though it 
is bounded experimentally to be 
$\bar\theta < 10^{-10}$~\cite{Griffith:2009zz}, assuming 
no other sources of \CP\ violation are present. 
The lack of an established explanation for the small size of $\bar\theta$ is
known as the ``strong \CP\ problem.'' Possible explanations must be compatible, too, 
with $\delta \sim {\cal O}(1)$, which experimental measurements 
of \CP-violating observables 
in $B$-meson decays demand \cite{Abe:2001xe,Aubert:2001nu,Aubert:2002ic}.  
The manner of its resolution can also impact the possible 
numerical size of non-CKM sources of \CP\ violation, see Ref.~\cite{Pospelov:2005pr}
for a discussion. If the Peccei-Quinn mechanism operates, so that there is indeed
a new continuous symmetry \cite{Peccei:1977hh}
which is spontaneously and mechanically broken at low energies,
then we could win on two counts. 
There would be 
a new particle, the \emph{axion} \cite{Weinberg:1977ma,Wilczek:1977pj}, which
we may yet discover \cite{Asztalos:2009yp,Graham:2011qk}, 
and non-CKM sources of \CP\ violation could also be of ${\cal O}(1)$ in size. 
This particular resolution of the strong \CP\ violation problem would also imply 
that a nonzero EDM speaks to the existence of physics beyond the SM. 
In Sec.~\ref{edm:sec:broad}, we consider how \PX\ capacities for EDM searches could 
be adapted to a new sort of axion search~\cite{Graham:2011qk}. 

The electric dipole moment $d$ and magnetic moment $\mu$ of a nonrelativistic 
particle with spin $\bm{S}$ is defined via 
\begin{equation}
{\cal H} = - d\frac{\bm{S}}{S} \cdot \bm{E}
- \mu\frac{\bm{S}}{S} \cdot \bm{B},
\label{edm:eqn:edmdef}
\end{equation}
noting $\bm{d}\equiv d\bm{S}/S$ as well as $\bm{\mu}\equiv\mu\bm{S}/S$.
This expression in itself suggests
an experimental method: a nonzero $d$ is present if the energy splitting of the spin states
in a magnetic field is altered upon the reversal of an applied electric field---and this
method has been the basis of EDM searches for decades \cite{Ramsey:1982td}. 
The ${\bm{S}} \cdot \bm{E}$ interaction for a spin 1/2 particle 
has the following relativistic generalization 
\begin{equation}
{\cal L} = - d\frac{i}{2} \bar\psi \sigma^{\mu\nu} \gamma_5 \psi F_{\mu\nu} ,
\end{equation}
if \CPT\ symmetry is assumed. Such a dimension-five operator can be generated in 
a variety of well-motivated extensions of the SM, giving rise to 
EDMs substantially in excess of the predictions of the CKM model \cite{Pospelov:2005pr,Engel:2013lsa}. 
We suppose that the SM is the low-energy limit of a more fundamental theory in which 
new particles appear at some energy scale $\Lambda$.  At energies below that scale, the new 
degrees of freedom no longer appear, but their presence can still be felt through the
appearance of effective operators of dimension $D$, with $D>4$, which augment the SM. The new
effective operators, constructed from SM fields, are suppressed by a factor 
of $\Lambda^{D-4}$, and, moreover, 
respect the SU(3)${}_C \times$SU(2)${}_L \times$U(1) gauge symmetry of the SM. 
Upon imposing SU(2)${}_L\times$U(1) gauge invariance this chirality-changing, dimension-five operator becomes
of dimension-six in numerical effect.
Under naive dimensional analysis, the EDM of a fermion with mass $m_f$ can be estimated as $d_f\sim e
\sin\phi_{\CP} \, m_f/\Lambda^2$, where $\phi_{\CP}$ is a \CP-violating phase \cite{DeRujula:1990db}.
To give a sense of the sensitivity of the existing experiments, we note that the currently best measured
limit of the neutron EDM is $|d_n| < 2.9\times 10^{-26}$~e-cm \cite{Baker:2006ts}, whereas that of the
electron is $|d_e| < 1.05\times 10^{-27}$~e-cm \cite{Hudson:2011zz}---we report both limits at 90\%~CL.
If $\sin\phi_{\CP} \sim 1$, as $\sin\delta$ is in the CKM mechanism, then the current experimental limits
on the electron and neutron imply that $\log_{10} [\Lambda ({\rm GeV})] \sim 5$, where we employ a light
quark mass $m_f= m_q \sim 10\,{\rm MeV}$ in the neutron case.
Including a loop suppression factor of $\alpha/4\pi \sim 10^{-3}$, we estimate, crudely, that energy scales
of some 3 TeV are probed by current experiments, with the next generation of EDM experiments, anticipating a
factor of 100 in increased sensitivity, improving the energy reach by a factor of 10.

\begin{figure}
    \centering
    \includegraphics[width=\textwidth]{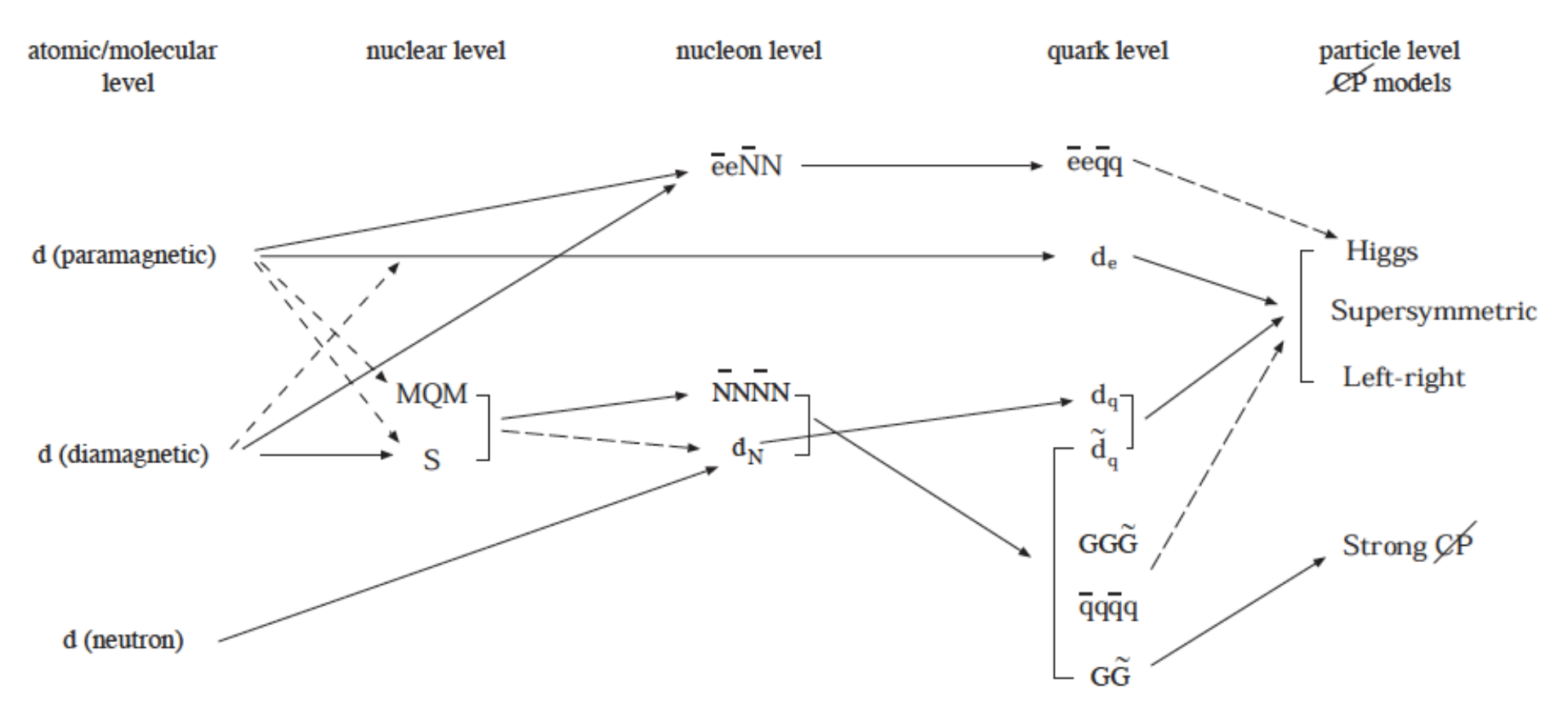}
    \caption[A flow chart for the analysis of EDMs]{A flow chart for the analysis of EDMs, connecting 
% Read from the right, N
        new sources of \CP\ violation at the TeV scale through the parameters of effective Lagrangians at 
        ever lower energy scales to give rise finally to nonzero lepton, nucleon, nuclear, atomic, 
        molecular, and solid-state EDMs.
        Empirical limits on EDMs in various systems in turn constrain different \CP-violating sources, as 
        indicated by arrows; dashed, as opposed to solid, lines note the existence of weaker constraints.
        From Ref.~\cite{Ginges:2003qt}.}
\label{edm:fig:flow}
\end{figure}

We have considered the new-physics reach of an EDM measurement in simple 
systems, but the greatest experimental sensitivities can be found in complex
systems, most notably in atoms, molecules, and solids (crystals). 
The connection between the empirical EDM 
limits in such systems and new sources of \CP\ violation at the TeV scale is made
indirectly, through multiple theoretical frameworks, each with its own range of
validity tied to a particular energy scale. 
Tracking the manner in which TeV-scale sources of \CP\ violation 
emerge in the low-energy theoretical frameworks 
appropriate to the descriptions of 
nuclei, atoms, and molecules is a richly complex task; 
we illustrate it, 
schematically and incompletely, 
in Fig.~\ref{edm:fig:flow} and refer to Ref.~\cite{Engel:2013lsa} for a recent
review. 
In the following discussion, we start at the energy scales of the systems
in question and evolve upward, ending with a discussion of the models which 
evince new TeV-scale sources of \CP\ violation.

\subsection{EDMs of Atoms and Molecules}
\label{edm:subsec:atoms}

Atoms and molecules differ from fundamental particles such as electrons, muons, and taus, as well as
from neutrons, protons, deuterons, and indeed nuclei, 
in that their composite nature guarantees 
that their EDMs vanish in the point-like, nonrelativistic limit 
even if their constituents have nonzero EDMs---this is the so-called 
Schiff theorem \cite{Schiff:1963zz}. This effect 
suppresses the visibility of an EDM in an experimental measurement, but enhancements
also arise because the cancellation can be strongly violated by relativistic
and finite-size effects. The former effect can give such an atomic experiment sensitivity to $d_e$, whereas 
the latter effect can give rise to a nonzero atomic 
EDM through  $P$-odd, $T$-odd nuclear moments, of which the ``Schiff moment'' is typically the driving
contribution. If the electrons have nonzero total spin, then
a magnetic quadrupole moment (MQM) can also contribute~\cite{Ginges:2003qt}. 
The precise role of the mechanisms in realizing a nonzero EDM 
depends on the nature of the particular atom in question. 
Atoms are broadly classified as either paramagnetic or diamagnetic. 
In paramagnetic atoms, the atomic electrons have an unpaired spin, and in this case 
relativistic effects are most 
important, making EDMs  in paramagnetic atoms or molecules sensitive to $d_e$, strikingly
so in heavy atoms with atomic number $Z$, scaling as $Z^3\alpha^2$ \cite{Sandars:1965,Sandars:1966}. 
This enhancement can be interpreted in terms of an enhanced, ``effective'' electric field. 
Polar diatomic molecules, such as TlF or YbF, can evince even larger enhancements of the effective
electric field \cite{Sandars:1967,Ginges:2003qt}. 
Paramagnetic systems are also sensitive to $P$-odd, $T$-odd electron-nucleon interactions. 
In contrast, diamagnetic atoms have only paired electronic spins, so that the EDMs of
these systems are particularly sensitive to the tensor $P$-odd, $T$-odd electron-nucleon interactions, which
can act to unpair electrons in a closed shell. However, 
once hyperfine effects are included, other $P$-odd, $T$-odd electron-nucleon interactions, as well as 
$d_e$, can also contribute to the EDM \cite{Ginges:2003qt,KhripLamor}. 
As we have noted, 
atomic EDMs can also be induced by $P$-odd, $T$-odd nuclear moments, 
whose effects operate in both diamagnetic and paramagnetic
atoms, 
though the effects are much less difficult to probe in diamagnetic systems. 
Moreover, nuclear deformation 
and atomic state mixing can give rise to marked enhancements. For example, 
in a heavy diamagnetic atom of a rare isotope, for which the nucleus has 
octupole strength \cite{Engel:1999np,Flambaum:2002mv} 
or a permanent octupole deformation \cite{Auerbach:1996zd,rf:Spevak1997,Dobaczewski:2005hz}, 
the $T$-odd, $P$-odd 
charge distribution in the nucleus, 
characterized by the Schiff moment, is predicted to be significantly enhanced relative to $^{199}$Hg. 
We refer to Refs.~\cite{Ginges:2003qt,Dzuba:2012bh,Engel:2013lsa} for detailed treatments of
all these issues.

%\subsection{EDMs of rare atoms with rare isotopes}
\subsection{Rare Atom EDMs}
\label{edm:subsec:rare}

In what follows we focus 
on the systems of immediate relevance to experiments at the first stage of \PX: atoms
with rare nuclear isotopes, for which their EDMs can be markedly enhanced, and neutrons.
In connecting experimental limits on atomic EDMs to fundamental sources of \CP\ violation, 
multiple layers of theoretical analysis are required, as we have
noted. Specifically, 
it is the task of atomic theory to compute an atomic EDM in terms of nuclear inputs, 
such as the $P$-odd, $T$-odd nuclear moments, and it is the task of nuclear theory to 
compute these nuclear moments in terms of the nucleon EDMs and 
$P$-odd, $T$-odd nucleon-nucleon interactions. We postpone, for the moment, discussion of
the connection of these hadronic inputs to quantities realized in terms of quark and gluon degrees of
freedom---and return to them in the next section, in which we discuss the neutron EDM as well. 
To illustrate the connections explicitly, yet concisely, we consider
a \CP-odd, effective Lagrangian in hadron degrees of freedom, at the nuclear scale, 
which can generate both the
Schiff moments and the germane $P$-odd, $T$-odd electron-nucleon interactions. Namely \cite{Pospelov:2005pr}, 
\begin{equation}
{\cal L}_{\rm eff}^{\rm nuclear} = {\cal L}_{e-{\rm edm}} + {\cal L}_{eN} + {\cal L}_{\pi NN}, 
\end{equation} 
where ${\cal L}_{e-{\rm edm}} = -i (d_e/2)\bar e (F\sigma) \gamma_5 e$, 
\begin{eqnarray}
{\cal L}_{eN} &=&
C_S^{(0)} \bar e i\gamma_5 e \bar N N + 
C_P^{(0)} \bar e e \bar N i\gamma_5 N + 
C_T^{(0)} \epsilon_{\mu\nu\alpha\beta} \bar e \sigma^{\mu\nu} e \bar N \sigma^{\alpha\beta} N 
\nonumber \\ 
&&+ 
C_S^{(1)} \bar e i\gamma_5 e \bar N \tau^3 N + 
C_P^{(1)} \bar e e \bar N i\gamma_5 \tau^3 N + 
C_T^{(1)} \epsilon_{\mu\nu\alpha\beta} \bar e \sigma^{\mu\nu} e \bar N \sigma^{\alpha\beta} \tau^3 N,
\end{eqnarray} 
and 
\begin{eqnarray}
{\cal L}_{\pi NN} &=& 
-(i/2) \bar N (F\sigma) (d^{(0)} + d^{(1)} \tau^3) \gamma_5 N + 
{\bar g}_{\pi NN}^{(0)} \bar N \tau^a N\pi^a + 
{\bar g}_{\pi NN}^{(1)} \bar N  N\pi^0 
\nonumber \\ 
&&+ 
{\bar g}_{\pi NN}^{(2)} (\bar N \tau^a N\pi^a - 3 \bar N \tau^3 N \pi^0 ) + \dots \;,
\label{edm:eqn:piN}
\end{eqnarray} 
noting that the superscripts indicate interactions of isoscalar (0), 
 isovector (1), or isotensor (2) character and that 
$d^{(0)} + d^{(1)}$ and $d^{(0)} - d^{(1)}$ contribute to 
$d_p$ and $d_n$, respectively. 
In recent years the construction
of ${\cal L}_{\pi NN}$ has been revisited within the context of heavy-baryon chiral perturbation 
theory (HBChPT), 
a low-energy, effective field theory in which 
the nucleons are nonrelativistic and a momentum expansion effected in the context of interactions
which respect the chiral symmetry of QCD serves as an organizing 
principle, where we refer to 
Ref.~\cite{Bernard:2007zu} for a review. 
The result of this analysis \cite{Mereghetti:2010tp,Maekawa:2011vs,deVries:2012ab} 
yields terms which map to those articulated in 
Eq.~(\ref{edm:eqn:piN}), as well as explicit $T$-odd and $P$-odd contact interactions which 
capture the most important of the short-range $NN$ interactions \cite{Engel:2013lsa}. 
Employing Eq.~(\ref{edm:eqn:piN}), we note that the EDM of thallium atom, which is 
 paramagnetic, is given by \cite{Pospelov:2005pr} 
\begin{equation}
d_{^{205}{\rm Tl}} = -585 d_e - e (43 {\rm GeV})\times (C_S^{(0)} - 0.2 C_S^{(1)})\,, 
\end{equation}
though we caution the reader that the ultimate relative role of the terms is sensitive to
the precise BSM model. The experimental limit of 
$|d_{^{205}{\rm Tl}}| \le 9.4 \times 10^{-25}$ \ecm~at 90\% C.L., which is currently the 
most stringent limit in any paramagnetic atom, yields 
$|d_e| \le 1.6 \times 10^{-27}$ \ecm~at 90\% C.L. \cite{Regan:2002}
if one assumes the atomic EDM is saturated by $d_e$. 
Table \ref{edm:tab:limits} reveals that the recent EDM limit from YbF \cite{Hudson:2011zz} 
is somewhat stronger, but the possibility that the study of $^{211}$Fr at \PX, which we 
discuss in  Sec.\ref{edm:sec:FrEDM}, could yield an improved sensitivity of $\sim 10^3$ to $d_e$ 
tantalizes. 

For diamagnetic atoms, such as $^{129}$Xe, $^{199}$Hg, $^{223}$Rn, or $^{225}$Ra, 
there are two main contributions to the atomic EDM, as we have mentioned---a 
tensor electron-nucleon interaction~\cite{rf:Martensson1985}, controlled by 
$C_T^{(0)}$, and the $P$-odd and $T$-odd nuclear moments, of which the Schiff moment ${\cal S}$ 
appears in leading order in an expansion about the point-like limit. Indeed, it is the only 
$T$-odd, $P$-odd nuclear moment which generates an EDM in the current context. 
Typically we can characterize 
the EDM of a diamagnetic atom, $d_{\rm dia}$, in the following parametric way: 
\begin{equation}
d_{\rm dia} = d_{\rm dia}({\cal S}[{\bar g}_{\pi NN}^{(i)},d^{(i)}], C_S^{(i)},C_P^{(i)},C_T^{(i)},d_e). 
\label{edm:eqn:ddia}
\end{equation}
The Schiff moment $\mathbf{{\cal S}}$ tends to play a driving role, and this can be understood in 
the following way. The contribution of ${\mathbf{\cal S}}$ 
to the $T$-odd, $P$-odd nuclear electrostatic potential generates, in essence, 
an effective electric field in the nucleus that has a permanent projection along $\bm{I}$, 
the total nuclear angular momentum, and which is naturally $T$-odd and $P$-odd. 
This effective electric field polarizes the atomic electrons and thus gives rise to an 
atomic EDM \cite{Ginges:2003qt}. 
For an effective pion-mediated nucleon-nucleon interaction, the contributions to 
the Schiff moment can be decomposed into isospin components given by~\cite{rf:Engel2005}
\begin{equation}
S = g(a_0{\bar g}_{\pi NN}^{(0)}+a_1{\bar g}_{\pi NN}^{(1)}+a_2{\bar g}_{\pi NN}^{(2)}),
\label{edm:eqn:Schifmomentgpi}
\end{equation}
where $g$ is the usual, \CP-conserving $\pi NN$ coupling constant, $g\equiv 13.5$,
and the $P$-odd, $T$-odd physics is contained in $\bar g_{\pi NN}^{(0,1,2)}$, which are dimensionless.
We note that the latter contribute to $d_{n,p}$ as well. 
In general the $a_i$, in units of $e~{\rm fm}^3$, 
 represent the polarization of 
the nuclear charge distribution by a specific isospin component 
of the $P$-odd, $T$-odd interaction and can reflect intricate cancellations. 
The Schiff moment in $^{199}$Hg has been computed by different groups~\cite{Ginges:2003qt,Ban:2010ea}, 
 employing Skyrme 
effective interactions and, most recently, fully self-consistent mean field (Hartree-Fock-Bogoliubov) 
computations with core-polarization effects included in a unified way~\cite{Ban:2010ea}. 
A dispersion of the results with different Skyrme interactions is reflective of
the theoretical systematic error~\cite{rf:deJesus2005}. Beyond this, 
some dispersion in the collected results exists~\cite{Ban:2010ea,Ginges:2003qt}. 
It is crucial to note, however, that large collective enhancements of the 
Schiff moment, relative to the single-particle contributions, can occur under special
conditions. 
For nuclei with strong octupole collectivity, 
the Schiff moment 
may be significantly enhanced relative to $^{199}$Hg due to
the large intrinsic dipole moment and, 
for permanently deformed nuclei, 
the closely spaced, opposite parity levels that arise. 
In this picture, the enhanced Schiff moment for deformed systems can be written~\cite{rf:Spevak1997}
\begin{equation}
S\approx 0.05e\frac{\beta_2\beta_3^2ZA^{2/3}r_0^3\eta}{ E_+-E_-},
\label{edm:eqn:SchiffSchematic}
\end{equation}
where $E_+$ and $E_-$ are the energies of opposite parity states and
 $\eta$ is the matrix element 
of the effective  $T$-odd and $P$-odd interaction 
between nucleons. In the presence of rigid octupole deformation, the computation of
this latter quantity is expected to be more robust~\cite{Engel:2013lsa}. 
Here $\beta_2$ and $\beta_3$ are the quadrupole and octupole deformation parameters---and
are experimentally accessible as we shall detail.

\subsubsection{Octupole Deformation and Schiff Moment Enhancements}
\label{edm:subsubsec:OctupoleMotivation}

 Experimental programs in two important octupole-enhanced systems, 
$^{225}$Ra and $^{221/223}$Rn are underway, 
and the experimental details are presented in section~\ref{edm:sec:expt}. 
For $^{225}$Ra, with a half-life of 14.9 days, a great deal has been studied 
regarding its nuclear structure, including the 55 keV spacing of the ground $1/2^+$ state 
and the lowest $1/2^-$ state of the negative parity band, suggesting that $^{225}$Ra 
is octupole deformed~\cite{rf:Helmer1997}. Calculations including work by Engel 
and collaborators, who estimate the $a_{0,1,2}$, confirm that these quantities are 
indeed enhanced~\cite{rf:Engel2005}. 
Experimental studies also indicate that $^{226}$Ra is octupole deformed~\cite{rf:Cocks1999}. 
Recently the first direct evidence of octupole deformation with a 
determination of $\beta_3$ in $^{224}$Ra has been established 
through measurements of Coulomb excitation of 2.85 MeV/a.m.u. 
rare-isotope beams at REX-ISOLDE (CERN)~\cite{rf:Gaffney2013}, 
strengthening the confidence in the size of the Schiff moment. 
Though a precise estimate of the enhancement 
relative to $^{199}$Hg or $^{129}$Xe is hampered by the difficulty 
of accurately calculating the $a_i$, particularly for $^{199}$Hg~\cite{rf:deJesus2005,Ban:2010ea}, 
as a rough estimate we take $a_0=0.01$ for $^{199}$Hg and $a_0=5$ for $^{225}$Ra 
indicating an enhancement of 500 for the isoscalar contributions. 
Similar enhancements are expected for $a_1$ and $a_2$. 
In work at REX-ISOLDE~\cite{rf:Gaffney2013}, 
octupole collectivity was also determined for $^{220}$Rn 
indicating a similar $\beta_3$ compared to $^{224}$Ra, 
but as evidence of octupole vibrations and not of permanent octupole deformation. 
In this case, the formula of Eq.~(\ref{edm:eqn:SchiffSchematic}) would not apply; nevertheless, 
enhancements of the Schiff moment may occur \cite{Engel:1999np,Flambaum:2002mv}. 
Though the  spins and parities for $^{221}$Rn have not 
been determined for any states, three new gamma-ray lines 
between 200 keV and 300 keV excitation were identified 
in a subsequent experiment at REX-ISOLDE. 
Further measurements are necessary to determine the nature of $^{223}$Rn.

\begin{figure}
    \centering
    \includegraphics[scale=0.45]{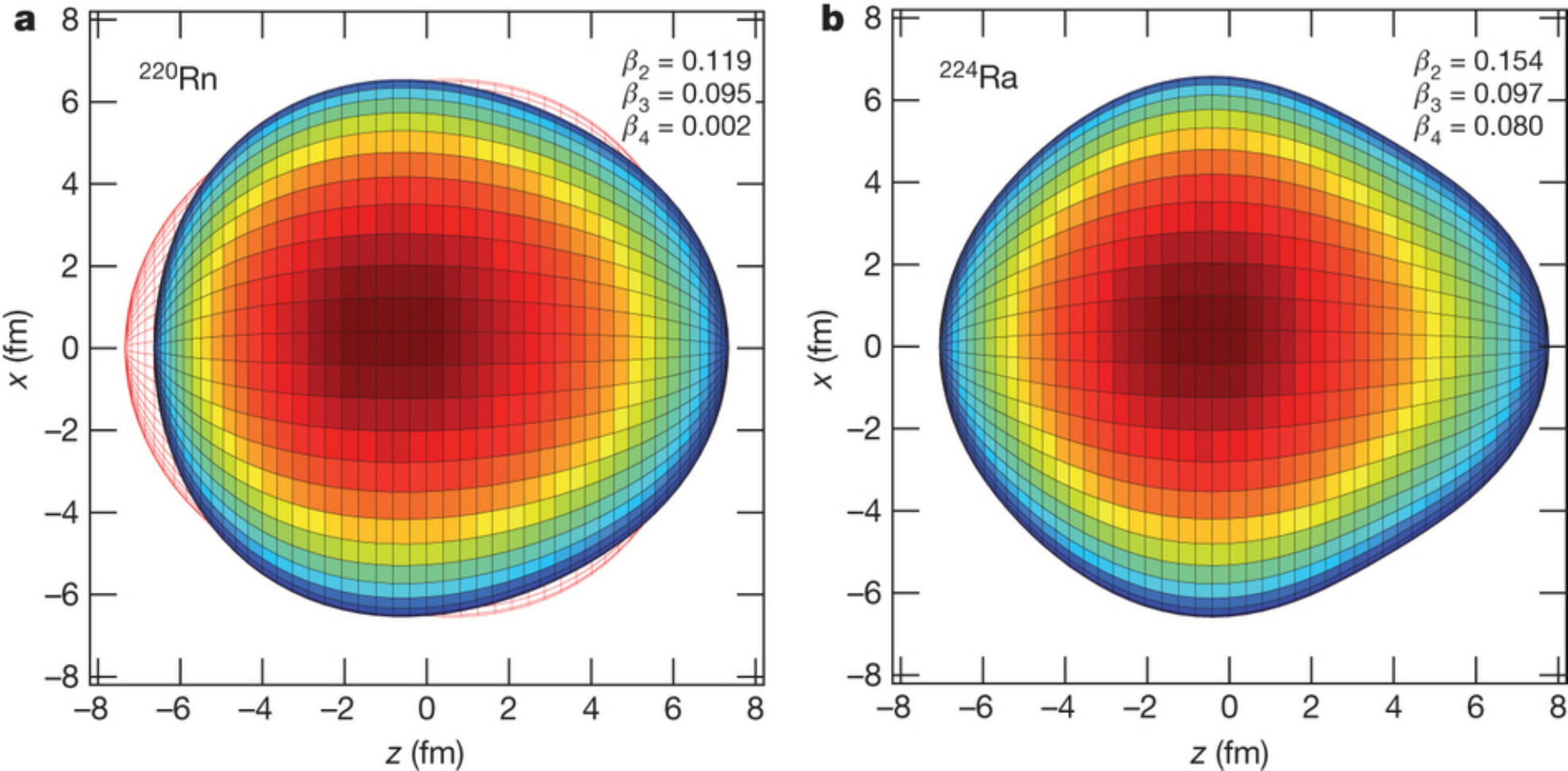}
    \caption[Representation of the shapes of $^{220}$Rn and $^{224}$Ra]{Representation of the shapes of 
        $^{220}$Rn and $^{224}$Ra. 
        The left panel depicts vibrational motion between two surfaces, the second indicated by the red 
        hatched line.
        The right panel denotes static deformation in the intrinsic frame.
        The color scale, from blue to red, represents the $y$-values of the surface.
        From Ref.~\cite{rf:Gaffney2013}, to which we refer for all details.}
    \label{edm:fig:pear}
\end{figure}

Although the EDMs of atomic systems, notably $^{199}$Hg \cite{Griffith:2009zz}, 
can be measured with much higher precision than that of the neutron, 
these results currently probe the underlying physics
at a level crudely commensurate to that of the neutron EDM limit, 
to the extent that they are comparable. This is simply a concrete consequence of
the Schiff theorem in $^{199}$Hg and need not hold generally. Indeed, one
can be optimistic in regards to the prospects for EDM studies in $^{225}$Ra and
other deformed systems. A measurement of the EDM in $^{225}$Ra of much less sensitivity
than that of $^{199}$Hg 
can probe the underlying physics to a comparable level, with 
improvements to the sensitivity of the current $^{199}$Hg limit yielding new-physics
sensitivity well beyond that. 

\subsection{EDMs of Light Nuclei}
\label{edm:subsec:few}

Advances in storage ring technology make sensitive EDM experiments of electrically charged particles possible.
The possible candidate systems include not only the proton and the muon, but also light nuclei, such as the
deuteron and $^3$He.
At \PX\ the first two
possibilities are more readily realizable, though there are plans afoot to realize the
latter elsewhere.
We refer to Sec.~\ref{edm:subsec:pEDM} for a description of the basic empirical concepts as well as an
overview of the possibilities.

The study of light nuclei appeal because the combination of 
chiral and isospin symmetry serve as powerful tools in distinguishing
the various possible \CP-violating interactions 
which appear in Eq.~(\ref{edm:eqn:piN}). 
It has been known for some time that the deuteron EDM is 
a particularly sensitive discriminant of its 
CP-violating source, notably $g_{\pi NN}^{(1)}$~\cite{Flambaum:1984fb}, 
where we refer to Ref.~\cite{Engel:2013lsa} for a review. 
Recently the deuteron EDM has been revisited \cite{Lebedev:2004va,deVries:2011re}, 
affirming the earlier arguments.
In HBChPT, the EDMs of the proton, neutron, and light nuclei can be analyzed within a single framework.
Consequently, exploiting the distinct way the various \CP-violating sources appear in leading-order HBChPT,
it has been shown that 
a systematic program of EDM measurements in these systems 
 could potentially disentangle their \CP-violating sources~\cite{deVries:2011re,deVries:2011an}. 
A proton storage ring EDM experiment would be a significant step along this path. 

We discuss the physics implications of a muon EDM experiment of improved sensitivity
at the close of Sec.~\ref{edm:subsec:tev}. 

\subsection{\CP-Violating Sources at $\sim$1 GeV}
\label{edm:subsec:had}

In this section we consider the connection between the low-energy constants of 
${\cal L}_{\pi NN}$, detailed in Eq.~(\ref{edm:eqn:piN}), to underlying quark and gluon 
degrees of freedom, where we wish to consider low-energy sources of \CP\ violation beyond the SM. 
Thinking 
broadly and systematically we organize the expected contributions
in terms of the mass dimension of the possible \CP-violating operators appearing in an effective field
theory with a cutoff of $\sim 1$ GeV~\cite{Pospelov:2005pr}: 
\begin{eqnarray}  
{\cal L}_{\Lambda}&=&
\frac{\alpha_s\bar\theta}{8\pi} 
\epsilon^{\alpha\beta\mu\nu}F_{\alpha\beta}^a F_{\mu\nu}^a - \frac{i}{2} \sum_i d_i \bar \psi_i F_{\mu\nu}\sigma^{\mu\nu} \gamma_5 \psi_i 
- \frac{i}{2} 
\sum_{i\in u,d,s} {\tilde d}_i \bar \psi_i F_{\mu\nu}^a t^a\sigma^{\mu\nu}\gamma_5 \psi_i  \nonumber\\
&&+ \frac{1}{3} w f^{abc} F_{\mu\nu}^a \epsilon^{\nu\beta\rho\delta}F_{\rho\delta}^b
F_\beta^{\,\,\mu,c}
+ \sum_{i,j} C_{ij} (\bar \psi_i \psi_i)(\bar \psi_j i\gamma_5\psi_j) + \dots 
\label{edm:eqn:Leff}
\end{eqnarray}
with $i,j\in u,d,s,e,\mu$ unless otherwise noted---all heavier degrees of freedom have been integrated out.
The leading term is the dimension-four strong~\CP\ term, proportional to the parameter $\bar\theta$.
Even in the presence of axion dynamics, a higher dimension operator could \emph{induce} a nonzero value of
$\bar\theta$~\cite{Bigi:1990kz,Pospelov:2005pr}; thus we retain it explicitly.
The balance of the terms are the nominally dimension-five fermion EDMs $d_i$ and quark chromo-EDMs (CEDM)
$\tilde d_i$, though they are effectively of dimension 6
once SU(2)$_{L}\times$U(1) symmetry is imposed.
Moreover there are the dimension-six Weinberg three-gluon operator, $w$, and \CP\ violating 4 fermion
operators, $C_{ij}$.
This list is not exhaustive even within the restricted operator dimensions we have considered.
To see this we consider the leading dimension set of operators in SM fields under SU(2)$_L\times$U(1) gauge
invariance at the electroweak scale, prior to electroweak symmetry breaking.
Turning to Ref.~\cite{Grzadkowski:2010es},
one finds that there are in total 
19 dimension-six operators in terms of gauge, Higgs, and 
fermion degrees of freedom which can contribute to an 
EDM.\footnote{In Ref.~\cite{Grzadkowski:2010es},
certain operators are of the same form for $f\in u,d,e$, making our 
tally consistent with Ref.~\cite{Engel:2013lsa}.} 
After electroweak symmetry breaking, 
certain of the terms
becomes those enumerated in Eq.~(\ref{edm:eqn:Leff}); the balance are largely four-fermion
operators which functionally become contributions of dimension 8 under 
SU(2)$_L\times$U(1) gauge invariance. We refer to Ref.~\cite{Engel:2013lsa} 
for an exhaustive analysis. 
Various extensions of the SM can generate 
the low-energy constants which appear, 
so that, in turn, EDM limits thereby constrain the new sources of \CP\ violation which appear
in such models. In connecting the Wilson coefficients of 
these operators and hence models of new physics
to the low-energy constants of
${\cal L}_{\pi NN}$, Eq.~(\ref{edm:eqn:piN}), requires the computation of nonperturbative 
hadron matrix elements. Parametrically, we have~\cite{Pospelov:2005pr}
\begin{eqnarray}
d_n &=& d_n({\bar \theta},d_i, {\tilde d}_i,w,C_{ij}) \nonumber \\
{\bar g}_{\pi NN}^{(i)} 
&=& 
{\bar g}_{\pi NN}^{(i)}({\bar \theta},d_i, {\tilde d}_i,w,C_{ij}). 
\label{edm:eqn:match}
\end{eqnarray} 
Several computational aspects must be considered in connecting a model of new physics at the TeV scale to the
low-energy constants of Eq.~(\ref{edm:eqn:Leff}).
After matching to an effective theory in SM degrees of freedom,
there are QCD evolution and operator
mixing effects, as well as flavor thresholds, 
involved in realizing the Wilson coefficients at a scale of $\sim$1 GeV. 
Beyond this, the hadronic matrix elements must be computed. 
We refer the reader to a detailed review of all these issues, including 
recent technical developments in this area \cite{Engel:2013lsa}. 
Typically QCD sum rule methods, or a SU(6) quark model, 
have been employed in the computation of the matrix elements \cite{Pospelov:2005pr}. 
For the neutron, 
we note Ref.~\cite{ARXIV:0806.2618} for a comparative review of different methods.
Lattice gauge theory can also be used to compute the needed proton and neutron matrix elements, and the
current status and prospects for lattice-QCD calculations are 
presented in Sec.~\ref{lqcd:subsec:EDM}.
So far lattice-QCD methods have only been 
used to compute the matrix element associated with $\bar\theta$,
but calculations of the 
dimension-six operators needed to make predictions for BSM theories are also underway.

To give a concrete yet simple example, for the neutron we note the estimate \cite{Pospelov:2005pr}
\begin{equation}
d_n^{\rm est} = \frac{8\pi^2 | \langle {\bar q} q \rangle|}{M_n^3} 
\left[ \frac{2\chi m_\ast}{3} e(\bar \theta - \theta_{\rm ind}) 
+ \frac{1}{3}(4d_d - d_u) + \frac{\chi m_0^2}{6}(4e_d {\tilde d}_d - e_u {\tilde d}_u) \right], 
\end{equation}
where terms which are naively of dimension 6 and higher have been neglected 
and $\theta_{\rm ind}$ is given in terms of $\tilde d_q$~\cite{Bigi:1990kz,Pospelov:2005pr}. 
The study of the EDM of $^{225}$Ra, in contrast, brings in sensitivity to 
${\bar g}_{\pi NN}^{(1)}$ and thus to the combination ${\tilde d}_u - {\tilde d}_d$. 
We refer to Ref.~\cite{Pospelov:2005pr} for all details. We note in passing
that $d_n$ and $d_p$ have also been analyzed 
in chiral perturbation theory employing the sources of Eq.~(\ref{edm:eqn:piN}), where 
we refer to Ref.~\cite{Engel:2013lsa} for a review, 
as well as in light-cone QCD~\cite{Brodsky:2006ez}. 
					
The electron-nucleon couplings, $C_{S,P,T}^{(i)}$, also play a role in atomic EDMs; they 
receive contributions from semileptonic, four-fermion couplings $C_{qe}$.
The hadronic matrix element which connects these quantities
can be computed using low-energy theorems for the matrix element of quark bilinears 
in the nucleon \cite{Pospelov:2005pr}.

\subsection{EDMs and New, TeV-Scale Sources of \CP\ Violation}
\label{edm:subsec:tev}

A variety of well-motivated extensions of the SM can generate EDMs substantially 
in excess of the predictions of the CKM model \cite{Pospelov:2005pr,Engel:2013lsa}. 
This includes  models with an 
extended Higgs sector, with manifest left-right symmetry at sufficiently
high energy scales, with extra spacetime dimensions, and  with weak-scale
supersymmetry, that  can generate EDMs through dimension-five operators, though, as we have noted, 
they are of dimension-six in numerical effect. 
Models with weak-scale
supersymmetry are particularly appealing in that they can potentially
resolve a variety of theoretical problems at once, 
yielding a cosmic baryon asymmetry through an electroweak phase transition
more efficiently than in the SM \cite{Carena:1996wj,Delepine:1996vn}, as well 
as providing a dark-matter candidate \cite{Jungman:1995df,Balazs:2004ae}. 
These models have and have had
significant implications for flavor physics. Furthermore,  limits from the 
nonobservation of EDMs and, more generally, of 
new interactions, constrain the appearance 
of new degrees of freedom \cite{Buchmuller:1985jz,Isidori:2010kg}. 
In the LHC era, it has been possible to search for the predicted new degrees of freedom 
directly, and all searches
have yielded null results thus far---though the campaign is far from over. 
As we have noted, EDMs retain their interest 
even if no new physics signals are observed at the LHC 
since,
modulo theoretical uncertainties and assumptions, a discovery would reveal the energy scale of new physics 
 beyond LHC reach.

\begin{figure}
    \centering
    \includegraphics[width=0.5\textwidth]{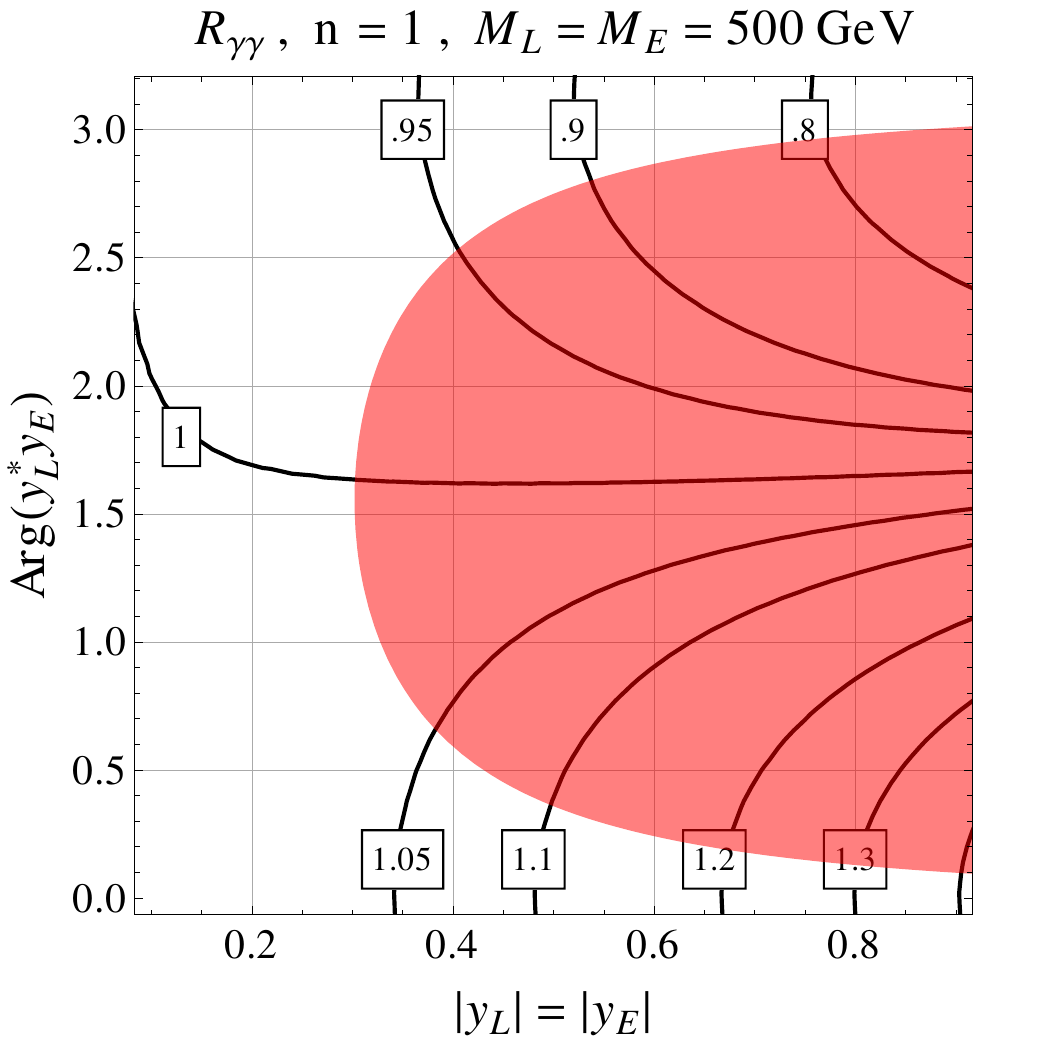}
    \caption[Electron EDM and $h\to\gamma\gamma$ rate in a model with vector-like leptons of charge~2]{%
        Electron EDM and $h\to\gamma\gamma$ rate in a model with exotic vector-like leptons of charge~2. 
        Possible modifications of the $h \to \gamma\gamma$ rate are shown in the plane of the Yukawa 
        couplings of the exotic leptons versus the relevant \CP-violating phase.
        Note that curves of fixed $R_{\gamma\gamma}$, defined as the modified $h\to\gamma\gamma$ rate in 
        units of the SM $h\to\gamma\gamma$ rate, are shown as solid black lines.
        The red region is excluded by the experimental limit on the electron EDM.
        From Ref.~\cite{Altmannshofer:HiggsEDMs}.}
\label{edm:fig:higgs}
\end{figure}

The discovery of a Higgs-like boson at the LHC~\cite{Aad:2012tfa,Chatrchyan:2012ufa} is a milestone in our
understanding of the mechanism of electroweak symmetry breaking, and provides a consistent mathematical
formulation of the SM of particle physics.
Given the absence of any direct signals of new physics at the LHC, as yet, attention is being focused on the
study of the properties of the Higgs-like boson.
At current sensitivities, the accessible production modes and decay rates are overall in reasonable agreement
with the predictions of the SM Higgs.
In particular, there is strong experimental evidence that the newly discovered particle decays with an
appreciable branching fraction to ZZ*, in spite of the very strong phase space suppression.
Hence, that indicates that the particle couples to the Z gauge boson at tree level as expected.
Previous hints of an enhanced $h \to \gamma \gamma$ rate persist in the ATLAS data~\cite{ATLAS:2013oma}, but
are not confirmed by the latest CMS analysis~\cite{CMS:ril}.
Detailed studies of the decays of the newly discovered particle into 4 leptons show kinematic distributions
consistent with a spin zero particle.
An assignment of spin 2 cannot be conclusively excluded but would demand a tuning of the tensor couplings to
fit current data.
Moreover, first studies of the \CP\ properties of the Higgs-like boson in the $h \to ZZ$ channel strongly
favor the scalar over the pseudoscalar hypothesis~\cite{ATLAS:2013nma,CMS:xwa}, as predicted by the SM.

It is of great importance to use all possible experimental handles to test possible departures from SM
properties of the newly discovered particle.
Observing evidence of departures would conclusively show that, even if the new Higgs-like particle is the one
responsible for electroweak symmetry breaking, the SM is an effective theory that requires extensions.
EDM experiments give complementary indirect information on the \CP\ properties of the Higgs-like boson.
In particular, the current experimental limits put strong constraints on possible \CP\ violation in the $h \to
\gamma\gamma$ and also the $h \to Z\gamma$ decays assuming that the couplings of the Higgs-like boson to
light fermions are SM-like~\cite{McKeen:2012av,Fan:2013qn,Altmannshofer:HiggsEDMs}.
EDM limits imply---barring accidental cancellations---that possible new physics which modifies the $h \to
\gamma\gamma$ and $h \to Z\gamma$ rates has to be approximately \CP\ conserving.
Simple extensions of the SM, which can modify the diphoton rate, are models with extra vector-like fermions.
Such models contain a physical \CP\ violating phase that can induce fermion EDMs at the two-loop level, through
Barr-Zee diagrams.
In regions of parameter space that lead to visible new physics effects in $h \to \gamma\gamma$, this phase is
constrained to be below $\lesssim 0.1$ (see Fig.
\ref{edm:fig:higgs}).
Moreover, EDMs give also the opportunity to obtain indirect information on the couplings of the Higgs-like
boson to the first generation of SM fermions, which are not directly experimentally accessible.
For example, current EDM limits already constrain possible imaginary parts in the couplings to electrons, up
quarks, and down quarks to be at most one order of magnitude below the corresponding SM Yukawa couplings.

Despite the absence of direct evidence for supersymmetric particles at the LHC, models of supersymmetry
(SUSY) remain among the most well-motivated and popular extensions of the Standard Model.
Besides direct searches, there exist various ways to probe SUSY models indirectly with low energy observables.
The minimal supersymmetric extension of the Standard Model (MSSM) generically contains numerous new sources
of \CP\ violation.
Parameterizing the soft SUSY breaking terms in the most general way, one finds O(50) new \CP\ violating phases
in the MSSM.
Most of them are connected to new sources of flavor violation, but even in the limit of completely
flavor-blind soft SUSY breaking terms, one is still left with 6 physical \CP\ phases that can be probed with
EDMs.
We discuss first the flavor blind case and come back to ``flavored EDMs'' at the end of the section.

\begin{figure}\centering
\includegraphics[width=0.6\textwidth]{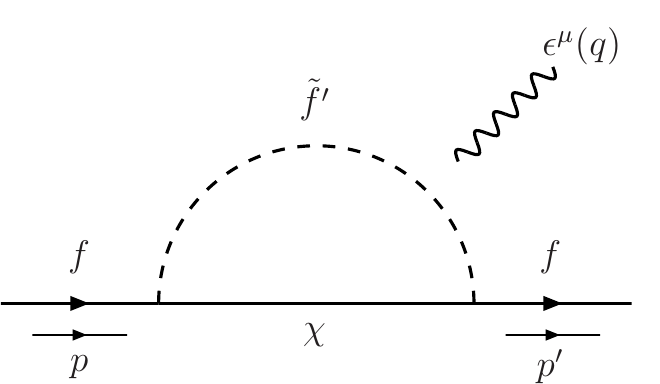}
    \caption[Generic one-loop SUSY diagram giving rise to a fermion EDM or CEDM]{Generic one-loop SUSY 
        diagram giving rise to a fermion EDM or CEDM.
        From Ref.~\cite{Ellis:2008zy}.}
    \label{edm:fig:diagram1}
\end{figure}

\begin{figure}
    \centering
    \includegraphics[width=0.4\textwidth]{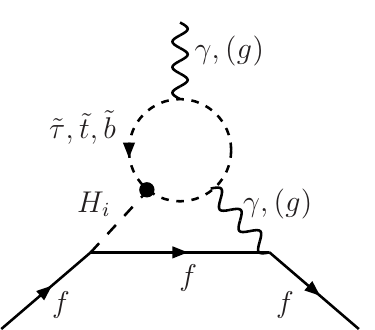} \hfill
    \includegraphics[width=0.4\textwidth]{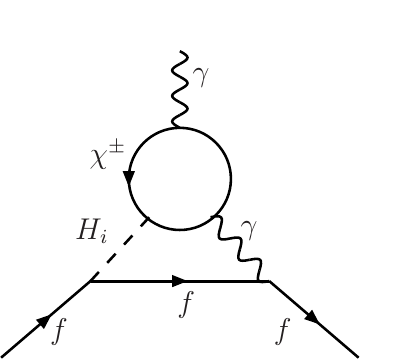}
    \caption[Example two-loop Barr-Zee diagrams giving rise to a fermion EDM or CEDM]{Example two-loop 
        Barr-Zee diagrams giving rise to a fermion EDM or CEDM.
        From Ref.~\cite{Ellis:2008zy}.}
    \label{edm:fig:diagram2}
\end{figure}

\begin{figure}
    \centering
    \includegraphics[width=0.925\textwidth]{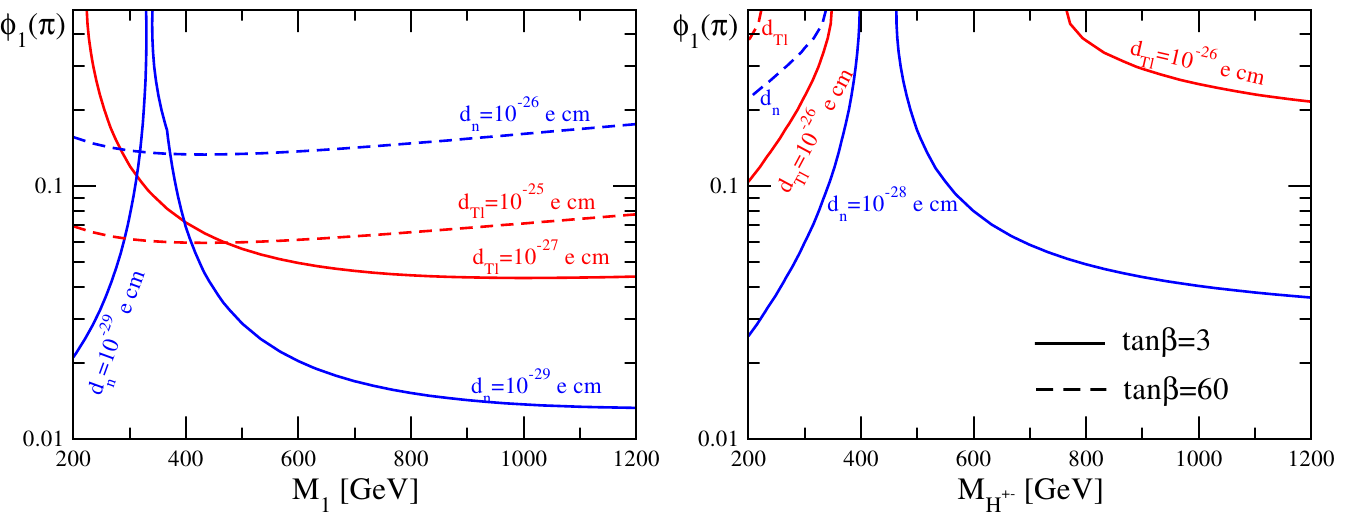}
    \caption[EDM contouts for thallium and the neutron in supersymmetry]{Curves of constant values for the 
        thallium (red) and neutron (blue) EDM as function of the bino mass~$M_1$ and the charged Higgs 
        mass~$M_{H^\pm}$ in the case of heavy first two generations for sfermions.
    From Ref.~\cite{Li:2010ax}.}
    \label{edm:fig:2loop}
\end{figure}

In the flavor blind case, the relevant \CP\ phases are the invariants $\arg(M_i \mu B_\mu^*)$ and 
$\arg(A_f\mu B_\mu^*)$, where $\mu$ is the Higgsino mass, $B_\mu$ is the soft Higgs mixing parameter, 
$M_{1,2,3}$ are the bino, wino, and gluino masses, and $A_{u,d,\ell}$ are universal trilinear couplings of
the up-type squarks, down-type squarks, and sleptons, respectively.
These phases can induce contributions to all the operators in
Eq.~(\ref{edm:eqn:Leff})~\cite{Pospelov:2005pr,Ellis:2008zy,Li:2010ax}.
In the following, we assume that a Peccei-Quinn symmetry takes care of the $\bar\theta$ term which appears in
dimension-four, so that we need not consider that operator further.
However, a nonzero $\bar\theta$ term can be induced through the appearance of an appropriate higher-dimension
operator; though we will not consider this possibility further as we have no explicit axion
dynamics~\cite{Bigi:1990kz,Pospelov:2005pr}.
Generically, the largest SUSY effect comes from fermion EDMs, $d_i$, and CEDMs ${\tilde d}_i$, that can be
induced at the one-loop level by sfermion-gaugino and sfermion-higgsino loops.
A generic Feynman diagram is shown in Fig.~\ref{edm:fig:diagram1}.
In the illustrative case of a degenerate SUSY spectrum at the scale $M_{\rm SUSY}$, and only two \CP\ phases
$\theta_\mu$ (the phase of the higgsino mass), and $\theta_A$ (the universal phase of the trilinear
couplings) one finds~\cite{Pospelov:2005pr}
\begin{eqnarray}
 \frac{d_e}{e \kappa_e} &\simeq& \frac{g_1^2}{12} \sin\theta_A + \left(\frac{5g_2^2}{24} + \frac{g_1^2}{24} \right) \sin\theta_\mu \tan\beta ~, \\
 \frac{d_q}{e_q \kappa_q} &\simeq& \frac{2g_3^2}{9} \Big( \sin\theta_\mu R_q - \sin\theta_A \Big)  ~, \\
 \frac{\tilde d_q}{\kappa_q} &\simeq& \frac{5g_3^2}{18} \Big( \sin\theta_\mu R_q - \sin\theta_A \Big)  ~,
\end{eqnarray}
where $R_q = \tan\beta$ for down quarks, and $R_q = \cot\beta$ for up quarks, and 
\begin{equation}
    e \kappa_i \simeq \frac{m_i}{1~{\rm MeV}} 
    \left(\frac{1~{\rm TeV}}{M_{\rm SUSY}}\right)^2 \times 1.3 \times 10^{-25}\, e~\mathrm{cm}.
\end{equation}
If the SUSY \CP\ phases are assumed to be generically of O(1), the masses of first generation scalar fermions
have to be at least several TeV to avoid the constraints from the current EDM limits.
If the first two generations of sfermions have mass far above the TeV scale and, hence 
are decoupled, as,
e.g., in the "more minimal supersymmetric Standard Model" scenario~\cite{Cohen:1996vb}, two-loop
contributions to the fermion (C)EDMs can become very important.
In particular, two-loop Barr-Zee type diagrams~\cite{Barr:1990vd} can access the light third generation of
sfermions (see the left diagram in Fig.~\ref{edm:fig:diagram2}) and can typically lead to strong constraints
on the \CP\ phases.
Only in scenarios where the only sizable phase comes from the bino mass $M_1$, the constraints are still
rather weak as shown in Fig.~\ref{edm:fig:2loop}.
In such frameworks, electroweak baryogenesis remains possible in the MSSM, but demands very large splitting
between the two top scalar partners, and that the lighest supersymmetric particle be a neutralino with mass
close to half of the Higgs mass value to avoid LHC constraints on the production and decay rates of the
125~GeV Higgs-like boson~\cite{Carena:2008vj,Carena:2012np}.
An explanation of the matter-antimatter asymmetry (BAU) in this framework leads to lower bounds on EDMs that
are only about 2 orders of magnitude below the current experimental limits
\cite{Morrissey:2012db,Cirigliano:2006dg,Cirigliano:2009yd}.
The sensitivities that can be achieved with \PX\ will allow one to probe essentially the entire parameter
range of that framework.
Even if all SUSY scalars are completely decoupled, as in models of split
SUSY~\cite{ArkaniHamed:2004fb,Giudice:2004tc,ArkaniHamed:2004yi}, two-loop Barr-Zee type diagrams that
contain gauginos and Higgsinos (see the right diagram in Fig.~\ref{edm:fig:diagram2}) can lead to effects in
EDMs that will be accessible with improved experimental sensitivity~\cite{Giudice:2005rz}.

In SUSY scenarios with large $\tan\beta$, very important contributions to EDMs can also come from
\CP-violating 4 fermion operators~\cite{Lebedev:2002ne,Demir:2003js}.
The coefficients $C_{ij}$ can be generated at the one-loop level by neutral Higgs boson mixing and by complex
Yukawa threshold corrections and scale with the third power of $\tan\beta$.
It is important to note that these contributions do not decouple with the scale of the SUSY particles, but
with the mass of the heavy Higgs bosons.
Finally, the Weinberg three-gluon operator $w$ receives contributions at the two-loop level by squark gluino
loops, and at the three-loop level by diagrams involving the Higgs bosons.
For a TeV-scale SUSY spectrum, the induced effects on EDMs are typically small.

Going beyond the MSSM, EDMs remain highly sensitive to additional sources of \CP\ violation.
In particular, SUSY models with extended Higgs sectors, like the next-to-minimal supersymmetric standard
model (NMSSM)~\cite{Ellwanger:2009dp} or the effective beyond the MSSM (BMSSM) setup~\cite{Dine:2007xi}, can
contain \CP\ phases in the Higgs sector already at tree level.
Existing and expected limits on EDMs lead to strong constraints on such
phases~\cite{Blum:2010by,Altmannshofer:2011rm,Altmannshofer:2011iv}.
Nonetheless, electroweak baryogenesis can be made compatible with EDM constraints in such
models~\cite{Huber:2006wf,Blum:2010by,Pietroni:1992in,Davies:1996qn,Huber:2000mg,Kang:2004pp,Menon:2004wv,%
Kumar:2011np,Huber:2006ma,Profumo:2007wc}, and at the same time a light dark matter candidate can be
viable~\cite{Carena:2011jy}.

EDMs are highly sensitive probes of additional sources of \CP\ violation beyond those present already in the
SM.
On the other hand, new \CP\ violating phases could in principle also modify the SM predictions for \CP\
violation in meson mixing or in rare $B$-meson decays.
Given the strong constraints from flavor observables on possible new sources of flavor violation at the TeV
scale~\cite{Isidori:2010kg}, an often adopted assumption is the principle of minimal flavor violation
(MFV)~\cite{D'Ambrosio:2002ex}, which states that the SM Yukawa couplings remain the only sources of flavor
violation even in extensions of the SM.
It is important to stress, that MFV does not forbid the existence of additional flavor diagonal sources of
\CP\ violation~\cite{Altmannshofer:2008hc,Mercolli:2009ns,Paradisi:2009ey}.
While MFV ensures that flavor constraints are generically under control for new physics at the TeV scale, EDM
experiments probe flavor diagonal \CP\ phases of O(1), up to tens of TeV, as discussed above.
In concrete new physics models, however, both EDMs and \CP\ asymmetries in rare $B$-meson decays can have
comparable sensitivity to new physics at the TeV scale and give complementary information on the model
parameters (see \cite{Altmannshofer:2008hc} for a corresponding study in the MSSM with MFV).

If new physics introduces generic new sources of flavor violation, then flavor constraints push the new
physics spectrum far above the TeV scale.
Nonetheless, EDMs can provide important constraints in such frameworks.
Due to the generic flavor mixing, the EDMs of first generation fermions can become proportional to the masses
of the third generation.
These large enhancements allow to probe scales of 1000 TeV with EDMs.
An explicit example of such a case is given by the mini-split SUSY
framework~\cite{Hall:2011jd,Ibe:2011aa,Arvanitaki:2012ps,ArkaniHamed:2012gw}, where squarks and sleptons in a
range from $\sim$ 100--10,000 TeV allow to accommodate a 125~GeV Higgs mass in an ``effortless'' way.
Current EDM bounds already probe 100~TeV squarks in this
framework~\cite{McKeen:2013dma,Altmannshofer:minisplit}.
Future improved sensitivities will allow to probe squarks at 1000 TeV and above.

In the future, if a nonzero EDM is discovered in one particular system, the measurement of EDMs of other
types is essential to resolving the underlying mechanisms of \CP\ violation.
The neutron and heavy atom EDMs are thought to be most sensitive to different \CP-violating
sources~\cite{Ginges:2003qt}.
In addition, the recently proposed storage ring EDM experiments of the proton and deuteron 
aim to probe combinations of \CP-violating contributions which differ from those in the 
neutron EDM. In contrast, experiments with paramagnetic atoms or molecules are sensitive 
to the EDM of the 
electron and a possible new \CP-violating electron-quark interaction. 

On a separate tack, a muon EDM of improved sensitivity probes other aspects of 
new-physics models. Recalling the simple dimensional estimate 
$d_f\sim e \sin\phi_{\CP} \, m_f/\Lambda^2$, one might think that a muon
EDM $d_\mu$ would need to be no more than a factor of 100 less stringent than 
the electrom EDM to probe interesting new physics. This is not the case; several authors
have discussed the size of $d_\mu$ within supersymmetric scenarios~\cite{Romanino:2001zf,Hiller:2010ib},
finding that $d_\mu$ could be as large as $d_\mu \sim 10^{-22}$ \ecm, which is easily
within the reach of planned, future dedicated initiatives. Such a large EDM could
speak to the structure of flavor breaking at the Planck scale~\cite{Hiller:2010ib}.

%%%%%%%%%%%%%%%%%%%%%%%%%%%%%%%%%%%%%%%%%%%%%%%%%%%%%%%%%%%%
\section{Experiments}
\label{edm:sec:expt}
%%%%%%%%%%%%%%%%%%%%%%%%%%%%%%%%%%%%%%%%%%%%%%%%%%%%%%%%%%%%

The high-energy, high-intensity, high-power proton beam envisioned for \PX\ provides a number of exciting
experimental opportunities for EDM measurements that can be separated into three categories distinguished by
system and technical aspects:
\begin{enumerate}
\item
Rare-atom EDM experiments with $^{225}$Ra, $^{223}$Rn, 
and $^{211}$Fr that would make use of the Th-target
concept for rare-isotope production.
\item
A neutron EDM experiment, for example using the apparatus being developed for the SNS experiment, which 
would take advantage of the superior neutron flux from a spallation target powered by the high-power proton 
beam.
\item
Storage ring EDM experiments, specifically a dedicated muon experiment and a proton-EDM experiment that uses 
developments in polarized proton beams and polarimetry.
\end{enumerate}

EDM experiments are a unique mix of atomic, low-energy-nuclear and accelerator-physics techniques, which, for
the most part, are stand-alone efforts that only require the isotope or neutron production facilities
anticipated at \PX.
Unlike other \PX\ physics activities, no facility-wide detectors would be employed.
For example, the rare-isotope EDM experiments derive beam from the Th target concept for isotope production,
which uses the 1 GeV beam exclusively.
Without doubt, \PX, already at Stage~1, would be a tremendous enabler for the planned nuclear studies.
The Project~X Injector Experiment (PXIE) with proton beams of 40 MeV at 1 mA may also provide useful yields
of isotopes for fundamental physics research, but this is still under study.
This set of results could potentially unravel the various sources of \CP\ violation encoded in the low-energy
constants associated with \CP-violating effective operators at low energies.
Such systematic studies would be key were an EDM discovered in any system; the highly leveraged nature of the
enhancements in $^{225}\rm{Ra}$ make this system an excellent example of the opportunities presented by \PX.

\subsection{Rare Atom EDM Experiments}

Rare-atom experiments use short-lived isotopes with half lives varying from 3 min for $^{211}$Fr to 24
min for $^{223}$Rn to 14 days for $^{225}$Ra.
For the two-week $^{225}$Ra, it is practical to use separated isotope derived from a $^{229}$Th (7340 yr)
source for development at the first EDM measurements; however in order to reach the $10^{-29}$ level, an
isotope production faculty must be used.
FRIB is a promising source of $^{225}$Ra, however the potential of \PX\ would ultimately provide the
largest sample.
For the shorter lived Fr and Rn, \PX\ would also provide unprecedented rates for rare-isotope
production; however these would require an on-line style experiment.
As discussed in section~\ref{edm:subsubsec:OctupoleMotivation}, octupole collectivity of nuclei with
$Z\approx 88$ and $N\approx 134$ leads to a charge distribution in the body frame which is polarized or
aligned with the spin by the isospin-dependent \CP-violating contributions.
Permanent octupole deformation also leads to closely spaced opposite parity levels that enhance the
polarizability, for example in $^{225}$Ra and possibly $^{223}$Rn.
The recent REX-ISOLDE work~\cite{rf:Gaffney2013} suggests that the enhancement in $^{225}$Ra may be an order
of magnitude greater than for $^{221}$Rn.
For francium, the unpaired electron is subject to $P$-odd, $T$-odd forces due to the interaction of the electron
EDM with the electric field of the nucleus as a consequence of relativistic effects as well as of the
existence of a contact interaction with the nucleus mediated by a $P$-odd, $T$-odd scalar current.

\subsubsection{Radon-221, 223}
\label{edm:sec:RnEDM}

The promise of  an EDM experiment in radon arises for several reasons.  Most importantly, precision measurements with  polarized noble gases in cells  have demonstrated the feasibility of an EDM experiment. For $^{129}$Xe, it was measured that $d=0.7\pm 3.4\times 10^{-27}$ \ecm~\cite{Rosenberry:2001zz}.
A number of techniques have been developed including spin-exchange-optical-pumping (SEOP) using rubidium, construction of EDM cells and wall coatings that reduce wall interactions, in particular for spins 
greater than 1/2. 
The Radon-EDM collaboration has developed an experiment (S-929) at TRIUMF's ISAC, an on-line isotope
separator-facility, which has been approved with high priority.
The experimental program includes development of on-line techniques including collection of rare-gas isotopes
and transfer to a cell, and techniques for detection of spin precession based on gamma-ray anisotropy, beta
asymmetry and laser techniques.

For polarized rare-isotope nuclei, the excited states of the daughter nucleus populated by beta decay are
generally aligned, leading to a $P_2(\cos\theta)$ distribution of gamma-ray emission.
The gamma anisotropy effect has been used to detect nuclear polarization in
$^{209}$Rn~\cite{Kitano:1988zz,Tardiff:2008zz} and $^{223}$Rn \cite{Kitano:1988zz}.
At TRIUMF, the large-coverage HPGe gamma-detector array TIGRESS or the new GRIFFIN array may be used.
Alternatively, the beta asymmetry can be used to detect nuclear polarization with a higher efficiency.
Both the gamma-anisotropy and beta-asymmetry detection techniques have an analyzing power expected to be
limited to 0.1--0.2.
The sensitivity of the EDM measurement is proportional to the analyzing power, thus laser-based techniques
are also under investigation.
A newly conceived two-photon magnetometry for $^{129}$Xe, which may also be useful as a co-magnetometer in
neutron-EDM measurements, is under development.
The analyzing power for two-photon transitions can be close to unity as long as the density is sufficient.

EDM measurements in radon isotopes will ultimately be limited by production rates.
The \PX\ isotope separator scenario is projected to produce 1--2 orders of magnitude more than current
facilities and provides a promising alternative to extracting rare-gas isotopes from the FRIB beam dump as
indicated in Table~\ref{edm:tb:RadonProjections}.
\begin{table}
    \centering
    \caption[Projected sensitivities at TRIUMF, FRIB, and \PX]{Projected sensitivities for $^{221/223}$Ra 
        and the corresponding sensitivities in $^{199}$Hg at TRIUMF, FRIB, and \PX.}
    \label{edm:tb:RadonProjections}
    \begin{tabular}{lccc}
    \hline\hline
    Facility & TRIUMF-ISAC  &  FRIB ($^{223}$Th source) & \PX\   \\
    \hline
     Rate           & $2.5\times 10^7$ s$^{-1}$ & $1\times 10^9$ s$^{-1}$     & $3\times 10^{10}$ s$^{-1}$    \\
    \# atoms        & $3.5\times 10^{10}$  & $1.4\times 10^{12}$     & $4.2\times 10^{13}$     \\
    EDM Sensitivity & $1.3\times 10^{-27}$ \ecm & $2\times 10^{-28}$ \ecm     & $5\times 10^{-29}$ \ecm    \\
     $^{199}$Hg equivalent & $1.3\times 10^{-29}$ \ecm & $2\times 10^{-30}$ \ecm     & $5\times 10^{-31}$ \ecm    \\
    \hline\hline
    \end{tabular}
\end{table}

\subsubsection{Radium-225 Atomic EDM}
\label{edm:sec:RaEDM}

The primary advantage of $^{225}$Ra is the large enhancement \cite{Spevak:1996tu,Ban:2010ea,% 
Dobaczewski:2005hz}, approximately 
%with up to 
a factor of 1000, of the atomic EDM over $^{199}$Hg that arises from both
the octupole deformation of the nucleus and the highly relativistic atomic electrons.
This favorable case is being studied at both Argonne National Laboratory \cite{Guest:2007zz} and Kernfysisch
Versneller Instituut (KVI) \cite{De:2009zz}.
The scheme at Argonne is to measure 
the EDM of $^{225}$Ra atoms in an optical dipole trap (ODT) as first
suggested in Ref.~\cite{Romalis:1999zz}.
The ODT offers the following advantages: 
${\bm{v}}\times{\bm{E}}$ and geometric phase effects are suppressed,
collisions are suppressed between cold fermionic atoms, vector light shifts and parity mixing induced shifts
are small.
The systematic limit from an EDM measurement in an ODT can be controlled at the level of
10$^{-30}$\ecm~\cite{Romalis:1999zz}.

The Argonne collaboration has completed the development of a multi-step 
process that prepares cold, trapped $^{225}$Ra atoms, 
and has observed the nuclear spin precession of $^{225}$Ra 
atoms in an optical dipole trap. In the next step of an EDM 
measurement, the precession frequency and its dependence on a 
strong electric field will be studied. A linear dependence would signify 
the existence of a nonzero EDM. In the experiment, $^{225}$Ra atoms 
are first chemically reduced in a hot oven and physically 
evaporated into a collimated atomic beam. Transverse cooling is 
applied to enhance the forward atomic beam flux by a factor of 100. 
The atoms are slowed and are captured into a magneto-optical trap 
(MOT)~\cite{Guest:2007zz}. The trapped atoms are then transferred 
to a movable optical dipole trap (ODT) that is controlled 
by a lens mounted on a translation stage. The ODT carries the cold 
$^{225}$Ra atoms into a neighboring measurement chamber, 
and hands the atoms off to a stationary, standing-wave ODT~\cite{Parker:2012}. 
With the observation of nuclear precession in the standing-wave ODT, 
the collaboration is poised to begin the first phase of 
the EDM measurement at the sensitivity level of 10$^{-26}\ecm$, 
which should be competitive with 10$^{-29}\ecm$~for $^{199}$Hg 
in terms of sensitivity to $T$-violating physics. For phase 2 
of this experiment, the collaboration plans to upgrade 
the optical trap. In the present MOT, the slower and 
trap laser operate at 714 nm where there is a relatively 
weak atomic transition rate. In phase 2, they would upgrade 
the trap to operate at 483 nm where a strong transition can be exploited for slowing and trapping.

In the first and second phases, a typical experimental run will use 1-10 mCi of $^{225}$Ra presently 
available.
The next-generation isotope facility, such as FRIB after upgrade or \PX, is expected to produce more than
10$^{13}$ $^{225}$Ra atoms/s~\cite{Mustapha:2004}.
In this case it should be possible to extract more than 1~Ci of $^{225}$Ra for use in the EDM apparatus.
This would lead to a projected sensitivity of $10^{-28}\ecm$~for $^{225}$Ra, competitive with
$10^{-31}\ecm$~for $^{199}$Hg.
Table~\ref{edm:table2} summarizes the projected sensitivities.

\begin{table}
\centering
\caption[Projected sensitivities for three scenarios]{Projected sensitivities for $^{225}$Ra and their 
$^{199}$Hg-equivalent values for three scenarios.}
\label{edm:table2}
\begin{tabular}{lccc}
\hline\hline
Phase & Phase 1 & Phase 2 & FRIB after upgrade, \PX\ \\
\hline
Ra (mCi) & 1-10 & 10 & $>$ 1000 \\
$d(^{225}\mathrm{Ra})$ ($10^{-28}\ecm$) & 100 & 10 & 1 \\
equiv. $d(^{199}\mathrm{Hg})$ ($10^{-30}\ecm$) & 10 & 1 & 0.1 \\
\hline\hline
\end{tabular}
\end{table}

\subsubsection{Electron EDM with Francium}
\label{edm:sec:FrEDM}

For paramagnetic systems including alkali atoms, 
the EDM of the atom arises predominantly due to the 
electron EDM and due to \CP-violating components of the electron-nuclear 
interaction. In particular, the electron EDM induces an atomic EDM 
that is approximately proportional to $Z^3\alpha^2$, 
and for heavy atoms, the atomic EDM is enhanced relative to the electon EDM. 
Francium is an extremely promising system in which to study the electron EDM~\cite{Hewett:2012ns}, 
and for $^{211}$Fr, the large nuclear spin and magnetic 
dipole moment allow efficient laser cooling. 
Systematic effects, including a magnetic field that arises 
due to leakage currents resulting from the applied  electric field 
can couple to the magnetic moment producing a false EDM, 
and the  motional magnetic field $\bm{B}_\text{mot}=({\bm v}\times\bm{E})/c^2$ leads 
to systematic effects linear in $\bm{E}$. 
For an experiment in zero magnetic field, the atom 
is quantized along the electric field, and 
these effects can be removed in first order and residual magnetic fields are small. 
Remaining systematic effects scale as $1/E^n$, with $n>2$. 
Consequently, the ratio of systematic effect sensitivity 
to electron EDM sensitivity in $^{211}$Fr is 
two orders of magnitude smaller than in any lighter alkalis. 
An ISOL source at \PX\ would have proton beam currents 
about two orders of magnitude larger than TRIUMF and ISOLDE, 
and may produce $10^{13}$ Fr s$^{-1}$, 
which would be sufficient to lower the electron EDM upper limit by a factor of up to 1000.

\subsection{SNS Neutron EDM}
\label{edm:sec:USnEDM}

A large effort is underway to develop the SNS nEDM experiment with the goal of achieving a sensitivity $< 3
\times 10^{-28}$ \ecm, two orders of magnitude beyond the current experimental limits.
The SNS is a dedicated accelerator-based neutron source utilizing a high-powered 1 GeV proton beam at 
1.4-3~MW incident on the liquid-mercury spallation target.
A cold moderator provides neutrons to the Fundamental-Neutron-Physics-Beamline (FNPB).
The nEDM experiment will use 8.9~\AA~neutrons which are converted to ultra-cold neutrons (velocity less
than $\approx$ 8 m/s) in superfluid helium in the nEDM experiment.
The rate of neutron production is limited by the power of the proton beam and practical considerations of the
target and moderators.
Since the FNPB is a multipurpose cold beam, the nEDM experiment will take what it can get.
The possibility of developing a significantly more intense source of 8.9~\AA~neutrons for the nEDM experiment
by utilizing the higher powered \PX\ proton beam may provide unprecedented sensitivity to the neutron
EDM.

The nEDM experiment, based on Ref.~\cite{Golub:1994cg}, uses a novel
polarized $^3$He co-magnetometer and will
detect the neutron precession via the spin-dependent neutron capture on
$^3$He. The capture reaction produces energetic proton and triton, which
ionize liquid helium and generate scintillation light that can be detected.
Since the EDM of $^3$He is strongly suppressed by electron screening in the
atom it can be used as a sensitive monitor of the  $\sim 30$ mGauss magnetic field.
High densities of trapped UCNs are produced via
phonon production in superfluid $^4$He which can also support large
electric fields, and  $\sim 70$ kV/cm is anticipated.

The nEDM technique allows for a number of independent
checks on systematics including:
\begin{enumerate}
\item Studies of the temperature dependence of false EDM signals in the $^3$He.
\item Measurement of the $^3$He precession frequency using SQUIDs.
\item Cancellation of magnetic field fluctuations by matching the
effective gyromagnetic ratios of neutrons and $^3$He with the ``spin dressing'' 
technique~\cite{Golub:1994cg}.
\end{enumerate}
Key R\&D developments underway  in preparation of the full experiment include:
\begin{enumerate}
\item Maximum electric field strength for large-scale electrodes made of
appropriate materials in superfluid helium below a temperature of 1 K.
\item Magnetic field uniformity for a large-scale magnetic coil and a
superconducting Pb magnetic shield.

\item Development of coated measurement cells that preserve both neutron and
$^3$He polarization along with neutron storage time.

\item Understanding of polarized $^3$He injection and transport in the superfluid.

\item Estimation of the detected light signal from the scintillation
in superfluid helium.

\end{enumerate}
The experiment will
be installed at the FNPB (Fundamental Neutron Physics Beamline) at the
SNS and construction is likely to take at least five years, followed by
hardware commissioning and data taking. Thus first results could
be anticipated by the end of the decade.

Regarding next generation neutron EDMs, a spallation source optimized 
for ultra-cold neutrons (UCNs) could substantially improve 
the sensitivity of next generation neutron EDM experiments, 
which in the case of the US nEDM experiment, 
will be statistically limited at the SNS. 
A key metric for these experiments is the density 
of UCNs provided to or in the experiment, 
which can be much higher per incident beam proton in an optimized UCN source. 
There are several paths \PX\ can provide to improve the statistical 
reach of the nEDM after completion of running at FNPB.  
For example, a cold neutron source envisioned 
for a neutron-antineutron oscillation 
experiment early in \PX\ (NNbarX) 
could provide increased cold neutron flux 
for UCN production in the nEDM experiment as well. 
Alternatively, one can implement an optimized UCN 
source at Project X and couple it ``externally'' 
to a nEDM experiment. Such a source should 
also be compatible with NNbarX and 
is projected to provide substantial gains in available UCN density for nEDM.

\subsection{Storage Ring EDMs}
\label{edm:subsec:pEDM}

The EDM of the muon was measured as part of the muon 
$g-2$ measurement~\cite{Bennett:2008dy} in a magnetic
storage ring and has led to the idea of measuring the EDM of 
charged particles in storage rings with
magnetic, electric, or a combination of fields.
With stable particles combined with proton polarization and polarimetry, the storage ring method brings a
revolution in statistics to the field.
\PX\ will fulfill the need for intense low emittance beams that enable longer storage times,
narrower line widths and therefore the promise to extend EDM sensitivity to the 10$^{-30}$ \ecm~range.
But perhaps more crucial is adding the direct measurement of the proton-EDM to the set of EDM results, which
will provide new information, specifically on the isovector contributions to the nucleon EDMs.
In addition, the expertise in accelerator physics that will concentrate with \PX\ will provide the
storage ring method with a highly instrumented ring with the most sensitive equipment, and it will be a
testing ground for many notions regarding accelerators and storage rings.
In turn, innovative solutions can be applied to \PX\ accelerators to understand it better and improve
their performance.

Stage~1 of \PX\ can be configured in a straightforward manner to provide the necessary polarized protons
with the required energy, emittance, and flux.
High power is not required, and it is true that the existing Fermilab linac could be reconfigured, at
substantial expense, to drive the experiment.
If Stage~1 of \PX\ is built to drive many new experiments, then this is definitely the most cost
effective approach to include the elements of the accelerator required for the proton EDM experiment (e.g.
polarized proton source).
There are ``frozen spin'' muon EDM 
concepts proposed at JPARC and developed for PSI that are dedicated configurations
for EDM sensitivity that could operate in Stage~1 of \PX, and can thus have much higher sensitivity
than the EDM measurement parasitic to the $g-2$~measurement.

Storage Ring EDM methods also provide a unique set of systematic errors and solutions.
Since the particles are stored in a ring, information regarding their position can reveal systematic-error
sources.
Storing particles in clock-wise (CW) and counter-clock-wise (CCW) directions, in alternating 
fills, reveals
the main systematic error source, i.e., a net radial magnetic field (a menace in EDM experiments).
The difference in particle positions and EDM-like signals is a powerful tool against the main systematic
errors.
In addition, for particles with a positive anomalous magnetic moment, it is possible to use an
all-electric-field ring to store the particles for an EDM experiment.
Storing particles in CW and CCW directions simultaneously (possible in an all-electric ring), further
simplifies combating the main systematic errors.
Finally, another systematic error, the so-called geometrical phase, can also be revealed by a) looking at the
EDM-like signal as a function of the azimuthal location of the ring, b) compare the beam position around the
azimuth to its ``ideal'' closed orbit, etc.
These standard techniques for a storage ring eliminate the geometrical phase error as well.
Clearly, the storage ring EDM methods can substantially advance the quest for ever greater sensitivity with
the high intensity beams available even today, e.g., protons and deuterons and by applying well developed
beam storage techniques, e.g., for protons, deuterons, muons, etc.

\subsubsection{Measurement Principle}

The interaction energy for a particle at rest with a magnetic dipole moment (MDM), $\bm{\mu}$, and
electric dipole moment (EDM),  $\bm{d}$, in magnetic and electric fields is given by
Eq.~(\ref{edm:eqn:edmdef}), 
i.e., magnetic dipole moments couple only to magnetic fields 
and electric dipole moments couple only to electric fields.  
The spin precession rate for rectilinear motion is given by
\begin{equation}
\frac {d\bm{s}} {dt^\prime}  =   \bm{\mu} \times \bm{B^\prime} + \bm{d} \times \bm{E^\prime}\,,
\end{equation} 
where $t^\prime$, $\bm{E^\prime}$, and $\bm{B^\prime}$ are evaluated in the particle rest frame. 
We note $\bm{s}$ is the rest-frame particle spin vector, with $\bm{\mu} = g(q/2m) \bm{s}$, 
yielding $\mu=(1+a)q\hbar/2m$ with $a = (g-2)/2$ the anomalous magnetic moment.
Correspondingly, $\bm{d} = \eta (e/2mc) \bm{s}$, with $\eta$ a dimensionless 
parameter that plays a similar 
role to the EDM that $g$ plays for the MDM.
In storage rings, the momentum precession is given by
\begin{equation}
\frac {d \bm{\beta}} {dt}  =  \frac {q} {m\gamma}  \left[  
    \frac {\bm{E}} {c} + \bm{\beta} \times \bm{B} - 
    \frac {\left( \bm{\beta} \cdot \bm{E} \right) \bm{\beta}} {c} \right] \,,
\end{equation}
where $\bm{\beta} = \bm{v} / c$ is the particle velocity divided by the speed of light.
We emphasize that $t$, $\bm{E}$, and $\bm{B}$ are all laboratory frame quantities. 
frame and consider the Thomas precession, if applicable.
The spin precession rate due to the magnetic and electric fields 
including Thomas precession is given by~\cite{jdjackson,Bargmann:1959gz}
\begin{eqnarray}
\frac {d \bm{s}} {dt}  = \frac{q}{m} \bm{s} &\times&
    \Bigg[ \left( \frac{g}{2} - 1 + \frac{1}{\gamma} \right)  \bm{B} - \left(\frac{g}{2} -1 \right)
    \frac {\gamma} {\gamma + 1} \left( \bm{\beta} \cdot \bm{B} \right) \bm{\beta} - \left(\frac {g} {2}
    -  \frac {\gamma} {\gamma + 1}  \right) \frac {\bm{\beta} \times \bm{E}} {c} \nonumber \\
&+&\frac {\eta} {2}   \left( \frac {\bm{E}} {c} -  \frac{\gamma} {1 + \gamma} \frac
{\bm{\beta} \cdot \bm{E}} {c} +  \bm{\beta} \times \bm{B}   \right) \Bigg],
\end{eqnarray}
% \color{red}
% \begin{eqnarray}
% \frac {d \bm{s}} {dt}  = \frac {q} {m} \bm{s} &\times& 
%     \left[ \left( \frac {g} {2} -1 + \frac {1} {\gamma} \right)  \bm{B} - \left(\frac {g} {2} -1 \right) 
%     \frac {\gamma} {\gamma + 1} \left( \bm{\beta} \cdot \bm{B} \right) \bm{\beta} - \left(\frac {g} {2} 
%     -  \frac {\gamma} {\gamma + 1}  \right)  \frac {\bm{\beta} \times \bm{E}} {c} \right] \nonumber \\  
% &+&\frac {\eta} {2} \frac {q} { m}  \left[ \frac {\bm{E}} {c} -  \frac{\gamma} {1 + \gamma} \frac 
% {\bm{\beta} \cdot \bm{E}} {c} +  \bm{\beta} \times \bm{B}   \right], 
% \end{eqnarray}
% \color{black}
where the last term is due to the EDM.

One way to determine the so-called $g-2$ precession rate is to compute the time-dependence of 
the scalar product between the velocity unit vector,
$\bm{\hat{\beta}}$, and spin vector, $\bm{s}$. 
When $\bm{\beta} \cdot \bm{B} =\bm{\beta} \cdot \bm{E} = 0$, it can also be estimated using
\begin{equation}
\bm{\Omega} = \bm{\omega}_a + \bm{\omega}_{\rm EDM}  = - \frac {q} {m}  \left[ 
a \bm{B} - \left( a - \left( \frac {mc} p \right)^2 \right)  \frac {\bm{\beta} \times \bm{E}} {c} \right] 
- \frac {\eta} {2} \frac {q} { m}  \left[ \frac {\bm{E}} {c}  +  \bm{\beta} \times \bm{B}   \right]
\end{equation}
showing that while the so-called $g-2$ precession rate ($\omega_a$) 
is mostly in the horizontal plane there is also a small
precession in the out of plane direction for a nonzero EDM.\footnote{In the absence of electric
fields, $\omega_a$ is independent of $\gamma$.} 
This causes a small tilt in the $g-2$ precession plane.
The dedicated EDM method, developed by the Storage Ring EDM Collaboration, optimizes the EDM sensitivity by
minimizing the horizontal precession and maximizing the vertical one. 
Since the EDM precession is in the vertical direction, where there is no acceleration, there is no Thomas
precession involved and the EDM precession is directly proportional to $\eta$ and not to 
$\eta - 2$ as it is
correspondingly in the horizontal plane.

\subsubsection{EDM Optimization}

As we have noted, 
the traditional way to search for an EDM 
is to place a neutral system in a weak magnetic field region and
observe the interaction energy of the magnetic moment of the system with the magnetic field.
Then apply a very strong electric field 
and look for a change in the interaction energy when the electric
field direction is reversed.
If there is an energy shift proportional 
to the applied electric field, it would signal a permanent electric dipole
moment along the MDM of the system.
A charged system would be accelerated out of the electric 
field region and get lost in a very short time.
However, charged particles in a storage ring 
are regularly stored for hours in very large numbers without any
special difficulty.
This fact provides a special opportunity 
to look for an optimization process in probing the charged particle
EDMs.

The out-of-plane or ``vertical'' polarization change as a function of time is an indication of an EDM signal.
The vertical polarization is given by
\begin{equation}
\Delta P_V = P_L \frac {\omega_{\rm EDM}} \Omega \sin(\Omega t + \theta_0)
\end{equation}
with $\Omega = \sqrt{(\omega_a^2 + \omega_{\rm edm}^2)}$, and $P_L$ the longitudinal polarization.
Setting $\omega_a = 0$ maximizes the sensitivity to the EDM of the stored particles.
Here are two distinct ways we can achieve this goal:
\begin{enumerate}
\item Use a special combination of dipole magnetic and radial electric fields to cancel the horizontal spin
precession.
The radial electric field required is equal to
\begin{equation}
E_r = \frac {a B c \beta \gamma^2} {1-a \beta^2 \gamma^2} \approx aBc \beta \gamma^2
\end{equation}
\begin{figure}
    \centering
    \includegraphics[width=.7\textwidth]{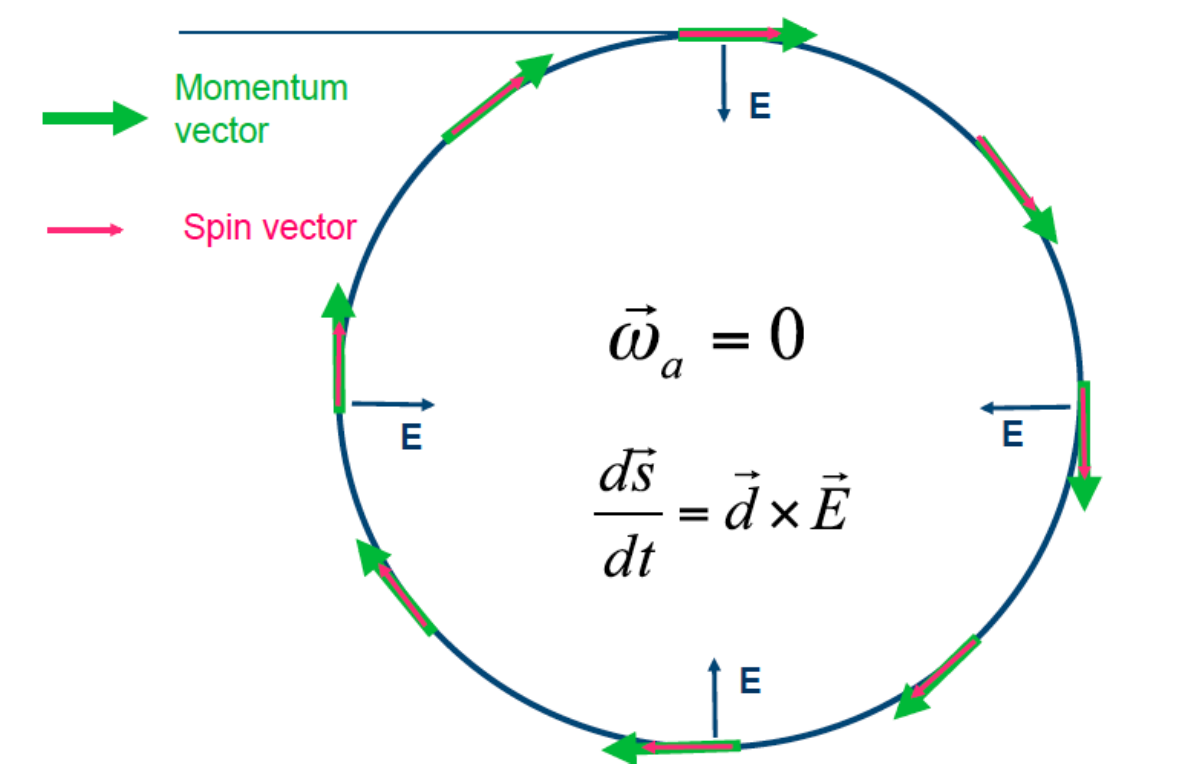}
    \caption[Proton EDM storage ring]{At the magic momentum the particle spin and momentum vectors precess 
        at the same rate in an electric storage ring, i.e., the so-called $g-2$ precession rate is zero.
        That allows the spin to precess in the vertical direction if the particle EDM is nonzero.}
    \label{edm:fig:spin_mom}
\end{figure}

This method was used for the muon EDM LOI to J-PARC and the deuteron EDM proposal at BNL.
The advantage of this method is that the ring can be quite small, and the sensitivity quite large since the
equivalent electric field in the particle rest frame is equal to $\bm{v} \times \bm{B}$, equal to 300~MV/m
for a relativistic particle in a 1~T magnetic field.
The disadvantage is that it requires the development of a combined system with a dipole magnetic field and a 
radial electric field in the same region.
Systematic error sources are any net vertical electric field average around the ring, requiring frequent
clock-wise (CW) and counter-clock-wise (CCW) storage.
During that rotation, the dipole magnetic field is flipped but the radial electric field remains constant.

\item Use $\bm{B}=\bm{0}$, by eliminating every possible source of magnetic fields.
Then use a special momentum value at which the horizontal spin precession is equal to the momentum
precession, i.e., $\omega_a = 0$, as is shown in Fig.~\ref{edm:fig:spin_mom}.
The special particle momentum is equal to $p = mc/\sqrt{a}$.
The particle momentum is equal to 3.08~GeV/$c$ for the muon and 0.7~GeV/$c$ for the proton.
The advantage of this method is the simplicity of the electric ring.
In addition, simultaneous CW and CCW beam storage is possible, 
enabling the detection of spurious $B$~fields by
probing the \emph{relative} displacements of the counter-rotating beams.
The disadvantage is that the ring real estate is large because the maximum electric fields available are
still small compared to the equivalent magnetic fields.
Furthermore, the ring needs to be shielded to first order from spurious magnetic fields adding to the cost.
Obviously this method cannot be applied to particles with negative anomalous magnetic moment values, e.g.,
the deuteron, $^3$He, etc.
\end{enumerate}
The required parameters for the polarized beams are given in Table~\ref{edm:tab:sr_edm}.

\begin{table}
\centering
\caption{The required beam parameter values and the projected sensitivities}
\label{edm:tab:sr_edm}
\begin{tabular}{llllll}
\hline\hline
Particle & Beam intensity,  & Horizontal, vertical  &  Momentum  & Projected Sensitivity \\
               & polarization,  & emittance $95\%$,       &   [GeV/$c$]              &  [\ecm]\\
               &  $N P^2$          &  normalized [mm-mrad], &                       &            \\
               &                        &  dp/p &                       &            \\
\hline
Protons    & $4 \times 10^{10}$, $>80\%$    &   2, 6, $2\times 10^{-4}$ rms   &   0.7  & $10^{-29}$,  $10^{-30}$   \\
Deuterons    & $2 \times 10^{11}$, $>80\%$    &   3, 10, $10^{-3}$   &   1  & $10^{-29}$   \\
$^3$He    & TBD, $>80\%$    &   TBD   &   TBD       & $<10^{-28}$   \\
Muons    & $NP^2=5 \times 10^{16}$ total   &   800, 800 , 2\% max  &   0.5         & $10^{-24}$   \\
\hline\hline
\end{tabular}
\end{table}

\subsubsection{Plans}

The Storage Ring EDM Collaboration has proposed a proton EDM experiment sensitive to \linebreak
$10^{-29}$~\ecm \cite{Farley:2003wt,edmweb}.
With an upgrade, applying stochastic cooling to the stored proton beam, it may be possible to achieve another
order of magnitude in sensitivity for the proton, down to $10^{-30}$~\ecm.
The proposal requires highly polarized protons with an intensity of more than $10^{10}$ particles per cycle
of 15 minutes.
The method uses polarized protons at the so-called ``magic'' momentum of 0.7~GeV/$c$ in an all-electric
storage ring with a radius of $\sim 40$~m.
At this momentum, the proton spin and momentum vectors precess at the same rate in any transverse electric
field.
When the spin is kept along the momentum direction, the radial electric field acts on the EDM vector causing
the proton spin to precess vertically.
The vertical component of the proton spin builds up for the duration of the storage time, which is limited to
$10^{3}$~s by the estimated horizontal spin coherence time (hSCT) of the beam within the admittance of the
ring.
This spin coherence time can be further prolonged by various techniques, e.g., stochastic cooling or
sextupoles placed at special locations around the ring.

The strength of the storage ring EDM method comes from the fact that a large number of highly polarized
particles can be stored for a long time, a large hSCT can be achieved and the transverse spin components can
be probed as a function of time with a high sensitivity polarimeter.
The polarimeter uses elastic nuclear scattering off a solid carbon target placed in a straight section of the
ring serving as the limiting aperture.
The collaboration over the last few years has developed the method and improved their understanding and
confidence in it.
Some notable accomplishments are listed below:

\begin{enumerate}

\item Systematic errors, the efficiency and analyzing power of the polarimeter has been studied.
The polarimeter systematic errors, caused by possible beam drifting, are found to be much lower than the
statistical sensitivity~\cite{brantjes}.

\item A tracking program has been developed to accurately simulate the spin and beam dynamics of the stored
particles in the all-electric ring.
Several aspects of the beam and spin dynamics have been developed analytically, so it is possible to compare
with precision simulations~\cite{orlov}.
The required ring parameters are readily available at BNL with current capabilities.
At Fermilab we would need a polarized proton source.
There is no need for a siberian snake since the acceleration to the required kinetic energy (233 MeV) is done
in the LINAC.

\item The required radial E-fields ($\approx 100$~kV/cm) can be achieved using  technology developed as part of  the international
linear collider (ILC) and energy recovery linacs (ERL) R\&D efforts \cite{dunham}.
Tests at J-Lab indicate that much higher than 100 kV/cm across a 3 cm plate separation can be achieved~\cite{jlab}.

\item The geometrical phase effect can be reduced to a level lower than  the statistical sensitivity based on
a position tolerance of commonly achievable $\sim 0.1  {\rm mm}$ in
the relative positioning of the E-field plates around the ring.

\end {enumerate}

\section{Broader Possibilities}
\label{edm:sec:broad}

In this section we consider broader prospects in regards to the quest for new sources
of \CP\ violation at low energies. 

As we have discussed, one possible resolution of the strong \CP\ problem implies that 
a new particle, an axion, exists---this is well-motivated new physics which does not
in any way rely on the notion of naturalness and new physics at the weak scale. 
In particular, the dynamical relaxation of the axion potential solves the strong \CP\ problem, where 
the essential components of the axion Lagrangian
are \cite{Pospelov:2005pr}
\begin{equation}
{\cal L}_a = \frac{1}{2} \partial_\mu a \partial^\mu a + \frac{a(x)}{f_a} \frac{\alpha_s}{8\pi} 
F_{\mu\nu}^a {\tilde F}^{\mu\nu\,a}. 
\label{edm:eqn:axion}
\end{equation} 
The mass of the axion is controlled by its mixing with the neutral pion; current algebra 
techniques \cite{Bardeen:1977bd,Kandaswamy:1978wi} yield $m_a \approx f_\pi m_\pi /f_a$ 
where $f_\pi$ is the usual pion decay constant. Consequently, for large values of 
the Peccei-Quinn scale $f_a$, the axion is very light. 
Current searches place the 
limit $f_a > 10^{10}$ GeV \cite{Asztalos:2009yp}.  Cosmological constraints on 
$f_a$ have also been thought to operate, requiring that $f_a \lesssim 10^{12}$ GeV so that
the axion contribution to the closure energy density today, $\Omega_a$, is 
less than unity \cite{Preskill:1982cy,Abbott:1982af,Dine:1982ah}. 
The cosmological 
constraints, however, are not compelling; they can be evaded \cite{Fox:2004kb}, 
most notably, if $f_a$ is in excess of the inflation energy scale \cite{sthomas}. 
Moreover, much larger values of $f_a$ can have a ready origin: a string/M-theory QCD axion
in which gauge unification arises from four-dimension renormalization group running has
$f_a$ which is naturally of the grand unification scale, $f_a \sim 10^{16}$ GeV \cite{Fox:2004kb}.
This window of parameter space has not been probed directly, previously; however, 
as suggested in Refs.~\cite{Graham:2011qk,grahamtalk}, the axion can give
rise to a time-dependendent EDM, which potentially can be identified using NMR 
techniques. Current plans \cite{grahamtalk} consider such a search in the 
context of a solid-state EDM experiment, or, more precisely, of an experiment 
building on the technical 
capacities developed in such contexts. 
We offer a brief view of the status of such 
experiments in what follows. 

In solid-state systems, the EDM of the unpaired electrons is detectable either through the magnetic field
produced when the electron EDMs are aligned by the strong internal electric field ($\bm{B}_{\rm
ind}\cdot\bm{E}$) or through the electric field induced when the electron magnetic moments are polarized by a
strong magnetic field 
($\bm{E}_{\rm ind}\cdot\bm{B}$)~\cite{rf:Liu2004,rf:Shapiro1968,rf:Buhmann2002,rf:Sushkov2009,rf:Sushkov2010}.
For example, in PbTiO$_3$, a ferroelectric crystal, sensitivity to the electron EDM is enhanced due to the
large number of electrons in the solid and due to the strong internal electric field in a cooled
crystal~\cite{rf:Mukhamedjanov2005}.
A similar measurement in gadolinium-gallium garnet is under way~\cite{rf:Liu2004}.
Another approach using ferromagnetic gadolinium-iron garnet would detect the electric field produced by the
electron EDMs aligned with the magnetically polarized spins~\cite{rf:Heidenreich2005}.

It should be possible to employ the NMR techniques we have mentioned to the study of 
time-dependent EDMs in other systems, possibly in rare atom experiments at \PX. 

As a second possibility we consider the possibilty of probing spin-\emph{independent}
sources of \CP\ violation at low energies. 
Radiative $\beta$ decay, e.g.,  offers the opportunity of studying $T$-odd correlations
which do not appear in ordinary $\beta$ decay. 
That is, a triple-product momentum correlation among the final-state particles in that process 
is both parity 
$P$- and naively time-reversal $T$-odd but independent of the particle spin. 
Its spin independence renders it distinct from searches for 
permanent electric-dipole-moments (EDMs) of neutrons and nuclei. 
The inability of the Standard Model (SM) to explain the cosmic baryon asymmetry
prompts the search for sources of \CP~violation which do not appear within it
and which are not constrained by other experiments. 
Here, too, $T$-violation is linked to \CP~violation through the CPT-theorem. 
A decay correlation, however, can be, 
by its very nature, only ``pseudo'' $T$-odd, 
so that it can be mimicked by final-state
effects without fundamental $T$ or \CP\ violation; in beta decay, these final-state
interactions are electromagnetic in nature and wholly calculable 
at the needed levels of precision \cite{Gardner:2012rp,Gardner:2013}. 
Such a calculation is crucial to establishing a baseline in the search 
for new sources of \CP-violation in such processes. 
It is motivated in large part by the determination, due to Harvey, Hill, and Hill, 
that pseudo-Chern-Simons 
terms appear in SU(2)${}_{\rm L}\times$U(1) gauge theories at 
low energies---and that they can 
impact low-energy weak radiative processes involving baryons \cite{Harvey:2007rd,Harvey:2007ca}. 
In the SM such pseudo-Chern-Simons interactions are \CP-conserving, 
but considered broadly they are not, so that searching 
for the $P$- and $T$-odd effects that \CP-violating 
interactions of pseudo-Chern-Simons form would engender
offers a new window on physics beyond the SM, specifically 
on new sources of \CP\ violation mediated by the weak vector current, 
probing the \CP\ structure of particular hidden sector models \cite{Gardner:2013aiw}. 
The ultraheavy atoms we have considered in the context of EDM searches would
not do here; the value of $Z$ cannot be too large, or the computation of the SM
background from FSI is not controlled. Rather, lighter nuclei, such as 
$^{35}$Ar, are better choices~\cite{Gardner:2013aiw,Gardner:2013}, so that these studies
may be better suited to the FRIB facility.

%%%%%%%%%%%%%%%%%%%%%%%%%%%%%%%%%%%%%%%%%%%%%%%%%%%%%%%%%%%%
\section{Summary}
\label{edm:sec:summary}
%%%%%%%%%%%%%%%%%%%%%%%%%%%%%%%%%%%%%%%%%%%%%%%%%%%%%%%%%%%%

EDM searches of enhanced experimental sensitivity give us 
sensitive probes of new physics beyond the SM, potentially 
giving insight on the energy scale of new physics, irrespective
of developments at the LHC. 

The first stage of the \PX\ accelerator gives us unpredented 
sensitivity to rare atom EDMs, particularly $^{225}$Ra and $^{211}$Fr,
giving us the opportunity to sharpen 
constraints on low-energy sources of \CP\ violation in chirality-changing
interactions by orders of magnitude. 
The experimental capacities 
involved can be extended to the study of broader 
prospects as well, using \CP\ violation to probe the nature of dark
matter and possibly leading to the discovery of the axion. 
Moreover, a storage ring EDM experiment offers the possibility of probing 
the proton EDM directly and with high sensitivity for the very first time.

At anticipated levels of experimental sensitivity,
there is little doubt that a nonzero EDM would be 
Nature's imprimatur of the existence of physics beyond the SM. However, 
to go beyond this, 
theory must make strides in order to take advantage of such experimental developments.
There are different issues to consider. The interpretation of the EDM of a complex system such as
$^{225}\rm{Ra}$---beyond discovery---involves hadronic, 
nuclear, and atomic computations. 
Lattice gauge theory methods may well be able to redress limitations in current computations
of hadronic matrix elements. 
Ultimately the interpretability of possible EDMs in terms of underlying sources of \CP
violation may prove sharpest in simple systems such as the neutron and proton, and
in these systems as well \PX\ can open new decades of sensitivity.

\bibliographystyle{apsrev4-1}
\bibliography{edm/refs}

 % Lu, Tim C. & Susan G.

%%%%%%%%%%%%%%%%%%%%%%%%%%%%%%%%%%%%%%%%%%%%%%%%%%%%%%%%%%%%
% \chapter[$\protect\nnb$ Oscillations with \PX]{Neutron-Antineutron Oscillations with \PX}
\chapter{Neutron-Antineutron Oscillations with \PX}
%\chapter[Neutron-Antineutron Oscillations with \PX]{Neutron-Antineutron Oscillations with \PX}
\label{chapt:nn}
%%%%%%%%%%%%%%%%%%%%%%%%%%%%%%%%%%%%%%%%%%%%%%%%%%%%%%%%%%%%

\authors{Yuri Kamyshkov, Chris Quigg, William M. Snow, Albert R. Young, \\
Usama~Al-Binni,
Kaladi~Babu,
Sunanda~Banerjee,
David~V.~Baxter,
Zurab~Berezhiani,
Marc~Bergevin,
Sudeb~Bhattacharya,
Stephen~J.~Brice,
Thomas~W.~Burgess,
Luis~A.~Castellanos,
Subhasis~Chattopadhyay,
Mu-Chun~Chen,
Christopher~E.~Coppola,
Ramanath~Cowsik,
J.~Allen~Crabtree,
Alexander~Dolgov,
Georgi~Dvali,
Phillip~D.~Ferguson,
Tony~Gabriel,
Avraham~Gal,
Franz~Gallmeier,
Kenneth~S.~Ganezer,
Elena~S.~Golubeva,
Van~B.~Graves,
Geoffrey~Greene,
Cory~L.~Griffard,
Thomas~Handler,
Brandon~Hartfiel,
Ayman~Hawari,
Lawrence~Heilbronn,
James~E.~Hill,
Christian~Johnson,
Boris~Kerbikov,
Boris~Kopeliovich,
Vladimir~Kopeliovich,
Wolfgang~Korsch,
Chen-Yu~Liu,
Rabindra~Mohapatra,
Michael~Mocko,
Nikolai~V.~Mokhov, 
Guenter~Muhrer,
Pieter~Mumm,
Lev~Okun,
Robert~W.~Pattie~Jr.,
David~G.~Phillips~II,
Erik~Ramberg,
Amlan~Ray,
Amit~Roy,
Arthur~Ruggles,
Utpal~Sarkar,
Andy~Saunders,
Anatoly~Serebrov,
Hirohiko~Shimizu,
Robert~Shrock,
Arindam~K.~Sikdar,
Aria~Soha,
Stefan~Spanier,
Sergei~Striganov,
Zhaowen~Tang,
Lawrence~Townsend,
Robert~S.~Tschirhart,
Arkady~Vainshtein,
Richard~J.~Van~Kooten,
and Bernard~Wehring}

\section{Introduction}
\label{nnbar:sec:intro}

An observation of neutron-antineutron (\nnb) transformation
would constitute a discovery of fundamental importance for cosmology
and particle physics.  It would provide the first direct
experimental evidence for baryon number ($\mathcal{B}$) violation, and
would qualitatively change our ideas of the scales relevant for
quark-lepton unification and neutrino mass generation.  If seen at
rates achievable in next-generation searches, \nnb\ transformation
must be taken into account for any quantitative understanding of
the baryon asymmetry of the universe.  A discovery of this
process would also prove that all nuclei are ultimately unstable.
In fact, a search for \nnb\ oscillations using free neutrons
at \PX\ possesses excellent potential in exploring
the stability of matter. A limit on the free-neutron oscillation time
$\tau_{\nnb} > 10^{10}$ $\rm{s}$, which appears to be within
the range of the next generation of experiments described in this chapter,
would correspond to a limit on matter stability of $T_{A} =
1.6-3.1\times 10^{35}$ $\rm{yrs}$~\cite{Dover:1983cd,Friedman:2008ef}.

\PX\ presents an opportunity to probe \nnb\ transformation
with free neutrons with an unprecedented improvement in sensitivity.  Improvements would
be achieved by creating a unique facility, combining a high-intensity
cold-neutron source \emph{dedicated} to particle physics experiments with
advanced neutron-optics technology and detectors which build on
the demonstrated capability to detect antineutron
annihilation events with zero background.  Existing slow-neutron sources at research reactors
and spallation sources possess neither the required space nor the degree of access
to the cold source needed to take full advantage of advanced neutron-optics technology which enables a greatly improved free \nnb\ transformation
search experiment. Therefore, a dedicated source devoted exclusively
to fundamental neutron physics, such as would be available at \PX, represents an exciting tool to explore not only \nnb\ oscillations, 
but also other Intensity Frontier questions accessible through slow neutrons.

The current best limit on \nnb\ oscillations comes from the
Super-Kamiokande experiment, which determined an upper bound on the
free-neutron oscillation time of $\tau_{\nnb} >$ 3.5$\times10^{8}$ s
from \nnb\ transformation in $^{16}$O nuclei~\cite{Friedman:2008ef,Abe:2011ka}.
An important point for underground detector measurements is that these experiments
are already limited in part by atmospheric
neutrino backgrounds.  Because only modest increments in detector mass
over Super-Kamiokande are feasible and the atmospheric neutrino backgrounds will scale
with the detector mass, dramatic improvements in the current limit
will be challenging for underground experiments.

Experiments that utilize free neutrons to search for \nnb\
oscillations have a number of remarkable features.  The basic idea for
these experiments (we go into much greater detail in Sections~\ref{nnbar:sec:physics} and~\ref{nnbar:sec:neutronpx})
is to prepare a beam of slow (below room temperature) neutrons which propagate
freely from the exit of a neutron guide to a distant annihilation target.  During
the time in which the neutron propagates freely, a $\mathcal{B}$-violating interaction can
produce oscillations from a pure ``$n$" state to one with an admixture of ``$n$" and ``${\bar n}$" amplitudes. Antineutron appearance is sought through
annihilation in a thin target, which generates a star pattern of several secondary
pions seen by a tracking detector situated around the target.  This signature
strongly suppresses backgrounds.  We note that, to observe this
signal, the ``quasi-free" condition must hold, in which the $n$ and $\bar{n}$
are effectively degenerate in energy.  This creates a requirement for low pressures
(below roughly $10^{-5}$~Pa for \PX) and very small ambient magnetic fields
(between 1 and 10~nT for \PX) in order to prevent level splittings between the
neutron and antineutron from damping the oscillations. Advantages of a new \nnb\ oscillation search experiment at \PX\ would include:
\begin{itemize}
\item detection of annihilation events with zero background (see discussion next paragraph), maximizing the discovery potential for these experiments,
\item a systematic cross-check of a non-zero \nnb\ signal is possible by
a modest increase in the magnetic field, which damps out oscillations,
\item and orders of magnitude improvement in sensitivity over the current free-neutron limit through the use of cutting-edge neutron optics, greatly increasing the neutron integrated flux and average transit time to the annihilation target.
\end{itemize}
These advantages provide a strong motivation to search for \nnb\
oscillations as a part of \PX.

\begin{figure}
    \centering
    \includegraphics[width=0.875\textwidth]{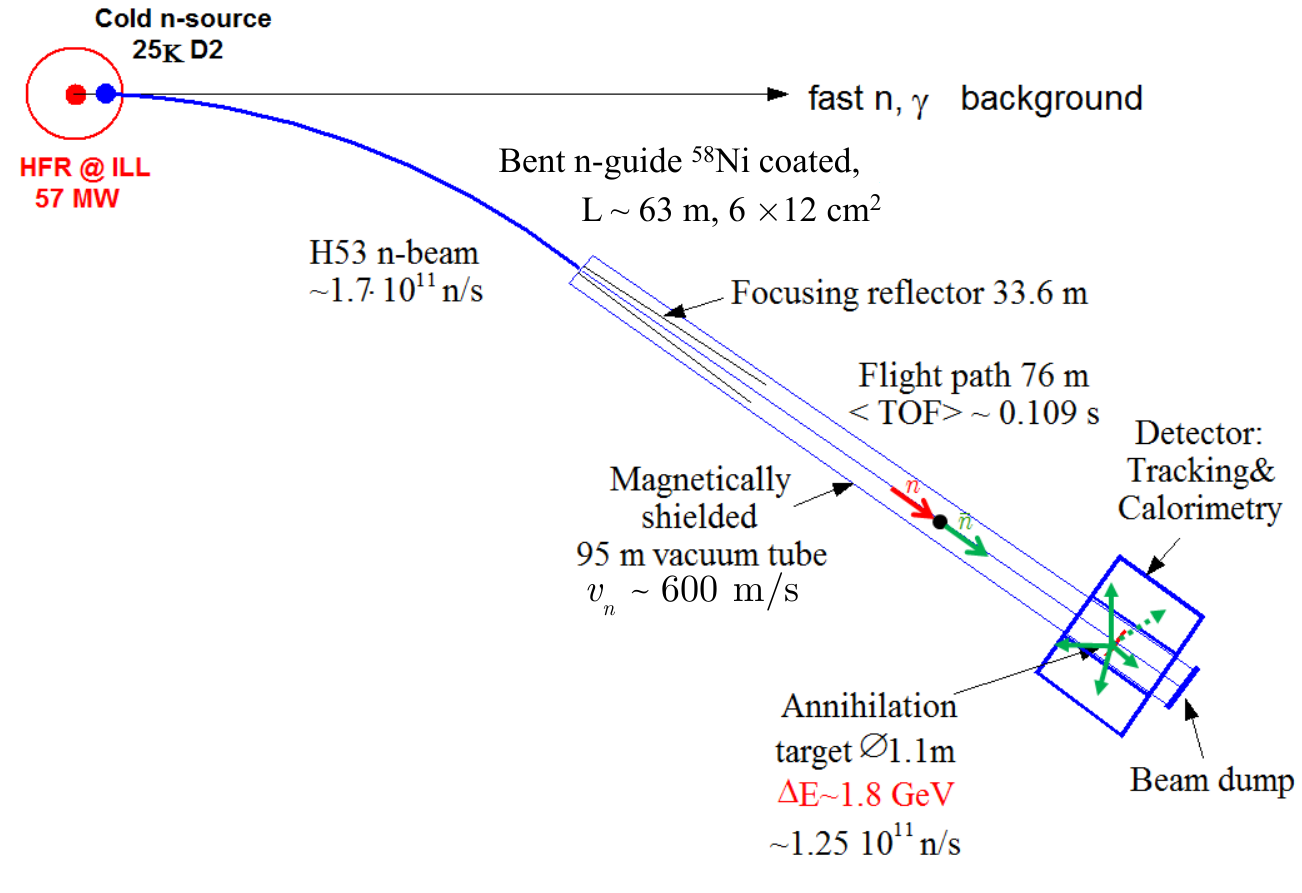} \vspace*{-6pt}
    \caption[Configuration of the horizontal \protect\nnb\ search at ILL/Grenoble]{Configuration of the 
        horizontal \nnb\ search experiment at ILL/Grenoble~\cite{Baldo:1994bc}, published in 1994.}
    \label{ill:fig:logo}
\end{figure}

The current best limit for an experimental search for free \nnb\
oscillations was performed at the ILL in
Grenoble in 1994~\cite{Baldo:1994bc} (see Fig.~\ref{ill:fig:logo}).
This experiment used a cold neutron beam from their 58 MW research reactor
with a neutron current of $1.25\times10^{11} n/\text{s}$
incident on the annihilation target and gave a limit of
$\tau_{\nnb} > 0.86\times10^{8}$ $\rm{s}$~\cite{Baldo:1994bc}.
The average velocity of the cold neutrons was $\sim600~\text{m/s}$
and the average neutron observation time was $\sim0.1~\text{s}$.
A vacuum of $P\simeq 2\times10^{-4}~\text{Pa}$ maintained in the
neutron flight volume and a magnetic field of $|\bm{B}| < 10$~nT
satisfied the quasi-free conditions for oscillations to occur.
Antineutron appearance was sought through annihilation with a $\sim130$-$\mu\text{m}$ thick carbon-film 
target that generated at least 
two tracks (one due to a charged particle) in the tracking detector with
a total energy above 850 MeV in the surrounding calorimeter. In one year of
operation the ILL experiment saw zero candidate events with zero
background.

An \nnb\ oscillation search experiment at \PX\ (NNbarX) is
conceived of as a two-stage experiment. The neutron spallation
target/moderator/reflector system and the experimental apparatus need
to be designed together in order to optimize the sensitivity of the
experiment. The target system and the first-stage experiment can be
built and start operation during the commissioning of the first-stage
of \PX, which is based on a 1-GeV proton beam Linac
operating at 1~mA. The first-stage of NNbarX will be a horizontal
experiment with configuration similar to the ILL experiment performed
in the 1990s, but employing
modernized technologies that include an optimized slow-neutron
target/moderator/reflector system and an elliptical supermirror
neutron-focusing reflector. Our very conservative baseline goal for a
first-stage experiment is a factor of 30 improvement of the
sensitivity (probability of appearance) for \nnb\ oscillations
beyond the limits obtained in the ILL experiment.
This level of sensitivity would also surpass the \nnb\
oscillation limits obtained in the Super-Kamiokande, Soudan-II, and SNO intranuclear
searches~\cite{Abe:2011ka,Chung:2002jc,Bergevin:2010mb}.  In fact, although still in progress, our
optimization studies indicate that this horizontal geometry is capable
of improvements of a factor of 300 or more in 3 years of operation at
\PX.  A future, second stage of an NNbarX experiment can achieve higher
sensitivity by exploiting a vertical layout and a moderator/reflector
system that can make use of colder neutrons and ultracold neutrons
(UCN) for the \nnb\ search.  This experimental arrangement
involves new technologies that will require a dedicated R$\&$D
campaign, but the sensitivity of NNbarX should improve by another
factor of $\sim$ 100 with this configuration, corresponding to limits
for the oscillation time parameter $\tau_{\nnb} > 10^{10}~\rm{s}$.

In what follows we present a more detailed analysis of the theoretical
formalism and motivation for measurements of \nnb\ oscillations.
We then proceed to a more detailed description of our experimental program
to measure \nnb\ oscillations at \PX.

%%%%%%%%%%%%%%%%%%%%%%%%%%%%%%%%%%%%%%%%%%%%%%%%%%%%%%%%%%%%
\section{Physics Motivation for $\protect\nnb$  Oscillation Searches}
\label{nnbar:sec:physics}
%%%%%%%%%%%%%%%%%%%%%%%%%%%%%%%%%%%%%%%%%%%%%%%%%%%%%%%%%%%%
The search for neutron--antineutron oscillations\cite{Kuzmin:1970nx,Glashow:1979kx,Mohapatra:1980qe} may illuminate  two of the great mysteries of particle physics and cosmology: the great stability of ordinary matter and the origin of the preponderance of matter over antimatter in the universe. Processes that violate baryon number and lepton number must be highly suppressed, but they must be present if the observed matter excess evolved from an early universe in which matter and antimatter were in balance~\cite{Sakharov:1967as,Dolgov:1992ad,Dolgov:1998ad}. The primitive interactions of quantum chromodynamics and the electroweak theory conserve baryon number $\mathcal{B}$ and lepton number $\mathcal{L}$, but we have not identified a dynamical principle or symmetry that compels conservation of either baryon number or lepton number. 

Indeed, grand unified theories (GUTs)~\cite{Georgi:1974hg,Raby:2008sr} and
nonperturbative effects in the Standard Model itself 
lead to baryon number violation~\cite{Hooft:1976gh,Hooft:1978gh,Kuzmin:1985vk}.  The baryon-number--violating effects in all these models appear with a very weak strength so that
stability of atoms such as hydrogen, helium, etc., is not significantly
affected on the time scale of the age of the universe. The discovery that neutrino species mix, which demonstrates that individual ($e, \mu, \tau$) lepton numbers are not conserved, leaves open the possibility that overall lepton number is conserved. The observation of neutrinoless double-beta decay would establish $\mathcal{L}$ nonconservation. Once we accept the
possibility that baryon number is not a good symmetry of nature, there are many
questions that must be explored to decide the nature of physics associated with
$\mathcal{B}$-violation:

\begin{itemize}

\item

Is (a nonanomalous extension of) baryon number, $\mathcal{B}$, a global or local symmetry?

\item 

Does baryon number occur as a symmetry by itself or does it appear in
combination with lepton number, $\mathcal{L}$, i.e. $\mathcal{B}-\mathcal{L}$, as the Standard Model (SM)
would suggest?

\item 

What is the scale of baryon-number violation and the nature of the associated
physics that is responsible for it? For example, is this physics characterized
by a mass scale not too far above the TeV scale, so that it can be probed in
experiments already searching for new physics in colliders as well as low
energy rare processes?

\item 

Are the details of the physics responsible for baryon-number violation such
 that they can explain the origin of matter ?

\end{itemize}

Proton-decay searches probe baryon-number violation due to physics at a grand
unified scale of $\sim 10^{15}$-$10^{16}$ GeV.  In contrast, the baryon-number--violating
process of \nnb\ oscillation, where a free neutron spontaneously
transmutes itself into an anti-neutron, has very different properties and
probes quite different physics; for one thing it violates baryon number by two
units and is caused by operators that have mass dimension nine so that it
probes new physics at mass scales $\sim 1$~TeV and above. Therefore it can be probed by
experiments searching for new physics at this scale. Secondly, it may be deeply
connected to the possibility that neutrinos may be Majorana fermions, a natural
expectation.  A key question for experiments is whether there are theories that
predict \nnb\ oscillations at a level that can be probed in
currently available facilities such as reactors or in contemplated ones such as
\PX\ at Fermilab, with intense neutron fluxes.  Equally important to know
is what conclusion can be drawn about physics beyond the Standard Model if no signal appears after the
free-neutron oscillation time is improved by two orders of magnitude above the current limit
of $\sim 10^8$ s.

% ========================================================================

\subsection{Some Background Concerning Baryon Number Violation} 
\label{nnbar:subsec:theorybkgd}

Early on, it was observed that in a model with a left-right symmetric
electroweak group, $G_{\mathrm{LR}} = \text{SU}(2)_{\mathrm{L}} \times \text{SU}(2)_{\mathrm{R}} \times {\rm
U}(1)_{\mathcal{B-L}}$, baryon and lepton numbers in the combination $\mathcal{B}-\mathcal{L}$ can be
gauged in an anomaly-free manner. The resultant U(1)$_{\mathcal{B-L}}$ can be combined
with color SU(3) in an SU(4) gauge group \cite{Pati:1974jp}, giving rise to the group
$G_{422} = \text{SU}(4) \times \text{SU}(2)_{\mathrm{L}} \times \text{SU}(2)_{\mathrm{R}}$
\cite{Pati:1974jp,Mohapatra:1975rm,Mohapatra:1975jp}. A higher degree of unification involved models that embed either
the Standard Model gauge group $G_{\mathrm{SM}} = \text{SU}(3)_{\mathrm{c}} \times \text{SU}(2)_{\mathrm{L}}
\times \text{U}(1)_Y$ or $G_{422}$ in a simple group such as SU(5) or SO(10)
~\cite{Georgi:1974hg,Raby:2008sr}.  The motivations for grand unification theories
are well known and include the unification of gauge interactions and their
couplings, the related explanation of the quantization of weak hypercharge and
electric charge, and the unification of quarks and leptons. While the gauge
couplings do not unify in the Standard Model, they do unify in a minimal
supersymmetric extension of the Standard Model.  Although supersymmetric
particles have not been discovered in the 7-TeV and 8-TeV data at the Large
Hadron Collider, they may still be observed at higher energy.  Supersymmetric
grand unified theories thus provide an appealing possible ultraviolet
completion of the Standard Model. The unification of quarks and leptons in
grand unified theories (GUTs) generically leads to the decay of the proton and
the decay of neutrons that would otherwise be stably bound in nuclei. These
decays typically obey the selection rule $\Delta \mathcal{B} = -1$ and $\Delta \mathcal{L} = -1$.
However, the general possibility of a different kind of baryon-number violating
process, namely the $|\Delta \mathcal{B}|=2$ process of \nnb\ oscillations, was
suggested~\cite{Kuzmin:1970nx} even before the advent of GUTs.  This was further
discussed and studied after the development of GUTs in~\cite{Glashow:1979kx,Mohapatra:1980qe} and in a number of subsequent models~\cite{Kuo:1980tk,Chang:1980lc,Mohapatra:1980rn,Cowsik:1981rc,Rao:1982sr,Misra:1983sm,Rao:1984sr,Huber:2001sh,Babu:2001kb,Nussinov:2002sn,Mohapatra:2005rm,Babu:2006kb,Dutta:2006bd,Berezhiani:2006zb,Babu:2009kb,Mohapatra:2009rm,Gu:2011pg,Babu:2013kb,Arnold:2013ja}.
Recently, a number of models have been constructed that predict \nnb\ oscillations
at levels within reach of an improved search, e.g.~\cite{Babu:2001kb,Nussinov:2002sn,Dutta:2006bd,Babu:2013kb}.
We proceed to discuss some of these.

% =====================================================================

\subsection{Some Models with \nnb\ Oscillations} 
\label{nnbar:subsec:models}

It was pointed out in 1980 that a class of unified theories for Majorana
neutrino mass in which the seesaw mechanism operates in the TeV mass range
predicts \nnb\ oscillation transition times that are in the
accessible range being probed in different experiments~\cite{Mohapatra:1980rn}. This model
was based on the idea that $\mathcal{B}-\mathcal{L}$ is a local rather than a global symmetry. This
idea is incorporated in the electroweak gauge group $G_{\mathrm{LR}}$ and accommodates
right-handed neutrinos and an associated seesaw mechanism.  The Majorana
neutrino mass terms are $|\Delta \mathcal{L}|=2$ operators and hence, in the context of
the U(1)$_{\mathcal{B-L}}$ gauge symmetry, it is natural that they are associated with
baryon number violation by $|\Delta \mathcal{B}|=2$.  Thus, \nnb\ oscillations are
an expected feature of this model.  Detailed analysis of the model shows that
it naturally predicts the existence of TeV-scale color-sextet Higgs particles
that can be probed at the LHC.

The question of how restrictive the range of neutron oscillation time is in
this class of models has recently been investigated by requiring that the model
also explain the observed matter-anti-matter asymmetry. The basic idea is that
since \nnb\ oscillations are a TeV-scale $\mathcal{B}$-violating phenomenon, they
will remain in equilibrium in the thermal plasma down to very low temperatures
in the early universe.  Hence, in combination with Standard-Model
baryon-number--violating processes they will erase any pre-existing baryon
asymmetry in the universe. Therefore in models with observable \nnb\
oscillation, one must search for new ways to generate a matter-antimatter
asymmetry near or below the weak scale. Such a mechanism was proposed in a
few recent papers~\cite{Babu:2006kb,Babu:2009kb,Gu:2011pg}, where it was shown that high-dimensional operators
that lead to processes such as neutron oscillation can indeed generate a baryon
asymmetry via a mechanism called post-sphaleron baryogenesis. This mechanism
specifically applies to the class of $G_{422}$ models for neutron
oscillation discussed in Ref.~\cite{Mohapatra:1980rn}, as well as to other models for
neutron oscillation.

Because of quark-lepton unification, the field responsible for the seesaw mechanism
now has color-sextet partnerss. The neutral scalar field,
which breaks $\mathcal{B}-\mathcal{L}$ gauge symmetry to generate neutrino masses has couplings to
these colored scalars and decays slowly to the six-quark states via the
exchange of virtual color sextet fields. This decay in combination with \textsf{CP}
violation is ultimately responsible for baryogenesis. Due to its slowness, the
decay cannot, however, compete with the Hubble expansion until the universe
cools below the weak scale. The cosmological requirements for baryogenesis then
impose strong constraints on the parameters of the model and predict that there
must be an \textit{upper limit} on the free-neutron oscillation time of $5\times 10^{10}$
s~\cite{Babu:2013kb}, while for most of the parameter range it is below
$10^{10}$ s.  Essentially what happens is that if the neutron oscillation
time exceeds this bound, then the magnitude of the baryon asymmetry becomes
smaller than the observed value or the color symmetry of the model breaks down,
neither of which is acceptable for a realistic theory.  It may therefore be
concluded that if the search for \nnb\ oscillation up to a
transition time of $10^{10}$~s comes out to be negative, this class of
interesting neutrino mass models will be ruled out.

A different type of model that predicts \nnb\ oscillations at a rate close to current limits involves an
extra-dimensional theoretical framework~\cite{Nussinov:2002sn}.
Although current experimental data are fully consistent with a four-dimensional Minkowski spacetime, it is
useful to explore the possibility of extra dimensions, both from a purely phenomenological point of view and
because the main candidate theory for quantum gravity---string theory---suggests the existence of higher
dimensions.
Ref.~\cite{Nussinov:2002sn} focuses on theories in which standard-model fields can propagate in extra
dimensions and the wave functions of standard-model fermions have strong localization at various points in
this extra-dimensional space.
The effective size of the extra dimension(s) is denoted $L$; the associated mass parameter $\Lambda_L=L^{-1}$
can be $\sim 50$--100~TeV.
Such models are of interest partly because they can provide a mechanism for obtaining a hierarchy in fermion
masses and quark mixing.
In generic models of this type, excessively rapid proton decay can be avoided by arranging that the
wavefunction centers of the $u$ and $d$ quarks are separated far from those of the $e$ and $\mu$.
However, as was pointed out in Ref.~\cite{Nussinov:2002sn}, this does not guarantee adequate suppression of
\nnb\ oscillations.
Indeed, for typical values of the parameters of the model, it was shown that \nnb\ oscillations occur at
levels that are in accord with the current experiment limit but not too far below this limit.
One of the interesting features of this model is that it is an example of a theory in which proton decay is
negligible, while \nnb\ oscillations could be observable at levels close to current limits.
Other models of this type have recently been studied in~\cite{Arnold:2013ja}.
These models have scalar fields in two representations of $\text{SU}(2)\times\text{SU}(2)\times\text{U}(1)$
and violate baryon number by two units.
Some of the models give rise to \nnb\ oscillations, while some also violate lepton number by two units.
The range of scalar masses for which \nnb\ oscillations are measurable in the next generation of experiments
is also discussed in~\cite{Arnold:2013ja}.
In extra dimensional models with low scale gravity, neutron-antineutron oscillations are predicted to occur 
1--2 orders of magnitude less frequently than current experimental limits~\cite{Bambi:2006mi}.

We conclude that there is strong motivation to pursue
a higher-sensitivity \nnb\ oscillation search experiment that can achieve a lower
bound of $\tau_{\nnb} \gtrsim 10^9$--$10^{10}~\text{s}$.  

%We next present a general phenomenological discussion of \nnb\
%oscillations, as background to the later discussion of current experimental
%limits and future plans. 

% =======================================================================

\subsection{General Formalism for Analyzing \nnb\ Oscillations}
\label{nnbar:subsec:formalism}

% ======================================================================

\subsubsection{Oscillations in a Field-Free Vacuum} 
\label{nnbar:subsubsec:free}

We denote the effective Hamiltonian that is responsible for \nnb\
oscillations as $H_{\eff}$.  This has the diagonal matrix elements 
\begin{equation}
\langle n | H_{\eff} | n \rangle \ = \ 
\langle \bar n | H_{\eff} | \bar n \rangle \ = m_n - \frac{i \lambda}{2} \ ,
\label{diagonal}
\end{equation}
where $\lambda^{-1} = \tau_n = 0.88 \times 10^3$ s is the mean life of a free
neutron. Here we assume \textsf{CPT} invariance, so that $m_n = m_{\bar{n}}$.
The transition matrix elements are taken to be real and are denoted 
\begin{equation}
\langle \bar n | H_{\eff} | n \rangle = 
\langle n | H_{\eff} | \bar n \rangle \equiv \delta m \ .
\label{nnbtransition}
\end{equation}
Consider the $2 \times 2$ matrix

\begin{equation}
\cal{M}_{F}=\left(\begin{array}{cc}
m_n - i\lambda/2 & \delta m \\
\delta m         & m_n - i\lambda/2 \end{array}\right)
\label{nnbmatrix}
\end{equation}
Diagonalizing this matrix ${\cal M_F}$ yields the mass eigenstates
\begin{equation}
|n_\pm \rangle =\frac{|n \rangle \pm |\bar n \rangle}{\sqrt{2}}
\end{equation}
with mass eigenvalues
\begin{equation}
m_{\pm} = (m_n \pm \delta m) - \frac{i\lambda}{2} \ .
\label{meigfree}
\end{equation}
Hence, if one starts with a pure $|n\rangle$ state at $t=0$, then there is a 
finite probability $P$ for it to be an $|\bar n\rangle$ at $t \ne 0$ given by
\begin{equation}
P_{\bar{n}\leftarrow n}(t) = |\langle \bar n|n(t) \rangle|^2 = \sin^2(t/\tau_{\nnb})\displaystyle{
e^{-\lambda t}}  , 
\label{pfree}
\end{equation}
where
\begin{equation}
\tau_{\nnb} = \frac{1}{|\delta m|} .
\label{tau}
\end{equation}
Neutron--antineutron oscillations would likewise be inhibited by a neutron--antineutron mass difference.  Should oscillations be observed, $\tau_{\nnb}$ can also be interpreted as a limit on $|m_n - m_{\bar{n}}|$, and so test \textsf{CPT} invariance~\cite{Abov:1984bt}. 

Current lower limits on the oscillation lifetime, $\tau_{\nnb} \gsim 10^8$ s, greatly exceed the lifetime for $\beta$-decay of the free neutron.

% ===================================================================

\subsubsection{Oscillations in a Magnetic Field}
\label{nnbar:subsubsec:bfield}

We next review the formalism for the analysis of \nnb\ oscillations in
an external magnetic field~\cite{Mohapatra:1980rn,Cowsik:1981rc}. This formalism is relevant for an experiment
searching for \nnb\ oscillations using neutrons that propagate some
distance in a vacuum pipe, because although one must use degaussing methods to
greatly reduce the magnitude of the magnetic field in the pipe, it still plays
an important role in setting the parameters of the experiment.  This formalism is relevant for both the ILL experiment at Grenoble and NNbarX.

The $n$ and $\bar n$ interact with the external $\bm{B}$ field through their
magnetic dipole moments, $\bm{\mu}_{n,\bar n}$, where 
$\mu_n = -\mu_{\bar n} = -1.9 \mu_N \approx 6 \times 10^{-14}$~MeV/Tesla.  Hence, the matrix $\mathcal{M}_B$ now takes the
form 
\begin{equation}
\mathcal{M}_B=\left(\begin{array}{cc}
m_n - \bm{\mu}_n \cdot \bm{B} - i\lambda/2 & \delta m \\
\delta m         & m_n + \bm{\mu}_n \cdot \bm{B} -  i\lambda/2
\end{array}\right)
\end{equation}
Diagonalizing this mass matrix yields mass eigenstates
\begin{equation}
|n_1 \rangle = \cos \theta \ |n \rangle + \sin\theta \ |\bar n \rangle 
\end{equation}
and
\begin{equation}
|n_2 \rangle = -\sin\theta \ |n \rangle + \cos\theta \ |\bar n \rangle \ , 
\end{equation}
where 
\begin{equation}
\tan(2\theta) = -\frac{\delta m}{\bm{\mu}_n \cdot \bm{B}} \ . 
\end{equation}
The eigenvalues are 
\begin{equation}
m_{1,2} = m_n \pm \sqrt{(\bm{\mu}_n \cdot \bm{B})^2 + (\delta m)^2} \ -
 \frac{i\lambda}{2} \ . 
\end{equation}
Experiments typically reduced the magnitude of the magnetic field to $|\bm{B}| \sim 10^{-4}~\text{G} =
10^{-8}$~T, so $|\bm{\mu}_n \cdot \bm{B}| \simeq 10^{-21}$~MeV. Since one knows from the
experimental bounds that $|\delta m| \lsim 10^{-29}~\text{MeV}$, which is much
smaller than $|\bm{\mu}_n \cdot \bm{B}|$, it follows that $|\theta| \ll1$.  Thus, 
\begin{equation}
\Delta E \equiv m_1 - m_2 = 2 \sqrt{(\bm{\mu}_n \cdot
\bm{B})^2 + (\delta m)^2} \simeq 2 |\bm{\mu}_n \cdot \bm{B}| \ . 
\end{equation}
The transition probability is then 
\begin{equation}
P_{\bar{n}\leftarrow n}(t) = \sin^2(2\theta) \, \sin^2 [(\Delta E)t/2] \, e^{-\lambda t} \
. 
\end{equation}

In a free propagation experiment, one arranges that the neutrons propagate
for a time $t$ such that $|\bm{\mu}_n \cdot \bm{B}|t \ll 1$ and also $t
\ll \tau_n$.  Then, 
\begin{equation}
P_{\bar{n}\leftarrow n}(t) \approx  
(2\theta)^2 \Big ( \frac{\Delta E t}{2} \Big )^2 \simeq
\bigg ( \frac{\delta m}{\bm{\mu}_n \cdot \bm{B}} \bigg )^2
\bigg (\bm{\mu}_n \cdot \bm{B} \, t \bigg )^2 = [(\delta m) \, t]^2 =
(t/\tau_{\nnb})^2 \ . 
\end{equation}
Then the number of $\bar n$'s produced by the \nnb\ oscillations is given
essentially by $N_{\bar n}=P_{\bar{n}\leftarrow n}(t)N_n$, where $N_n$ is the number of neutrons observed.  The sensitivity of the experiment is proportional to the square of the propagation time $t$, so, with adequate
magnetic shielding, one wants to maximize $t$, subject to the condition that 
$|\bm{\mu}_n \cdot \bm{B}|t \ll 1$.

% =======================================================================

\subsubsection{Oscillations in Matter} 
\label{nnbar:subsubsec:matter}

To put the proposed free propagation \nnb\ oscillation experiment in
perspective, it is appropriate to review limits that have been achieved in the
search for \nnb\ oscillations in matter, using large nucleon-decay
detectors. In matter, the matrix $\mathcal{M}_A$ takes the form 
\begin{equation}
\mathcal{M}_A=\left(\begin{array}{cc}
m_{n\,\eff}  & \delta m \\
\delta m         & m_{\bar n\, \eff} \end{array}\right)
\label{mat}
\end{equation}
with
\begin{equation}
m_{n\, \eff} = m_n + V_n \ , \quad m_{\bar n\, \eff} = m_n + V_{\bar n} \ .
\end{equation}
The nuclear potential $V_n$ is practically real, $V_n = V_{nR}$, but $V_{\bar n}$ has
an imaginary part representing the $\bar n N$ annihilation,
\begin{equation}
V_{\bar n} = V_{\bar n R} - i V_{\bar n I} \ ,
\end{equation}
with~\cite{Dover:1983cd,Friedman:2008ef,Misra:1983sm}
\begin{equation}
V_{n R}, \ V_{\bar n R}, \ V_{\bar n I} \sim \mathrm{O}(100) \ \text{MeV} \ . 
\end{equation}
The mixing is thus strongly suppressed; $\tan(2\theta)$ is determined by
\begin{equation}
\frac{2\delta m}{|m_{n\, \eff} - m_{\bar n\, \eff}|} =
\frac{2\delta m}{\sqrt{(V_{n R}-V_{\bar n R})^2 + V_{\bar n I}^2}} \ll 1 \ . 
\end{equation}
Using the upper bound on $|\delta m|$ from the ILL reactor experiment, this
gives $|\theta| \lsim 10^{-31}$. This suppression in mixing is compensated for
by the large number of nucleons in a nucleon decay detector such as Soudan-2~\cite{Chung:2002jc} or
Super-Kamiokande~\cite{Abe:2011ka} e.g., $\sim 10^{33}$ neutrons in the (fiducial part of the)
Super-Kamiokande detector.

The eigenvalues  of $\mathcal{M}_A$ are 
\begin{equation}
m_{1,2} = \frac{1}{2} \bigg [ m_{n\, \eff} +  m_{\bar n\, \eff} \pm
\sqrt{ (m_{n\, \eff} -  m_{\bar n\, \eff})^2 + 4(\delta m)^2 } \ \bigg ] \ . 
\end{equation}
Expanding $m_1$ for the mostly-$n$ mass eigenstate $|n_1\rangle \simeq
|n\rangle$, one obtains
\begin{equation}
m_1 \simeq m_n + V_n - i \frac{(\delta m)^2 \, V_{\bar n I}}
{(V_{n R}-V_{\bar n R})^2 + V_{\bar n I}^2} \ . 
\end{equation}
The imaginary part leads to matter instability
via annihilation of the $\bar n$, producing mainly pions (with mean 
multiplicity $\langle n_\pi \rangle \simeq 4-5$).  The rate for this is
\begin{equation}
\Gamma_m = \frac{1}{\tau_m} = \frac{2(\delta m)^2 |V_{\bar n I}|}
{(V_{n R} - V_{\bar n R})^2 + V_{\bar n I}^2} \ . 
\end{equation}
Thus, $\tau_m = 1/\Gamma_m \propto (\delta m)^{-2}$. Writing
\begin{equation}
\tau_m = R \, \tau_{\nnb}^2 \ ,
\end{equation}
one has
\begin{equation}
R \simeq 100 \ \text{MeV} \ ,
\end{equation}
 i.e.,
\begin{equation}
R \simeq 1.5 \times 10^{23} \ \text{s}^{-1} \ . 
\end{equation}
The lower bound on $\tau_{\nnb}$ from \nnb\ searches in reactor
experiments yields a lower bound on $\tau_m$ and vice versa. With estimated
inputs for $V_{n R}$, $V_{\bar n R}$, and $V_{\bar n I}$ from nuclear
calculations, $\tau_{\nnb} > 0.86 \times 10^8$ s yields $\tau_m \gsim 2
\times 10^{31}$ yr.

Limits on matter instability due to \nnb\ oscillations have been
reported by several nucleon decay experiments~\cite{Beringer:2012jb}.  The signature is the
emission of an energy of $2m_n \simeq 2$ GeV, mainly in the form of pions.
However, these are emitted from a point within the nucleus 
(oxygen in a water Cherenkov detector and mainly iron in the Soudan detector),
and interact as they propagate through the nucleus.  Thus, modeling this
process is complicated.  In 2002, the Soudan experiment reported the bound~\cite{Chung:2002jc} 
\begin{equation}
\tau_m > 0.72 \times 10^{32} \ \text{yr} \ (90 \% \  \text{CL}) \ . 
\end{equation}
Using the relation 
\begin{equation}
\tau_{\nnb} = \sqrt{ \frac{\tau_m}{R}} \ , 
\end{equation}
this is equivalent to $\tau_{\nnb} \gsim 1.3 \times 10^8$ s. 
In 2011, the Super-Kamiokande experiment reported a limit~\cite{Abe:2011ka} 
\begin{equation}
\tau_m > 1.9  \times 10^{32} \ \text{yr} \ (90 \% \ \text{CL}) \ , 
\end{equation}
equivalent to $\tau_{\nnb} \gsim 2.4 \times 10^8$ s~\cite{Dover:1983cd}, or $\tau_{\nnb} \gsim 3.5 \times 10^8$ s~\cite{Friedman:2008ef}. 

The envisioned free neutron propagation experiment has the potential to improve
substantially on these limits. Achieving sensitivities of $\tau_{\nnb} \sim 10^9$ s to $10^{10}$ s would be roughly equivalent to
\begin{equation}
\tau_m \simeq (1.6-3.1 \times 10^{33} \ \text{yr}) \Big ( \frac{\tau_{\nnb}}
{10^9 \ \text{s}} \Big )^2 .
\end{equation}
A field-theoretic approach to the $\nnb$ transition in nuclei yields results very close to the results of
the potential approach~\cite{Kopeliovich:2011aa}, and further studies of the suppression in matter are under
way~\cite{ArkadyPXPS}.
%A.~Vainshtein, ``Neutron-antineutron oscillations \textit{versus} nuclei stability,'' contribution to the \PX\ Physics Study, \url{http://j.mp/11eRBuj}.
% =====================================================================

\subsection{Operator Analysis and Estimate of Matrix Elements}
\label{nnbar:subsec:analysis}

At the quark level, the $n \to \bar n$ transition is 
$(u d d) \to (u^c d^c d^c)$.  This is
mediated by six-quark operators ${\cal O}_i$, so the effective Hamiltonian is
\begin{equation}
H_{\eff} = \int d^3x {\cal H}_{\eff}  , 
\end{equation}
where the effective Hamiltonian density is 
\begin{equation}
{\cal H}_{\eff} = \sum_i c_i {\cal O}_i \ . 
\end{equation}
In four-dimensional spacetime, this six-quark operator has Maxwellian 
dimension 9 in mass units, so the coefficients have dimension $-5$. We write
them generically as 
\begin{equation}
c_i \sim \frac{\kappa_i}{M_X^5}
\end{equation}
If the fundamental physics yielding the \nnb\ oscillation is
characterized by an effective mass scale $M_X$, then, with $c_i \sim \mathrm{O}(1)$
(after absorbing dimensionless numerical factors into the effective scale
$M_X$), then the transition amplitude is
\medskip
\begin{equation}
\delta m = \langle \bar n | H_{\eff} | n \rangle = \frac{1}{M_X^5}
\sum_i c_i \langle \bar n |{\cal O}_i  | n \rangle
\end{equation}
Hence,
\begin{equation}
\delta m \sim \frac{\kappa \Lambda_{\mathrm{QCD}}^6}{M_X^5} \ , 
\end{equation}
where $\kappa$ is a generic $\kappa_i$ and $\Lambda_{\mathrm{QCD}} \approx 200$~MeV
arises from the matrix element $\langle \bar n | {\cal O}_i | n \rangle$.  For
$M_X \sim \text{few} \times 10^5$ GeV, one has $\tau_{\nnb} \simeq 10^9~\text{s}$.

The operators ${\cal O}_i$ must be color singlets and, for $M_X$
larger than the electroweak symmetry breaking scale, also $\text{SU}(2)_{\mathrm{L}} \times
\text{U}(1)_Y$-singlets.  An analysis of these (operators) was carried out in~\cite{Rao:1982sr}
and the $\langle \bar n | {\cal O}_i | n \rangle$ matrix elements were 
calculated in the MIT bag model.  Further results were obtained varying MIT bag
model parameters in~\cite{Rao:1984sr}.  These calculations involve integrals over 
sixth-power polynomials of spherical Bessel functions from the quark 
wavefunctions in the bag model.  As expected from the general arguments above,
it was found that 
\begin{equation}
|\langle \bar n | {\cal O}_i | n \rangle | \sim O(10^{-4})
\ \text{GeV}^6 \simeq (200 \ \text{MeV})^6 \simeq \Lambda_{\mathrm{QCD}}^6 .
\end{equation}
A calculation of the \nnb\ transition matrix elements at $\sim10$--20\% precision would be highly
informative.
In the near future, lattice QCD can provide a first-principles calculation of the complete set of \nnb\
transition matrix elements with controlled uncertainties.
Exploratory results for \nnb\ matrix elements presented at the \PX\ Physics Study~\cite{Buchoff:2012mb} are
consistent, at the order-of-magnitude level, with dimensional expectations.
For more discussion, see Sec.~\ref{lqcd:subsec:nnbar}.

Another interesting, successful way to describe baryons is as Skyrmions, topological configurations that are permitted in the chiral Lagrangian once a stabilizing term is added~\cite{Skyrme:1961vq, Skyrme:1962vh}. Being topological objects, pure Skyrmions are forbidden from decaying and are therefore not a useful laboratory for studying baryon-number--violating processes.  The chiral bag model, in which the center of the Skyrmion is replaced with a volume of free massless quarks, joins the exact chiral symmetry of the Skyrme picture with a more accurate short-distance description of QCD. This modification relaxes the topological selection rule that would forbid proton decay or \nnb\ oscillations, but Martin and Stavenga have argued~\cite{Martin:2011nd,GerbenPXPS} that an important inhibition remains. They estimate a $\times 10^{-10}$ suppression of the \nnb\ oscillation rate. This line of reasoning requires further examination.
% =====================================================================

%%%%%%%%%%%%%%%%%%%%%%%%%%%%%%%%%%%%%%%%%%%%%%%%%%%%%%%%%%%%
\section{\protect\nnbx: A Search for $\protect\nnb$ Oscillations with \PX}
\label{nnbar:sec:neutronpx}
%%%%%%%%%%%%%%%%%%%%%%%%%%%%%%%%%%%%%%%%%%%%%%%%%%%%%%%%%%%%

As mentioned in Sec.~\ref{nnbar:sec:intro}, the search for \nnb\ oscillations
using free neutrons (as opposed to neutrons bound in nuclei) requires intense beams of
very low energy (meV) neutrons. Such neutron beams are available at
facilities optimized for condensed matter studies focused on neutron
scattering. These sources may be based on high flux reactors such as
the ILL or the High Flux Isotope Reactor (Oak Ridge) or on accelerator
based spallation sources such as the SNS, the JSNS in Japan, or SINQ
(Switzerland).  Indeed, as stated in Sec.~\ref{nnbar:sec:intro}, the best
limit to date for \nnb\
oscillation times was set at the ILL  in 1991. Existing neutrons
sources are designed and optimized to serve a large number of neutron
scattering instruments that each require a relatively small beam. A
fully optimized neutron source for an \nnb\ oscillation
experiment would require a beam having a very large cross section and
large solid angle. There are no such beams at existing sources as
these attributes would preclude them from providing the resolution
necessary for virtually all instruments suitable for materials
research. The creation of such a beam at an existing facility would
require very major modifications to the source/moderator/shielding
configuration that would seriously impact the its efficacy for neutron
scattering. In point of fact, the reason there has been no improvement
in the limit on free neutron \nnb\ oscillations since the ILL
experiment of 1991 is that no substantial improvement is possible
using existing sources (or any likely future source devoted to
materials research).

From Sec.~\ref{nnbar:subsubsec:bfield}, the figure of merit for the sensitivity of a free
\nnb\ search experiment is $N_{n}\cdot t^{2}$, where $N_{n}$
is the number of free neutrons observed and $\emph{t}$ is the neutron
observation time. A schematic of the ILL \nnb\ experiment~\cite{Baldo:1994bc} is
shown in Fig.~\ref{ill:fig:logo}. The initial intensity of the
neutron source was determined in the ILL experiment by the brightness
of the liquid deuterium cold neutron source and the transmission of
the curved neutron guide.  Although, in principle, one expects the
sensitivity to improve as the average velocity of 
neutrons is reduced, it is
not practical to use very cold (velocity below 200~m/s) and ultracold
neutrons UCN (below 7~m/s) with a horizontal layout for the \nnb\ search due to
effects of Earth's gravity, which will not allow free transport of very
slow neutrons over significant distances in the horizontal direction.

Only modest improvements in the magnetic
field and vacuum levels reached for the ILL experiment would still
assure satisfaction of the quasi-free condition for the horizontal
experiment planned at \PX, but in our ongoing optimizations we
will investigate limits of $|\bm{B}|\leq 1$~nT in the whole free
flight volume and vacuum better than $P\sim 10^{-5}$~Pa in
anticipation of the more stringent requirements for the vertical
experiment. The costs of realizing these more stringent goals will be
considered in our ongoing optimization of the experimental design.

\begin{figure}
    \centering
    \includegraphics[width=0.99\textwidth]{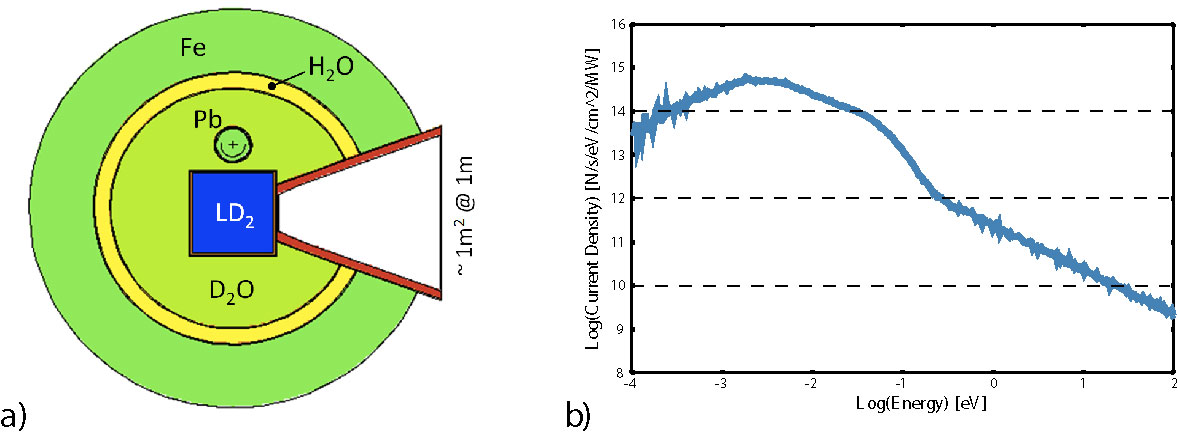} 
    \caption[Initial NNbarX source design]{Initial NNbarX source design.
        Panel (a) depicts the layout of a baseline cold neutron source geometry and (b) depicts an MCNP 
        simulation of the cold neutron spectrum entering the neutron optical system.}
    \label{nnbarx:fig:source}
\end{figure}

The \PX\ spallation target system will include a cooled
spallation target, reflectors and cold source cryogenics, remote
handling, nonconventional utilities, and shielding.  The delivery
point of any high-intensity beam is a target which presents
technically challenging issues for optimized engineering design, in
that optimal neutron performance must be balanced by effective
strategies for heat removal, radiation damage, remote handling of
radioactive target elements, shielding, and other aspects and
components of reliable safe operation.

The NNbarX baseline design incorporates a spallation target core which
can be cooled by circulating water or heavy water and will be coupled
to a liquid deuterium cryogenic moderator with optimized size and
performance (see Fig. \ref{nnbarx:fig:source}).  As we point out
below, existing, operating spallation sources provide an excellent
starting point for an optimized target design, as several such sources
exist and would be perfectly adequate for the NNbarX experiment at
Fermilab. In the next three sections, we review some of the
specifications for operating 1 MW spallation neutron sources, our
strategy to increase the number of neutrons we direct to the
annihilation target, and the sensitivity improvements relative to
the ILL experiment.

%%%%%%%%%%%%%%%%%%%%%%%%%%%%%%%%%%%%%%%%%%%%%%%%%%%%%%%%%%%%
\subsection{Currently Existing Spallation Sources}
\label{nnbar:subsec:current}
%%%%%%%%%%%%%%%%%%%%%%%%%%%%%%%%%%%%%%%%%%%%%%%%%%%%%%%%%%%%

Domestic and international 1 MW spallation sources include the Spallation Neutron Source (SNS) at ORNL and
the PSI SINQ~\cite{Blau:2009bb,Fischer:1997wf} source in Villigen, Switzerland.
The SNS at ORNL~\cite{Mason:2006tm} uses a liquid mercury target running at 1.0 MW with a proton energy of
825 MeV, and a frequency of 60 Hz.
The time-averaged flux of neutrons with kinetic energies below 5 meV at a distance of 2 m from the surface of
the coupled moderators is $1.4\times 10^{9}~n\,\text{cm}^{-2}\,\text{s}^{-1}$ at
1~MW~\cite{Iverson:2003ei,Iverson:2013ei}.
A similar source, JSNS, is running at JPARC in Japan~\cite{Maekawa:2010fk}.

The SINQ source is currently the strongest operating continuous mode spallation neutron source in the world.
It receives a continuous (51 MHz) 590 MeV proton beam at a current up to 2.3~mA.
Under normal operation the beam current is typically 1.5~mA.
The SINQ source uses a \emph{cannelloni} target made of an array of Zircaloy clad lead cylinders.
The cold neutron beam contains a flux of $2.8\times10^{9}~n\,\text{cm}^{-2}\,\text{s}^{-1}$ at 1MW and a
distance of 1.5~m from the surface of the Target 8 coupled moderators~\cite{Wohlmuther:2011mw,Wagner:1997ww}.
These facilities demonstrate that the substantial engineering challenges of constructing a 1 MW spallation
target/moderator/reflector (TMR) system can be overcome.
However, as noted earlier, none of these existing multipurpose facilities is a suitable host for the next
generation \nnb\ experiment due to constraints imposed on their TMR designs by their materials research
missions.

%%%%%%%%%%%%%%%%%%%%%%%%%%%%%%%%%%%%%%%%%%%%%%%%%%%%%%%%%%%%
\subsection{Increased Sensitivity of the \protect\nnbx\ Experiment}
\label{nnbar:subsec:sensitivity}
%%%%%%%%%%%%%%%%%%%%%%%%%%%%%%%%%%%%%%%%%%%%%%%%%%%%%%%%%%%%

A higher sensitivity in the NNbarX experiment compared to the previous
ILL experiment~\cite{Baldo:1994bc}, can be achieved by employing
various improvements in neutron optics and
moderation~\cite{Snow:2009ws}.  Conventional moderator designs can be
enhanced to increase the yield of cold neutrons through a number of
neutronics techniques such as a reentrant moderator
design~\cite{Ageron:1989pa}, use of
reflector/filters~\cite{Mocko:2013mm},  supermirror
reflectors~\cite{Swiss:2013sn}, and high-albedo materials such as
diamond nanoparticle
composites~\cite{Nezvizhevsky:2008vz,Lychagin:2009el,Lychagin:2009em}.
Although potentially of high positive impact for an \nnb\
experiment, some of these techniques are not necessarily suitable for
multipurpose spallation sources serving a materials science user
community (where sharply defined neutron pulses in time may be
required, for example).

Supermirrors based on multilayer coatings can greatly increase the
range of reflected transverse velocities relative to the nickel guides
used in the ILL experiment. In the following discussion, $m$,
denotes the increased factor for near-unity reflection above
nickel. Supermirrors with $m = 4$, are now mass-produced and
supermirrors with up to $m = 7$, can be produced~\cite{Swiss:2013sn}.

\begin{figure}
    \centering
    \includegraphics[width=0.99\textwidth]{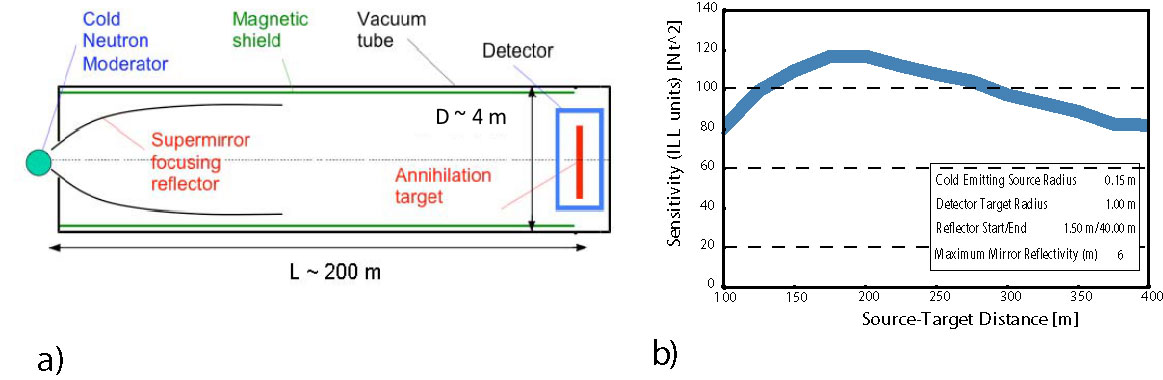}
    \caption[NNbarX layout and sensitivity]{The NNbarX layout and a sensitivity calculation.
        Panel~(a) is a schematic diagram of a candidate NNbarX geometry, depicting the relative location of 
        the cold neutron source, reflector, target, annihilation detector and beam dump.
        Panel~(b) depicts a calculation of the \nnb\ oscillation sensitivity for a geometry similar to that 
        in panel~(a), where all parameters are fixed except for the source-target distance $L$.
        The semi-major axis of the elliptical reflector is equal to $L/2$, so one focus is at the source and 
        the other is at the target.}
    \label{nnbarx:fig:sensitivity}
\end{figure}

To enhance the sensitivity of the \nnb\ search the
supermirrors can be arranged in the shape of a truncated focusing
ellipsoid~\cite{Kamyshkov:1995yk} as shown in
Fig.~\ref{nnbarx:fig:sensitivity}a.  The focusing reflector with a large
acceptance aperture will intercept neutrons within a fixed solid angle
and direct them by single reflection to the target. The cold neutron
source and annihilation target will be located in the focal planes of
the ellipsoid. The geometry of the reflector and the parameter ${\it
m}$ of the mirror material are chosen to maximize the sensitivity
$N_{n}\cdot t^{2}$ for a given brightness of the source and a given
size of the moderator and annihilation target.  Elliptical
concentrators of somewhat smaller scale have already been implemented
for a variety of cold neutron
experiments~\cite{Boni:2010pb}. Critically, the plan to create a ${\it
dedicated}$ spallation neutron source for particle physics experiments
creates a unique opportunity to position the NNbarX neutron optical
system to accept a huge fraction of the neutron flux, resulting in
large gains in the number of neutrons directed to the annihilation
target. Because such a strategy makes use of such a large fraction of
the available neutrons for a single beamline, it would be incompatible
with a typical multi-user materials science facility.  The NNbarX
collaboration contains specialists in neutronics design, moderator
development and spallation target construction and design (including
leaders of the design and construction team for the SNS and the Lujan
Mark III systems).  Initial steps towards an optimized design have
been taken, with an NNbarX source design similar to the SINQ source
modeled and vetted vs. SINQ source performance (see
Fig. \ref{nnbarx:fig:source}), and a partially optimized elliptical
neutron optics system shown in Fig. \ref{nnbarx:fig:sensitivity}(a).

MCNPX~\cite{Mcnpx:1981mc} simulation of the performance of the cold
source shown in Fig.~\ref{nnbarx:fig:source} produced a flux of
cold neutrons emitted from the face of cryogenic liquid  deuterium
moderator into forward hemisphere with the spectrum shown in
Fig.~\ref{nnbarx:fig:source}.  Only a fraction of the integrated flux is
accepted by the focusing reflector to contribute to the sensitivity at
the annihilation target.

For sensitivity ($N_n\cdot t^{2}$) calculations, neutrons emitted from
the surface of neutron moderator were traced through  the detector
configuration shown in Fig.~\ref{nnbarx:fig:source} with gravity
taken into account and with focusing reflector  parameters that were
adjusted by a partial optimization procedure. The flux of cold
neutrons impinging on the annihilation detector target located at the
distance $L$ from the source was calculated after reflection (mostly
single) from the focusing mirror. The time of flight to the target
from the last reflection was also recorded in the simulation
procedure. Each traced neutron contributed its $t^{2}$ to the total
sensitivity figure $N_n\cdot t^{2}$ that was finally normalized to the
initial neutron flux from the moderator. Sensitivity as function of
distance between neutron source and target ($L$) is shown in
Fig.~\ref{nnbarx:fig:sensitivity}(b). The simulation has several parameters
that affect the sensitivity: emission area of the moderator, distance
between moderator and annihilation target, diameter of the
annihilation target, starting and ending distance for truncated
focusing mirror reflector, minor semi-axis of the ellipsoid, and the
reflecting value ``$m$" of the mirror. Sensitivity is a complicated
functional in the space of these parameters. An important element of
our ongoing design work is to understand the projected cost for the
experiment as a function of these parameters.

A sensitivity in NNbarX in units of the ILL experiment larger than 100
per year of  (i.e. a 300-fold gain over the anticipated three-year
run) seems feasible from these simulations.  Configurations of
parameters that would correspond to even larger sensitivities are
achievable, but for the baseline simulation shown in the
Fig.~\ref{nnbarx:fig:sensitivity} we have chosen a set of parameters that
we believe will be reasonably achievable and economical after
inclusion of more engineering details than can be accommodated in our
simulations to date.

As emphasized above, the optimal optical configuration for an
\nnb\ search is significantly different from anything that has
previously been built, so the full impact on the sensitivity of cost
and other engineering considerations is not straight-forward to
predict at this early stage of the project. To demonstrate that the
key parameters contributing to the sensitivity predicted by these
simulations do not dramatically depart from existing engineering
practice, we include below a table identifying the value of these same
parameters at existing MW-scale spallation neutron sources for the
source and optical parameters, and the 1991 ILL experiment for the
overall length $L$.

\begin{table}
    \centering
    \begin{threeparttable}
    \caption{Comparison of parameters in NNbarX simulations with
            existing practice.}
    \label{edm:tab:lqcd}
    \begin{tabular}{ccccc}
       \hline\hline 
       Parameter & Units & NNbarX & Existing MW    & Ref. \\ 
                 &       & Simulations    & Facility Value & \\ 
       \hline
       Source brightness & $n$/(s cm$^{2}$ sterad MW) &
       3.5$\times$10$^{12}$ & 4.5$\times$10$^{12}$ &
       \cite{Maekawa:2010fk} \\  ($E <$ 400 meV) &  &  &  & \\
       Moderator viewed area & cm$^{2}$ & 707 & 190 &
       \cite{Maekawa:2010fk} \\ Accepted solid
       angle\tnote{1} & sterad & 0.2 & 0.034 & \cite{Kai:2005tk} \\ Vacuum
       tube length & m & 200 & 100 & \cite{Baldo:1994bc} \\ $^{12}$C
       target diameter & m & 2.0 & 1.1 & \cite{Baldo:1994bc} \\
       \hline\hline
    \end{tabular}
    \begin{tablenotes}
        \footnotesize
        \item[1] The solid angle quoted from JSNS is the total for a coupled parahydrogen moderator feeding 
            five neighboring beamlines (each of which would see a fifth of this value), whereas at NNbarX 
            the one beam accepts the full solid angle.
    \end{tablenotes}
    \end{threeparttable}
\end{table}

%%%%%%%%%%%%%%%%%%%%%%%%%%%%%%%%%%%%%%%%%%%%%%%%%%%%%%%%%%%%
\subsection{Requirements for an Annihilation Detector}
\label{nnbar:subsec:detector}
%%%%%%%%%%%%%%%%%%%%%%%%%%%%%%%%%%%%%%%%%%%%%%%%%%%%%%%%%%%%

As mentioned in Sec.~\ref{nnbar:sec:intro}, a free \nnb\
transformation search NNbarX experiment could require a vacuum of
10$^{-5}$~Pa and magnetic fields of $|\bm{B}| <1$~nT along the flight path
of the neutrons. The target vacuum is achievable with standard vacuum
technology, and the magnetic fields could be achieved with an
incremental improvement on the ILL experiment through  passive
shielding and straight-forward active field
compensation~\cite{Baldo:1994bc,Pxps:2012pp,Pxps:2012px}.

\begin{figure}
    \centering \includegraphics[width=0.66\textwidth]{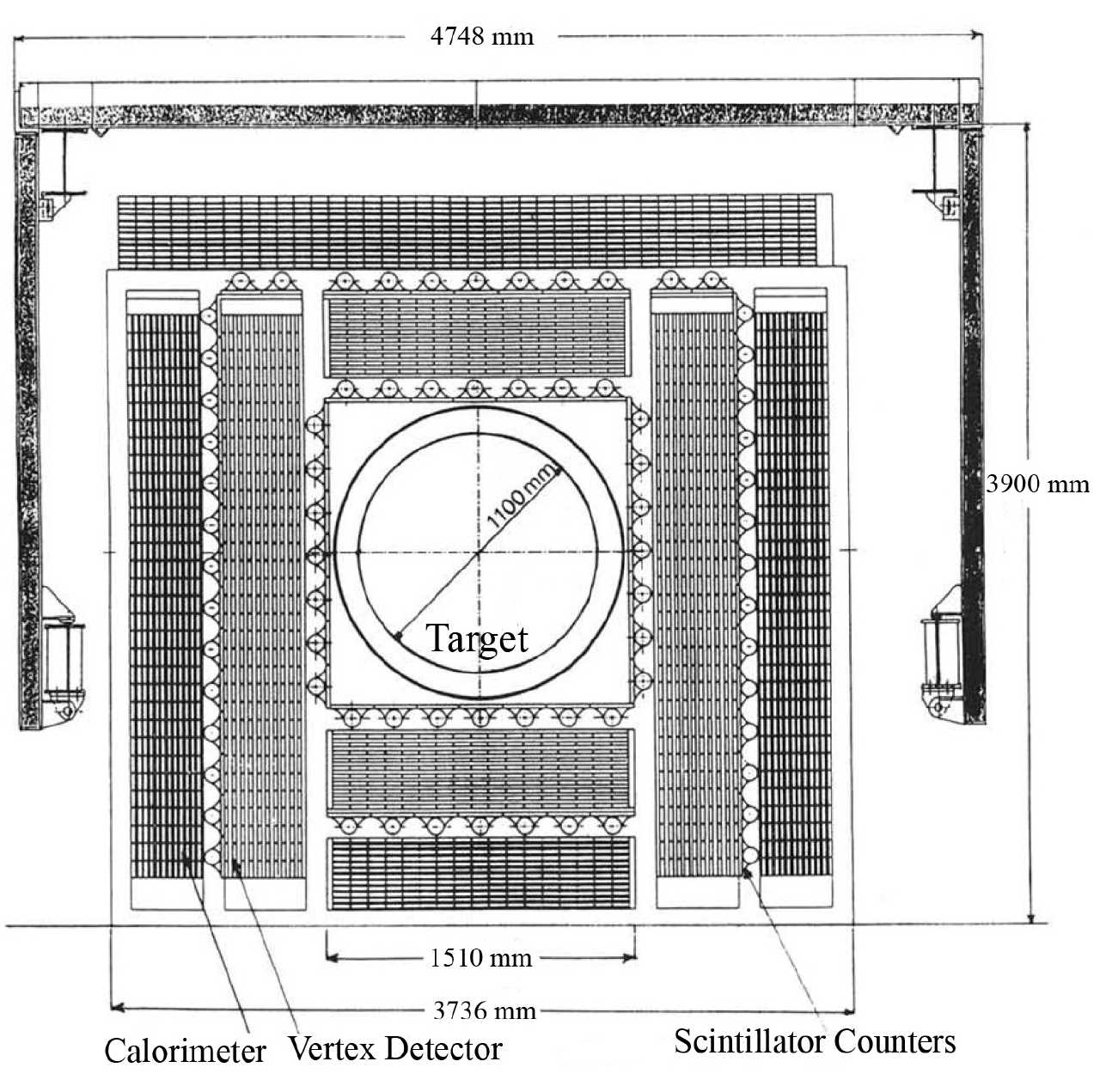} 
    \caption[ILL/Grenoble \protect\nnb\ detector]{Cross-sectional drawing of the ILL/Grenoble \nnb\
        annihilation detector apparatus~\cite{Baldo:1994bc}.}
    \label{nnbarx:fig:illdetector}
\end{figure}

In the design of the annihilation detector, our strategy is to develop
a state-of-the-art realization of the detector design used in the ILL
experiment~\cite{Baldo:1994bc}; see Fig. \ref{nnbarx:fig:illdetector}. 
Major subsystems of the NNbarX
annihilation detector (radially in the outward direction) will
include:  (i) the annihilation target; (ii) the detector vacuum
region; (iii) the tracker; (iv) the time of flight systems (before and
after the tracker); (v) the calorimeter; and (vi) the cosmic veto
system. Requirements for these subsystems are formulated below. In
general, the \nnb\ detector doesn't require premium performance, but
due to relatively large size needs rather careful optimization of the
cost. The detector should be built along the detector vacuum region
with several layered detection subsystems (sections (iii) - (vi)) and
should cover a significant solid angle (in $\theta$-projection from
$\sim$20$^{\circ}$ to 160$^{\circ}$ corresponding to the solid angle
coverage of $\sim$94$\%$). In the $\phi$-projection, the detector
configuration can be cylindrical, octagonal, hexagonal, or square
(similar to the ILL experiment~\cite{Baldo:1994bc}).

The spallation target geometry of NNbarX introduces a new
consideration in the annihilation detector design, because of the
possible presence of fast neutron and proton backgrounds.  These
backgrounds were effectively completely eliminated from the ILL
experiment, which produced fewer high energy particles in the reactor
source and eliminated the residual fast backgrounds using a curved
guide system to couple the cold source to the \nnb\ guide.  For
NNbarX, we utilize a strategy of integrating our shielding scheme for
fast particles into the design of the source and beamline, and
optimizing the choice of tracker detectors to differentiate between
charged and neutral tracks. We note that the residual fast backgrounds
at the detector are a strong function of the guide tube length,
detector threshold, and pulse structure for the proton beam.  In
particular, if needed, we can perform a slow chopping of the proton
beam (1 ms on, 1 ms off) to effectively eliminate fast backgrounds
completely.

%%%%%%%%%%%%%%%%%
\subsubsection{Annihilation Target}
\label{nnbar:subsubsec:target}
%%%%%%%%%%%%%%%%%

A uniform carbon disc with a thickness of $\sim$ 100 $\mu$m and
diameter $\sim$ 2 m would serve as an annihilation target. It would be
stretched on a low-$Z$ material ring and installed in the center
of the detector vacuum region. The choice of carbon is dictated by low
capture cross section for thermal neutrons $\sim$ 4 mb and high
annihilation cross-section $\sim$ 4 kb.  The fraction of hydrogen in
the carbon film should be controlled below $\sim$ 0.1$\%$ to to reduce
generation of capture $\gamma$s.

%%%%%%%%%%%%%%%%%%%
\subsubsection{Detector Vacuum Region}
\label{nnbar:subsubsec:vacuum}
%%%%%%%%%%%%%%%%%%%

The detector vacuum region should be a tube with inner diameter $\sim$
4 m and wall thickness $\sim$ 1.5 cm.  The wall should be made of
low-$Z$ material (Al) to reduce multiple scattering for tracking
and provide a low $(n,\gamma)$ cross-section.  Additional
lining of the inner surface of the vacuum region with $^{6}$LiF pads
will reduce the generation of $\gamma$s by captured neutrons.  The
detector vacuum region is expected to be the source of $\sim$ 10$^{8}$
$\gamma$s per second originating from neutron capture. Unlike in the
neutron beam flight vacuum region, no magnetic shielding is required
inside the detector vacuum region. As mentioned before, the vacuum
level should be better than 10$^{-4}$~Pa via connection with the neutron beam
vacuum region.  We plan to have a section of the vacuum tube in the
detector recessed.  This area will have no support or detector
elements in the neutron beam, which will reduce the rate of neutron
captures.

%%%%%%%%%%%%%%%
\subsubsection{Tracker}
\label{nnbar:subsubsec:tracker}
%%%%%%%%%%%%%%%

The tracker should be radially extended from the outer surface of the
detector vacuum tube by $\sim$ 50 cm and should have solid angle
coverage of $\sim$20$^{\circ}$ to 160$^{\circ}$.  It should provide
rms $\leq$ 1 cm accuracy of annihilation vertex reconstruction to the
position of the target in the $\theta$-projection (compared to 4 cm in
ILL experiment). This is a very important resource for the control of
background suppression in the detector. Reconstruction accuracy in the
$\phi$-projection can be a factor of 3 - 4 lower. Vertex information
will be also used for the total momentum balance of annihilation
events both in the $\theta$- and $\phi$-projections. Relevant tracker
technologies can include straw tubes, proportional and drift
detectors. Limited Streamer Tubes (LST), as used in the ILL
experiment, are presumed to be worse than proportional mode detectors
due to better discrimination of the latter to low-energy capture
$\gamma$s.

A system similar to the ATLAS transition radiation tracker (TRT) is
currently under consideration for the tracking system.  The ATLAS TRT
covers a pseudorapidity range less than 2 and has a measured barrel
resolution of 118 $\mu$m and an end-cap resolution of 132 $\mu$m.  The
ATLAS TRT is capable of providing tracking for charged particles down
to a transverse momentum of $p_{T} =$ 0.25 GeV with an efficiency
above 90$\%$, but typically places a cut of $p_{T} >$ 1.00 GeV due to
combinatorics on the large number of tracks in collision events.  For
tracks that have at least 15 TRT hits, a transverse momentum $p_{T} >$
1.00 GeV, and are within 1.3 mm of the anode, the efficiency was found
to be 94.4$\%$ for the 7 TeV ATLAS data with similar results for the
0.9 TeV ATLAS data
set~\cite{Animma:2011an,Stahlman:2011js,Boldyrev:2012ab,Vogel:2013av}.
For a cut of $p_{T} >$ 0.25 GeV, the efficiency drops down to
93.6$\%$.  For higher momentum tracks (e.g. $p_{T} >$ 15.00 GeV), the
efficiency increases to 97$\%$ and is more indicative of the
single-straw efficiency~\cite{Vankooten:2013rv}.  The efficiency drops
at the edges of the straw due to geometric and reconstruction effects.
The straw tubes in the TRT have a diameter of 4 mm and are made from
wound kapton reinforced with thin carbon fibers.  The anode at the
center of each straw is gold plated tungsten wire with a diameter of
31 $\mu$m.  The cathodes were kept at -1.5 kV, while the anodes were
kept at ground.  The tubes are filled with a gas mixture of 70$\%$ Xe,
27$\%$ CO$_{2}$, and 3$\%$ O$_{2}$, however we will have to optimize
our gas mixture for a different set of backgrounds in this experiment,
particularly fast $n$-backgrounds and proton backgrounds.  If it
will be determined that the tracker should be moved inside the
detector vacuum region for better accuracy (also giving rise to the
problem of gas and electrical vacuum feedthroughs), then the
requirements on the detector tube material and thickness should be
revisited.

%%%%%%%%%%%%%%%%%%
\subsubsection{Time of Flight System}
\label{nnbar:subsubsec:tof}
%%%%%%%%%%%%%%%%%%

The time of flight (TOF) systems should consist of two layers of fast
detectors (e.g. plastic scintillation slabs or tiles) before and after
the tracker with solid angle coverage of $\sim$20$^{\circ}$ to
160$^{\circ}$.  With appropriate segmentation, TOF should provide
directional information for all tracks found in the tracker. The TOF
systems could also be a part of the trigger. With two layers separated
by $\sim$50 cm - 60 cm, the TOF systems should have timing accuracy
sufficient to discriminate the annihilation-like tracks from the
cosmic ray background originating outside the detector volume.

%%%%%%%%%%%%%%%
\subsubsection{Calorimeter}
\label{nnbar:subsubsec:calorimeter}
%%%%%%%%%%%%%%%

The calorimeter will range out the annihilation products and should
provide trigger signal and energy measurements in the solid angle
$\sim$20$^{\circ}$ to 160$^{\circ}$.  The average multiplicity of
pions in annihilation at rest equals 5, so an average pion can be
stopped in $\sim$20 cm of dense material (like lead or iron). For low
multiplicity (but small probability) annihilation modes, the amount of
material can be larger. Calorimeter configuration used in the ILL
experiment with 12 layers of Al/Pb interspersed with gas detector
layers (LST in ILL experiment) might be a good approach for the
calorimeter design. Detailed performance for the measurement of total
energy of annihilation events and momentum balance in $\theta$- and
$\phi$-projections should be determined from simulations. The
proportional mode of calorimeter detector operation possibly can be
less affected by copious low-energy $\gamma$-background than the LST
mode.  An approach using MINER$\nu$A-like wavelength shifting fibers
coupled to scintillating bars is also being
considered~\cite{Mcfarland:2006km}.

%%%%%%%%%%%%%%%%%%%
\subsubsection{Cosmic Veto System}
\label{nnbar:subsubsec:veto}
%%%%%%%%%%%%%%%%%%%

The cosmic veto system (CVS) should identify all cosmic ray
background. All annihilation products should be totally stopped in the
calorimeter. Large area detectors similar to MINOS scintillator
supermodules~\cite{Michael:2008dm} might be a good approach to the
configuration of the CVS. Possible use of timing information should be
studied in connection with the TOF system. CVS information might not
be included in the trigger due to high cosmogenic rates, particularly
in the stage-one horizontal \nnb\ configuration on the
surface, but should be recorded for all triggers in the off-line
analysis.

\begin{figure}
    \centering
    \includegraphics[width=0.92\textwidth]{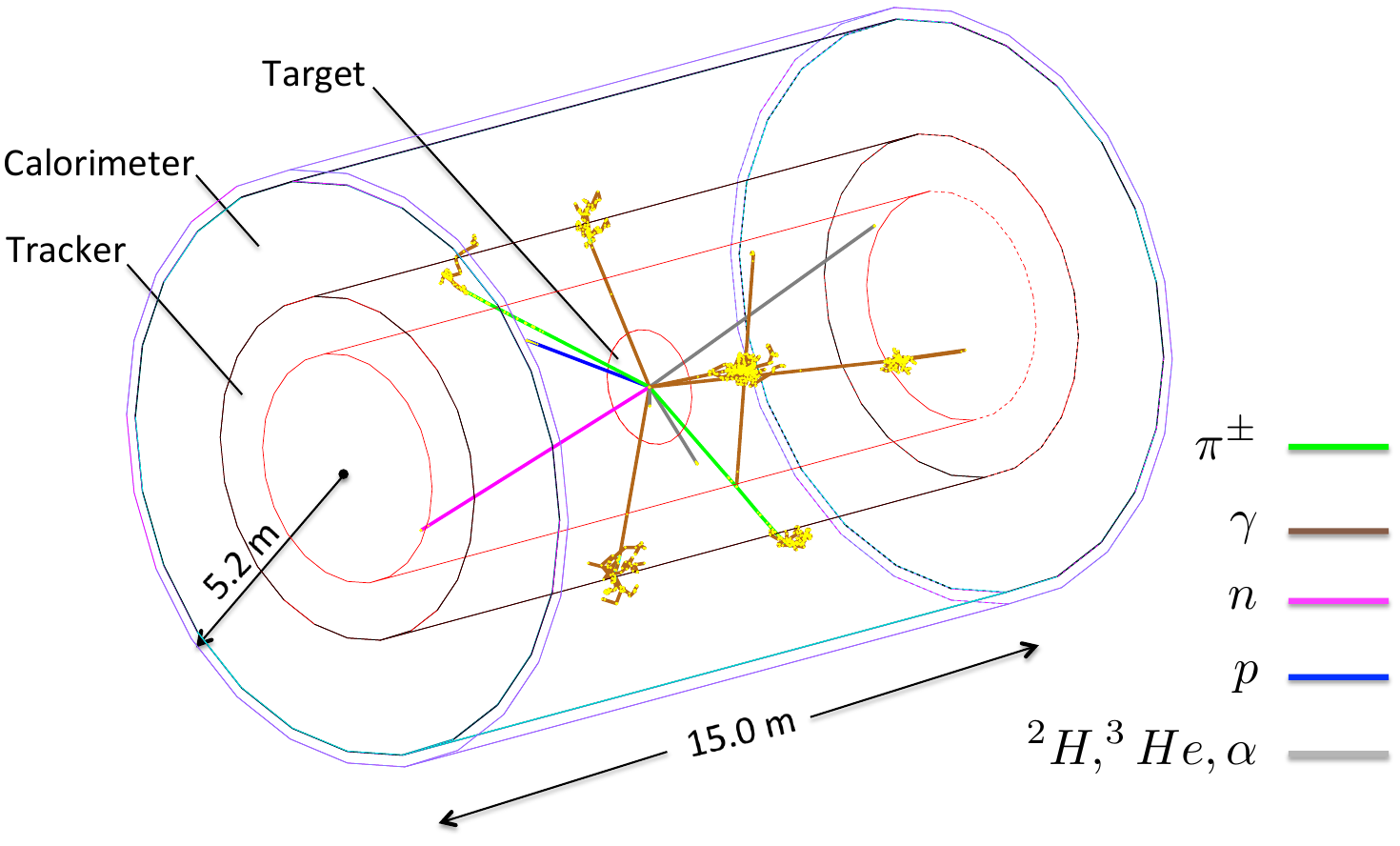} 
    \caption[Event display in Geant4 for a $\pi^{+}\pi^{-}2\pi^{0}$ annihilation event]{Event display 
        generated in our preliminary Geant4~\cite{Geant:2013ge} simulation for a $\pi^{+}\pi^{-}2\pi^{0}$ 
        annihilation event in a generalized NNbarX detector geometry. 
        Given the short lifetime of the $\pi^{0}$, they decay immediately to 2$\gamma$, as shown above.}
    \label{nnbarx:fig:nnbarxdetector}
\end{figure}

%%%%%%%%%%%%%%%%%%%%%%%%%%%%%%%%%%%%%%%%%%%%%%%%%%%%%%%%%%%%
\section{\protect\nnbx\ Simulation}
\label{nnbar:sec:simulation}
%%%%%%%%%%%%%%%%%%%%%%%%%%%%%%%%%%%%%%%%%%%%%%%%%%%%%%%%%%%%

Developing a detector model through simulation that allows us to reach our goal of zero background and
optimum signal event detection efficiency is the primary goal of our simulation campaign, which is currently
underway.
We are using Geant 4.9.6~\cite{Geant:2013ge} to simulate the passage of annihilation event products through
the annihilation detector geometry with concurrent remote development coordinated through
GitHub~\cite{Github:2013gi}.
A detailed treatment of \nnb\ annihilation modes in $^{12}$C is under development, however for this report,
we present a list of \nnb\ annihilation modes in $^{16}$O~\cite{Abe:2011ka} (see Table 6.2), which we expect
to be similar to the physics of NNbarX.
The event generator for \nnb\ annihilation modes in $^{12}$C uses programs developed for the IMB experiment
and Kamiokande II collaborations~\cite{Jones:1984tj,Takita:1986mt} validated in part by data from the LEAR
experiment~\cite{Golubeva:1996eg}.
The branching ratios for the \nnb\ annihilation modes and fragmentation modes of the residual nucleus were
taken from Ref.~\cite{Abe:2011ka,Berger:1990cb,Fukuda:2003yf,Botvina:1990ab}.
The cross sections for the $\pi$-residual nucleus interactions were based on extrapolation from measured
$\pi$-$^{12}$C and $\pi$-Al cross sections.
Excitation of the $\Delta$(1232) resonance was the most important parameter in the nuclear propagation phase.
Nuclear interactions in the event generator include $\pi$ and $\omega$ elastic scattering, $\pi$ charge
exchange, $\pi$-production, $\pi$-absorption, inelastic $\omega$-nucleon scattering to a $\pi$, and $\omega$
decays inside the nucleus.
Fig.~\ref{nnbarx:fig:nnbarxdetector} shows an event display from our preliminary Geant4 simulation of a
$\pi^{+}\pi^{-}2\pi^{0}$ annihilation event in a detector geometry with a generalized tracker and calorimeter.

\begin{table}
\centering
    \caption[List of \nnb\ annihilation modes]{List of \nnb\ annihilation modes and branching ratios from 
        the Super-Kamiokande simulation study~\cite{Abe:2011ka}.}
    \label{edm:tab:sim}
        \begin{tabular}{cc}
        \hline\hline \nnb\ Annihilation Mode & Branching Ratio \\ \hline
        $\pi^{+}\pi^{-}3\pi^{0}$ & 28$\%$ \\
        $2\pi^{+}2\pi^{-}\pi^{0}$ & 24$\%$ \\
        $\pi^{+}\pi^{-}2\pi^{0}$ & 11$\%$ \\
        $2\pi^{+}2\pi^{-}2\pi^{0}$ & 10$\%$ \\
        $\pi^{+}\pi^{-}\omega$ & 10$\%$ \\
        $2\pi^{+}2\pi^{-}$ & 7$\%$ \\
        $\pi^{+}\pi^{-}\pi^{0}$ & 6.5$\%$ \\
        $\pi^{+}\pi^{-}$ & 2$\%$ \\
        $2\pi^{0}$ & 1.5$\%$ \\
        \hline\hline
    \end{tabular}
\end{table}

% \clearpage

%%%%%%%%%%%%%%%%%%%%%%%%%%%%%%%%%%%%%%%%%%%%%%%%%%%%%%%%%%%%
\section{The \protect\nnbx\ Research and Development Program}
\label{nnbar:sec:randd}
%%%%%%%%%%%%%%%%%%%%%%%%%%%%%%%%%%%%%%%%%%%%%%%%%%%%%%%%%%%%

In October of 2012, the Fermilab Physics Advisory Committee strongly
supported the physics of NNbarX and recommended that ``R$\&$D be
supported, when possible, for the design of the spallation target, and
for the overall optimization of the experiment, to bring it to the
level required for a proposal to be prepared.''   At the core of this
activity is integrating models for the source, neutron optics and
detectors into a useful tool for evaluating overall sensitivity to
annihilation  events and fast backgrounds, and developing a cost
scaling model.  In addition to this activity, the NNbarX collaboration
has identified several areas where research and development may
substantially improve the physics reach of the experiment: target and
moderator design, neutron optics optimization and the annihilation
detector design.

As touched on Sec.~\ref{nnbar:subsec:sensitivity}, for the target and
moderator, there exist a number of improvements which have already
been established as effective that might be applied to our baseline
conventional source geometry.  For example, one can shift from a ${\it
cannelloni}$ target to a lead-bismuth eutectic (LBE)
target~\cite{Wohlmuther:2011mw}, utilize a reentrant moderator
design~\cite{Ageron:1989pa}, and possibly use
reflector/filters~\cite{Mocko:2013mm}, supermirror
reflectors~\cite{Swiss:2013sn}, and high-albedo materials such as
diamond nanoparticle
composites~\cite{Nezvizhevsky:2008vz,Lychagin:2009el,Lychagin:2009em}.
At present, the collaboration envisions a program to perform neutronic
simulations and possibly benchmark measurements on several of these
possibilities, with high-albedo reflectors as a priority.  At present,
we envision at least a factor of two improvement arising from some
combination of these improvements.

For neutron optics, members of the collaboration are currently
involved in the production of high $m$ supermirror guides.  Although
the basic performance is established, optimizing the selection of
supermirror technology for durability ($vs$ radiation damage) and cost
could have a very large impact on the ultimate reach of the experiment.

Finally, for the detector, the collaboration is using the WNR facility
at LANSCE to determine the detection efficiency and timing properties
of a variety of detectors from 10 MeV to 800 MeV neutrons.  Detectors
under evaluation include proportional gas counters with various gas
mixtures, straw tubes and plastic scintillators.  Evaluating different
available detector options and modernizing the annihilation detector
should improve the background rejection capability and permit reliable
scaling to more stringent limits for \nnb\ oscillations.  The
main technical challenges in for NNBarX is to minimize the cost of
critical hardware elements, such as the large-area super-mirrors,
large-volume magnetic shielding, vacuum tube, shielding of the
high-acceptance front-end of the neutron transport tube, and
annihilation detector components.  These challenges will be addressed
in the R$\&$D phase for the NNBarX experiment.

%%%%%%%%%%%%%%%%%%%%%%%%%%%%%%%%%%%%%%%%%%%%%%%%%%%%%%%%%%%%
\section{Summary}
\label{nnbar:sec:summary}
%%%%%%%%%%%%%%%%%%%%%%%%%%%%%%%%%%%%%%%%%%%%%%%%%%%%%%%%%%%%

Assuming beam powers up to 1 MW on the spallation target and that 1
GeV protons are delivered from the \PX\ linac, the goal of NNbarX
will be to improve the sensitivity of an \nnb\ search
($N_{n}\cdot t^{2}$) by at least a factor 30 (compared to the previous
limit set in ILL-based experiment~\cite{Baldo:1994bc}) with a
horizontal beam experiment; and by an additional factor of $\sim$ 100
at the second stage with the vertical layout. The R$\&$D phase of the
experiment, including development of the conceptual design of the cold
neutron spallation target, and conceptual design and optimization of
the performance of the first-stage of NNbarX is expected to take 2-3
years.  Preliminary results from this effort suggest that an
improvement over the ILL experiment by a factor of more than 100 may
be realized even in this horizontal mode, but more work is needed to
estimate the cost of improvements at this level.  The running time of
the first stage of NNbarX experiment is anticipated to be three years. The
second stage of NNbarX will be developed depending upon the
demonstration of technological principles and techniques of the first
stage.

%%%%%%%%%%%%%%%%%%%%%%%%%%%%%%%%%%%%%%%%%%%%%%%%%%%%%%%%%%%%

\bibliographystyle{apsrev4-1} \bibliography{nnbar/refs}
 % Albert Y. & Chris Q.

%%%%%%%%%%%%%%%%%%%%%%%%%%%%%%%%%%%%%%%%%%%%%%%%%%%%%%%%%%%%
% \chapter[New, Light, Weakly-coupled Particles with \PX]{Searching for New, Light, Weakly-coupled Particles 
% with \PX}
\chapter{New, Light, Weakly-coupled Particles with \PX}
\label{chapt:nlwcp}
%%%%%%%%%%%%%%%%%%%%%%%%%%%%%%%%%%%%%%%%%%%%%%%%%%%%%%%%%%%%

% \title{Proton Beams to Search for \\ New Light Weakly-Coupled Particles}
\author{Brian Batell, William Wester, \\
Patrick deNiverville, 
Ranjan Dharmapalan,
Athanasios~Hatzikoutelis,
David McKeen, \\
Maxim Pospelov,
Adam Ritz,
    and
Richard Van de Water}

\section{Introduction}

The empirical evidence for new physics, such as dark matter and neutrino mass, does not necessarily point to
a specific mass scale, but instead to a hidden sector, weakly-coupled to the Standard Model (SM).
This point has recently been amplified by the LHC's exploration of the weak scale which, despite the
impressive discovery of a SM-like Higgs boson, has yet to uncover new physics.
Hidden sectors containing light degrees of freedom, with mass in the MeV--GeV range, are motivated by various
questions about dark matter, neutrinos, and early universe cosmology as we discuss below.
An intense proton source such as \PX, with a fixed target and rare meson decay program, would provide an
ideal setting in which to explore this new physics landscape.

If we focus on the compelling evidence for dark matter, a number of anomalies in direct and indirect
detection have led recently to a broader theoretical perspective, beyond the characteristic weakly
interacting massive particle (WIMP) with a weak-scale mass.
The simple thermal relic scenario, with abundance fixed by freeze-out in the early universe, allows a much
wider mass range if there are light (dark force) mediators which control the annihilation rate.
Current direct detection experiments lose sensitivity rapidly once the mass drops below a few GeV, and
experiments at the intensity frontier provide a natural alternative route to explore this dark matter regime.
Moreover, dark matter may not be a thermal relic at all, and could be composed of sub-MeV very weakly
interacting slim particles (WISPs), e.g., axions, sterile neutrinos, gravitinos, dark photons, etc.
Possible inconsistencies of the $\Lambda$CDM picture of structure formation on galactic scales, and the
advent of precision CMB tests of light degrees of freedom at the era of recombination have also focussed
attention of the possibility of new light degrees of freedom.

These empirical (or bottom-up) motivations for exploring new light weakly-coupled particles (NLWCPs) can also
be placed within a more systematic framework.
As we discuss in the next subsection, a general effective field theory perspective of the interaction between
new gauge singlet fields with the SM points to a specific set of operators, known as {\it portals}.
These extend the usual right-handed neutrino coupling, which provides a natural explanation for neutrino
mass, to include interactions of dark singlet scalars with the Higgs, kinetic mixing of a new $\text{U}(1)$ dark
photon (or $Z'$) with the hypercharge gauge boson, and the coupling of axion-like pseudoscalars to the axial
vector current.
These couplings are also quite generic in top-down models of new physics.
Light pseudo-Nambu-Goldstone bosons, such as axions, are generic in scenarios where new symmetries are broken
at a high scale, and scalars and pseudoscalars can also arise from compactification of extra dimensions.
Extensions of the SM gauge group to include new $\text{U}(1)$ sectors are also quite generic in string theory.

Portal interactions naturally describe the generic coupling of light degrees of freedom in a hidden sector
with the SM.
The combination of relatively light sub-GeV mass, along with a weak (but not super-weak) coupling, lends
itself to production at high luminosity accelerator-based facilities.
In many cases the suppressed interaction rate also requires large volume detectors to search for rare
scattering events.
These features point to the intensity frontier, and the high luminosity proton source at \PX\ as ideally
suited to host an experimental program exploring this sector.

\subsection{Hidden Sectors}

A conventional parametrization of the interactions between the SM and a hidden sector assumes that any light
hidden sector states are SM gauge singlets.
This automatically ensures weak interactions, while the impact of heavier charged states is incorporated in
an effective field theory expansion of the interactions of these light fields at or below the weak scale,
\begin{equation}
    {\cal L} \sim \sum_{n=k+l-4} \frac{c_n}{\Lambda^n} {\cal O}^{(k)}_{\rm SM} {\cal O}^{(l)}_{\rm hidden},
\end{equation}
where the two classes of operators are made from SM and hidden fields, respectively.
The generic production cross section for hidden sector particles via these interactions scales as $\sigma
\sim E^{2n-2}/\Lambda^{2n}$.
It follows that the lower dimension interactions, namely those that are unsuppressed by the heavy scale
$\Lambda$, are preferentially probed at lower energy.
Such hidden sectors are natural targets for the intensity frontier.
Given the LHC's discovery of a SM-like Higgs boson, it is appropriate to delineate these interactions in a
form which builds in the SM electroweak gauge group structure.
In this case, the set of low-dimension interactions, usually termed {\it portals}, is quite compact.
Up to dimension five ($n\leq 1$), assuming SM electroweak symmetry breaking, the list of portals includes:
\begin{center}
\begin{tabular}{rl}
Dark photons & $-\frac{\kappa}{2}B_{\mu\nu}V^{\mu\nu}$ \\
Dark scalars & $ (A S + \lambda S^{2})H^{\dagger}H $ \\
Sterile neutrinos & $ y_N LHN $ \\
Pseudoscalars & $ \frac{\partial_{\mu}a}{f_{a}} \overline{\psi}\gamma^{\mu}\gamma^{5}\psi$ \\
\end{tabular}
\end{center}
On general grounds, the coupling constants for these interactions are either unsuppressed, or, for
pseudoscalars, minimally suppressed by any heavy scale of new physics, and thus it would be natural for new
weakly-coupled physics to first manifest itself via these portals.
Indeed, we observe that the right-handed neutrino coupling is amongst this list, which provides the simplest
renormalizable interpretation for neutrino mass and oscillations.
It is natural to ask if the other portals are also exploited in various ways, and many have been discussed
recently in the dark matter context.

\subsubsection{Light Dark Matter}
\label{nlwcp:sec:lightDM}

Dark matter provides one of the strongest empirical motivations for new particle physics, with a vast array
of evidence coming from various disparate sources in astrophysics and cosmology.
While the vast majority of the particle physics community has focused on the possibility of WIMPs with a mass
at the weak scale and interaction strength similar to the SM weak interactions, this is certainly not the
only possibility.
With the lack of evidence for new states at the weak scale from the LHC, a broader approach to the physics of
DM and new experimental strategies to detect its non-gravitational interactions are called for.
In particular, the particle(s) that comprise dark matter may be much lighter than the weak scale.
Crucially, in the regime of sub-GeV dark matter, direct searches looking for the nuclear recoil of DM
particles in the halo lose sensitivity.
High intensity proton beams offer a new opportunity to search for light DM particles.

An important requirement of light thermal relic dark matter is the presence of new mediators which connect
the SM to the dark sector, which open up new annihilation channels.
The same mediators can then be utilized as a bridge to the SM and give signatures in proton beam fixed target
experiments.
Simple models involving dark matter coupling through a dark photon that kinetically mixes with the SM have
been constructed in Refs.~\cite{Pospelov:2007mp,deNiverville:2011it}, and these models pass all terrestrial,
astrophysical and cosmological constraints.

\subsubsection{Dark Photons}

A new $\text{U}(1)$ vector gauge boson $V_\mu$ can couple via kinetic mixing~\cite{Holdom:1985ag} with the
hypercharge gauge boson of the SM: ${\cal L} \supset -({\kappa}/{2}) V^{\mu\nu} F_{\mu\nu}$, providing one of
the few renormalizable interactions between the SM and a hidden sector.
In terms of the physical mass eigenstates, the interaction above generates a coupling between the dark photon
and ordinary matter, ${\cal L}\supset e \kappa V_\mu \bar \psi_{SM} \gamma^\mu \psi_{SM}$.
The strength of the kinetic mixing can range over many orders of magnitude depending on how it is generated
at the high scale.
For example, in supersymmetric models, it is quite naturally a loop factor below the scale of the
electromagnetic coupling.
Interest in dark photons in recent years has been motivated by a variety of experimental and observational
data.
The observation of a rise in the cosmic ray positron spectrum~\cite{Adriani:2008zr,Aguilar:2013qda} is
suggestive of TeV-scale dark matter interacting through a new dark force mediated by the dark
photon~\cite{ArkaniHamed:2008qn,Pospelov:2008jd}.
Furthermore, a dark photon with a mass in the range of several MeV to a few~GeV gives a positive
contribution to the anomalous magnetic moment of the muon~\cite{Fayet:2007ua,Pospelov:2008zw}, potentially
resolving the $3 \sigma$ discrepancy between theory and experiment~\cite{Bennett:2006fi}.
Indeed new experimental programs to search for such dark photons decaying to SM final states has commenced
at Thomas Jefferson National Laboratory and at the Institute for Nuclear Physics of the Johannes Gutenberg
University of Mainz using electron-beam fixed-target
experiments~\cite{Bjorken:2009mm,Essig:2010xa,Merkel:2011ze,Abrahamyan:2011gv}.
 
\subsubsection{Dark Scalars}

Given the discovery of the Higgs boson by the LHC experiments, the possibility of a Higgs portal to a hidden
sector has become a reality.
The Higgs portal couples new scalars to the SM via the operator ${\cal L}\supset (A S+\lambda S^2)H^\dag H$.
Higgs mediated interactions between light fermions are the amongst the weakest in the SM, and characterize
the sensitivity of the current generation of direct detection experiments looking for WIMP dark matter;
indeed $S$ provides a simple WIMP candidate if $A=0$.
The small SM width of the Higgs, combined with the existence of the low dimension portal, makes probes of
Higgs couplings a primary test of new physics.
The LHC limits on the Higgs invisible width impose constraints on light scalars coupled through the Higgs
portal, but precision tests through rare decays of $B$ and $K$ mesons at the intensity frontier can provide
greater sensitivity in the relevant kinematically accessible mass range.
Producing these states via proton beams is more difficult at low energy due to the Yukawa suppression of the
Higgs coupling to light quarks, and the small parton densities for sea quarks.
It should be noted though that a dark $\text{U}(1)$ rendered massive via the Higgs mechanism in the hidden sector
naturally allows a Higgs portal coupling to the {\it dark Higgs}, and this can be probed more efficiently via
dark Higgs-strahlung.

\subsubsection{Singlet Neutrinos}

A conventional weakly coupled particle that falls within this classification is the sterile neutrino.
While the right-handed neutrinos of a type I see-saw may be too heavy to mediate interactions of interesting
strength, light sterile neutrinos could help to resolve neutrino oscillation anomalies, and are another light
dark matter candidate.
These scenarios can be tested at long-baseline facilities, either by precise tests of the neutrino
oscillation pattern, or for heavier mass via precise measurements of neutrino scattering in the near detector.
It is important to note that larger-than-weak couplings of new singlet neutrinos to the baryon current are
not well constrained by other experiments.
Such baryonic neutrinos \cite{Pospelov:2011ha,Harnik:2012ni} could play a role in various low mass anomalies
in direct detection, and could be searched for at high luminosity proton fixed target experiments.

\subsubsection{Axion-like Particles}

The QCD axion is a highly motivated dark matter candidate, as it derives naturally in the context of quantum
chromodynamics via spontaneous breaking of a new symmetry that forces \CP-conservation by the strong
interaction.
While the parameter space for which axions may contribute significantly to dark matter is best probed with
resonant cavities such as the ADMX experiment, other pseudoscalars produced in high scale symmetry breaking
can also naturally be light and mediate interactions with a hidden sector via the pseudoscalar portal.
Axion-like particles (ALPs), for which the mass is not tied to the symmetry breaking scale which solves the
strong \CP\ problem, may therefore be probed at the intensity frontier \cite{Ringwald:2012hr}.

\subsubsection{Other Possibilities}

We have summarized some of the scenarios analyzed in the recent literature, which involve couplings to the lowest dimension singlet portals. There are of course many other possibilities, for which an intense proton source could provide sensitivity. We should mention the possibility of allowing for parity-violation in the mediator couplings, as initially studied in some generality for dark $\text{U}(1)$ vectors \cite{Fayet:1980ss,Fayet:1981rp}, and couplings to flavor-dependent lepton and baryon currents. While such scenarios are generally more complex, and may require additional states for anomaly cancelation, we note that the existing sensitivity can be comparatively weak. 

\subsection{Current Experimental Sensitivity}

The past five years has seen a renewed interest in experimental probes of light weakly-interacting particles, with a focus on testing the portal couplings. In this subsection, we briefly summarize the current landscape.

\bigskip
\noindent{\it Neutrino beams and proton fixed targets}: There is already a significant infrastructure of short and long-baseline neutrino beam experiments. Most utilize intense proton sources impacting a target, with a decay volume in which charged pions, kaons and muons decay to produce neutrinos. Facilities such as MINOS, NO$\nu$A, T2K, and MiniBooNE already provide significant sensitivity to sterile neutrinos and non-standard interactions (NSI's). These facilities  can also  exploit the large volume (near-)detectors to study light states coupled through the other portals. However, the need to suppress the large neutrino background actually favors running in beam-dump mode, without the large decay volume. A number of constraints have been deduced from existing data, but as yet the only dedicated analysis for light dark matter coupled via the vector portal is being explored at MiniBooNE \cite{Dharmapalan:2012xp}.

\bigskip \noindent{\it Rare meson decays}: The search for rare decays has for many years imposed stringent
constraints on models of new physics.
The kaon physics program at \PX\ could play an important role in searches for hidden sectors.
The ORKA experiment, aiming to measure the $K^+ \rightarrow \pi^+\nu\bar\nu$ rate, will have sensitivity to
suppressed decays to light dark matter coupled via a dark photon.
Similarly decays of kaons and $B$ mesons are sensitive tests of suppressed couplings via the scalar and
pseudoscalar portals.

\bigskip
\noindent{\it Electron fixed targets}: A number of experiments at JLab and MAMI/Mainz, e.g., APEX, HPS, DarkLight, and others \cite{Bjorken:2009mm,Essig:2010xa,Merkel:2011ze,Abrahamyan:2011gv,Boyce:2012ym} have recently been developed to search primarily for light dark photons through their decays to electrons and muons. Electron beams are generally lower in energy than the existing proton beams, but have the advantage of a cleaner electromagnetic production process. 

\bigskip \noindent{\it Meson factories}: Significant sensitivity to various portals is available via heavy
meson factories such as BaBar, Belle, Kloe, BES-III, and in the future Belle-II.
The ultimate luminosity is lower than for proton fixed targets, but the precision detectors allow significant
sensitivity up to higher mass for e.g.
light scalars, pseudoscalars and dark photons decaying primarily to SM states.
While some analyses are still underway, future progress in this area may come from Belle-II.

\bigskip \noindent{\it Direct detection}: Current direct detection experiments, searching for nuclear
recoils, are now probing the threshold of Higgs-mediated scattering for weak-scale WIMPs.
However, sensitivity drops rapidly to zero for masses below a few GeV.
There are proposals to extend this reach with alternate technologies, e.g., CCD's at DAMIC
\cite{Barreto:2011zu}, and also the analysis of very low energy electron recoils \cite{Essig:2012yx}.
However, currently experiments at the intensity frontier have significantly greater reach.

\bigskip
\noindent{\it LHC}: The energy frontier of course also provides sensitivity to light hidden sector states, primarily through unusual jet structures, eg. lepton jets in the case where SM decays are unsuppressed \cite{Strassler:2006im,ArkaniHamed:2008qp}, or through missing energy events \cite{Goodman:2010ku,Fox:2011pm,Shoemaker:2011vi}. However, as noted above, for light mediators coupled through the renormalizable portals  the production rates go down and radiation of hidden sector states is suppressed. Therefore,  the energy frontier does not currently provide the strongest sensitivity to hidden sectors with light mediators.

\bigskip
In the next section, we focus on the specific sensitivity and advantages of proton fixed target experiments.

\section{Opportunities at Neutrino Facilities}

Neutrino experiments provide an excellent opportunity to search for light weakly-coupled particles due to the large number of protons on target (currently reaching $\sim 10^{21}$ POT), the position of a near or single detector within a kilometer of the target, which in turn has a large mass with low energy thresholds and sensitive event characterization and background rejection. This potential was pointed out in Ref.~\cite{Batell:2009di}, which explored the sensitivity of the LSND experiment to probe a variety of hidden sector particles such as dark photons, dark Higgs bosons, and dark matter.  Refs.~\cite{deNiverville:2011it,deNiverville:2012ij} explore the potential of the LSND, MiniBooNE, NuMi/MINOS, and T2K experiments to search for light dark matter. Ref.~\cite{Essig:2010gu} further explores the sensitivity of LSND, MiniBooNE, MINOS, and CHARM to axions and dark photons. 

Neutrino experiments are designed to produce neutrinos at a sizable hadronic rate via meson decays, and then detect their weak-scale scattering at the sub percent level.  Thus, new weakly coupled states, produced with a rate at or below that of neutrinos,  and with  interaction strengths on the order of $G_F$ or possibly below can be probed with these experiments.
Such states can be produced in the primary proton-target collisions through a variety of physics processes, travel to a detector due to their weak coupling to SM matter, and then leave a signature in the detector through their decay to SM particles or by scattering with nucleons or electrons. 
 
There are many hidden sector particle candidates that can be probed using neutrino experiments.
There is still a great deal of work to be done to understand the sensitivity of these experiments over the
full model and parameter ranges in these scenarios, beyond the investigations in the references presented
above.
Rather than discuss all of the possibilities here, we will highlight in detail a specific proposal to search
for light dark matter with the MiniBooNE experiment~\cite{Dharmapalan:2012xp}.
This proposal represents the most detailed and precise investigation on the physics potential of neutrino
experiments to search for hidden sector physics to date.

Indeed, as emphasized above, a unique advantage of these large neutrino detectors is the ability to search
for new weakly-coupled particles via scattering.
This opens up dark sector searches to `invisible modes', where the dark photon decays to light NLWCPs that
then travel the distance to the detector and scatter.
The MiniBooNE experiment is sensitive to a model of light dark matter, which achieves the required relic
abundance via thermal freeze-out through a dark photon mediator coupled to the SM via kinetic mixing.
Such portal couplings render these models the least restricted by other terrestrial and astrophysical
constraints.
At MiniBooNE energies, the dark photon can be produced in the decays of the neutral pseudoscalar bosons
$\phi^0, \eta \rightarrow \gamma V$ of which MiniBooNE has produced a huge sample.
These dark photons subsequently decay to a pair of dark matter particles, which then travel to the detector
and scatter.
These searches therefore nicely compliment those being being done at JLAB and MAMI/Mainz that look for dark
photon decay to Standard Model particles.

Generically, beyond MiniBooNE the mass range that can be covered is dictated by the proton beam energy and
the production mechanism involved.
In the case of dark sector models with portal couplings to the visible sector, the accessible DM mass range
is from a few MeV up to a few GeV for typical proton machines used for neutrino production (e.g.
the FNAL Booster and Main Injector).
It is important to emphasize that this covers a region at low DM masses that cannot currently be explored in
underground direct detection experiments.

\begin{figure}
\centering
\includegraphics[scale=0.425]{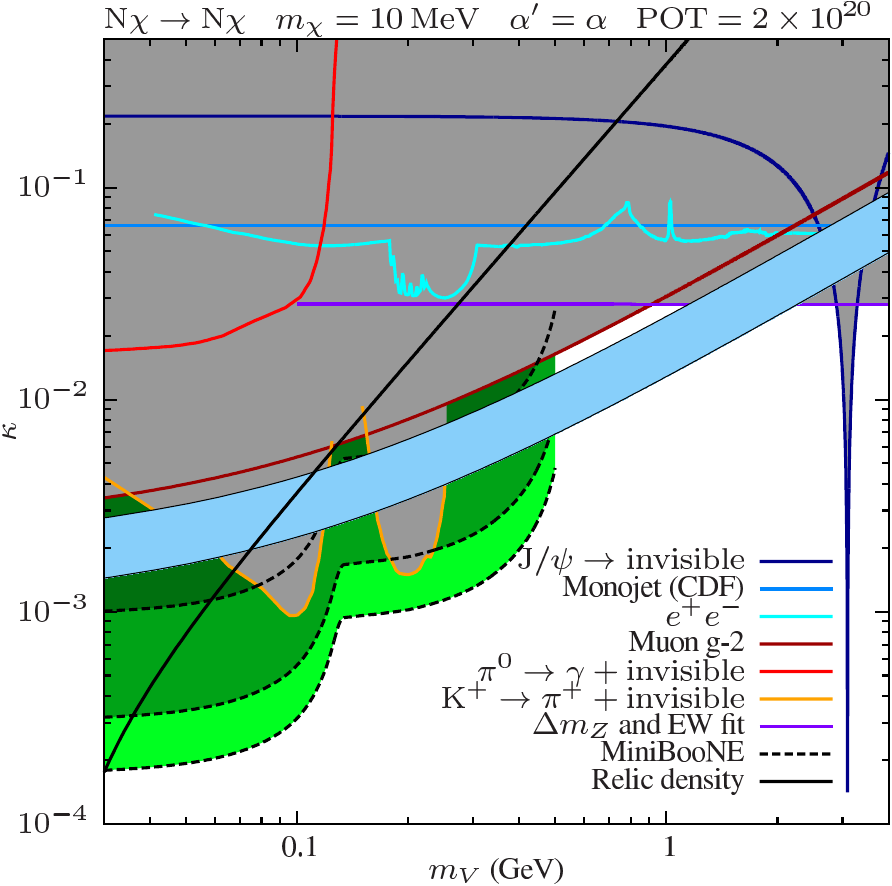} \hfill
\includegraphics[scale=0.425]{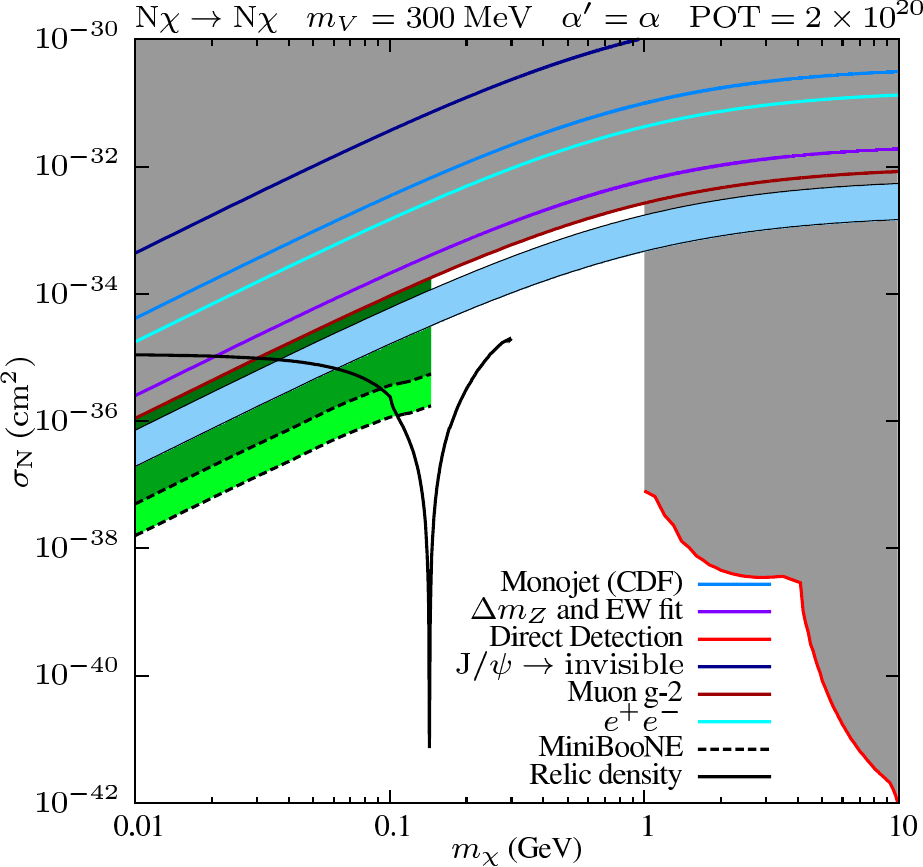}
\caption[MiniBooNE cross section versus mass sensitivity to WIMP production]{The MiniBooNE sensitivity to
light dark matter scattering.
This is shown in the plane of kinetic mixing versus vector mass (left), assuming a WIMP mass of 10~MeV, and
in the plane of non-relativistic per-nucleon scattering cross section versus dark matter mass (right), using
a vector mediator mass of 300 MeV.
The light green band indicates greater than 10 events at MiniBooNE given a $2\times10^{20}$~POT
beam-off-target run.
The various constraints and sensitivities are shown in the legend, the light blue band is the muon $g-2$
signal region, and the required thermal relic density is shown as the black line.
See Ref.~\cite{Dharmapalan:2012xp} for more details.}
\label{nlwcp:fig:MB}
\end{figure}

The proposal \cite{Dharmapalan:2012xp} describes the the potential to search for light sub-GeV WIMP dark
matter at MiniBooNE.
An important aspect of the proposal is to take advantage of the ability to steer the proton beam past the
target and into an absorber, leading to a significant reduction in the neutrino background and allowing for a
sensitive search for elastic scattering of WIMPs off nucleons or electrons in the detector.
Additional background reduction strategies involve utilizing precision timing to account for the small delay
of massive dark matter propagating to the detector, as compared to neutrinos, and also the distinct
kinematics of the scattering \cite{Dharmapalan:2012xp}.
Dark matter models involving a dark photon mediator can be probed in a parameter region consistent with the
required thermal relic density, and which overlaps the region in which these models can resolve the muon 
$g-2$ discrepancy.
The expected number of signal events is shown in Fig.~\ref{nlwcp:fig:MB} for a range of parameter points.
The signal significance for various operational modes is described in more detail in
\cite{Dharmapalan:2012xp}.

The experimental approach outlined for MiniBooNE to search for light NLWCPs is applicable to other neutrino
facilities.
For instance the MicroBooNE LAr detector can also make a search comparable to that outlined for MiniBooNE
with a long enough beam-off-target run.
Other neutrino experiments such as MINOS, NO$\nu$A, and T2K have potential to search for low mass NLWCPs.
Refs.~\cite{deNiverville:2011it,deNiverville:2012ij} have demonstrated the potential of such experiments to
probe light dark matter, but more detailed studies by the experimental collaborations would be required to
precisely determine the reach in the parameter space of these models.

\section{\PX\ Beam Parameters}

In the following we briefly summarize the \PX\ beam parameters relevant to new particle searches.
Such searches will benefit from increased intensity at each step of the staged approach, as this translates
into greater search sensitivity given detectors able to provide sufficient background rejection.
\PX\ is also ideally suited to a broad search for a range of possible NLWCPs, via flexibility in beam
energies, targets, timing structures, and other configurations.
It is expected that optimization over many different possible beam parameters will target searches for
specific models or for specific areas of unexplored parameter space.

a) Stage~1 of \PX\ can provide 1000 kW at 1~GeV from a spallation target, which is essentially a high-Z,
high-mass beam-dump.
The beam timing is a continuous train of 50-ps wide proton pulses separated by 25,000~ps (25~ns).
Pion production per Watt of beam power is essentially flat between 1~GeV and 8~GeV, so the Stage-1 beam dump
would have a neutral pion flux of $\sim20$ times that of the 8-GeV Booster beamline (this includes the lack
of focusing relative to the BNB).
The sharp pulse train timing will be useful in rejecting prompt neutrino backgrounds and searching for
particles with mass below the pion threshold.
Stage~1 would provide significantly more protons to both the BNB and MI, and hence any beam dump experiments
on these beamlines.

b) Stage~2 of \PX\ will provide Stage-1 resources and another 1000 kW at 3~GeV from low-medium
(carbon-gallium) targets, with the same Stage-1 time structure.
The higher energy would allow higher particle mass searches due to the production of $\eta$ mesons.

c) Stage~3 of \PX\ will continue the resources of Stages~1 and~2, and in addition replace the 8-GeV Fermilab
Booster beam with an 8-GeV, 200-kW source pulsed at 10~Hz.

A future search with a dedicated beam dump experiment on the Main Injector would benefit from the higher
energy 120-GeV protons.
For an experiment such as the MiniBooNE proposal, this would allow allow searches for dark photons and dark
matter up to masses of a few GeV, covering the gap in the muon $g-2$ region up to the mass range at which
current direct detection limits apply.

\section{New Detector Technologies}

The opportunity of searching for motivated new physics from a
possible dark sector lends itself not only to exploiting existing
and planned accelerator facilities but also to exploiting newly developed
and improving detector technologies. In the following discussion, we assume
a proton beam dump configuration where an intense proton beam is incident
upon a target that may allow for the production of new light weakly coupled
particles. Such particles will travel through shielding material until
they either scatter or decay in a downstream detector. Presumably, beam
related backgrounds will be dominated by neutrinos as other
particles will be absorbed by shielding. Unlike a conventional neutrino beam
facility that would try to concentrate the parent $\pi$ and $K$ mesons that
decay into muons and neutrinos, 
an incident proton beam can specifically be directed
onto a target without such focusing such that the neutrinos are spread out
and have a lower density compared with possible directly-produced NLWCPs.
In addition, having the ability to adjust the incident beam
energy will change the composition of any neutrino background with lower 
energy neutrinos having a smaller interaction cross section in a downstream
detector. Other possible handles for reducing the neutrino background include
precise timing information, as discussed in \cite{Dharmapalan:2012xp}, to 
distinguish the travel time to scatter off GeV-mass NLWCPs from neutrinos traveling at essentially the speed 
of light.

The different strategies for reducing backgrounds to near negligible levels
make the overall approach similar to a direct detection dark matter experiment,
that has the greatest sensitivity when backgrounds are negligible. With 
negligible backgrounds, experiments gain in sensitivity faster with
increased intensity. Beyond the
simple counting type experiment, new detector technologies that are sensitive
either to low energy scatters off the nucleus or surrounding electrons, or
are sensitive to possible final state Standard Model decay products, can
be exploited for a particular search. For example, detectors based upon
liquid noble elements like liquid argon can be used to look for
new weakly interacting particle scatters with or without a TPC that might 
allow for particle identification in case the new particle
also decays into Standard Model particles. Detectors with low
energy thresholds (such as DAMIC) could also be employed 
in such an experiment.

In short, there are a number of directions for using new detector technologies
in the search for new light weakly coupled particles. Like the worked out
example of the MiniBooNE proposal, we see that existing detectors can cover
interesting regions of parameter space. The parameter space depends on
the initial beam parameters and \PX\ would naturally allow for
a broad range of possibilities. In addition, the ways in which 
hidden sector particles may either scatter off detector materials or
perhaps decay into detectable Standard Model particles also gives a large
range of possibilities and optimizations. 

\section{Summary}

The possibility of new physics in the form of light weakly coupled particles from a hidden sector is
motivated in various ways, from both bottom-up and top-down arguments.
A general effective field theory perspective points to the minimal set of renormalizable portal interactions
with the Standard Model as the primary couplings to probe for the existence of a neutral hidden sector.
It turns out that existing neutrino experiments are particularly well suited to make the first measurements
that cover interesting regions of parameter space, such as those motivated by the muon $g-2$ anomaly or by
astrophysical observations.
A wider, more systematic, exploration would be possible at a high intensity facility such as \PX\ having many
different configurations for initial beam energy, timing, and other parameters.
Detectors can be optimized for generic searches or for more specific well-motivated searches.
Who knows, but it may be a novel beam dump experiment at \PX\ that just might be the first to reveal a new
level of understanding of New Physics!

\bibliographystyle{apsrev4-1}
\bibliography{nlwcp/refs}
 % Brian B., Richard VdW, William C.W.3

\chapter{Hadronic Structure with \PX}
\label{chapt:hadron-dy}

\authors{Markus~Diefenthaler,
Xiaodong~Jiang,
Andreas~Klein,
Wolfgang~Lorenzon, \\
Naomi~C.~R.~Makins,
and
Paul~E.~Reimer}

\section{Introduction}

The proton is a unique bound state, unlike any other yet confronted by
physics.  We know its constituents, quarks and gluons, and we have a
theory, QCD, to describe the strong force that binds these
constituents together, but two key features make it a baffling system
that defies intuition: the confining property of the strong force, and
the relativistic nature of the system. Real understanding of the
proton can only be claimed when two goals are accomplished: precise
calculations of its properties from first principles, and the
development of a meaningful picture that well approximates the
system's dominant behavior, likely via effective degrees of freedom.

The excitement and challenge of the quest for this intuitive picture
is well illustrated by the ongoing research into the
spin structure of the proton, and in particular, into the contribution
from quark orbital angular momentum (OAM).
As experiment provides new clues
about the motion of the up, down, and sea quarks, theory continues to make
progress in the interpretation of the data, and to confront fundamental
questions concerning the very definition of $L$ in this context.
Yet crucial pieces are still missing on the experimental side.
One substantial missing piece is the 
the lack of any spin-dependent data from one of the most powerful
probes of hadronic substructure available, the Drell-Yan process.

%-------------------------------------------------------------------
% \section{The Proton Spin Puzzle and Orbital Angular Momentum}
\section{Proton Spin Puzzle and Orbital Angular Momentum}

In its simplest form, the proton spin puzzle is the effort to
decompose the proton's total spin into its component parts
\begin{equation}
	\frac{1}{2} = \frac{1}{2} \Delta\Sigma + \Delta G + L_q + L_g\, .
  \label{hadron-dy:eq:spinsum}
\end{equation}
$\Delta\Sigma$ is the net polarization of the quarks, summed over
flavor, and is known to be around 25\% \cite{deFlorian:2008mr,
  deFlorian:2009vb}. The gluon polarization, $\Delta G$, is currently
under study at the RHIC collider; the data collected to date favor a
positive but modest contribution.  What remains is the most mysterious
contributions of all: the orbital angular momentum of the partons.

\begin{figure}
    \centering
    \includegraphics[width=0.2\textwidth]{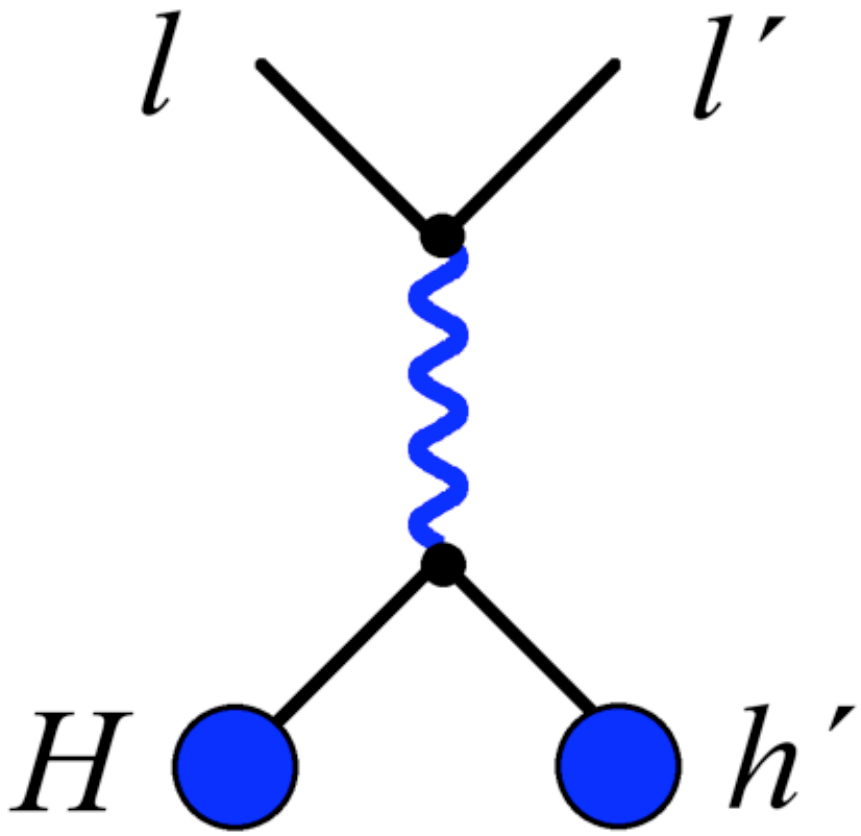} 
    \hspace*{3em}
    \includegraphics[width=0.2\textwidth]{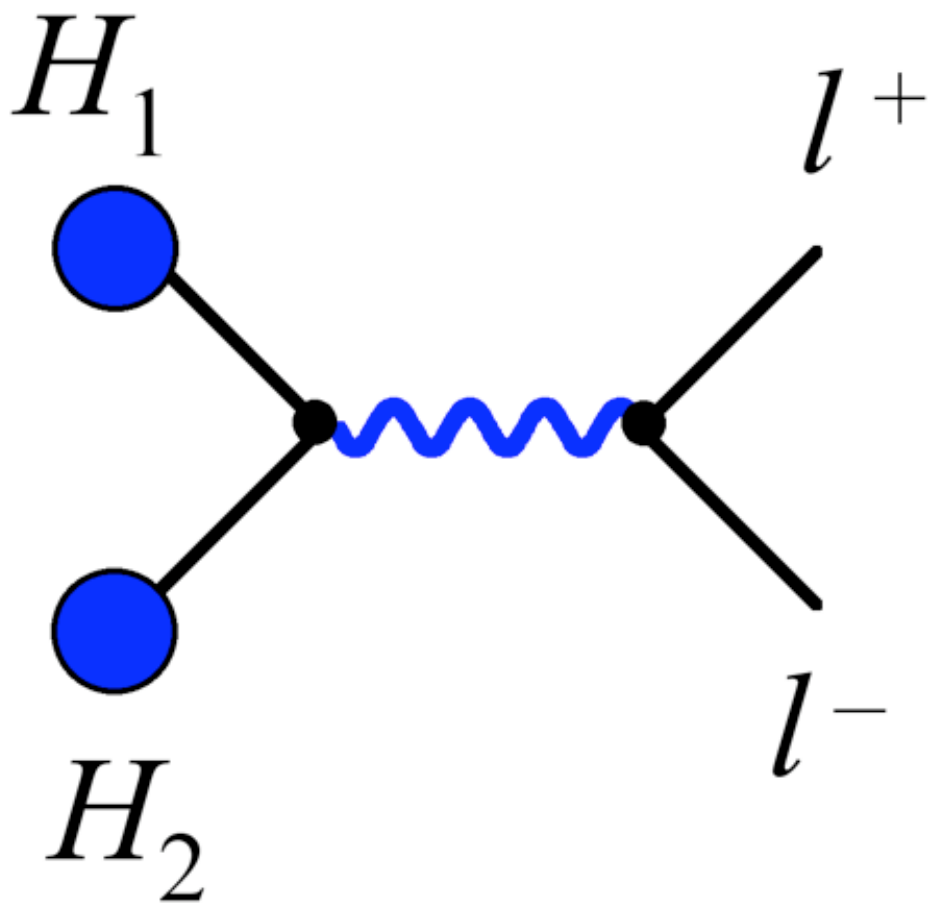} 
    \hspace*{3em}
    \includegraphics[width=0.18\textwidth]{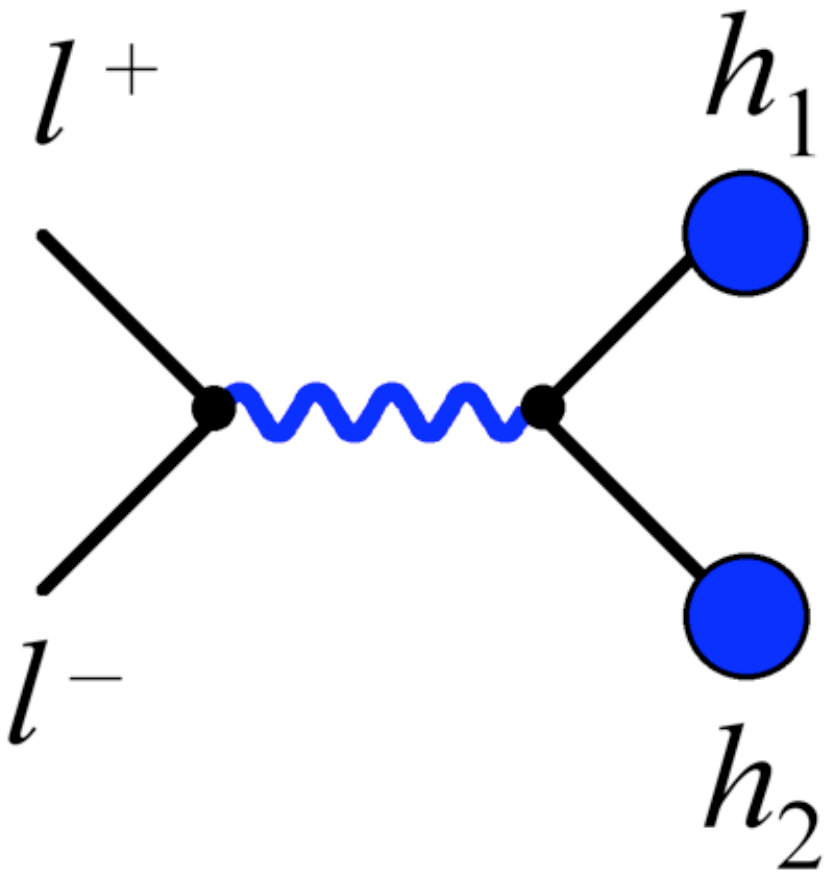}
    \vspace*{0.15in}
    \caption[Tree-level hard-scattering processes relevant to TMD universality]{Tree-level hard-scattering processes of the three
    reactions where TMD universality has been established (a)
    semi-inclusive DIS (b) Drell-Yan / $W$-production (c) $e^+ e^-$
    annihilation.}
    \label{hadron-dy:fig:processes}
\end{figure}

With the spin sum above as its capstone goal, the global effort
in hadronic spin structure seeks to map out the proton's
substructure at the same level of scrutiny to which the atom and the nucleus
have been subjected.
To this end, experiments with high-energy beams map out the proton's
parton distribution functions (PDFs):
the number densities of quarks and gluons as a function of
momentum, flavor, spin, and, most recently, space. Deep-inelastic scattering
(DIS) has yielded the most precise information on the unpolarized and
helicity-dependent PDFs $f_1^q(x)$ and $g_1^q(x)$ for quarks. Here $q$
represents quark flavor and includes the gluon, $g$, while
$x$ is the familiar Bjorken scaling variable
denoting the fraction of the target nucleon's momentum carried by the
struck quark. (The logarithmic dependence of the PDFs on the hard scattering
scale has been suppressed for brevity.)
For antiquarks, these distributions are accessed most cleanly by
the Drell-Yan and $W$-production processes in proton-nucleon scattering.
As with semi-inclusive DIS (SIDIS) or deep-inelastic jet production,
both of these processes are purely leptonic in one half
of their hard-scattering diagrams (see Fig.~\ref{hadron-dy:fig:processes}),
which facilitates clean interpretation and enables the
event-level determination of the parton kinematics.
The unique sensitivity of Drell-Yan and $W$-production to sea quarks
is clearly shown: an antiquark is needed at the annihilation vertex
in both cases. The Fermilab E866 experiment used Drell-Yan scattering
to make its dramatic determination
of the pronounced $\bar d(x) / \bar u(x)$ excess in the sea;
the PHENIX and STAR experiments at RHIC are currently measuring $W$-production
with polarized proton beams to determine the antiquark helicity PDFs
$\Delta \bar u(x)$ and $\Delta \bar d(x)$ with new precision.

%...................FIGURES 1 & 2............................
%.............the big three processses.......................

%.............the TMD stoplights.......................

\begin{figure}
    \centering
    \includegraphics[width=0.8\textwidth]{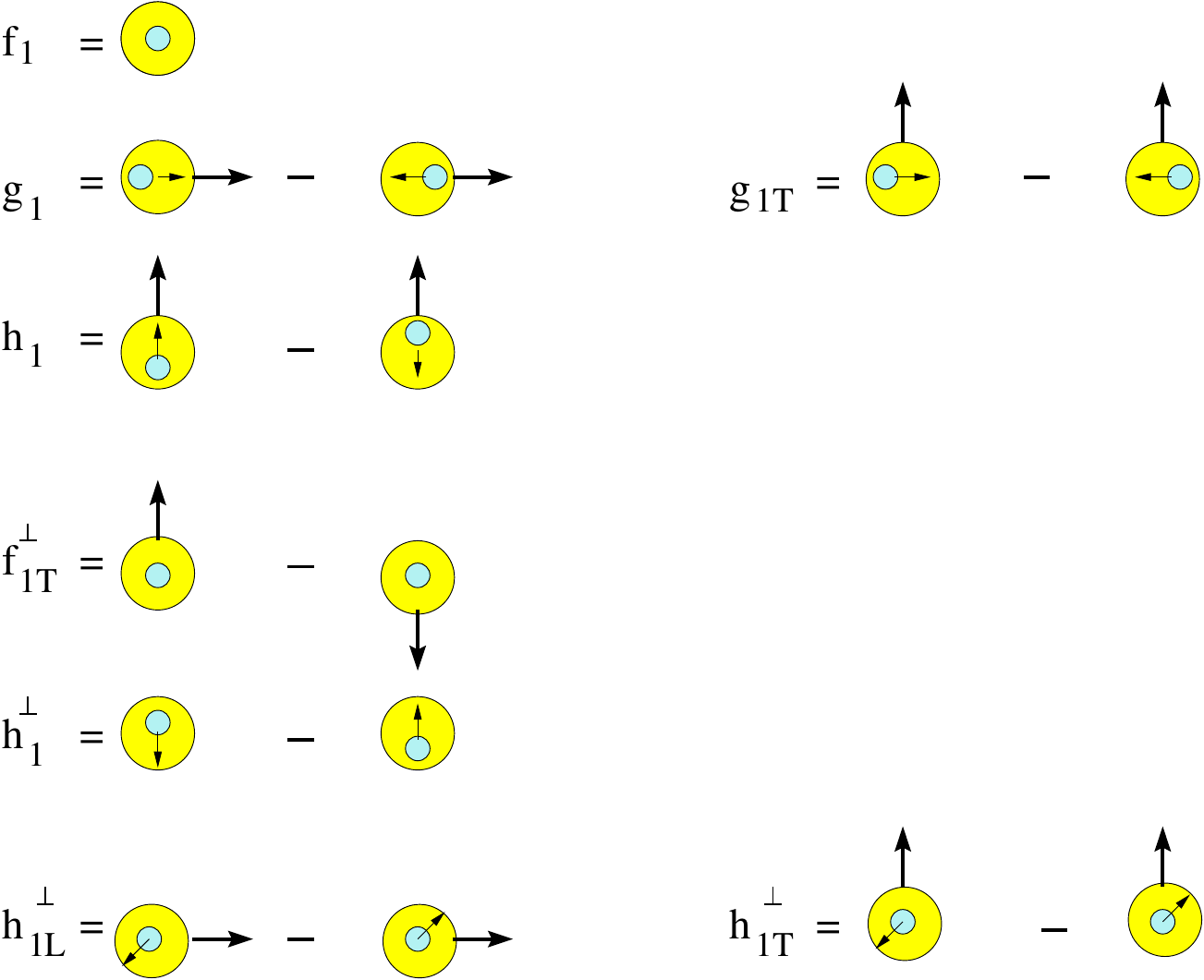}
    \caption[Operator structures of the the eight leading-twist TMDs]{Operator structures of the the eight 
    leading-twist TMDs. 
	The horizontal direction is that
	of the virtual boson probing the distribution.
	The large and small circles represent the proton and quark
	respectively, while the attached arrows indicate their
	spin directions.}
    \label{hadron-dy:fig:stoplights}
\end{figure}
%.............................................................

Over the past decade, attention has shifted to two new classes of
parton distribution functions that offer a richer description of the
proton's interior than $q(x)$ and $\Delta q(x)$. These are the TMDs
(transverse momentum dependent PDFs) and the GPDs (generalized parton
distributions). The two descriptions are complementary: they correlate the
partons' spin, flavor, and longitudinal momentum $x$ with transverse
momentum $\euvec{k}_T$ in the TMD case and with transverse position
$\euvec{b}_T$ in the GPD case. Both offer access to $L$, via different
experimental approaches. The GPD approach relies on the measurement of
exclusive photon and meson production with lepton beams at large
$Q^2$.  This proposal focuses on the TMDs, which are accessed most
cleanly via the azimuthal distributions of the final-state products of
the SIDIS and Drell-Yan processes with polarized beams and/or
targets. The details of these ``single-spin azimuthal asymmetries''
are presented in Sec~\ref{hadron-dy:sec:spin-L}.

When parton transverse momentum $\euvec{k}_T$ is included---i.e.,
momentum transverse to that of the $s$- or $t$-channel virtual
boson---one obtains the \textit{transverse momentum distributions}.
Theoretical analysis of the SIDIS process has led to the
identification of eight such TMDs at leading
twist~\cite{Mulders:1995dh, Bacchetta:2006tn}.  Their operator
structure is shown schematically in Fig.~\ref{hadron-dy:fig:stoplights}.  Three
of these survive on integration over $\euvec{k}_T$: the transverse
extensions $f_1^q(x,\euvec{k}^2_T)$ and $g_1^q(x,\euvec{k}^2_T)$ of
the familiar PDFs and a third distribution, $h_1^q(x,\euvec{k}^2_T)$
termed transversity. The remaining five TMDs bring $\euvec{k}_T$ into
the picture at an intrinsic level, and vigorous theoretical work has
been devoted to deciphering their significance. The most intensely
studied are the Sivers~\cite{Sivers:1989cc, Sivers:1990fh}
distribution $f_{1T}^{\perp,q}(x,\euvec{k}^2_T)$ and the
Boer-Mulders~\cite{Boer:1997nt} distribution
$h_1^{\perp,q}(x,\euvec{k}^2_T)$. As shown in
Fig.~\ref{hadron-dy:fig:stoplights}, they describe the correlation of the
quark's momentum with the transverse spin of either the proton
(Sivers) or the quark itself (Boer-Mulders).  At first sight, the
operator differences depicted in the figure seem absurd: in the Sivers
case, for example, how can the quark's momentum distribution change if
one simply rotates the proton's spin direction by 180 degrees?  A
solution is presented when one considers the \emph{orbital angular
  momentum} of the quarks, $L_q$. If the up quarks' OAM is aligned
with the proton spin, then the quarks will be
oncoming---blue-shifted---on \emph{different sides} of the proton
depending on its spin orientation. The search for a rigorous,
model-independent connection between the Sivers distribution and quark
OAM is ongoing (see Refs.~\cite{Brodsky:2002cx, Burkardt:2003uw,
  Bacchetta:2011gx} for examples of recent approaches). The connection
is as yet model-dependent, but what is clear is that the existence of
the Sivers function \emph{requires} nonzero quark OAM.

%-------------------------------------------------------------------------
% \subsection{Spin, $\boldsymbol{L}$, and QCD}
\subsection{\protect Spin, Orbital Angular Momentum, and QCD}
\label{hadron-dy:sec:spin-L}

Orbital angular momentum provides one of the most dramatic illustrations
of the challenge of understanding the most fundamental bound state of QCD,
the proton. In atomic and nuclear physics, $L$ is a conserved quantity:
a good quantum number that leads to the shell structure
of these familiar systems. Not so with the proton. As the masses of the
light quarks are so much smaller than the energy-scale of the system
(e.g., the mass of the proton itself: 938\,MeV compared with the 3--5~MeV
of the up and down quarks), the system is innately relativistic.
In relativistic quantum mechanics, $L$ is \emph{not} a conserved quantity:
neither it nor spin commute with even the free Dirac Hamiltonian, and
so a shell structure within the proton is excluded. A simple calculation
of the ground state of a light Dirac fermion bound in a central potential, for
example, shows that the ground-state spinor is in a mixed state of $L$:
$L = 0$ for the upper components and $L = 1$ for the lower components~\cite{Liang:1992hw}.

Furthermore, the definition of quark OAM is under active
dispute. The simple spin sum of Eq.~(\ref{hadron-dy:eq:spinsum}) conceals a wealth
of complexity in the definition of its components.  Two versions of
this decomposition have dominated the discussion to date.  They are
colloquially referred to as the Jaffe~\cite{Bashinsky:1998if} and
Ji~\cite{Ji:1996ek} decompositions, though they have been addressed by
numerous authors; see Refs.~\cite{Wakamatsu:2010cj,Wakamatsu:2010qj,Wakamatsu:2010cb, Burkardt:2005km} for 
elegant summaries of the issues.

The Ji decomposition can be expressed as
\begin{equation}
  \euvec{J}_\text{proton}
	= \int \psi^\dagger \frac{1}{2}\euvec{\Sigma} \psi \,d^3x
	+ \int \psi^\dagger \euvec{x} \times \frac{1}{i} \euvec{D} \psi \,d^3x
	+ \int \euvec{x} \times (\euvec{E}^a \times \euvec{B}^a) \,d^3x ,
  \label{hadron-dy:eq:Ji}
\end{equation}
where $a$ is a color index.
It has three gauge-invariant terms which, in order, represent the quark
spin $\Delta\Sigma$, quark OAM $L_q$, and total angular momentum $J_g$
of the gluons. The advantage of this decomposition is its
rigorous connection to experiment via the Ji sum rule \cite{Ji:1996ek}, which
relates $J_q$ for each quark flavor $q$ to the second moment of two GPDs
\begin{equation}
  \euvec{J}_q = \lim_{t\rightarrow 0} \int \euvec{x} \left[ H_q(x,\xi,t) + E_q(x,\xi,t) \right] d^3x .
\end{equation}
The actual measurement of these GPDs is an enormous experimental task; it
was initiated at HERMES and will be continued with greater precision at
Jefferson Laboratory and COMPASS. The Ji decomposition can also be addressed
by lattice QCD, which has already been used to compute moments of the GPDs
under certain approximations (e.g., Ref.~\cite{Gockeler:2006zu}).
One disadvantage of this decomposition is the lack of a gauge-invariant
separation of the gluon $J_g$ into spin and orbital pieces.
A second disadvantage is the problem of interpreting
its definition of $L_q$ as $\euvec{x} \times \euvec{D}$.
The appearance of the covariant derivative $\euvec{D} = \euvec\nabla + i \euvec{g}$
brings gluons into the definition. This is not the familiar, field-free OAM,
$\euvec{x} \times \euvec{p}$, that is addressed by quark models of the proton.

The Jaffe decomposition is
\begin{equation}
  \euvec{J}_\text{proton}
	= \int \psi^\dagger \frac{1}{2}\euvec{\Sigma} \psi \,d^3x
	+ \int \psi^\dagger \euvec{x} \times \frac{1}{i}\euvec{\nabla}\psi \,d^3x
	+ \int \euvec{E}^a \times \euvec{A}^a \,d^3x
	+ \int E^{ai} \euvec{x} \times \euvec{\nabla} A^{ai} \,d^3x \, .
  \label{hadron-dy:eq:Jaffe}
\end{equation}
It has four gauge-invariant terms, which in order represent the
quark spin, quark OAM, gluon spin, and gluon OAM. Here, $L_q$ \emph{is}
the field-free, canonical operator $\euvec{x} \times \euvec\nabla$.
The gluon spin and OAM are separated in a gauge-invariant way, and
in the infinite-momentum frame, parton distribution
functions for the four pieces can be defined. The disadvantage of the
Jaffe decomposition is that it is unclear how to measure its $L_q$
and $L_g$ terms, either in the lab or on the lattice, as they are nonlocal
operators unless one selects a specific gauge (the lightcone gauge, $A^+ = 0$).

At present, we are thus confronted with one definition of $L_q$ that can
be measured but not interpreted, and another that can be interpreted but
not measured. The ``dynamical'' OAM, $\euvec{x} \times \euvec{D}$, of the Ji
decomposition brings us face-to-face with the confining nature of QCD:
we cannot avoid interactions in a theory where quarks cannot be freed.
Can we learn to interpret this quantity? This remains an open question, as
only the ``canonical'' OAM definition, $\euvec{x} \times \euvec\nabla$, obeys
the commutation relations of angular momentum algebra.

%-------------------------------------------------------------------------

%-------------------------------------------------------------------------
\section{Polarized Drell-Yan: The Missing Spin Program}

If we are to resolve the puzzle of quark spin in general and quark OAM in
particular, it is vital to make a direct measurement of the Sivers distribution for antiquarks.
The only process with which this task can be cleanly accomplished is Drell-Yan, with its innate sensitivity 
to antiquarks. 
A~potential alternative, $W$-production, cannot be used in this endeavor
as the unobserved neutrino blurs the final-state azimuthal distributions.

The need for a spin-dependent Drell-Yan program has become an urgent priority
for the hadron-structure community world-wide. The three processes depicted in
Fig.~\ref{hadron-dy:fig:processes} are the only ones where the TMD formalism
has been theoretically shown to yield universal functions: PDFs and
fragmentation functions that are process-independent.
Of the three, only Drell-Yan has not yet been explored with
polarized beams and/or targets. It is the missing component in the
ultimate goal of a global analysis of TMD-related data.
The crucial nature of this missing spin program arises from three facts:
the innate sensitivity of Drell-Yan to antiquarks,
its freedom from fragmentation functions, and the unique possibility
it affords to test the TMD formalism.

%-------------------------------------------------------------------------
\subsection{Measurement of the Sivers Sign Change with a Polarized Proton Beam}
\label{hadron-dy:sec:sign-change}

The previous sections have framed the context in which polarized
Drell-Yan experiments would be placed, and described its crucial place
in the spin puzzle. We now turn to the specific motivation for 
using a polarized proton beam.

For Drell-Yan kinematics, $x_f \approx x_b - x_t$, where $x_b$ and
$x_t$ are the longitudinal momentum fractions of the annihilated
quarks from the beam and target, respectively. As with E906/SeaQuest,
E866, and their predecessor experiments, the high $x_b$ values
selected by the forward $x_f > 0$ spectrometer mean that the partons
from the beam will almost certainly be quarks, with the antiquark
coming from the target.  Taking $u$-quark dominance into consideration
(due to the charge-squared weighting of the cross section and the
preponderance of up quarks in the proton at high $x$), the measurement
will be heavily dominated by valence up quarks from the polarized
proton beam. The proposed measurement will thus be sensitive to Sivers
function for up quarks, $f_{1T}^{\perp,u}(x,\euvec{k}^2_T)$, times the
familiar unpolarized PDF for anti-up quarks, $\bar u(x)$.

Given the unique access to sea quarks afforded by the Drell-Yan
process, the reader may wonder why this proposal aims to measure the
Sivers function for valence quarks, and valence up
quarks at that---the flavor most precisely constrained by SIDIS data
from HERMES and COMPASS.

The goal of this first spin-dependent Drell-Yan measurement is exactly
to compare Drell-Yan and SIDIS, in order to test the 10-year-old
prediction of a sign change in the Sivers function from SIDIS
to Drell-Yan.  Given the theoretical definition of the Sivers
function~\cite{Boer:1997nt}, this sign change follows directly from
field theory and \CPT\ invariance~\cite{Collins:2002kn}. Observing the
sign change is essential to our interpretation of present and future
TMD data in terms of angular momentum and spin. The sign change also
offers a rigorous test of QCD in the nonperturbative regime---a rare
thing indeed.  Observation of the Sivers sign change is one of the DOE
milestones for nuclear physics and is the first step for any
spin-dependent Drell-Yan program~\cite{NPMilestone2008}.

Beyond the verification of the TMD framework and the tantalizing access it affords to OAM in the proton,
there is rich physics behind the Sivers sign change itself.
This physics lies in the definition of the Sivers function.
The function was first proposed as a possible explanation of the ``E704 effect'': the large left-right
analyzing power observed in inclusive pion production from a transversely polarized proton beam of 200 GeV
incident on a beryllium target.
The polarized beam at Fermilab~E704 was a tertiary beam obtained from the production and subsequent decay of
hyperons.
(Its intensity was thus far below that required for Drell-Yan measurements.) As has happened repeatedly when
spin degrees of freedom are introduced for the first time in experimental channels, new effects were observed
at E704 that provoked rich new areas of study.
The measured analyzing power was $A_N \propto \euvec{S}_\mathrm{beam} \cdot (\euvec{p}_\mathrm{beam} \times
\euvec{p}_\mathrm{pion})$.
This single-spin asymmetry is odd under so-called ``naive time-reversal'', the operation that reverses all
vectors and pseudo-vectors but does not exchange initial and final states.
The only way to produce such an observable with a $T$-even interaction is via the interference of 
$T$-even amplitudes.
The interfering amplitudes must have different helicity structures---one spin-flip and one non-spin-flip
amplitude are required---and they must differ by a nontrivial phase.
Both of these requirements are greatly suppressed in the perturbative hard-scattering subprocess, so the
source of the E704 effect must be soft physics~\cite{Kane:1978nd}: an interference in either the initial or
final state.
The original Sivers idea was of an initial-state interference~\cite{Sivers:1989cc, Sivers:1990fh}.
A~complementary proposal from Collins suggested a spin-orbit effect within the fragmentation
process~\cite{Collins:1992kk}.

The breakthrough that led to our modern understanding of the E704 analyzing power occurred many years later
when the HERMES collaboration measured pion single-spin asymmetries for the first time in deep-inelastic
scattering, i.e., using a lepton rather than proton beam~\cite{Airapetian:2001eg,Airapetian:1999tv}.
Unlike inclusive $p p \rightarrow \pi$, the SIDIS process $e p \rightarrow e^\prime \pi$ allows complete
kinematic determination of one side of the hard scattering diagram and involves two distinct scattering
planes (as do all three processes in Fig.~\ref{hadron-dy:fig:processes}).
With this additional control, HERMES was able to separate single-spin effects arising from initial- and
final-state interactions \cite{Airapetian:2004tw}.
An electron beam interacts much more weakly than a hadron beam.
It was widely assumed that initial-state interactions would be excluded in SIDIS, thereby isolating the
final-state ``Collins mechanism'', but the data showed otherwise: both initial- and final-state effects were
found to be sizable.
The explanation was provided in 2002 by Brodsky, Hwang, and Schmidt~\cite{Brodsky:2002cx}.
They revisited the QCD factorization theorems and discovered that previously-neglected gauge~links
between the struck quark and target remnant---soft gluon reinteractions necessary for gauge invariance---had
to be included in the very definition of the parton distribution functions.
Their paper presented a proof-of-principle calculation showing how a naive-$T$-odd distribution function
could be generated at leading twist, and therefore observable in lepton SIDIS at high $Q^2$: by interfering
two diagrams within the PDF's definition, one with no gauge-link rescattering and an $L=0$ quark, and one
with a single gluon exchanged and an $L=1$ quark.

\begin{figure}
  \hspace*{\fill}
\includegraphics[width=0.4\textwidth]{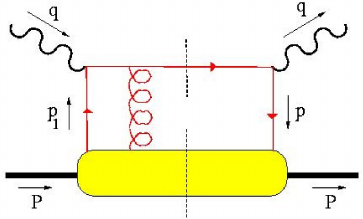}
  \hspace*{\fill}
  \includegraphics[width=0.4\textwidth]{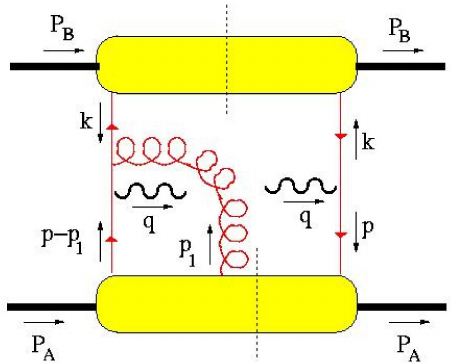}
  \hspace*{\fill}
  \caption[Gauge-link topology of the one-gluon exchange forward scattering amplitudes]{
	Gauge-link topology of the one-gluon exchange forward scattering
	amplitudes involved in the Sivers function in the
	(a) semi-inclusive DIS and (b) Drell-Yan scattering processes.}
  \label{hadron-dy:fig:topology}
\end{figure}

This PDF is what is now called the Sivers function, $f_{1T}^{\perp,q}$.
Its definition and its very existence at leading twist are intimately related to gauge invariance and our
understanding of QCD as a gauge theory.
Its universality has been demonstrated---to within a sign---only for SIDIS and Drell-Yan
(Fig.~\ref{hadron-dy:fig:processes}).
The sign change arises from the different topology of the gauge links in these two hard-scattering processes
(Fig.~\ref{hadron-dy:fig:topology}).
In the SIDIS case, the reinteraction is attractive as it occurs between the struck quark and the target
remnant.
For the Drell-Yan case, the reinteraction is repulsive as it connects the parton from the beam to the remnant
from the target (and vice versa).
As Dennis Sivers has put it, the Sivers function and its sign change teach us about the gauge 
structure of QCD itself.

%...................FIGURE 3: gauge-link topology.........................
%.............gauge-link topology.......................

\pagebreak
Testing the Sivers sign change is vital to the ongoing study of TMDs.
It is the inevitable first step for any Drell-Yan spin program
and is the key goals of this proposal. By polarizing the Main Injector
beam, Fermilab will be able to continue its long and distinguished history
of landmark Drell-Yan measurements and take the first step toward
becoming the site of the missing piece of the global spin program.

\subsection{Polarized Beam Drell-Yan Measurements at Fermilab} 
\label{hadron-dy:sec:polBeam}

The physics goals described in Sec.~\ref{hadron-dy:sec:sign-change} can only be
achieved with a combination of a large acceptance spectrometer for the
correct kinematics, beam energy and, most importantly, high
luminosity.  With the addition of a polarized source and polarization
maintaining Siberian snakes~\cite{SPINFermilab:2011aa}, Fermilab will
offer a rare convergence of these three conditions at one facility for
a Drell-Yan determination of the valence-quark Sivers distribution.

\begin{figure}
    \centering
    \includegraphics[width=0.875\textwidth]{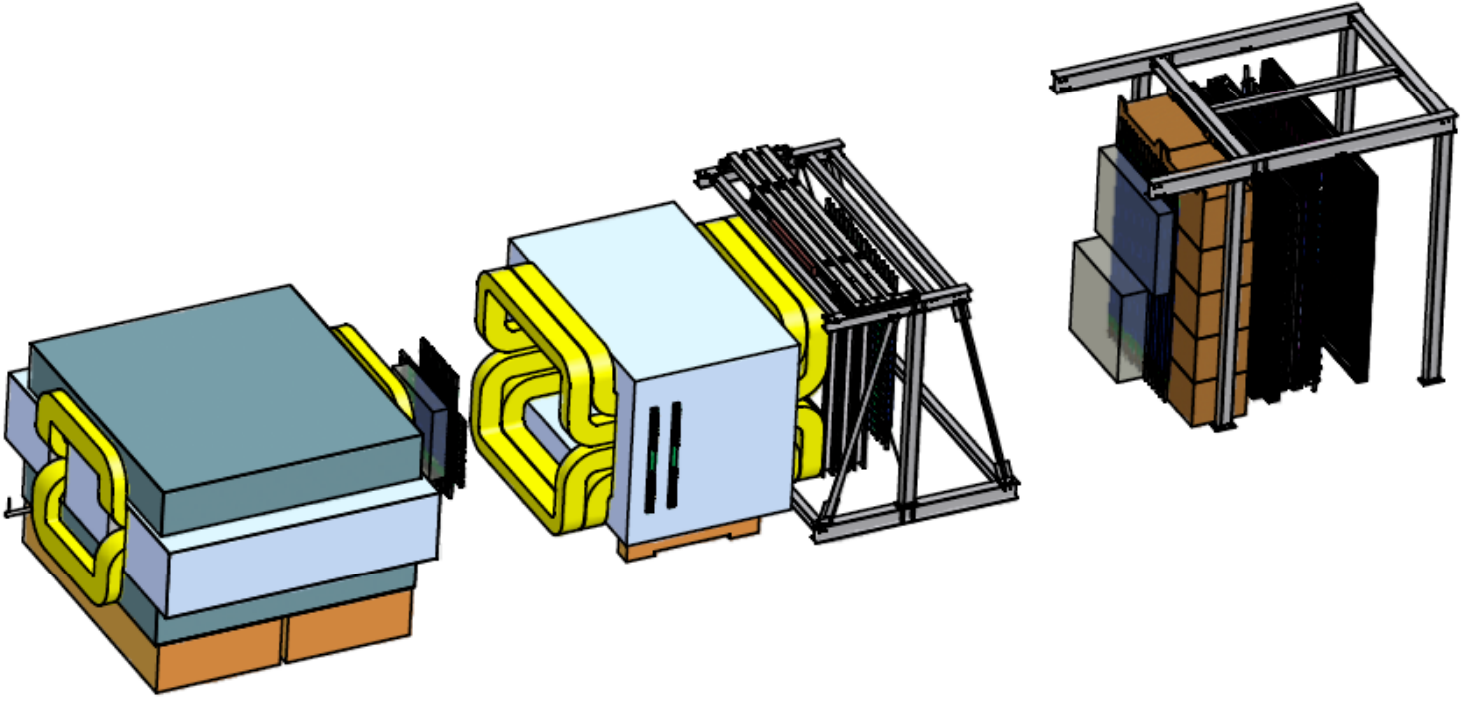}
    \vspace{6pt}
    \caption[SeaQuest Spectrometer during the 2012 commissioning run]{Schematic view of the SeaQuest 
    Spectrometer as it was during the 2012 commissioning run.} 
    \label{hadron-dy:fig:SeaQuest}
\end{figure}

The SeaQuest spectrometer, illustrated in Fig.~\ref{hadron-dy:fig:SeaQuest}, was
specifically designed to achieve the desired large, forward
acceptance.  This acceptance is critical to obtaining the proper range
in parton momentum fraction $x$, i.e., $x_b = 0.35-0.85$ covering
the valence quark region, and $x_t = 0.1-0.45$ covering the sea quark
region. This coverage dictates an event sample primarily from events
in which a target antiquark and beam quark interact.
  
In order to be certain that the di-lepton pair that is detected is
from a Drell-Yan interaction, the invariant mass of the virtual photon
must, in general, be above $M_{J/\psi}$. The available phase space for
a di-lepton pair falls as the center-of-mass energy, $\sqrt{s}$,
falls.  At the same time, backgrounds from uncorrelated pion decay
in-flight will increase with decreasing $\sqrt{s}$.  These two
considerations make it difficult to envision a fixed target Drell-Yan
measurement with a beam energy less than approximately 50 GeV.  On the
other hand, the Drell-Yan cross section for fixed $x_t$ and $x_b$
scales as $1/s$ implying that a smaller beam energy is desirable.  The
combination of these two factors places an extracted beam from the
Fermilab Main Injector near the ``sweet spot'' for this type of
measurement.

The measurement of the Sivers distribution sign change and the
connection of the Sivers distribution with OAM has generated great
interest around the globe.  There are now plans for a wide variety of
experiments to measure polarized Drell-Yan either with a polarized
beam or a polarized target (see Table~\ref{tab:dy:pol-DY}). While each
of these experiments can contribute to the overall picture, none
offer the sensitivity over a wide kinematic range that can be achieved
at Fermilab.  COMPASS at CERN and Panda at GSI plan to perform fixed
target experiments with either pion, proton or anti-proton beams,
whereas PAX at GSI, and NICA at JINR plan collider experiments with
polarized proton beams.  NICA and the polarized Drell-Yan programs at
RHIC will be sensitive to the interaction between valence quarks and
sea antiquarks. PAX and COMPASS plan to measure the interaction
between valence quarks and valence antiquarks, and are not sensitive
to sea antiquarks. And Panda is designed to study $J/\psi$ formation
rather than Drell-Yan physics due to the low antiproton beam
energy. The only  experiment scheduled to run in the near future is 
COMPASS, which will
  measure $A_N$ in one $x_f$-bin centered at $x_f=0.2$ in the
  invariant mass region $4<M<9$\,GeV.
COMPASS is scheduled to take data in 2014 for one
year and expects to measure the sign of the Sivers function in the
same kinematics as semi-inclusive DIS
with a statistical precision on $\delta A_N/A_N$ of 1--2\%.

\begin{table}
  \caption[Planned polarized Drell-Yan experiments]{Planned polarized Drell-Yan experiments. 
      $x_b$ and $x_t$
    are the parton momentum fractions in the beam and target,
    respectively.\label{tab:dy:pol-DY}}
\centering
%  \begin{small}
    \begin{tabular}{lccccc}
    \hline
    \hline
    Experiment & Particles & Energy & $x_b$  or  $x_t$ & Luminosity & Expected \\[-3pt]
& & (GeV) &  & (cm$^{-2}$s$^{-1}$) & start \\
    \hline
COMPASS~\cite{COMPASSII-prop}& \multirow{2}[2]{*}{$\pi^\pm + p^\uparrow$} & 160 & 
    \multirow{2}[2]{*}{$x_t=0.2$--0.3} & \multirow{2}[2]{*}{$1 \times 10^{32}$ } & 
    \multirow{2}[2]{*}{2014} \\*[-0.05in]
(CERN) & & $\sqrt{s}$ = 17.4 & & & \\*[0.05in]
PAX~\cite{Barone:2005pu}& \multirow{2}[2]{*}{$p^\uparrow + \overline{p}$} & collider & 
    \multirow{2}[2]{*}{$x_b$ = 0.1--0.9 } & \multirow{2}[2]{*}{$2 \times 10^{30} $} & 
    \multirow{2}[2]{*}{$>$2017} \\*[-0.05in]
(GSI) & & $\sqrt{s}$ = 14 & & & \\*[0.05in]
%    \hline
PANDA~\cite{Panda-prop}& \multirow{2}[2]{*}{$\overline{p} + p^\uparrow $} & 15 & 
    \multirow{2}[2]{*}{$x_t=0.2$--0.4 } & \multirow{2}[2]{*}{$2 \times 10^{32} $} & 
    \multirow{2}[2]{*}{$>$2016} \\*[-0.05in]
(GSI) & & $\sqrt{s}$ = 5.5 & & & \\*[0.05in]
%    \hline
NICA~\cite{NICA-prop}& \multirow{2}[2]{*}{$p^\uparrow + p $} & collider & 
    \multirow{2}[2]{*}{$x_b$ = 0.1--0.8 } & \multirow{2}[2]{*}{$1 \times 10^{30} $} & 
    \multirow{2}[2]{*}{$>$2014} \\*[-0.05in]
(JINR) & & $\sqrt{s}$ = 20 & & & \\*[0.05in] 
PHENIX~\cite{RHIC-Spin}& \multirow{2}[2]{*}{$p^\uparrow + p$} & collider &
    \multirow{2}[2]{*}{$x_b$ = 0.05--0.1 } & \multirow{2}[2]{*}{$2 \times 10^{32} $} &
    \multirow{2}[2]{*}{$>$2018} \\*[-0.05in] 
(BNL) & & $\sqrt{s}$ = 200 & & & \\*[0.05in]
%    \hline
%    RHIC internal~\cite{RHIC-int}& \multirow{3}[3]{*}{$p^\uparrow  + p$} & 250 &  \multirow{3}[3]{*}{$x_b$ = 0.25 -- 0.4 }  & &  \\*[-0.05in]
%    \hspace*{0.2in}target phase-1  &       & $\sqrt{s}$ = 22 &  &  $2 \times 10^{33} $ & $>$2015  \\*[-0.05in]
%    \hspace*{0.2in}target phase-2 &       & $\sqrt{s}$ = 22 &  & $3 \times 10^{34} $  &  $>$2018 \\*[0.05in]
Pol. Fermilab$^\ddag$ & \multirow{2}[2]{*}{$p^\uparrow  + p$} & 120 & 
    \multirow{2}[2]{*}{$x_b$ = 0.35--0.85 } & \multirow{2}[2]{*}{$2 \times 10^{35} $} & 
    \multirow{2}[2]{*}{{$>$2015}} \\*[-0.05in]
(Fermilab)  &       & $\sqrt{s}$ = 15 &  &  &  \\*[0.05in]
    \hline
    \hline
    \end{tabular}
%  \end{small}
\begin{flushleft}\small\hspace*{3em}
    $^\ddag$ $L = 1 \times 10^{36}\; \text{cm}^{-2}\text{s}^{-1}$ (SeaQuest LH$_2$ target limited),\\
    \hspace*{3em}\phantom{$^\ddag$ } 
    $L = 2 \times 10^{35}\; \text{cm}^{-2}\text{s}^{-1}$ (10\% of Main Injector beam limited).
\end{flushleft}
\end{table}

\begin{figure}
    \centering
    \includegraphics[width=0.75\textwidth]{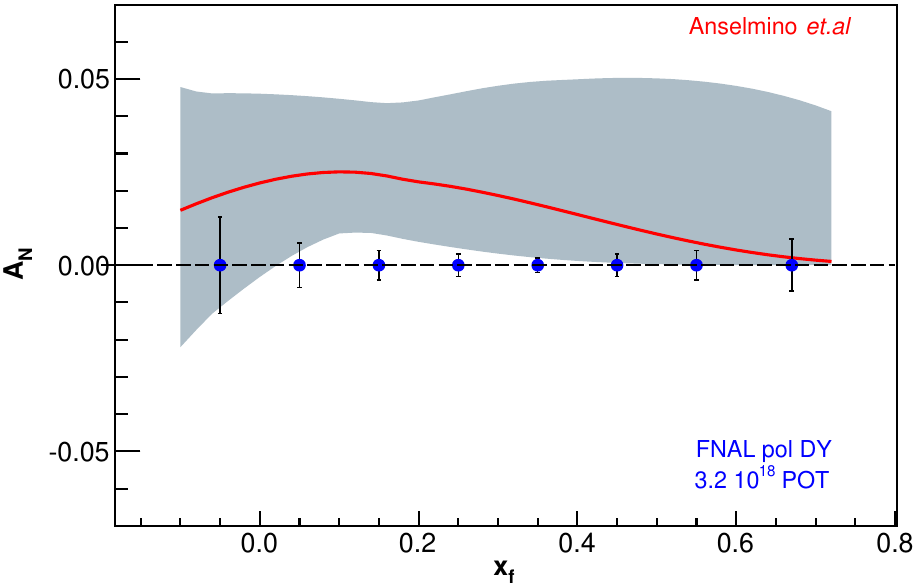}
    \caption[Single-spin asymmetry $A_N$ as a function of $x_f$]{Single-spin asymmetry $A_N$ as a function 
        of~$x_f$.
        $A_N$ (red line) is related to the Sivers SSA amplitude by 
        $A_N=(2/\pi)A_{\text{UT}}^{\sin{\phi_b}}$. 
        The gray shaded area represents the $\sqrt{20}\sigma$ error 
        band~\cite{Anselmino:2009st,AnselminoPrivate}.
        The expected statistical uncertainties (blue solid circles) for a 70\% polarized beam on an 
        unpolarized target and $3.2 \times 10^{18}$ protons on target are (arbitrarily) plotted on the zero 
        line.}
      \label{hadron-dy:fig:prediction}
\end{figure}

With the SeaQuest spectrometer and the Fermilab Main Injector beam
energy, the sensitivity of a measurement is limited by statistical
precision.  A quick examination of the proposed experiments in
Table~\ref{tab:dy:pol-DY} shows that Fermilab can achieve three orders
of magnitude more integrated luminosity than other facilities.  The
sensitivity that could be achieved at Fermilab is illustrated in
Fig.~\ref{hadron-dy:fig:prediction}, compared with a fit of existing SIDIS
Sivers distribution data by Anselmino, \emph{et al.}~\cite{Anselmino:2009st,AnselminoPrivate}.

The combination of high luminosity and large $x$-coverage makes
Fermilab arguably the best place to measure single-spin asymmetries in
polarized Drell-Yan scattering with high precision. At Fermilab, the
only ingredient that is missing is It would allow for the first time
to perform a measurement of the sign, the magnitude, and the shape of
the Sivers function with sufficient precision to verify this
fundamental prediction of QCD conclusively.

\subsection{OAM in the Sea}

As theory continues to wrestle with fundamental questions of the
origins of the proton's spin and OAM, experiment continues to
measure. An enticingly coherent picture of quark OAM has emerged from
the measurements of the Sivers function made via polarized SIDIS by
the HERMES and COMPASS collaborations~\cite{Airapetian:2004tw,
  Alexakhin:2005iw}. When subjected to a global
fit~\cite{Anselmino:2009st, Anselmino:2008sg, Anselmino:2008sga,
  Bacchetta:2011gx} and combined with the chromodynamic lensing model
of Ref.~\cite{Burkardt:2003uw}, they indicate $L_u > 0$ and \mbox{$L_d
  < 0$~\cite{Makins2012INT}}.

This agrees with the most basic prediction of the meson-cloud model of
the proton. In this model, the proton is described as a superposition
of a zeroth-order bare proton state of three constituent $uud$ quarks
and a first-order cloud of nucleon-pion states.  The seminal idea
behind this model is that hadrons, not quarks and gluons, are the best
degrees of freedom with which to approximate the essential features of
the proton. The pion cloud has two components: $n\,\pi^+$ and
$p\,\pi^0$, weighted by the Clebsch-Gordan coefficients of these two
isospin combinations. Immediately, we have an explanation for the
dramatic excess of $\bar d$ over $\bar u$ observed by Fermilab
E866/NuSea~\cite{Hawker:1998ty,Towell:2001nh}: with the sea quarks
wrapped up in the lightest hadronic states, the $\pi^0$ cloud
contributes $\bar d$ and $\bar u$ in equal measure but the $\pi^+$
contributes only $\bar d$.  Further, as the pions have zero spin, the
antiquarks should be unpolarized. This agrees with the HERMES SIDIS
data on $\Delta \bar u(x)$ and $\Delta \bar
d(x)$~\cite{Airapetian:2004zf, Airapetian:2003ct}, both of which were
found to be consistent with zero.

The meson cloud's picture of orbital angular momentum is dramatic.
As the constituents are heavy in this picture, nonrelativistic quantum
mechanics applies and $L$ is once again a good quantum number. In what state
of $L$ is the pion cloud? The pions have negative parity while
the nucleons have positive parity. To form a positive-parity proton
from $n\,\pi^+$ or $p\,\pi^0$, the pions must carry $L = 1$. The lowest-order
prediction of the meson cloud model is thus of an orbiting cloud;
application of Clebsch-Gordan coefficients yields $L_u>0$ and 
$L_d<0$~\cite{Makins2012INT,Thomas:2008ga}.

Unfortunately, this apparently coherent picture is at odds with
lattice-QCD calculations, which give $L_u < 0$ and $L_d > 0$ at the $Q^2$
scales of the Sivers measurements~\cite{Hagler:2007xi,Richards:2007vk}.
Recent work from a number of directions suggests
that the resolution of this puzzle lies in the proton sea. As
the sea quarks' spin polarization is near zero, and as the sea quarks'
disconnected diagrams are difficult to treat on the lattice (they were omitted in 
Refs.\cite{Hagler:2007xi,Richards:2007vk}), a 
tendency to neglect them has emerged in the spin community. As a
result, the simple fact has eluded us that the $L_u$ and $L_d$
determined from quark models and from SIDIS data refer to quarks only,
while the lattice-QCD calculations include both quarks and
antiquarks of the given flavor. Several recent developments have
highlighted the perils of this bias. First, data from HERMES and
BRAHMS on single-spin azimuthal asymmetries for kaon production have
shown mild-to-dramatic differences between them
~\cite{Airapetian:2004tw, Airapetian:2012yg, Arsene:2008aa}. A fast,
final-state $\pi^+$ meson ``tags'' $u$ and $\bar d$ quarks (i.e.,
enhances their contribution to the cross section), while a $K^+$ tags
$u$ and $\bar s$. The only difference between the two is the
antiquark; if it is causing pronounced changes in Sivers or
Boer-Mulders asymmetries, it may be indicative of antiquark OAM.
(Alternative explanations, such as higher-twist effects, also exist.)
Second, Wakamatsu~\cite{Wakamatsu:2009gx} has confronted the baffling
negative sign of $L_u - L_d$ from lattice QCD by calculating $L_u$ and
$L_d$ in the chiral quark soliton model, using both the Jaffe and Ji
definitions.  The paper shows not only the stark difference between
the two definitions, but also separates the sea and valence quark
contributions. In both definitions, the $\bar u$ and $\bar d$
antiquarks are the dominant players, and in the Jaffe
definition, are entirely responsible for the negative sign of
this quantity. Third, the $\chi$QCD Collaboration~\cite{Liu:2012nz} has, for the first time,
succeeded in including disconnected diagrams in a lattice calculation
of $L$.  They find the same: the sea quarks carry as much or more OAM
as the valence quarks. Finally, we return to the meson cloud
picture. Its orbiting cloud of $L=1$ pions gives as much OAM to the
antiquarks as to the quarks.

\subsection{Polarized Target Drell-Yan Measurements at Fermilab} 
\label{sec:dy:polTar}

\begin{figure}
    \centering
    \includegraphics[width=0.5\textwidth]{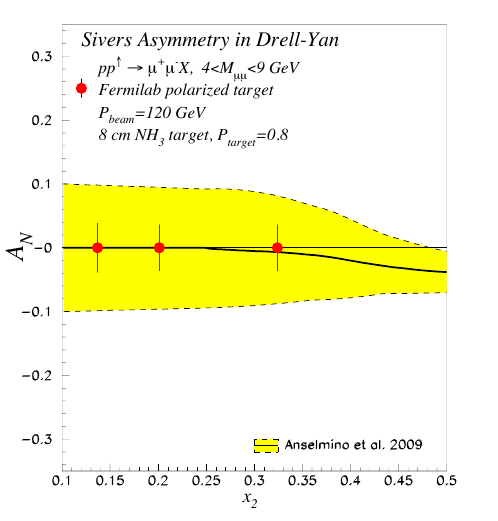}
    \caption[Estimated statistical precision for the Drell-Yan sea-quark Sivers asymmetry]{Estimated 
        statistical precision for the Drell-Yan sea-quark Sivers asymmetry \vs~$x_2$.
        Also shown is the prediction from Anselmino~\cite{Anselmino:2008sga,AnselminoPrivate} for the 
        magnitude of the asymmetry based on a fit of existing data.
        Note that we have extended the estimate below its valid minimum of $x_2$ of 0.2, in order to guide 
        the eye. 
        There is currently no good prediction available for the asymmetry below that value.
        The statistical uncertainties are based on $2.8\times 10^{18}$ protons on target.}
    \label{hadron-dy:fig:polTar}
    \end{figure}

The same combination of spectrometer acceptance, proton beam energy
and available luminosity that enabled the polarized beam measurement
is at play for polarized target measurements.  In this case, the
missing piece in the implementation is a polarized hydrogen target.
This is currently being developed by modifying and refurbishing an
existing, superconducting magnet and polarized target system.  With
this target and an integrated $2.7\times 10^{18}$ protons delivered,
the experiment expects to record and reconstruct $1.1\times 10^6$
Drell-Yan events.  The statistical precision on the asymmetry as a
function of $x_t$ from these events is shown in
Fig.~\ref{hadron-dy:fig:polTar}.

\subsection{Improvements with \PX\ Luminosity}

The measurements outlined in Secs.~\ref{hadron-dy:sec:polBeam} and
\ref{sec:dy:polTar} are statistically limited.  An examination of
Figs.~\ref{hadron-dy:fig:prediction} and \ref{hadron-dy:fig:polTar} quickly reveals
that these are both initial measurements.  
A true exploration of
  the Sivers distribution for both valence and sea quarks will benefit
  greatly from increased integrated luminosity.

The present luminosity is limited by the duration of the ``slow
spill'' during the Main Injector cycle; by the number of protons in
the ``slow spill''; and by the spectrometer's rate capabilities.
\PX\ Stage~1 and~2 will allow for a factor of 7 more 
  unpolarized protons per Main Injector spill.  The proposed
polarized target and associated utilities are likely capable of
handling a factor of 2 increase in proton intensity.  Additional
investment would be required to take full advantage of the \PX\
luminosity.

The ability of the SeaQuest spectrometer and analysis to record and
reconstruct events in this situation depends critically on the duty
factor of the the proton beam.  Improvements to the spectrometer's
rate capabilities in the \PX\ era can easily be foreseen,
including: finer segmentation in the tracking chamber; the use of
GEMs to replace the tracking chambers with the highest rate; finer
segmentation in the triggering system; and an open aperture magnet
that would allow for better triggering and track reconstruction.
Later stages of \PX\ will allow for even greater increases in
 unpolarized proton beam intensity. Utilizing this additional
luminosity would require additional investments in the spectrometer.

For the polarized beam experiment, the increase in proton beam
intensity must start with an improved  polarized proton source.
The present plan~\cite{SPINFermilab:2011aa} is to use a 1 mA polarized
proton source to eventually deliver $1\times 10^{13}$ protons/spill to
the experiment.  Although the polarized source is not in the baseline
for \PX, there are already foreseen improvements to the source
proposed for Fermilab E1027 that could lead to up to a 5 mA polarized
source and a corresponding increase in proton delivery to the
experiment with the additional \PX\ Stage~I improvements.

\bibliographystyle{apsrev4-1}
\bibliography{hadron-dy/refs}
 % Paul R., N. Makins, and W. Lorenzon

%%%%%%%%%%%%%%%%%%%%%%%%%%%%%%%%%%%%%%%%%%%%%%%%%%%%%%%%%%%%
\chapter{Hadronic Spectroscopy with \PX}
\label{chapt:hadron-s}
%%%%%%%%%%%%%%%%%%%%%%%%%%%%%%%%%%%%%%%%%%%%%%%%%%%%%%%%%%%%

\authors{J\"urgen Engelfried and Stephen Godfrey}

\section{Hadron Spectroscopy}

\subsection{Introduction}
\label{had:sec:intro}

Hadron spectroscopy is the manifestation of QCD in the soft, low $Q^2$, regime.
There have been significant developments in theory in recent years, particularly as the result of improved
and more complete results from lattice QCD; see Chapter~\ref{chapt:lqcd} and Refs.~\cite{Fodor:2012gf,%
Dudek:2009qf,Dudek:2011tt,Dudek:2011bn,Dudek:2010wm,Morningstar:1999rf,Chen:2005mg}.
Lattice QCD has established the existence of non quark model states in the physics QCD spectrum
\cite{Dudek:2009qf,Dudek:2011tt,Dudek:2011bn,Dudek:2010wm,Morningstar:1999rf,Chen:2005mg}.
These non--quark-model states would represent a new form of hadronic matter with explicit gluonic degrees of
freedom, the so called hybrids and glueballs, and multi-quark states beyond the quark-model $q\bar{q}$ mesons
and $qqq$ baryons.
However, despite searching for these states for over twenty years, these states have yet to be unambiguously
established experimentally.
Reviews on the subject are given in Refs.~\cite{Godfrey:1998pd,Meyer:2010ku,Crede:2008vw,Klempt:2007cp}.
See also the Particle Data Group \cite{Beringer:1900zz}.
There remains considerable interest in unambiguously identifying such states, as demonstrated by the high
interest in the reports by the CLEO, BaBar and Belle collaborations for evidence of possible exotic states,
the so called $X$ $Y$ $Z$ states \cite{Godfrey:2008nc}, which are among the most cited publications from
these experiments.
The discovery of hybrids is the motivation for the GlueX experiment at Jefferson Lab and a primary motivation
for the CEBAF 12~GeV upgrade \cite{Dudek:2012vr}.
GlueX uses high energy photons to excite mesons which many models predict will excite the gluonic degree of
freedom to produce hybrids \cite{Close:2003fz,Close:2003ae}.
The GlueX program can only explore a limited mass range due to the photon beam energy so while it may be able
to discover hybrid mesons and unambiguously establish their existence, it would not be able to fully map out
the hybrid spectrum.

Here, we outline an idea for an experiment at \PX, which we call the Fermilab Exotic Hadrons Spectrometer
(FEHS).
Its purpose is to map out the hybrid meson spectrum, complete the light meson spectrum and resolve some long
standing puzzles in hadron spectroscopy using high energy kaon beams that \PX\ has the unique capability
of producing.
The prototype for this experiment is the Large Aperture Superconducting Solenoid (LASS) experiment at SLAC
\cite{Aston:1987uc}, which advanced our understanding of strange and strangeonium spectroscopy to a degree
that has yet to be surpassed.

\subsection{Physics Motivation}
\label{had:sec:physics}

\begin{figure}
    \centering
    \includegraphics[width=1.0\textwidth]{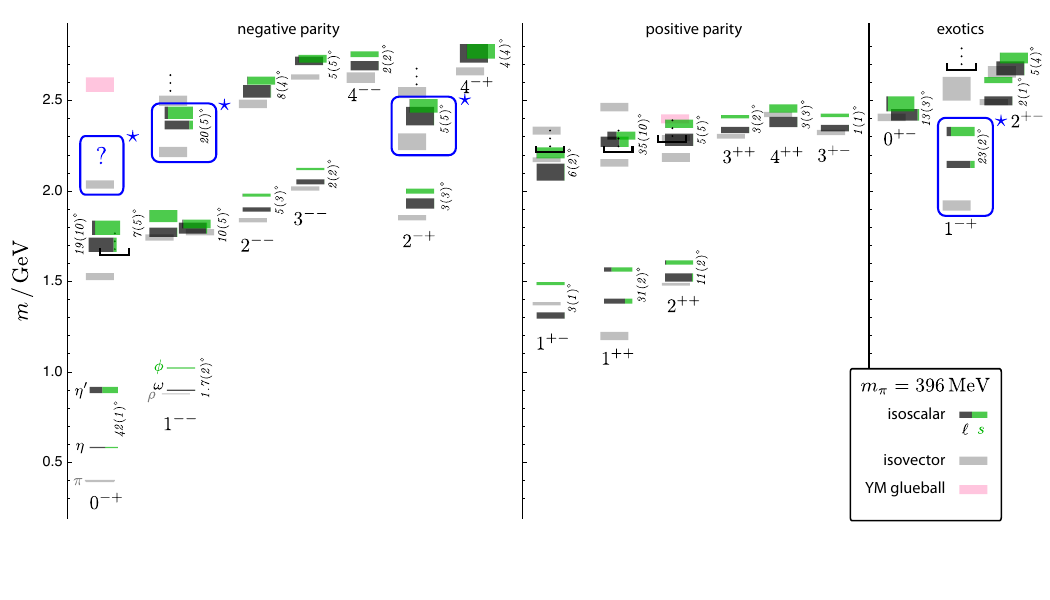} % provide eps and pdf and omit extension
    \vspace{-36pt}
    \caption[Light isoscalar meson spectrum from lattice QCD]{The light isoscalar meson spectrum as 
        calculated using lattice QCD labelled by $J^{PC}$. 
        The box height indicates the one sigma statistical uncertainty above and below the central value.
        The light ($u$, $d$) strange ($s$) quark content of each state ($\cos^2\alpha$, $\sin^2 \alpha$) is 
        given by the fraction of (black, green) and the mixing angle is also shown.
        Grey boxes indicate the positions of isovector meson states extracted on the same lattice 
        \cite{Dudek:2010wm}. 
        Pink boxes indicate the position of glueballs in the quarkless Yank-Mills theory 
        \cite{Morningstar:1999rf}.
        The candidate states for the lightest hybrid mesons are indicated by the blue boxes and stars
        \cite{Dudek:2011bn}.
        From Refs.~\cite{Dudek:2011tt,Dudek:2011bn}.}
    \label{had:fig:spectrum}
\end{figure}

There have been great strides in quantitatively mapping out the hadron spectrum using lattice QCD
\cite{Fodor:2012gf,Dudek:2009qf,Dudek:2011tt,Dudek:2011bn,Morningstar:1999rf,Chen:2005mg,Dudek:2010wm}.
Recent results indicate the existence of states with explicit gluonic degrees of freedom
\cite{Dudek:2009qf,Dudek:2011tt,Dudek:2011bn,Dudek:2010wm,Morningstar:1999rf,Chen:2005mg}.
These gluonic degrees of freedom manifest themselves as ``glueballs,'' which are hadrons without valence
quark content~\cite{Morningstar:1999rf,Chen:2005mg}, and ``hybrids,'' which are states with both valence
quarks and explicit gluonic degrees of freedom~\cite{Dudek:2009qf,Dudek:2011tt,Dudek:2011bn,Dudek:2010wm}.
Because the excited gluonic field could carry $J^{PC}$ quantum numbers other that $0^{++}$, the gluonic
quantum numbers can couple to $q\bar{q}$ quantum numbers resulting in $J^{PC}$ quantum numbers that are not
accessible to a $q\bar{q}$ pair alone.
Observation of a state with such exotic quantum numbers, $0^{--}$, $0^{+-}$, $1^{-+}$, $2^{+-}$, $3^{-+}
\ldots$, is considered the smoking-gun signature for states beyond the simple $q\bar{q}$ quark-model
states.
Lattice QCD predicts a rich spectrum of both isovector and isoscalar exotic hybrids, and the hadron spectrum
from one set of recent calculations \cite{Dudek:2011tt,Dudek:2011bn,Dudek:2010wm} is shown in
Fig.~\ref{had:fig:spectrum}, along with the glueball spectrum \cite{Morningstar:1999rf}.
It is crucial that these calculations be verified by experiment.

Over the years a number of candidate glueball states have been reported but due to the dense spectrum of
conventional hadrons it has been difficult to unambiguously identify a glueball candidate and rule out
conventional explanations \cite{Godfrey:1998pd}.
It is expected that the lowest lying glueballs are scalar mesons ($J^{PC}=0^{++}$) which are difficult to
disentangle from $q\bar{q}$ states with the same quantum numbers \cite{Crede:2008vw}.
Furthermore, the physical hadronic states are expected to be some linear combination of $q\bar{q}$, glueballs
and higher Fock space components rather than pure $q\bar{q}$ or glueballs.
Other glueballs with conventional quantum numbers are expected in the 2~GeV mass region but they are also
expected to be difficult to distinguish from conventional states \cite{Morningstar:1999rf}.
The lowest lying glueballs with exotic quantum numbers are expected to lie $\simeq 2.5$~GeV and will be more
difficult to produce.
As a consequence of the expected glueball properties they are likely to be difficult to unambiguously
identify as unconventional non-$q\bar{q}$ states.

Predictions of hybrid meson properties suggest that they are likely to be easier to discover than glueballs
for a number of reasons.
Primarily, it is expected that hybrid states with exotic quantum numbers exist low enough in mass that they
should be abundantly produced.
The phenomenological properties of hybrids have been studied using several different models
\cite{Isgur:1985vy,Page:1998gz,Close:1994hc} and hybrid properties such as quantum numbers, masses and decays
can be used to help in their discovery.
There are two approaches to distinguish hybrids from conventional states.
In the first, one looks for an excess of observed states compared to quark model predictions.
The drawback to this approach is that it depends on a good understanding of hadron spectroscopy in a mass
region that remains rather murky.
The other approach is to search for exotic quantum numbers that are not consistent with quark model
predictions.
The discovery of such exotic quantum numbers would be irrefutable evidence of something new.
Predicted properties of the lowest lying isovector and isoscalar hybrids are given in
Table~\ref{had:tab:hybrids}.
Lattice QCD predicts that the lowest hybrid excitations are expected at approximately 1.9~GeV for the
isovector $1^{-+}$ state and $\sim 2.1$~GeV and $\sim 2.3$~GeV for the mainly light and $s\bar{s}$ isoscalar
$1^{-+}$ states respectively \cite{Dudek:2011bn,Dudek:2011tt,Dudek:2010wm}.
Note that the isoscalars have mixed light and $s\bar{s}$ content.
We also expect strange hybrids.
However, strange mesons have a denser spectrum described by the the more limited set of $J^P$ quantum numbers
due to the flavored states not being eigenstates of charge-conjugation.
The immediate consequence is that there are no exotic strange mesons and therefore no smoking gun signature
for hybrid strange mesons.

\vspace{-3pt}
Observation of the $0^{+-}$ and $2^{+-}$ multiplets as well
as measuring the mass splittings with the $1^{+-}$ states would validate the lattice QCD calculations.
The decay properties probe the internal structure of the parent state so the predictions of a specific
model are very sensitive to the details of the model.  The decay predictions presented in 
Table~\ref{had:tab:hybrids} \cite{Isgur:1985vy,Page:1998gz,Close:1994hc}
are obtained using flux tube description of the gluonic degrees of freedom.
%which appear to be supported by lattice QCD.  
There appears to be 
a general property of hybrids that gluonic excitations 
cannot transfer angular momentum to the final states as relative angular momentum but rather, it 
must appear as internal angular momentum of the $q\bar{q}$ pairs. This results in 
an important selection rule of 
these models is that low-lying hybrids do not decay to identical mesons and that the preferred decay
channels are to $S+P$-wave mesons.   A consequence is that hybrids
tend to not couple strongly to simple final states making them in many cases difficult to reconstruct.
However, these results should not be taken as gospel, 
which is why we need experimental measurements to test these ideas.

\vspace{-3pt}
Using the hybrid decay properties given in Table~\ref{had:tab:hybrids}, we give 
examples of final states that a successful experiment should be able to study:
$b_2 \to a_1^+\pi^- \to (\rho^0 \pi^+)\pi^- \to \pi^+\pi^-\pi^+\pi^-$ where the final state
particles are charged, 
$h_2\to b_1^0 \pi^0 \to (\omega \pi^0)\gamma\gamma\to \pi^+\pi^-\gamma\gamma\gamma\gamma\gamma$, where
there are multiple final state photons, and 
$h_2' \to K_1^+K^- \to \rho^0 K^+K^- \to \pi^+\pi^- K^+ K^-$, requiring the
identification of strange particles.  
The $s\bar{s}$ hybrids, $\eta_1'$ and $h_2'$, are predicted by some models to be relatively narrow and 
are expected to decay to well-established strange resonances.
The decay of $s\bar{s}$ states to strange final states are enhanced relative to non-strange decays.  
To map out the $s\bar{s}$ hybrids will require measuring charged, neutral, and strange mesons
in the final state.  More generally, clearly identifying a 
large number of low lying hybrids would provide indisputable evidence for the 
the existence of exotic hybrid mesons.  To do so requires systematically studying the
strange and non-strange decay modes. 

\vspace{-3pt}
In addition to exotic hybrids one also expects hybrids with conventional $J^{PC}$ quantum numbers.
They will appear among conventional states with the same quantum numbers, so identifying the hybrids requires
having a good understanding of the meson spectrum.
To do so requires the ability to systematically study many final states, which will require considerable
statistics to be able to advance our knowledge of these states.

\begin{table}
    \centering
    \caption[Properties of exotic hybrid mesons]{Properties of exotic hybrid mesons.  
Masses are taken from lattice QCD calculations \cite{Dudek:2011bn,Dudek:2011tt,Dudek:2010wm}.
The estimates for widths and decay modes are taken from Ref.~\cite{Page:1998gz} for
the  PSS (Page, Swanson and Szczepaniak) \cite{Page:1998gz}
and IKP (Isgur, Kokoski and Paton) \cite{Isgur:1985vy} models.}
    \label{had:tab:hybrids}
\medskip
    \begin{tabular}{cccccc}
        \hline\hline
	& Mass (MeV) & $J^{PC}$ & \multicolumn{2}{c}{Total Width (MeV)} & Decays \\
	&	     &          & PSS & IKP & \\
\hline
$\pi_1$ & 1900 & $1^{-+}$ & 80-170 & 120 & $b_1 \pi$, $\rho \pi$, $f_1\pi$, $a_1 \eta$, $\eta(1295)\pi$ \\
$ \eta_1$ & 2100 & $1^{-+}$ & 60-160 & 110 & $a_1\pi$, $f_1 \eta$, $\pi(1300) \pi$ \\
$ \eta_1'$ & 2300 & $1^{-+}$ & 100-220 & 170 & $K_1(1400)K$, $K_1(1270)K$, $K^* K$ \\
\hline
$b_0$ & 2400 & $0^{+-}$ & 250-430 & 670 & $\pi(1300)\pi$, $h_1 \pi$ \\
$h_0$ & 2400 & $0^{+-}$ & 60-260 & 90 & $b_1\pi$, $h_1 \eta$, $K(1460)K$ \\
$h_0'$ & 2500 & $0^{+-}$ & 260-490 & 430 & $K(1460)K$, $K_1(1270) K$, $h_1 \eta$ \\
\hline
$b_2$ & 2500 & $2^{+-}$ & 10 & 250 & $a_2 \pi$, $a_1 \pi$, $h_1 \pi$ \\
$h_2$ & 2500 & $2^{+-}$ & 10 & 170 & $b_1 \pi$, $\rho \pi$ \\
$h_2'$ & 2600 & $2^{+-}$ & 10-20 & 80 & $K_1(1400) K$, $K_1(1270) K$, $K_2^* K$ \\
        \hline\hline
    \end{tabular}
\end{table}

%\begin{eqnarray}
%\tilde{\rho}^{-+}_1 & \to [B \pi ]_S & (\Gamma \approx 130 \hbox{MeV}) \\
%		    & \to [D \pi ]_S & (\Gamma \approx 50  \hbox{MeV}) \\
%\tilde{\omega}^{+-}_2 & \to [B \pi ]_P & (\Gamma \approx 500 \hbox{MeV}) \\
%\tilde{\omega}^{+-}_0 & \to [B \pi ]_P & (\Gamma \approx 250 \hbox{MeV}) \\
%\tilde{\phi}^{+-}_2 &\to [K^*(1420)\bar{K}]_P &(\Gamma\approx 250\hbox{MeV})\\
%		    &\to [\bar{K}Q_2 ]_P     &(\Gamma\approx 200\hbox{MeV})
%\end{eqnarray}

In addition to verifying the existence of these new forms of hadronic matter, there remain many issues in
conventional hadron spectroscopy.
The first issue is that the quark model predicts numerous states in the 1--2~GeV mass region that have not
been observed.
To fully understand conventional hadron spectroscopy, it is important that more of these missing states are
discovered and their properties measured.
A problem in improving our knowledge of mesons are that they are more difficult to produce via $t$-channel
exchange, and there is little control of the flavor quantum number.
In addition, these states are often broad and overlapping, the isospin zero states can and do mix, and there
is the possibility of glueballs, hybrids and multiquark states in the spectrum.
The LASS experiment at SLAC had considerable success in filling in some of the gaps in the strange and
$s\bar{s}$ meson sectors.
The leading $s\bar{s}$ states have been seen up to $J=5$, along with a few radial excitations.
However, some of the states have never been confirmed, with contradictory observations from other experiments,
and numerous states remain missing.
Furhemore, the LASS experiment was completed decades~ago.

To understand the physics, the LASS results should only be viewed as the start for unravelling the meson
spectrum.
It is time to find more of the radially excited states, fill in the orbitally excited multiplets, and proceed
to the more complicated $u\bar{u}$ and $d\bar{d}$ isoscalar and isovector mesons to test lattice-QCD
calculations.
Because of the complications of mixing between isoscalar states due to gluon annihilation and the
possibilities of glueballs, the strange mesons are a good starting point as they do not have these
complications.
In addition, because they are a heavy-light system, they probe glue dynamics in a different environment than
do mesons made out of equal mass quarks.
However, as mentioned previously, because they are not eigenstates of charge-conjugation there are no exotic
quantum numbers in the kaon sector and hence no smoking gun signal of hybrid states.
A detailed survey of the $s\bar{s}$ states would be a useful next step as they form a bridge between the
heavy quarkonia ($c\bar{c}$ and $b\bar{b}$) and the light quark mesons.

The second issue concerns puzzles and contradictory results in conventional hadron spectroscopy
\cite{Godfrey:1998pd,Beringer:1900zz}.
An example of such a puzzle is the $1^{++}$ $s\bar{s}$ state.
The $f_1(1420)$ is a candidate for the axial vector partner $s\bar{s}$ of the $a_1$ meson.
However, LASS also observed a $1^{++}$ $s\bar{s}$ state which fits in nicely as the $^3P_1$ $s\bar{s}$ state
but saw no evidence for the $f_1(1420)$.
The $\eta(1440)$ is another longstanding puzzle.
It is alleged by some to be a glueball although several conventional $q\bar{q}$ are expected in this mass
region.
Until the experimental situation clears up the glueball interpretation will remain suspect.
The FEHS experiment would be able to clear up these and many other puzzles that have festered for many years.

A final issue in conventional hadron spectroscopy is really how the first two issues affect the search for
gluonic hadrons and multiquark states.
The main impediment to finding exotic states with conventional quantum numbers is our incomplete
understanding of the conventional mesons.
To unambiguously recognize hybrids or glueballs will require a much better understanding of conventional
mesons.

We have focused primarily on the meson sector, because the baryon sector is denser and without exotic quantum
numbers.
Consequently, it will be difficult to distinguish non-quark model states from conventional baryons.
Also, in recent years the CLAS experiment at Jefferson Lab has improved our understanding of baryons.
Nevertheless, the \PX\ spectroscopy program can make important contributions to our knowledge of baryons.
The quark model predicts a very rich spectroscopy that has not been comprehensively tested.
Baryons have mainly been produced in $s$-channel $\pi N$ and $\bar{K} N$ formation experiments.
Quark model calculations predict that some states couple strongly to these channels and others will almost
completely decouple.
These features have been supported by experiment.
Thus, one way to find missing baryons is to study channels which couple more strongly to these missing states.
Another way is to produce baryons as decay states from higher states in the $\pi p$ and $K p$ channels.
Both approaches should be possible with the high intensity beams available at \PX.
In addition, because the number of baryons increases rapidly above 2~GeV, high statistics experiments will be
needed to disentangle the large number of states expected.
One sector that is relatively unexplored is the $sss$ $\Omega$ baryons.
A~suitable experiment at \PX\ should be able to observe and study these states.
It is important that the theoretical predictions for baryons be more completely tested by experiment by the
observation of many of the missing baryons and measurement of their properties.

Unravelling the spectroscopy will need high statistics experiments to perform partial wave analysis 
to filter by $J^{PC}$ quantum numbers.  To assist us in this process a guide to expected
properties will be useful.  There exist fairly complete calculations for expected masses and
decays of conventional states using the quark model and lattice QCD.  
While quantitative predictions might have large uncertainties,
qualitative predictions have proven to be reasonably reliable. 
The bottom line is that for many years progress in light hadron spectroscopy has been limited and
there is a compelling need for good quality data to advance the subject.

\subsection{Experimental Setup}
\label{had:sec:experiment}

% \subsubsection{Introductory}

As pointed out above, the LASS spectrometer at SLAC~\cite{Aston:1987uc} was the principal experiment
contributing to the physics of excited hadrons.
That said, the secondary beam-line at SLAC had a very poor duty factor, and LASS
ran with only 4--5 kaons per pulse to avoid pileup, 
and 100--180~pulses per second, resulting in fewer than 1000~kaons per
second~\cite{BlairRatcliff:2013}. 

A new experiment should aim to increase LASS's statistics by about a factor~50.
FEHS could run with a beam rate of about 50000~kaons per second, with a flat-top extraction this corresponds
to about $20~\mu\text{s}$ between beam particles and presents no problem whatsoever for the beam
instrumentation or the experimental setup.
For these reasons, a slightly updated copy of the LASS spectrometer would be the first approximation for the
experimental setup for FEHS.

\subsubsection{Beam Line and Target}

The LASS spectrometer featured a RF separated beam line of up to $16\,\mbox{GeV}/c$ momentum, and was usually
run at around $11\,\mbox{GeV}/c$.
At higher beam momentum, higher mass states can be produced, but
the length of the beam line needed for the RF separation increases with
the square of the momentum, while the decay losses decrease only linearly with
the momentum.
At higher momentum, say $\sim20\,\mbox{GeV}/c$, it will be necessary to use superconducting RF cavities to 
achieve a sufficient $p_t$ kick for the separation.  

One could also consider using an unseparated beam.
Depending on the momentum of the primary (proton-)beam and the secondary beam momentum, typically
the kaon:pion ratio is about 1:10, leading to a higher, but still tolerable,
flux in the spectrometer.
The final choice of the beam momentum will need a detailed
study of the all the above mentioned effects. 

In LASS, the beam particles were tagged with the help of two
threshold Cherenkov counters, the first $6\,\mbox{m}$ long filled 
with H$_2$ at $40\,\mbox{psia}$ (to count only
pions) and the second $1.28\,\mbox{m}$ long filled with CO$_2$ at 
$75\,\mbox{psia}$ (to count pions and kaons). The signals from the
two counters (anti-)coincidences were used to tag pions, kaon, and protons.
In FEHS, beam particle tagging can be performed in the same way.

% \subsubsection{The Target}

LASS featured a 33.5'' long liquid-hydrogen target.
FEHS could use the same without difficulty.

\subsubsection{The Spectrometer}

The experimental target at LASS was inside a solenoid magnetic field, surrounded by wire chambers, followed
downstream by a dipole magnet, again surrounded by multiwire proportional chambers.
This setup provided a nearly $4\pi$ coverage and proved to be very successful.
It was also adopted by the GlueX experiment~\cite{GlueX:2012me} at Jefferson Lab.

For FEHS this setup can also be used.
Conventional wire chambers equipped with modern readout electronics are sufficient for the flux conditions
described above.

Particle identification in LASS was performed via a pair of threshold Cherenkov detectors and a scintillator
hodoscope forming a TOF system.
Depending on the beam momentum, a similar PID system can be used for FEHS.

\subsubsection{Detector Summary}

In summary, the LASS spectrometer remains a suitable model for the FEHS experiment as well.
The detectors to be used (wire chambers, scintillator hodoscopes) are proven technology and optimizing the
designs for FEHS should not present serious problems.
Simulation studies have to be carried out to choose the beam momentum and to define the sizes and resolutions
of the different detector systems.

\subsection{Summary}
\label{had:sec:summary}

We have a long way to go before we can say that we understand hadron spectroscopy.
While there has been considerable progress made in lattice QCD, these results need to be reproduced by
experimental observation and measurements of the states' properties.
The unambiguous discovery of states with explicit glue degrees of freedom would be a major event as seen by
the excitement generated by recent candidate particles.
The details will provide important insights into quark and gluon dynamics.
They will help answer the question of how the glue degree of freedom manifests itself as collective
excitations or by some other description.

In addition, it is sobering to realize that we still do not understand conventional meson spectroscopy very
well.
Our knowledge and understanding of higher orbital and radial excitations is sparse at best.
It is worth remembering that there are many long-standing puzzles.
This poor understanding is hindering our ability to search for non $q\bar{q}$ states.
With a better understanding of conventional states it should be possible to distinguish hybrid states with
non-exotic quantum numbers from conventional states.
This would be especially important for strange mesons for which charge conjugation is not a good quantum
number.

The preferred approach is to search for hybrid states with exotic properties.  The least 
controversial characteristic is to look for states with exotic $J^{PC}$ quantum numbers with most
calculations predicting a $1^{-+}$ state to be accessible with mass less than 2~GeV.  The observation
of a resonance signal in this channel would be strong evidence for the discovery of a hybrid 
and is considered to be the starting point of any experimental search.  This is the approach taken
by the GlueX collaboration.  

To answer these questions and make progress in hadron spectroscopy a hadron spectrometer facility 
should be a part of the \PX\ physics program. The principal goal of the facility should be to search 
for gluonic excitations in hadrons and map out the spectroscopy of these states.  It is also 
important that the next generation of hadron spectroscopy experiments fill in as many of the missing
conventional meson and baryon states as possible. 

To make progress in this filed it is important that we study many properties of hadrons in many 
different channels to unravel the physics.  
The data will come from measurements at many different machines using different production 
mechanisms such as
$J/\psi$ radiative decays into light hadron final states studied by BESIII at IHEP in Beijing, 
photoproduction of states by GlueX at Jefferson Lab, 
$p\bar{p}$ annihilation by PANDA at GSI in Germany and high energy $\pi$, $K$ and $\bar{p}$ beams
by the COMPASS experiment at CERN.
However, the old LASS experiment has demonstrated that a dedicated high statistics 
hadroproduction experiment can make unique, important contributions. 

To address these questions the detector should have $4\pi$ acceptance for charged and neutral particles
with high detection efficiency, excellent tracking resolution and particle identification and be capable
of acquiring very high statistics.  The program will need $\pi$ and $K$ beams of 20~GeV
maximum energy with an appropriate sized experimental area to accommodate the detector.

The production mechanism in hadroproduction will complement other ongoing experiments such as GlueX
and BESIII as it will produce many different states in a wide variety of channels.

\bibliographystyle{apsrev4-1}
\bibliography{hadron-s/refs}
  % Steve G. and Jurgen

%%%%%%%%%%%%%%%%%%%%%%%%%%%%%%%%%%%%%%%%%%%%%%%%%%%%%%%%%%%%
% \chapter[Lattice QCD for \PX]{Lattice-QCD Calculations for \PX\ Experiments}
\chapter{Lattice-QCD Calculations for \PX}
\label{chapt:lqcd}
%%%%%%%%%%%%%%%%%%%%%%%%%%%%%%%%%%%%%%%%%%%%%%%%%%%%%%%%%%%%

\authors{Thomas Blum, Ruth S. Van~de~Water, \\
Michael Buchoff,
Norman H. Christ,
Andreas S. Kronfeld,
David G. Richards}

\section{Physics Motivation}
\label{lqcd:sec:physics}

As discussed in the previous chapters, the \PX\ accelerator complex will host a broad range of high-precision
measurements that probe quantum-mechanical loop effects and are sensitive to physics at higher energy scales
than are directly explored at the LHC.
Through the use of intense beams and sensitive detectors, the various \PX\ experiments will search for
processes that are extremely rare in the Standard Model (SM) and look for tiny deviations from Standard-Model
expectations.

In many cases, the comparison between the measurements and Standard-Model predictions are currently limited
by theoretical uncertainties from nonperturbative hadronic amplitudes such as decay constants, form 
factors, and meson-mixing matrix elements.
Lattice gauge theory provides the only known first-principles method for calculating these hadronic matrix
elements with reliable and systematically-improvable uncertainties, by casting the basic equations of QCD
into a form amenable to high-performance computing.
Thus, facilities for numerical lattice QCD are an essential theoretical compliment to the experimental
high-energy physics program.

The lattice-QCD community in the US and worldwide is expanding its program to meet the needs of the \PX\
physics program, as well as other upcoming intensity-frontier experiments.
In some cases, such as for the determination of CKM matrix elements that are parametric inputs to
Standard-Model predictions, improving the precision of existing calculations is sufficient, and the expected
increase in computing power due to Moore's law will enable a continued reduction in errors.
In other cases, like the muon $g-2$ and the nucleonic probes of non-SM physics, new hadronic matrix elements
are required; these calculations are typically computationally more demanding, and methods are
under active development.

The future success of the \PX\ physics program hinges on reliable Standard-Model predictions on the same
timescale as the experiments and with commensurate uncertainties.
In this chapter we discuss several key opportunities for lattice-QCD calculations to aid in the
interpretation of experimental measurements at \PX.
We focus on four general categories of calculations for which the technical issues are different: kaons, the
muon anomalous magnetic moment, nucleons, and hadron spectroscopy and structure.
We summarize the current status of lattice-QCD calculations in these areas; more detailed information can be
found in the talks on the \PX\ Physics Study website \cite{pxps:2012} and in the references.
We also discuss future prospects for lattice-QCD calculations in these areas, focusing on the computational
and methodological improvements needed to obtain the precision required by experiments at \PX.

This chapter is organized as follows.
In Sec.~\ref{lqcd:sec:intro}, we provide a brief introduction to numerical lattice QCD.
We summarize the dramatic progress in lattice-QCD calculations in the past decade, highlighting calculations
that validate the whole paradigm of numerical lattice-QCD.
This review sets the stage for Sec.~\ref{lqcd:sec:expt}, which describes a broad program of lattice-QCD
calculations that will be relevant for experiments at \PX, and that will be possible on the timescale of \PX.
Broadly, the lattice-QCD intensity-frontier effort has two main thrusts: (i) improving the precision of
present calculations and (ii) extending lattice gauge theory to new quantities relevant for upcoming
experiments.
Both require greater computational resources, and, where possible, we make forecasts for the expected
uncertainties in five years based on the assumption that computing resources continue to increase according
to Moore's law and that funding to support postdocs and junior faculty in lattice gauge theory does not
decrease.
In Sec.~\ref{lqcd:sec:resources}, we describe in some detail the computational resources needed to undertake
the calculations discussed earlier.
Finally, in Sec.~\ref{lqcd:sec:summary}, we recap the key lattice-QCD matrix elements needed to maximize the
scientific output of the \PX\ experimental physics program, and we summarize the case for continued support
of the US and worldwide lattice-QCD effort.

%%%%%%%%%%%%%%%%%%%%%%%%%%%%%%%%%%%%%%%%%%%%%%%%%%%%%%%%%%%%
\section{Introduction to Lattice QCD}
\label{lqcd:sec:intro}
%%%%%%%%%%%%%%%%%%%%%%%%%%%%%%%%%%%%%%%%%%%%%%%%%%%%%%%%%%%%

Lattice gauge theory formulates QCD on a discrete Euclidean spacetime lattice, thereby transforming the
infinite-dimensional quantum field theory path integral into a finite-dimensional integral that can be solved
numerically with Monte Carlo methods and importance sampling.
In practice, lattice-QCD simulations are computationally intensive and require the use of the world's most
powerful computers.
The QCD Lagrangian has $1 + N_f + 1$ parameters: the gauge coupling $g^2$, the $N_f$ quark masses $m_f$, and 
the \CP-violating parameter $\bar{\theta}$.
Because measurements of the neutron electric dipole moment (EDM) bound $\bar{\theta} < 10^{-10}$, most
lattice-QCD simulations set $\bar{\theta} = 0$.
The gauge-coupling and quark masses in lattice-QCD simulations are tuned by calibrating to $1 + N_f$
experimentally-measured quantities, typically hadron masses or mass-splittings.
Once the parameters of the QCD action are fixed, everything else is a prediction of QCD.

There are many ways to discretize QCD, particularly the fermions, but all of them recover QCD in the
continuum limit, i.e., when the lattice spacing $a\to 0$.
The various fermion formulations in use have different advantages (such as computational speed or exact
chiral symmetry) and different sources of systematic uncertainty; hence it is important to compute quantities
with more than one method for independent validation of results.
The time required for numerical simulations increases as the quark mass decreases (the condition number of
the Dirac operator, which must be inverted, increases with decreasing mass), so quark masses in lattice
simulations have usually been higher than those in the real world.
Typical lattice calculations now use quark masses such that the pion mass $m_\pi \lesssim 300$~MeV, while 
state-of-the art calculations for some quantities attain pions at or slightly below the physical mass of 
$m_\pi\sim140$~MeV. 
Over the coming decade, improvements in algorithms and increases in computing power will render chiral
extrapolations unnecessary.

Most lattice-QCD simulations proceed in two steps.
First one generates an ensemble of gauge fields with a distribution $\exp[-S_\text{QCD}]$; next one
computes operator expectation values on these gauge fields.
A major breakthrough in lattice-QCD occurred with the advent of gauge-field ensembles that include the
effects of the dynamical $u$, $d$, and $s$ quarks in the vacuum.
Lattice-QCD simulations now regularly employ ``$N_f = 2+1$" sea quarks in which the light $u$ and $d$
sea-quark masses are degenerate and somewhat heavier than the physical values, and the strange-sea quark mass
takes its physical value.
Further, ``$N_f = 2 + 1 + 1$" simulations that include a charm sea quark are now underway; dynamical charm
effects are expected to become important as precision for some quantities reaches the percent level.
During the coming decade, even $N_f=1+1+1+1$ simulations which include isospin-breaking in the sea are
planned.

The easiest quantities to compute with controlled systematic errors and high precision in lattice-QCD
simulations have only a hadron in the initial state and at most one hadron in the final state, where the
hadrons are stable under QCD (or narrow and far from threshold).
These quantities, often referred to as ``gold-plated,'' include meson masses and decay constants,
semileptonic and rare decay form factors, and neutral meson mixing parameters, and enable determinations of
all CKM matrix elements except $|V_{tb}|$.
Many interesting QCD observables are not gold-plated, however, such as resonances like the $\rho$ and $K^*$
mesons, fully hadronic decay matrix elements such as for $K \to \pi\pi$ and $B\to DK$, and long-distance
dominated quantities such as $D^0$-$\bar{D}^0$~mixing.
That said, lattice QCD with current resources is beginning to tackle such quantities, particularly in 
$K\to\pi\pi$ decay.

\begin{figure}
    \centering
    \includegraphics[width=\linewidth]{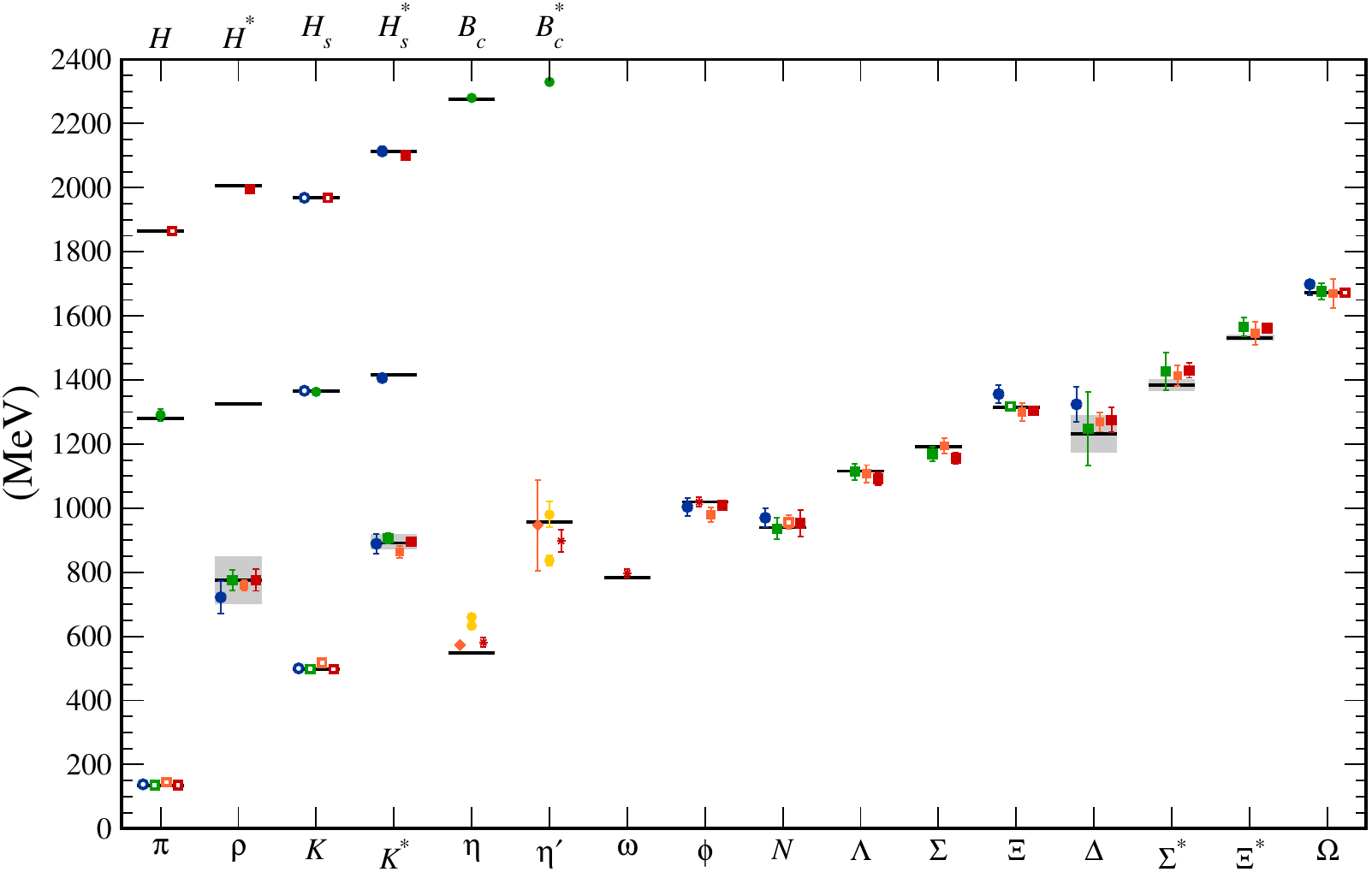}
    \caption[Hadron spectrum from many different lattice-QCD calculations]{Hadron spectrum from many 
        different lattice-QCD calculations~\cite{Aubin:2004wf,Bazavov:2009bb,Aoki:2008sm,Durr:2008zz,%
        Bietenholz:2011qq,Christ:2010dd,Dudek:2011tt,Gregory:2011sg,Bernard:2010fr,Gregory:2010gm,%
        Mohler:2011ke}.
        Open symbols denote masses used to fix bare parameters; closed symbols represent \emph{ab initio}
        calculations.
        Horizontal black bars (gray boxes) show the experimentally measured masses (widths).
        $b$-flavored meson masses ($B_c^{(*)}$ and $H_{(s)}^{(*)}$ near 1300~MeV) are offset by $-4000$~MeV.
        Circles, squares and diamonds denote staggered, Wilson and domain-wall fermions, respectively.
        Asterisks represent anisotropic lattices ($a_t/a_s<1$).
        Red, orange, yellow and green and blue signify increasing ensemble sizes (i.e., increasing range of 
        lattice spacings and quark masses).
        From Ref.~\cite{Kronfeld:2012uk}.}
    \label{lqcd:fig:spectrum}
\end{figure}

\begin{figure}
    \centering
    \includegraphics[width=0.495\textwidth]{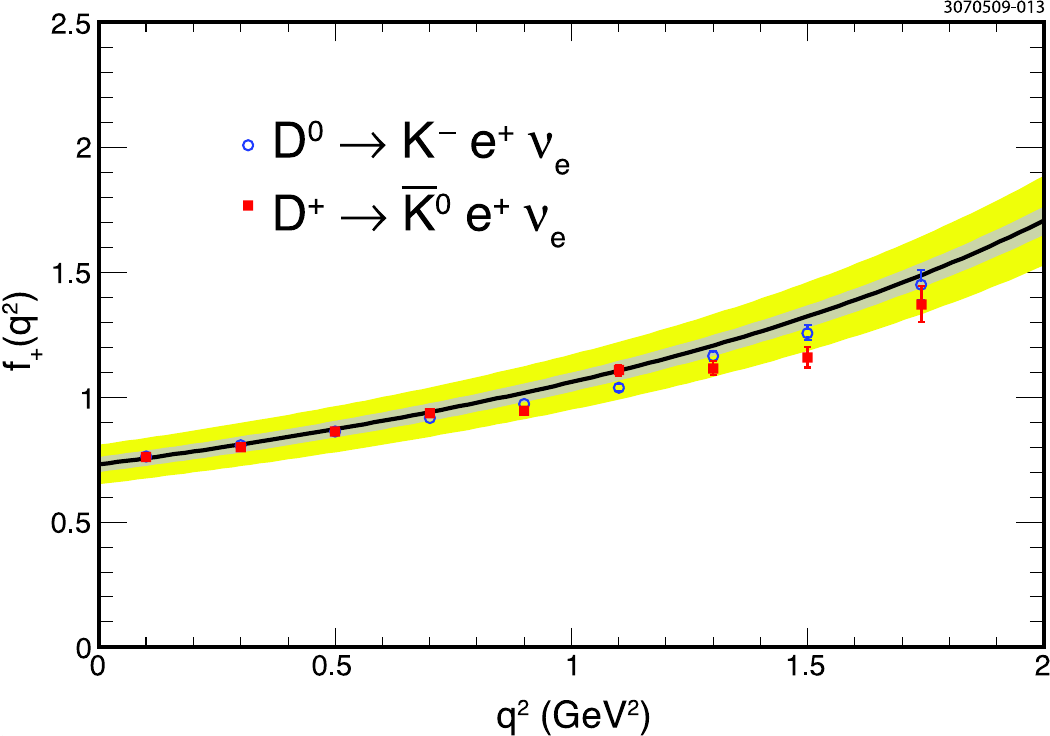}\hfill
    \includegraphics[width=0.495\textwidth]{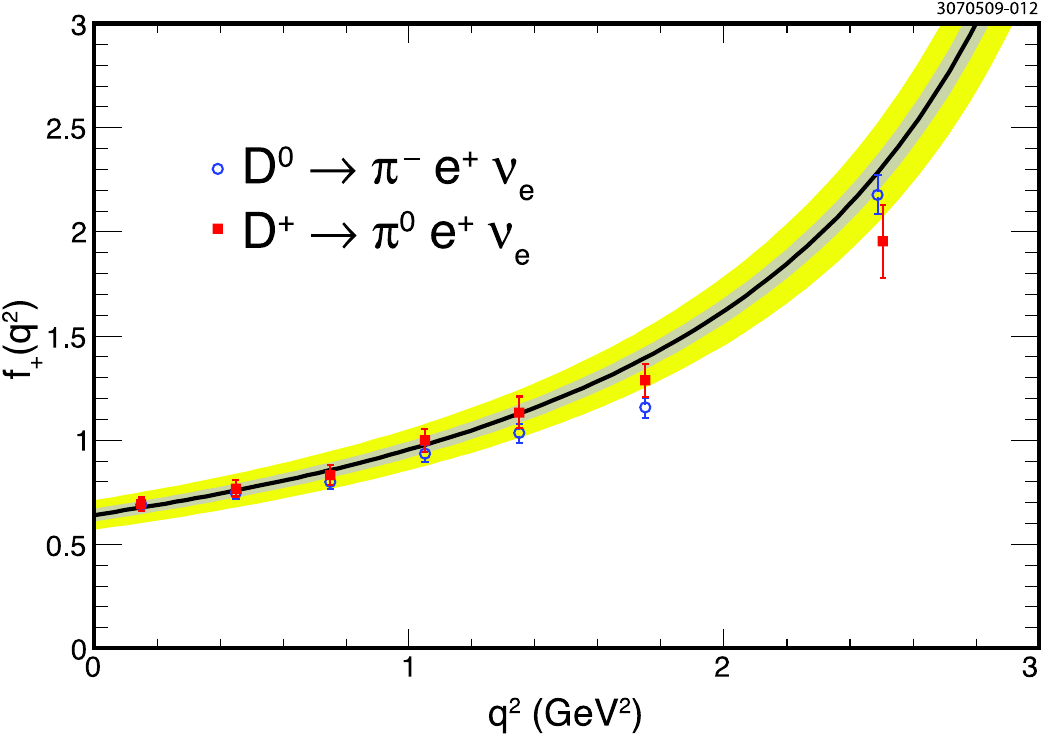}
    \caption[Lattice-QCD calculations of $D$-meson form factors compared with measurements]{Comparison of 
        $N_f = 2+1$ lattice-QCD calculations of $D$-meson form factors~\cite{Aubin:2004ej,Bernard:2009ke} 
        (curves with error bands) with measurements from CLEO~\cite{Besson:2009uv} (points with error bars).
        From Ref.~\cite{Besson:2009uv}.}
    \label{lqcd:fig:D2K}
\end{figure}

Many errors in lattice-QCD calculations can be assessed within the framework of effective field theory.
Lattice-QCD calculations typically quote the following sources of uncertainty:
\begin{itemize}
\item \emph{Monte Carlo statistics and fitting};
\item \emph{tuning lattice spacing and quark masses} by calibrating to a few experimentally-measured 
quantities such as $m_\pi$, $m_K$, $m_{D_s}$, $m_{B_s}$, $m_\Omega$, and $f_\pi$;
\item \emph{matching lattice gauge theory to continuum QCD} using fixed-order lattice perturbation theory, 
step-scaling, or other partly or fully nonperturbative methods;
\item \emph{chiral and continuum extrapolation} by simulating at a sequence of light (up and down) quark
masses and lattice spacings and extrapolating to $m_{\rm lat} \to m_{\rm phys}$ and $a\to0$ using functional
forms derived in chiral and weak-coupling QCD perturbation theory;
\item \emph{finite volume corrections}, which may be estimated using effective theory and/or studied directly 
by simulating lattices with different spatial volumes.
\end{itemize}
The methods for estimating uncertainties can be verified by comparing results for known quantities with
experiment.
Lattice-QCD calculations successfully reproduce the experimentally-measured low-lying hadron
spectrum~\cite{Aubin:2004wf,Bazavov:2009bb,Aoki:2008sm,Durr:2008zz,Bietenholz:2011qq,Christ:2010dd,%
Dudek:2011tt,Gregory:2011sg,Bernard:2010fr,Gregory:2010gm,Mohler:2011ke}, as shown in Fig.~\ref{lqcd:fig:spectrum}.
Lattice-QCD results agree with nonlattice determinations of the charm-and bottom-quark
masses~\cite{McNeile:2010ji,Beringer:1900zz,Chetyrkin:2009fv} and strong coupling~$\alpha_s$
\cite{Allison:2008xk,Davies:2008sw,Aoki:2009tf,McNeile:2010ji,Shintani:2010ph,Blossier:2012ef,%
Bethke:2011tr}, but now surpass the precision obtained by other methods.
Further, lattice-QCD calculations correctly predicted the mass of the $B_c$
meson~\cite{Allison:2004be,Abulencia:2005usa}, the leptonic decay constants $f_D$ and
$f_{D_s}$~\cite{Aubin:2005ar,Artuso:2005ym}, and the $D\to K \ell \nu$ semileptonic form
factor~\cite{Aubin:2004ej,Widhalm:2006wz} (see Fig.~\ref{lqcd:fig:D2K}) before the availability of precise
experimental measurements.
These successful predictions and postdictions validate the methods of numerical lattice QCD, and demonstrate
that reliable results can be obtained with controlled uncertainties.

We note that the huge strides made in lattice-QCD have been largely fueled by increased support for
lattice-QCD infrastructure and scientific staff in the United States, as well as similar efforts across the
globe.
Despite these considerable advances, however, for most quantities lattice errors remain significantly larger
than those in the corresponding experimental measurements.
Thus lattice QCD remains the bottleneck in these cases.
If we are to further squeeze the vise on the Standard Model with precise measurements at \PX\ and elsewhere,
we must continue to push forward with lattice~QCD.

%%%%%%%%%%%%%%%%%%%%%%%%%%%%%%%%%%%%%%%%%%%%%%%%%%%%%%%%%%%%
\section{Lattice QCD and \PX\ Experiments}
\label{lqcd:sec:expt}
%%%%%%%%%%%%%%%%%%%%%%%%%%%%%%%%%%%%%%%%%%%%%%%%%%%%%%%%%%%%

In this section we describe a broad program of lattice-QCD calculations that will be possible over the time
scale of \PX\ operations assuming that computer resources increase following Moore's law.
We organize this discussion according to physics topic or class of experiments for which the calculations are
needed.
In each subsection, we summarize the physics goals and their relationship to the experimental program,
describe the status of present lattice-QCD calculations, and explain what can be achieved over the next five
to ten years.

While the challenges to further reductions in errors depend on the quantity, there are many common features.
A key advance over the next five years will be the widespread simulation of physical $u$ and $d$ quark
masses, obviating the need for chiral extrapolations.
Such simulations have already been used for studies of the spectrum and several matrix elements including the
leptonic decay constant ratio $f_K/f_\pi$ and the neutral kaon mixing parameter
$\hat{B}_K$~\cite{Aoki:2009ix,Durr:2010vn,Durr:2010aw,Bazavov:2013cp,Dowdall:2013rya}.

A second advance will be the systematic inclusion of isospin-breaking
and electromagnetic (EM) effects. 
Once calculations attain percent-level accuracy, as is the 
case at present for quark masses,
 $f_K/f_\pi$, the $K\to\pi\ell\nu$ and $B\to D^*\ell\nu$ form factors, and $\hat B_K$,
one must study both of these effects.
A partial and approximate inclusion of such effects is already
made for light quark masses, $f_\pi$, $f_K$ and $\hat B_K$.
Full inclusion would require nondegenerate $u$ and $d$ quarks
and the incorporation of QED into the simulations.

A final across-the-board improvement that will likely become standard in the next five years is the use of
charmed sea quarks.
These are already included in two of the major streams of gauge-field ensembles being
generated~\cite{Baron:2009wt,:2012uw}.

%%%%%%%%%%%%%%%%%%%%%%%%%%%%%%%%%%%%
\subsection{Neutrino Experiments}
\label{lqcd:subsec:neutrino}
%%%%%%%%%%%%%%%%%%%%%%%%%%%%%%%%%%%%

Here we describe opportunities for lattice QCD to assist the \PX\ neutrino experimental program described in
Chapter~\ref{chapt:nu}.
\PX\ will provide intense neutrino sources and beams that can be used to illuminate nearby detectors at
Fermilab or far detectors at other facilities.

As discussed in Chapter~\ref{chapt:nu}, one of the largest sources of uncertainty in accelerator-based
neutrino experiments is from the determination of the neutrino flux.
This is because the beam energies are in the few-GeV range, for which the interaction with hadronic targets
is most complicated by the nuclear environment.
At the LBNE experiment, in particular, the oscillation signal occurs at energies where quasielastic
scattering dominates.
Therefore a measurement or theoretical calculation of the $\nu_\mu$ quasielastic scattering cross section as
a function of energy $E_\nu$ provides, to first approximation, a determination of the neutrino flux.
The cross section for quasielastic $\nu_\mu n \to \mu^-p$ and $\bar{\nu}_\mu p\to \mu^+n$ scattering is 
parameterized by hadronic form factors that can be computed from first principles with lattice QCD.

Once the LBNE far detector is large enough, and is shielded from cosmic rays either by an underground
location or an above-ground veto system, it will enable a proton-decay search that can improve upon the
projected reach of current facilities.
The interpretation of experimental limits on the proton lifetime as constraints on new-physics models depend
upon the expectation values $\langle \pi, K, \eta, \ldots | {\mathcal O}_{\rm BSM} | p \rangle$ of
non-SM operators; these can be computed with lattice QCD.

\subsubsection{Nucleon Axial-vector Form Factor}

The cross section for quasielastic scattering processes---$\nu_\ell n\to\ell^-p$ or 
$\bar{\nu}_\ell p\to\ell^+n$, where $\ell^\pm$ is a charged lepton---is a key element of many aspects of 
neutrino physics~\cite{RevModPhys.84.1307}.
The hadronic process is expressed via form factors, which must be known well to gain a full understanding of
neutrino scattering when the neutrino energy, $E_\nu$, on a fixed target is in the range $E_\nu<3$~GeV.
This knowledge is important both for using neutrinos to understand nuclear structure (which has ramifications
for many \PX\ experiments) and for understanding measurements of neutrino-oscillation parameters, in the 
precision era starting now with NO$\nu$A and T2K, and continuing on into \PX\ operations with LBNE.

The two most important form factors are the vector and axial-vector form factors, corresponding to the $V$ 
and $A$ components of $W^\pm$ exchange.
The vector form factor can be measured in elastic $ep$ scattering.
In practice, the axial-vector form factor has most often been modeled by a one-parameter dipole 
form~\cite{LlewellynSmith:1971zm}
\begin{equation}
    F_A(Q^2) = \frac{g_A}{(1+Q^2/M_A^2)^2},
    \label{lqcd:eq:dipole}
\end{equation}
although other parametrizations have been
proposed~\cite{Kelly:2004hm,Bradford:2006yz,Bodek:2007ym,Bhattacharya:2011ah}.
The normalization $g_A=F_A(0)=-1.27$ is taken from neutron $\beta$~decay \cite{Beringer:1900zz}.
The form in Eq.~(\ref{lqcd:eq:dipole}) matches the asymptotic behavior at large $Q^2$ (see, e.g.,
Ref.~\cite{Lepage:1980fj}), but in the low~$Q^2$ range relevant neutrino experiments, it does not rest on a
sound foundation.
It is worth noting that measurements of the vector form factor over a wide energy range do \emph{not} satisfy
the dipole form~\cite{Arrington:2006zm}.

Fits to Eq.~(\ref{lqcd:eq:dipole}) over different $Q^2$ ranges yield different results for the fit 
parameter $M_A$, e.g., $M_A\approx1.03~\textrm{GeV}$ from NOMAD and other higher-energy experiments 
($3~\textrm{GeV}<E_\nu<80~\textrm{GeV}$) \cite{Lyubushkin:2008pe}, 
but $M_A\approx1.35~\textrm{GeV}$ from MiniBooNE at lower energy
($0.4~\textrm{GeV}<E_\nu<2~\textrm{GeV}$) \cite{AguilarArevalo:2010zc}.
The difference may stem from nuclear effects, but without an \emph{ab initio} determination of the 
axial-vector form factor, one cannot know.
Indeed, fits employing a model-independent parametrization based on analyticity and 
unitarity~\cite{Bhattacharya:2011ah} find a consistent picture, obtaining $M_A=0.89^{+0.22}_{-0.07}$~GeV for 
a model-independent definition of~$M_A$.
The theoretical basis of Ref.~\cite{Bhattacharya:2011ah} is the same as that used successfully for meson 
decay form factors, say for the determination of $|V_{ub}|$ \cite{Bailey:2008wp,Ha:2010rf,Lees:2012vv}.

The lattice-QCD community has a significant, ongoing effort devoted to calculating $F_A(Q^2)$ 
\cite{Khan:2006de,Yamazaki:2009zq,Bratt:2010jn,Alexandrou:2010hf,Alexandrou:2013joa}.
Unfortunately, however, the results for the axial charge $g_A=F_A(0)$ have not agreed well with neutron 
$\beta$ decay experiments; see, e.g., Ref.~\cite{Hagler:2009ni} for a review.
Recently, however, two papers with careful attention to excited-state contamination in the lattice correlation functions and the chiral extrapolation~\cite{Capitani:2012gj} and 
lattice data at physical pion mass~\cite{Horsley:2013ayv} find results in agreement with experiment, 
$g_A\approx1.25$.
In addition to sensitivity to the chiral extrapolation, it is important to treat finite-volume effects 
more carefully than is often the case.
A~caveat here is that Refs.~\cite{Capitani:2012gj,Horsley:2013ayv} simulate with only $N_f=2$ sea quarks.
If these findings hold up with $2+1$ and $2+1+1$ flavors of sea quark, the clear next step is to compute the 
shape of the form factors with lattice QCD.
If the calculations of the vector form factor reproduce experimental measurements, then one could proceed to 
use the lattice-QCD calculation of the axial-vector form factor in analyzing neutrino data.

\subsubsection{Proton decay Matrix Elements}
\label{lqcd:subsubsec:protondecay}

Proton decay is forbidden in the Standard Model but is a natural prediction of grand unification.
% Thus, many theories of physics beyond the Standard Model predict proton decay.
Extensive experimental searches have, to date, found no evidence for proton decay, but future experiments
will continue to improve the limits.
To obtain constraints on model parameters requires knowledge of hadronic matrix elements
$\langle\pi,K,\eta,\ldots|\mathcal{O}_{\Delta B=1} | p \rangle$ of the baryon-number violating operators
$\mathcal{O}_{\Delta B=1}$ in the effective Hamiltonian.
Estimates of these matrix elements based on the bag model, sum rules, and the quark model vary by as much as
a factor of three, and lead to an $\text{O}(10)$ uncertainty in the model predictions for the proton
lifetime.
Therefore, \emph{ab initio} QCD calculations of proton-decay matrix elements with controlled systematic
uncertainties of even $\sim 20\%$ would represent a significant improvement, and be sufficiently precise for
constraining GUT theories.

Recently the RBC and UKQCD Collaborations obtained the first direct calculation of proton-decay matrix
elements with $N_f=2+1$ dynamical quarks~\cite{Aoki:2013yxa}.
The result is obtained from a single lattice spacing, and the total statistical plus systematic uncertainties
range from 20--40\%.
Use of gauge-field ensembles with finer lattice spacings and lighter pions, combined with a new technique to
reduce the statistical error~\cite{Blum:2012uh}, however, should enable a straightforward reduction of the
errors to the $\sim 10\%$ level in the next five years.

%%%%%%%%%%%%%%%%%%%%%%%%%%%%%%%%%%%%
\subsection{Kaon Physics}
\label{lqcd:subsec:kaon}
%%%%%%%%%%%%%%%%%%%%%%%%%%%%%%%%%%%%

Here we describe opportunities for lattice QCD to assist the \PX\ kaon physics program described in
Chapter~\ref{chapt:kaon}.
In many cases, hadronic matrix elements from lattice QCD are crucial for interpreting the experimental
measurements as tests of the Standard Model and constraints on new physics.
The ORKA experiment, which could well begin running before \PX, will measure the \CP-conserving rare decay
$K^+\to\pi^+\nu\bar{\nu}$ and collect $\sim$~200 events/year, assuming the Standard-Model rate.
With Stage~1 of \PX\ this rate will increase to $\sim$~340 events/year, enabling a measurement of the
branching fraction to $\sim$~3\% precision.
Stage~2 of \PX\ will enable a measurement of the branching fraction for the \CP-violating rare decay $K_L
\to \pi^0 \nu \bar{\nu}$ to $\sim$5\%, again assuming the Standard-Model rate.
The \PX\ kaon-physics experiments will also measure numerous other kaon observables such as
$\Gamma(K_{e2})/\Gamma(K_{\mu 2})$, $K^+ \to \pi^+ \ell^+ \ell^-$, and $K_L \to \pi^0 \ell^+ \ell^-$.
Correlations between these channels will allow discrimination between different new-physics scenarios,
provided sufficiently precise theoretical predictions from lattice QCD and elsewhere.

The worldwide lattice-QCD community has a well-established and successful kaon physics program.
The matrix elements needed to obtain pion and kaon leptonic decay constants, light-quark masses, the
$K\to\pi\ell\nu$ semileptonic form factor, and neutral kaon mixing are all gold-plated, and can therefore be
computed with lattice QCD to a few percent or better precision.
Many lattice-QCD collaborations are attacking these quantities with $N_f =
2+1$~\cite{Durr:2010hr,Follana:2007uv,Laiho:2011dy,Bazavov:2010hj,Aoki:2010dy,Lubicz:2009ht,Boyle:2010bh,Durr:2010vn,McNeile:2010ji,Bazavov:2009tw,Blum:2010ym,Kelly:2012uy,Bazavov:2012cd}
and now $N_f = 2+1+1$~\cite{Farchioni:2010tb,Bazavov:2011fh,Bazavov:2013cp,Dowdall:2013rya} gauge-field
ensembles, thereby providing independent cross checks and enabling global lattice-QCD
averages~\cite{Laiho:2009eu,Colangelo:2010et}.
A highlight of the lattice-QCD kaon physics effort is the calculation of the neutral-kaon mixing parameter
$B_K$, which enables a constraint on the apex of the CKM unitarity triangle when combined with experimental
measurements of indirect \CP-violation in the kaon system.
Until recently, the unitarity-triangle constraint from $\epsilon_K$ was limited by the $\sim$20\% uncertainty
in the hadronic matrix element $B_K$~\cite{Gamiz:2006sq}.
Several years ago, the lattice-QCD community identified $B_K$ as a key goal for lattice flavor physics, and 
devoted significant theoretical and computational effort to its improvement.
Now several independent lattice-QCD results for $B_K$ are in good
agreement~\cite{Durr:2011ap,Laiho:2011dy,Kelly:2012uy,Bae:2011ff}, and the error on the average is
$\lesssim1.5\%$~\cite{Laiho:2012ss}.
In fact, $B_K$ is now a sub-dominant source of uncertainty in the $\epsilon_K$ band, below the parametric
error from $A^4 \propto |V_{cb}|^4$ and the perturbative truncation errors in the Inami-Lim functions
$\eta_{cc}$ and $\eta_{ct}$~\cite{Brod:2011ty,BrodPXPS}.

Table~\ref{lqcd:tab:error} shows the status of lattice-QCD
calculations, comparing lattice errors in various matrix elements
to those in the corresponding  experimental measurements.
Where available, we also include forecasts made in 2007
for the expected errors in $\sim 2012$~\cite{whitepaper07},
which have proven to be quite accurate.
Given the maturity of these calculations, we expect the forecasts for 2018 to be reasonably accurate as well.  It is important to note that,
of the quantities in Table~\ref{lqcd:tab:error}, only for $f_K/f_\pi$
was a result available in 2007 with all errors controlled.
All other calculations have matured from having several
errors uncontrolled to all errors
controlled over the last five years.

\begin{table}
\centering
\caption[Forecasts for lattice QCD]{ History, status and future of selected lattice-QCD calculations needed
for the determination of CKM matrix elements relevant to the kaon sector.
Forecasts from the 2007 white paper (where available) assumed computational resources of 10--50 TF~years.
Present lattice errors are taken from Refs.~\cite{Bazavov:2013cp,Bazavov:2012cd,Durr:2011ap,Bailey:2010gb}.
Forecasts for 2018 assume that computer resources increase following Moore's law.}
\label{lqcd:tab:error}
% \begin{tabular}{c@{\;}c@{\quad}c@{\;\,}c@{\;}c@{\;\,}c@{\;\,}c}
\begin{tabular}{cccccc}
\hline\hline
Quantity  & CKM     & Present      & Forecast (2007) for & Present (2013) &      2018  \\[-1.5mm]
	      & element & expt.\ error & 2012 lattice  error & lattice  error & lattice error  \\  
%& \quad\quad QCD method \\
\hline
% 2014/2018 estimates discussed in notes in appendix from SS
%
% expt from Flavianet kaon WG 2010 
% 2007 forecast based on MILC2 ensembles
% current lat from MILC HISQ
$f_K/f_\pi$ & $|V_{us}|$ \rule[0mm]{0mm}{4mm} & 0.2\% &0.5\%&
 0.4\% & 0.15\%  \\ % & \\
%
% expt from Flavianet kaon WG 2010; 
% current lat from MILC HISQ-on-Asqtad
% 2014 consistent with Juettner at FNAL12/07
% 
$f_+^{K\pi}(0)$ & $|V_{us}|$ \rule[0mm]{0mm}{4mm} & 0.2\% & -- &
0.4\% & 0.2\%  \\ % & ChPT + quark model \\
%
% old forecast from Table in 2007 white paper
% experimental error is that in epsilon_K
% present error from BMW
% future guesses from text
$B_K$ \rule[0mm]{0mm}{4mm} & ${\rm Im}(V_{td}^2)$ & 0.5\% &3.5--6\% & 1.6\% & $< 1\%$ \\
% exp from HFAG end of 2011
% lat from FNAL/MILC 201? Lattice proc
% see text for forecasts (note that they are less optimistic
% than intensity frontier numbers)
$B\to D^{*}\ell\nu$ \rule[0mm]{0mm}{4mm} & $|V_{cb}|$ & 1.3\% & -- & 1.8\% & $<1\%$   \\ 
% & $B\to X_c\ell\nu$ + OPE + HQE \\
%
\hline\hline
\end{tabular}
\end{table}

The amplitudes listed so far all have one hadron in the initial state and zero or one in the final state and
are especially straightforward to determine for several reasons.
For example, the finite-volume errors are suppressed exponentially.
Recent advances in the methods of computational quantum field theory, numerical algorithms and computer
technology, however, are expanding the types of calculations that can be pursued and the experiments that can
be addressed.
For example, although the conceptual framework for computing $K\to\pi\pi$ amplitudes has been available for
twenty years, it was only in 2012 that the amplitude for $I=2$ was brought under
control~\cite{Blum:2011ng,Blum:2012uk}.
Progress is also being made in the calculation of long-distance amplitudes; methods are being tested for
$K_L$-$K_S$ mass difference $\Delta M_K$~\cite{Yu:2012xx} and will eventually be extended to rare
semileptonic kaon decays.

\subsubsection{$K\to\pi\pi$ Decays}

The advances in lattice-QCD calculations of weak interactions involving the strange quark open the exciting
possibility to search for physics beyond the standard model via experimental measurements of direct
\CP-violation in the kaon system.
The NA48 and KTeV experiments have measured ${\rm Re}(\epsilon^\prime/\epsilon)$ to around 10\%
precision~\cite{Batley:2002gn,Abouzaid:2010ny} , but the ability to constrain new physics with
$\epsilon^\prime$ has been handicapped by the uncertainty in the $K\to\pi\pi$ hadronic matrix elements.
Initial results suggest that calculation of the two complex decay amplitudes $A_0$ and $A_2$ describing the
decays $K\to(\pi\pi)_I$ for $I=0$ and 2 respectively are now realistic targets for large-scale lattice QCD
calculations.
This would allow a verification of the $\Delta I=1/2$ rule and a first-principles calculation of
$\epsilon'/\epsilon$ within the Standard Model.
Further, new physics in $\epsilon^\prime$ is tightly correlated with that in rare kaon decays; see,
e.g., Sec.~\ref{kaon:sect:BSM1} and Refs.~\cite{Buras:1999da,HaischPXPS}.
Thus the payoff of improved lattice-QCD calculations of $K\to\pi\pi$ decays with a precision comparable to
experiment will be significant.

The complex $I=2$, $K\to\pi\pi$ decay amplitude $A_2$ has now been computed in lattice QCD with 15\%
errors~\cite{Blum:2011ng,Blum:2012uk}.
Because the kaon mass is relatively small, the decay final states are dominated by two pions.
In such cases, QCD rescattering effects can be controlled using the method of Lellouch and
L{\"u}scher~\cite{Luscher:1986pf,Lellouch:2000pv}.
In the next two years, the addition of two smaller lattice spacings should reduce the dominant discretization
error, leading to a total error of $\sim 5\%$.
At this level, isospin violation must be included, which may be within reach on a five-year timescale.

The $I=0$ amplitude is considerably more challenging, and only trial calculations with unphysical kinematics
and $\sim 400$~MeV pions have been attempted to date~\cite{Blum:2011pu}.
The overlap between the $I=0$, $\pi\pi$ state and the vacuum results in quark-disconnected diagrams and a
noise to signal ratio that grows exponentially with time.
In addition, the simple quark-field boundary conditions used in the $I=2$ channel cannot give the correct
relative momentum to final-state pions with $I=0$.
A~promising solution is to impose $G$-parity boundary conditions on both the valence and sea quarks to
produce two-pion final states with $I=0$ and physical kinematics.
% Code has been written to generate $N_f = 2+1$ gauge-field ensembles with $G$-parity boundary
% conditions~\cite{Kelly:2012xx}, but a few more months of testing and additional coding effort are needed
% before the first full-QCD $G$-parity tests can be carried out.
The first results for $A_0$ from a single relatively coarse ensemble for an energy conserving decay with
physical pion and kaon masses are expected in 2014, and should reveal the method's ultimate effectiveness.
The systematic error associated with the nonzero lattice spacing, which was the dominant uncertainty for the
$I=2$ calculation ($\sim15\%$), will require simulations at a second lattice spacing and thus take longer to
control, but a 10\% error for $A_0$ appears possible by~2018.

In summary, a full calculation of $\epsilon^\prime$ with a total error at the 20\% level may be possible in 
two~years.
Given this precision, combining the pattern of experimental results for $K\to\pi\nu\bar\nu$ with
$\epsilon'/\epsilon$ can already help to distinguish between new-physics models, as discussed in
Sec.~\ref{kaon:sect:BSM1} and illustrated in Fig.~\ref{kaon:fig:1}; see also 
Refs.~\cite{Buras:1999da,HaischPXPS}.

\subsubsection{$K\to\pi\nu\bar\nu$ Decays}

The Standard-Model branching fractions for the rare kaon decays $K^+ \to \pi^+ \nu \bar{\nu}$ and $K_L \to
\pi^0 \nu \bar{\nu}$ are known to a precision unmatched by any other quark flavor-changing-neutral-current
process, so $K\to\pi\nu\bar\nu$ decays are promising channels for new-physics discovery.
The hadronic uncertainties are under good theoretical control because the form factors can be obtained using
experimental $K\to\pi\ell\nu$ data combined with chiral perturbation theory.
Further, long-distance contributions involving multiple operator insertions from the effective weak
Hamiltonian are subdominant due to quadratic GIM suppression.
The limiting source of uncertainty in the Standard-Model predictions for $\BR(K^+ \to \pi^+ \nu
\bar{\nu})$ and $\BR(K_L \to \pi^0 \nu \bar{\nu})$ is the parametric error from $|V_{cb}|^4$
and is approximately $\sim$10\%~\cite{Brod:2010hi,BrodPXPS}.
Therefore a reduction in the uncertainty on $|V_{cb}|$ is essential for interpreting the results of the
forthcoming measurements by NA62, KOTO, ORKA, and subsequent experiments at \PX\ as tests of the Standard
Model.
 
The CKM matrix element $|V_{cb}|$ can be obtained from exclusive $B \to D^{(*)} \ell\nu$ decays provided
lattice-QCD calculations of the hadronic form factors.
For the $B\to D^*\ell\nu$ form factor at zero recoil, the gap between experimental errors ($1.3\%$) and
lattice errors (presently $\sim 1.8\%$) has narrowed considerably over the last five
years~\cite{Bailey:2010gb}.
In the next five years, the lattice error is expected to drop below the experimental error, as shown in
Table~\ref{lqcd:tab:error}.
Particularly important for this will be the use of lattices with small lattice spacings and physical
light-quark masses, and the extension of the calculation to nonzero recoil~\cite{Qiu:2011ur}.
This projected improvement in the $B\to D^* \ell\nu$ form factor will reduce the error in $|V_{cb}|$ to
$\lesssim 1.5\%$, and thereby reduce the error on the Standard-Model $K\to\pi\nu\bar{\nu}$ branching
fractions to $\lesssim 6$\%.
With this precision, the theoretical uncertainties in the Standard-Model predictions will be commensurate
with the projected experimental errors in time for the first stage of \PX.

\subsubsection{Long-distance Amplitudes for Rare Kaon Decays}

Errors from long-distance contributions are subdominant in the Standard Model predictions for
$K\to\pi\nu\bar{\nu}$ due to quadratic GIM suppression, but are significant in other rare kaon decays such as
$K\to\pi\ell^+\ell^-$.
Currently the Standard-Model estimates for the $K \to \pi \ell^+ \ell^-$ branching fractions rely on chiral
perturbation theory and have large uncertainties that are not competitive with those on $K\to\pi\nu\bar{\nu}$.
If they can be brought under theoretical control, however, $K \to \pi \ell^+ \ell^-$ may afford
additional search channels that, through correlations with other observables, provide additional
handles to distinguish between new-physics scenarios.
See Sec.~\ref{kaon:sect:KLp0ll} and Ref.~\cite{Buras:1999da} for further details.
Thus the extension of lattice-QCD methods to compute long-distance weak amplitudes would have considerable
impact on the search for new physics.

The gold-plated kaon decays $K\to\ell\nu$ and $K\to\pi\ell\nu$, as well as the nonleptonic decay
$K\to\pi\pi$, are dominated by first-order weak processes in which a single $W^\pm$ is exchanged.
In the past, the only second-order quantities that were accessible to lattice~QCD were those which are
dominated by short distances, e.g., the \CP-violating parameter $\epsilon_K$ in 
$K^0$-$\overline{K}^0$ mixing.
These can be represented by matrix elements of local operators.
However, roughly 5\% of $\epsilon_K$~\cite{Buras:2010pza} and 30\% of the $K_L$-$K_S$ mass difference,
$\Delta M_K$,~\cite{Herrlich:1993yv,Brod:2010mj} come from long-distance contributions with two
flavor-changing interactions separated by distances of of order $\Lambda_{\rm QCD}^{-1}$.
Then both interactions, each represented by a four-fermion operator, must be explicitly included in a lattice
calculation, a challenge which may now be possible to meet with near-future computing resources.
Again, the effects of real intermediate states (rescattering effects) introduce finite-volume distortions.
It has recently been demonstrated, however, that, in the case of kaons, these distortions can be corrected in
a nonperturbative manner~\cite{Christ:2010zz,Christ:2012np}.

A pilot numerical study of $\Delta M_K$ using these methods is now
underway~\cite{Yu:2011gk,Yu:2012xx,Christ:2012se}.
The calculation is more challenging than those for the $K\to\pi\pi$ amplitudes, with a key issue being the
need to include charm quarks so as to enforce GIM cancellations.
First results from a single lattice spacing with unphysically heavy pions are due soon~\cite{Yu:2012xx}, and
a calculation at the physical light-quark masses may be finished in another year.
Because this calculation is still at an early stage in development, it is difficult to forecast the level of
resources that will be required to obtain an accurate, controlled result.
Pursuing this calculation will, however, be a major priority of the US lattice-QCD kaon physics program.

The lattice-QCD calculation of $\Delta M_K$ will pave the way for computations of the long-distance
contributions to neutral kaon mixing and rare kaon decays.
The difficulties here are similar to those for $\Delta M_K$, including the need for dynamical charm.
A method for calculating long-distance contributions to rare kaon decays such as
$K^+\to\pi^+\nu\overline{\nu}$, $K_L\to\pi^0\nu\overline{\nu}$ and $K\to\pi \ell^+\ell^-$ in lattice QCD has
been proposed in Ref.~\cite{Isidori:2005tv}.
These calculations are a higher priority for lattice QCD than the long-distance contribution to $\epsilon_K$,
in light of the ongoing NA62 experiment, the planned KOTO experiment, and the proposed high-sensitivity kaon
measurements at Fermilab.
Because lattice-QCD calculations of long-distance contributions to rare decays have not yet begun, it is
premature to forecast time scales for completion or uncertainties obtained.

%%%%%%%%%%%%%%%%%%%%%%%%%%%%%%%%%%%%
\subsection{Muon Experiments}
\label{lqcd:subsec:muon}
%%%%%%%%%%%%%%%%%%%%%%%%%%%%%%%%%%%%

Here we describe opportunities for lattice QCD to assist the \PX\ muon experimental program described in
Chapter~\ref{chapt:muon}.
The intense \PX\ beam with flexible time structure makes possible a range of muon experiments from searches
for charged-lepton flavor violation to a measurement of the muon electric dipole moment.

Stage~1 of \PX\ will enable the Mu2e experiment to improve the reach for $\mu \to e$ conversion on nuclei by
10--100 orders-of-magnitude.
The higher wattage of \PX\ Stage~2 will further improve the sensitivity of Mu2e, with an expected reach four
orders-of-magnitude better than current limits.
Stage~2 will also make possible other searches for charged-lepton flavor violation such as $\mu \to 3e$.
If $\mu \to e$ conversion is indeed discovered at \PX\, lattice-QCD calculations of the light- and
strange-quark contents of nucleon will be needed to make model predictions for the $\mu \to e$ conversion
rate and distinguish between possible new-physics theories.

The new Muon $g-2$ Experiment (E989) to improve the determination of the muon anomalous magnetic moment will
run at Fermilab before the \PX\ accelerator upgrade.
Although E989 is not part of \PX, a second-generation $g-2$ experiment would be possible with \PX\ if it
seemed warranted based on improvements in the theoretical calculation and the evolution of the discrepancy
with respect to the Standard Model; see Sec.~\ref{muon:sec:gm2}.
Lattice~QCD provides the only means to calculate the Standard-Model hadronic light-by-light contribution to
$g-2$ from first principles with controlled uncertainties that are systematically improvable.

\subsubsection{$\mu$-to-$e$ Conversion}
\label{lqcd:subsec:Mu2e}

Charged-lepton flavor violation (CFLV) is so highly suppressed in the Standard Model that any observation of
CLFV would be unambiguous evidence of new physics.
Many new-physics models, however, allow for CLFV and predict rates close to current limits; see
Sec.~\ref{mu:sec:theory} for examples.

Many experiments searching for charged-lepton flavor violation are running or are on the horizon.
The MEG experiment at PSI is currently searching for $\mu \to e \gamma$, and an improved search for $\mu \to
eee$ at PSI (the Mu3e Experiment) has also been proposed.
The Mu2e Experiment with \PX\ aims to search for $\mu N \to eN$ with a sensitivity four orders of magnitude
below the current best limit.
If CLFV is observed in these experiments, combining the measured rates of $\mu \to e \gamma$ and $\mu \to e$
conversion on different target nuclei can distinguish between models and reveal information on underlying
theory~\cite{Cirigliano:2009bz}.
Model predictions for the $\mu \to e$ conversion rate off a target nucleus depend upon the light- and
strange-quark contents of the nucleon; see Sec.~\ref{mu:sec:theory}.
These same quark scalar density matrix elements also needed to
interpret dark-matter detection experiments in which the dark-matter particle scatters off a
nucleus~\cite{Bottino:1999ei,Ellis:2008hf,Hill:2011be}.
Lattice-QCD can provide nonperturbative calculations of the scalar quark content of the nucleon with
controlled uncertainties.

Most lattice efforts on this front have focused on the determination of the strange-quark content of the
nucleon.
This is because the strange quark is least amenable to other perturbative approaches: it is too light for the
use of perturbative QCD, but too heavy for the reliable use of $SU(3)$ baryon chiral perturbation theory.
Calculations of $m_s \langle N| \bar{s}s|N \rangle$ have been performed with $N_f = 2+1$ and even $N_f =
2+1+1$ flavors using a variety of lattice-fermion
actions~\cite{Young:2009zb,Toussaint:2009pz,Durr:2011mp,Horsley:2011wr,Dinter:2012tt,Oksuzian:2012rzb,Engelhardt:2012gd,Freeman:2012ry,Shanahan:2012wh,Junnarkar:2013ac}.
Most groups compute the desired matrix element from direct simulation, but some exploit the Feynman-Hellmann
relation
\begin{equation}
	m_s \langle N| \bar{s}s|N \rangle = m_s\frac{\partial m_N}{\partial m_s} \,.
\end{equation}
The results obtained with different methods and lattice formulations agree at the 1--2$\sigma$ level, and a
recent compilation quotes an error on the average $m_s \langle N| \bar{s}s|N \rangle$ of about
25\%~\cite{Junnarkar:2013ac}.
With this precision, the current lattice results already rule out the much larger values of 
$m_s\langle N|\bar{s}s|N\rangle$ favored by early nonlattice
estimates~\cite{Nelson:1987dg,Kaplan:1988ku,Jaffe:1989mj}.
Lattice-QCD can also provide first-principles calculations of the pion-nucleon sigma
term~\cite{Young:2009zb,Durr:2011mp,Horsley:2011wr,Dinter:2012tt,Shanahan:2012wh} and the charm-quark content
of the nucleon~\cite{Freeman:2012ry,Gong:2013vja}.
A realistic goal for the next five years is to pin down the values of all of the quark scalar density matrix
elements for $q=u,d,s,c$ with $\sim$ 10--20\% uncertainties; even greater precision can be expected on the
timescale of a continuation of Mu2e at Stage~2 of \PX.

\subsubsection{Muon Anomalous Magnetic <oment}
\label{lqcd:subsec:gm2}

The muon anomalous magnetic moment provides one of the most precise tests of the SM and places important
constraints on extensions of it~\cite{Hewett:2012ns}.
The current discrepancy between experiment and the Standard Model has been reported in the range of 2.9--3.6
standard deviations~\cite{Aoyama:2012wk,Davier:2010nc,Hagiwara:2011af}.
With new experiments planned at Fermilab (E989) and J-PARC (E34) that aim to improve on the current 0.54 ppm
measurement at BNL~\cite{Bennett:2006fi} by at least a factor of four, it will continue to play a central
role in particle physics for the foreseeable future.

\begin{figure}
    \centering
    \includegraphics[width=0.35\linewidth]{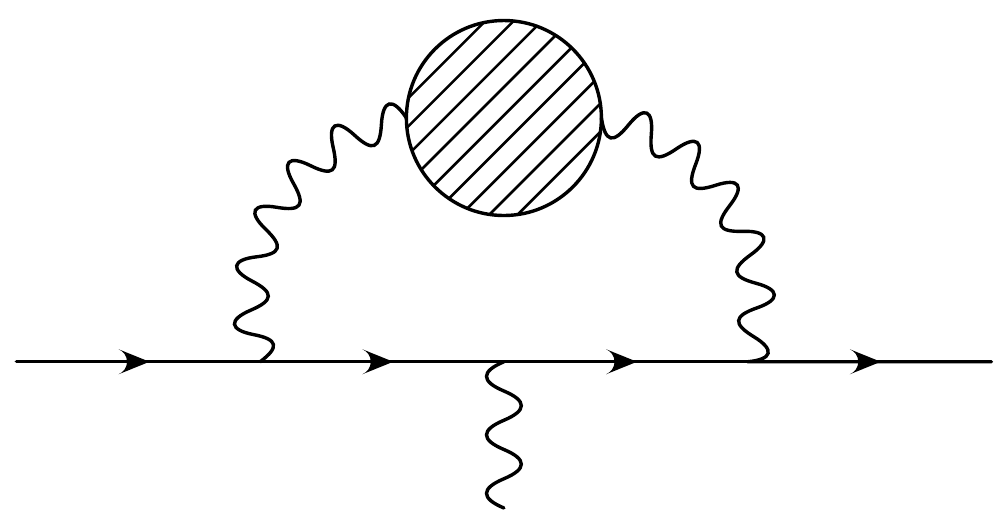} \hskip 1cm
    \includegraphics[width=0.35\linewidth]{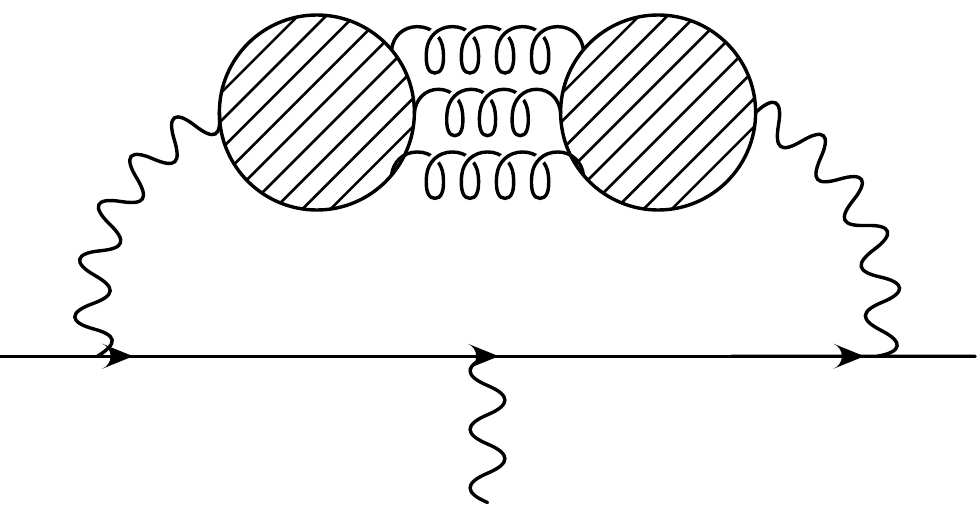} 
    \caption[Hadronic vacuum polarization diagrams for muon $g-2$]{Hadronic vacuum polarization diagrams 
        contributing to the Standard-Model muon anomaly.
        The horizontal lines represent the muon.
        The blob formed by the quark-antiquark loop represents all possible hadronic intermediate states.
        Right panel: disconnected quark line contribution in which the quark loops are connected by gluons.} 
    \label{lqcd:fig:hvp}
\end{figure}

\begin{figure}
    \centering
    \includegraphics[width=0.35\linewidth]{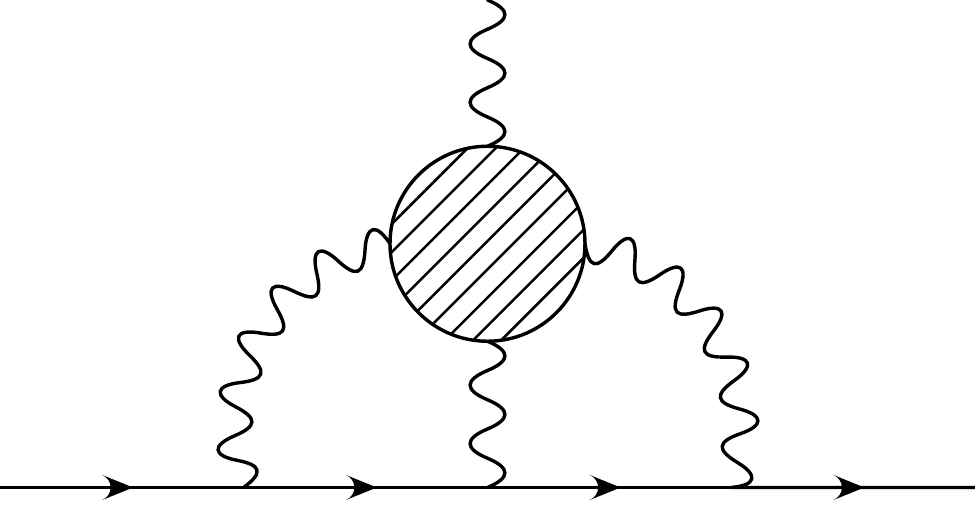} \hskip 1cm
    \includegraphics[width=0.35\linewidth]{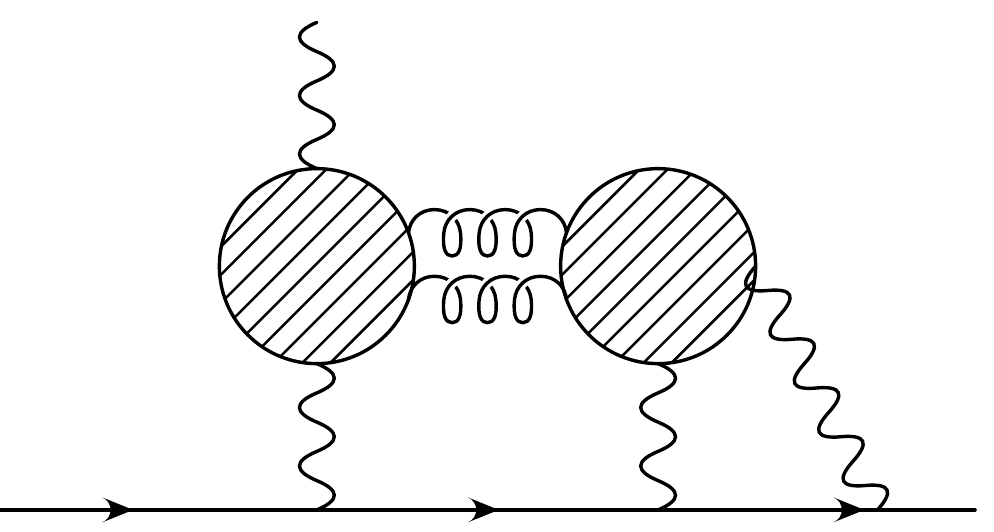} 
    \caption[Hadronic light-by-light diagrams for muon $g-2$]{Hadronic light-by-light scattering diagrams
        contributing to the Standard-Model muon anomaly.
        The horizontal lines represent the muon.
        The blob formed by the quark loop represents all possible hadronic intermediate states.
        Right panel: one of the disconnected quark line contributions in which the quark loops are connected 
        by gluons.}
    \label{lqcd:fig:hlbl}
\end{figure}

In order to leverage the improved precision on $g-2$ from the experiments, the theoretical uncertainty on the
Standard Model prediction must be shored-up, as well as be brought to a comparable level of
precision~\cite{Hewett:2012ns}.
The largest sources of uncertainty in the SM calculation are from the nonperturbative hadronic contributions.
The hadronic corrections enter at order $\alpha^{2}$ through the hadronic vacuum polarization (0.36 ppm),
shown in Fig.~\ref{lqcd:fig:hvp}, and $\alpha^{3}$ through hadronic light-by-light scattering (0.22 ppm), shown in
Fig.~\ref{lqcd:fig:hlbl}, as well as higher order hadronic vacuum polarization contributions.
Lattice QCD can provide calculations of the hadronic vacuum polarization (HVP) and hadronic light-by-light
(HLbL) contributions to the muon $(g-2)$ from QCD first principles with reliable uncertainties and,
ultimately, greater precision than currently available from nonlattice methods.

\paragraph{Hadronic vacuum polarization}
\label{lqcd:subsec:HVP}

The HVP contribution to the muon anomaly, $a_{\mu}(\rm HVP)$, has been obtained to a precision of 0.6\% using
experimental measurements of $e^{+}e^{-}\to\rm hadrons$ and $\tau\to\rm
hadrons$~\cite{Davier:2010nc,Hagiwara:2011af}.
The result including $\tau$ data is about two standard deviations larger than the pure $e^+e^-$
determination, and reduces the discrepancy with the Standard Model to below three standard
deviations~\cite{Davier:2010nc}.
The former requires isospin corrections which may not be under control.
Alternatively, $\rho$-$\gamma$ mixing may explain the difference and bring the $\tau$-based result in line
with that from $e^+e^-$~\cite{Jegerlehner:2011ti}.
A direct lattice-QCD calculation of the hadronic vacuum polarization with $\sim 1\%$ precision may help shed
light on the apparent discrepancy between $e^{+}e^{-}$ and $\tau$ data.
Ultimately, a lattice-QCD calculation of $a_{\mu}(\rm HVP)$ with sub-percent precision can circumvent these
concerns by supplanting the determination from experiment with one from first-principles QCD.

The hadronic vacuum polarization contribution is obtained by computing the two-point correlation function of the electromagnetic quark current, Fourier-transformed to momentum space, and then inserting the result into the one-loop QED integral for the interaction of the muon with an external photon field.   Lattice-QCD simulations enable a direct, nonperturbative computation of the renormalized vacuum polarization function $\Pi(Q^2)$ as a function of the Euclidean momentum-squared $Q^2$~\cite{Blum:2002ii}.  
 
The HVP contribution to the muon anomalous magnetic moment has been computed in lattice QCD by several groups~\cite{Blum:2002ii,Gockeler:2003cw,Aubin:2006xv,Feng:2011zk,Boyle:2011hu,DellaMorte:2011aa}, and statistical errors on lattice calculations of $a_{\mu}(\rm HVP)$ are currently
at about the 3--5\% level.  Important systematic errors remain, and these are being addressed through a combination of theoretical advances and brute-force computing.  Because simulating QCD on a computer requires a finite-sized lattice, lattice-QCD simulations can only access discrete momentum values in units of $2\pi/L$, where $L$ is the length of a side of the box.  As a consequence, lattice-QCD data are sparse and noisy in the low-$Q^2$ region.  The integral over $Q^2$ is dominated by momenta of order $m_\mu$, which is below the range directly accessible in current lattice simulations; thus the value of $a_{\mu}(\rm HVP)$ is sensitive to the functional form used to extrapolate $Q^2 \to 0$.  A new fitting approach based on Pad\'e approximants~\cite{Aubin:2012me} will eliminate this model dependence.  Further, smaller values of $Q^2$ can also be simulated directly using ``twisted" boundary conditions for the lattice fermions~\cite{Sachrajda:2004mi} and increasing the lattice box size, both of which are being pursued~\cite{DellaMorte:2011aa}.  Another significant source of uncertainty in $a_{\mu}(\rm HVP)$ is from the chiral extrapolation of the numerical simulation data to the physical light-quark masses.  Anticipated increases in computing resources will enable simulations directly at the physical quark masses, thereby eliminating this systematic.  The charm-quark contribution to HVP may be at the few-percent level (comparable to the hadronic light-by-light contribution), so calculations are underway using $N_f = 2+1+1$ gauge-field ensembles with dynamical charm quarks~\cite{Feng:2012gh}. A new method to extend the low momentum region to smaller values of momentum transfer, much like twisted boundary conditions, uses analytic continuation to access small time-like momenta~\cite{Feng:2013xsa}. It has the advantage that the energy, and therefore momentum transfer, can be varied continuously but at the expense of either introducing model-dependence into the calculation (to extend time to $\pm\infty$), or by truncating the time integral (sum), introducing an additional finite-size effect. While the authors do not expect the new method to increase the precision of HVP and similar calculations, it does avoid the difficulty of fitting the lattice data versus momentum transfer and its attendant problems (for low momenta), and provides an independent cross-check of the standard method to compute $a_{\mu}(\rm HVP)$ with different systematic uncertainties.

Given the above combination of theoretical improvements, plus increased computing resources, large error
reductions in lattice-QCD calculations of $a_{\mu}(\rm HVP)$ over the next one to two years are not only
possible, but likely.
The dominant quark-connected contribution $a_{\mu}(\rm HVP)$, shown on the left side of
Fig.~\ref{lqcd:fig:hvp}, will be calculated with few-percent errors on the timescale of the Muon $g-2$
experiment (E989).
This will provide a valuable cross-check of the semi-experimental determination from $e^+e^- \to {\rm
hadrons}$.
The quark-disconnected contribution, shown on the right side of Fig.~\ref{lqcd:fig:hvp}, is computationally
more demanding, but will be computed within the next five years.
Because the disconnected contribution is expected to contribute at the $\sim$1\% level, a rather large
uncertainty in this term can be tolerated.

\paragraph{Hadronic light-by-light}
\label{lqcd:subsec:HLbL}

The HLbL contribution to the muon anomaly cannot be extracted from experiment, as for the HVP contributions.
Thus present estimates of this contribution rely on models~\cite{Prades:2009tw,Nyffeler:2009tw}, and report
errors estimated to be 25--40\% range.
This uncertainty is neither fully controlled nor systematically improvable.
If not reduced, these errors will dominate over the HVP error as the latter is reduced via more experimental
data and lattice-QCD calculations.
Thus, there is a crucial need for an \emph{ab initio} calculation, and the HLbL contribution is the highest
theoretical priority for $(g-2)_\mu$.

Lattice-QCD can provide a calculation of $a_\mu({\rm HLbL})$ from QCD first principles with controlled uncertainties that are systematically improvable.  The importance of this calculation to the experimental program is well-known to the lattice-QCD community, and significant progress has been made on this topic.  The lattice-QCD calculation of the HLbL contribution is challenging, however, and still in early stages.  

The most promising strategy to calculate $a_\mu({\rm HLbL})$ is via lattice QCD plus lattice QED.
Then the muon and photons are treated nonperturbatively along with the quarks and
gluons~\cite{Hayakawa:2005eq}.
First results using this approach for the single quark-loop part of the HLbL contribution
(Fig.~\ref{lqcd:fig:hlbl}, left panel) have been reported recently~\cite{Blum:2013qu}.
The method can be checked in pure QED, where the LbL term has been calculated directly in perturbation
theory, allowing a benchmark for the procedure.
This test has been performed successfully~\cite{Chowdhury:2009}, showing the significant promise of the
method.
Much effort is still needed to reduce statistical errors, extrapolate to zero momentum transfer, and many
systematic errors (e.g., due to the finite lattice volume, unphysical heavy pions, and nonzero lattice
spacing) remain uncontrolled.
However, first signs of the HLbL contribution rising above the Monte-Carlo noise are encouraging.
The calculation is quenched with respect to QED, i.e., the sea quarks' electric charge is neglected,
so contributions from quark-disconnected diagrams with two separate quark loops connected only by a pair of
gluons are missing.
The disconnected contributions may be similar in size to the connected ones, so inclusion of these
contributions will be essential for a complete calculation with controlled errors.
This obstacle can be overcome simply by including of photons in the gauge field ensemble generation, by
reweighting the quenched ensembles to include the virtual sea quark contributions, or by brute-force
calculation of the disconnected diagrams.
All these approaches are under investigation.

Calculations are also being carried out that
will check both model and lattice-QCD calculations: for example, the
$\pi^{0}\to\gamma^{(*)}\gamma^{(*)}$ vertex
function~\cite{Feng:2012ck}, the axial-vector--vector--vector
three-point function~\cite{Melnikov:2003xd}, and the chiral magnetic
susceptibility~\cite{Ioffe:2010zz}.  The first of
these is also directly related to experimental measurements of the
Primakoff effect, $\gamma A\to\gamma\gamma$, which is dominated (like
HLbL) by the pion pole.  The four-point vector
correlation function in QCD needed for the HLbL amplitude, computed at
select fiducial values of the momenta at each vertex, is also under study.  

In order to bring the error on the HLbL contribution to, at, or below, the projected experimental uncertainty
on the time scale of the Muon $g-2$ experiment, one must reduce the error on $a_\mu({\rm HLbL})$ to
approximately 10\% or better by 2016--17.
Assuming this accuracy, a reduction of the HVP error by a factor of 2, and the expected reduction in
experimental errors, then the present central value would lie 7--8$\sigma$ from the SM prediction.
Reaching this milestone is certainly possible with sustained theoretical and computational effort from the
lattice-QCD community and continued advances in computer power, but is not guaranteed.
In particular, new theoretical developments, which are impossible to predict, may be needed to match the
target experimental precision.

%%%%%%%%%%%%%%%%%%%%%%%%%%%%%%%%%%%%
\subsection{Nucleon Matrix Elements and Fundamental Physics}
\label{lqcd:subsec:nucleons}
%%%%%%%%%%%%%%%%%%%%%%%%%%%%%%%%%%%%

Here we describe opportunities for lattice-QCD to assist the \PX\ physics program to study fundamental
physics with nucleons, as described in Chapters~\ref{chapt:edm} and~\ref{chapt:nn}.
The 1~GeV beam of \PX\ Stage~1 can power a spallation target facility optimized for particle physics, 
enabling experiments for ultracold neutrons and electric dipole moments (EDMs).
The interpretation of many of these experimental measurements as constraints on TeV-scale or GUT-scale new
physics requires knowledge of nucleon matrix elements that can be computed in lattice QCD.
The LBNE detector can be used to search for proton decay and will improve upon current limits,
once the far detector is sufficiently large and shielded from cosmic rays.
The proton-decay matrix elements needed to interpret experimental limits on the proton lifetime as
constraints on new-physics models are similar to those needed to interpret neutron $\beta$-decay experiments,
so we briefly mention them in this section; more details can be found in
Sec.~\ref{lqcd:subsubsec:protondecay}.

Although many nucleon matrix elements are gold plated, lattice-QCD calculations involving baryons are
generally more challenging than for mesons.
They are more computationally demanding because statistical noise in baryon correlation functions grows
rapidly with Euclidean time.
Further, the extrapolation to physical light-quark masses is more difficult because baryon chiral
perturbation theory converges less rapidly.

The most studied nucleon matrix element is that of the axial charge $g_A$.
Because it can be measured precisely in neutron $\beta$-decay experiments, $g_A$ provides a benchmark for of
the accuracy of lattice-QCD nucleon matrix element calculations.
Past lattice calculations of $g_A$ have quoted $\sim$ 6--10\%
uncertainties~\cite{Lin:2007ap,Lin:2008uz,Yamazaki:2008py,Bratt:2010jn}, but the central values have all been
systematically lower than the experimental measurement by about 10\%, indicating the presence of
underestimated uncertainties.
Two recent $N_f=2$ lattice-QCD calculations of $g_A$ have improved upon these calculations with a more
careful treatment of excited-state contamination in the three-point correlation
functions~\cite{Capitani:2012gj} and simulations at the physical pion mass~\cite{Horsley:2013ayv}, and obtain
results consistent with experiment.
These results, however, have yet to be confirmed with $N_f = 2+1$ flavors.
The expected increase in computing power over the next five years should allow simulations with larger
volumes and and more widespread use of physical light-quark masses, while new algorithms should greatly
reduce the statistical errors.
Percent-level lattice-QCD calculations of $g_A$ are therefore expected on this timescale.

Lattice-QCD calculations of proton and neutron electric dipole moments, proton and neutron decay matrix
elements, and $n$-$\bar{n}$ oscillation matrix elements are in earlier stages.
Percent-level precision is not needed, however, for these quantities to be of use to \PX.
Typically $\sim$~10 or 20\% accuracy is sufficient, which is an achievable target in the next five years.

\subsubsection{Proton and Neutron Electric Dipole Moments}
\label{lqcd:subsec:EDM}

Flavor physics experiments---aided, in part, by lattice-QCD calculations---have demonstrated that
Standard-Model \CP\ violation is not large enough to explain the baryon asymmetry of the universe.
Consequently, there must be as yet undiscovered \CP\ violating interactions beyond the Standard Model.
These could still show up in quark flavor-changing processes, but also elsewhere, such as in nonzero electric
dipole moments (EDMs) of leptons and nucleons \cite{Pospelov:2005pr}.

There are two possible sources of an electric dipole moment in the Standard Model.
Cabibbo-Kobayashi-Maskawa \CP\ violation makes a contribution to the nucleon EDM at the three-loop level and
lies well beyond experimental sensitivity.
The strong \CP-violating interaction, $\bar\theta G\tilde{G}$, directly makes a contribution.
Experimental limits on the size of the neutron EDM ($d_N$) constrain the size of $|\bar{\theta}| \lesssim
10^{-10}$, but this constraint is not known precisely because of uncertainties in model estimates for
$d_N/\bar{\theta}$.
Further, non-SM sources of \CP\ violation generate higher-dimension, EDM-inducing operators at low scales.
In some cases the BSM model predictions require nonperturbative hadronic matrix elements.
Interestingly, the strong-\CP\ contribution appears to flip sign between neutron and proton, while the BSM
contributions need not flip sign.

Lattice-QCD can provide first-principles QCD calculations of the strong-\CP\ contribution to
$d_N/\bar{\theta}$ with improved precision and controlled uncertainties, as well of matrix elements of non-SM
EDM-inducing operators.
Pilot lattice-QCD calculations have already been carried out for this strong-\CP contribution to the neutron
and proton EDMs using two methods: (i) calculating the energy difference between two spin states of the
nucleon in an external electric field~\cite{Shintani:2008nt}, and (ii) computing the form factor of the
electromagnetic current~\cite{Shintani:2005xg,Aoki:2008gv}.
Currently the statistical errors are still $\sim$30\%, both because of the general property that nucleon
correlation functions have large statistical errors and because the calculation involves correlations with
the topological charge density, which introduces substantial statistical fluctuations.
A lattice-QCD calculation of the matrix elements of dimension-6 operators needed for BSM theories is also
underway~\cite{Bhattacharya:2012bf}.
This research is still in an early phase, and a reasonable and useful goal for the coming five years is a
suite of matrix elements with solid errors at the 10--20\% level.

\subsubsection{Proton and Neutron Decays}

Experimental measurements of neutron $\beta$-decay can place constraints on TeV-scale new-physics models, in
particular those with scalar or tensor interactions, provided values for the nucleon scalar and tensor
charges $g_S$ and $g_T$.
The next generation of neutron $\beta$-decay experiments is expected to increase their sensitivity to scalar
and tensor interactions by an order of magnitude.
Model estimates of $g_S$ and $g_T$ disagree and provide only loose bounds, but lattice-QCD can provide
precise results for these quantities.

The calculation of neutron decay matrix elements is part of the lattice-QCD program to study nucleon
structure: see, e.g., Ref.~\cite{Green:2012ej}.
A realistic goal for lattice-QCD in the next five years is to pin the values of $g_S$ and $g_T$ down to
10--20\%.
Given this level of accuracy, experimental neutron $\beta$-decay measurements are more sensitive to scalar
and tensor contact terms than a 25~fb$^{-1}$ run at the 8~TeV LHC \cite{Lin:2011ab}.
Further, studies have shown this precision will be sufficient to exploit the anticipated experimental
sensitivity of the proposed UCN experiment at LANL~\cite{Bhattacharya:2011qm}.

The proton-decay matrix elements $\langle\pi, K, \eta, \ldots | {\mathcal{O}}_{\Delta B = 1} | p \rangle$
needed to interpret experimental limits on proton decay as constraints on GUT model parameters are similar to
the neutron-decay matrix elements discussed above.
Only a single small-scale lattice-QCD effort by the RBC and UKQCD Collaborations has been devoted to
calculating proton-decay matrix elements so far.
Recently they obtained the first direct calculation of these matrix elements with $N_f=2+1$ dynamical quarks
with uncertainties of $\sim$20--40\%.
They will include finer lattice spacings and lighter pion masses in a future work.
With these improvements, it should be straightforward to reduce the errors in proton-decay matrix elements to
the $\sim 10\%$ level in the next five years.

\subsubsection{Neutron-antineutron Oscillations}
\label{lqcd:subsec:nnbar}

A low-energy process that could provide distinct evidence for baryon number violation from BSM physics is the
transition of neutrons to antineutrons, which violates baryon number by two units~\cite{Mohapatra:1980qe}.
This process can be observed through the annihilation of the resulting antineutron.
Experimentally, this can be searched for with large scale proton decay detectors such as
Super-K~\cite{Super-K}, and also with experiments with nearly-free neutrons where flux and time of flight is
optimized~\cite{BaldoCeolin:1994jz}.
In particular, a neutron-antineutron oscillation experiment at \PX\ could improve the limit on the
$n$-$\bar{n}$ transition rate by a factor of $\sim 1000$.

For many grand unified theories (GUTs) with Majorana neutrinos and early universe sphaleron processes, the
prediction for the oscillation period is between $10^9$ and $10^{11}$ seconds~\cite{Nussinov:2001rb,%
Babu:2008rq,Mohapatra:2009wp,Winslow:2010wf,Babu:2012vc}.
However, this estimate is based on naive dimensional analysis, and could prove to be quite inaccurate when
the nonperturbative QCD effects are properly accounted for.
Calculations of these matrix elements with reliable errors anywhere below 50\% would provide valuable
guidance for new-physics model predictions.

Lattice-QCD calculations can provide both the matrix elements of the six-fermion operators governing this
process and calculate the QCD running of these operators to the scale of nuclear physics.
There are four independent operators, differing in their color and spin
structure~\cite{Rao:1982gt,Caswell:1982qs}.
Despite the fact that the operators involve more quark fields, the calculations are in many ways simpler than
those for the matrix elements discussed above, e.g., $\langle N|\bar s s|N\rangle$.
In particular, there are no quark-disconnected diagrams or spectator quarks.
Thus, we can ultimately expect very accurate results.

Initial work on these matrix elements is currently underway~\cite{Buchoff:2012bm}.
The main challenge at this stage is to make sufficient lattice measurements to obtain a statistically
significant signal.
A first result is expected in the next 1--2 years, with anticipated errors of $\sim 25\%$; results with
errors of $\sim 10\%$ or smaller should be achievable over the next five years.

%%%%%%%%%%%%%%%%%%%%%%%%%%%%%%%%%%%%
\subsection{Hadronic Physics}
\label{lqcd:subsec:hadron}
%%%%%%%%%%%%%%%%%%%%%%%%%%%%%%%%%%%%

Here we describe opportunities for lattice QCD to assist the \PX\ hadronic physics program described in
Chapters~\ref{chapt:hadron-dy} and~\ref{chapt:hadron-s}.
\PX\ will provide intense proton, pion, and kaon beams that enable measurements of the hadron spectrum and of
the proton structure; these will help address outstanding questions in QCD.
The spectroscopy experiment outlined in Chapter~\ref{chapt:hadron-s} will use a high-statistics kaon beam 
incident on a liquid hydrogen target to map out the hybrid meson spectrum and fill in the light-meson 
spectrum.
A comparison of the measured hadron spectrum with first-principles lattice-QCD calculations provides a
crucial test of our understanding of nonpertubative QCD dynamics.
The \PX\ hadron structure program will perform Drell-Yan measurements with a polarized proton beam to study
the the role of quark orbital angular momentum (OAM) in the fundamental structure of the proton.
The initial goal is to make the first spin-dependent Drell-Yan measurement and compare the measured Sivers
function for valence up quarks to that obtained from semi-inclusive deep inelastic scattering.
The subsequent goal is make a direct measurement of the Sivers distribution for antiquarks, which cannot be
accessed via semi-inclusive DIS.
Lattice QCD can provide first-principles calculations of nucleon structure quantities such as generalized
parton distribution functions and transverse momentum distribution functions.
Comparison of these theoretical predictions with experiment is needed to establish a complete and consistent
understanding of the proton (and neutron) spin, and ultimately resolve the proton-spin puzzle.

\subsubsection{Hadron Spectroscopy}
\label{lqcd:subsec:HadSpec}

The confrontation of experimental data on the spectrum with
high-precision calculations in lattice QCD is a vital test of our
understanding of QCD in the strong-coupling regime. Whilst the precise
calculation of the lowest-lying states represents an important
milestone in our ability to solve QCD, the calculation of the
excited-state spectrum at sufficient precision to delineate the states
provides an unrivaled opportunity to explore in detail the dynamics
of the theory, and to identify the collective degrees of freedom that
describe it.

The lattice calculations shown in
Figure~9.1\cite{Dudek:2010wm,Dudek:2011tt} provide a powerful
indication of the presence of mesons with exotic quantum numbers in
the energy regime accessible to the emerging generation of
experiments, and indeed the existence of ``hybrids,'' states in which
the gluons assume a structural role, with both exotic and non-exotic
quantum numbers.  The calculation of the spectrum of isoscalar mesons
reveals exotic states in the neighborhood of their isovector cousins,
and enables the flavor content to be
determined\cite{Christ:2010dd,Dudek:2011tt}.

\begin{figure}
    \centering
    \includegraphics[width=0.7\textwidth]{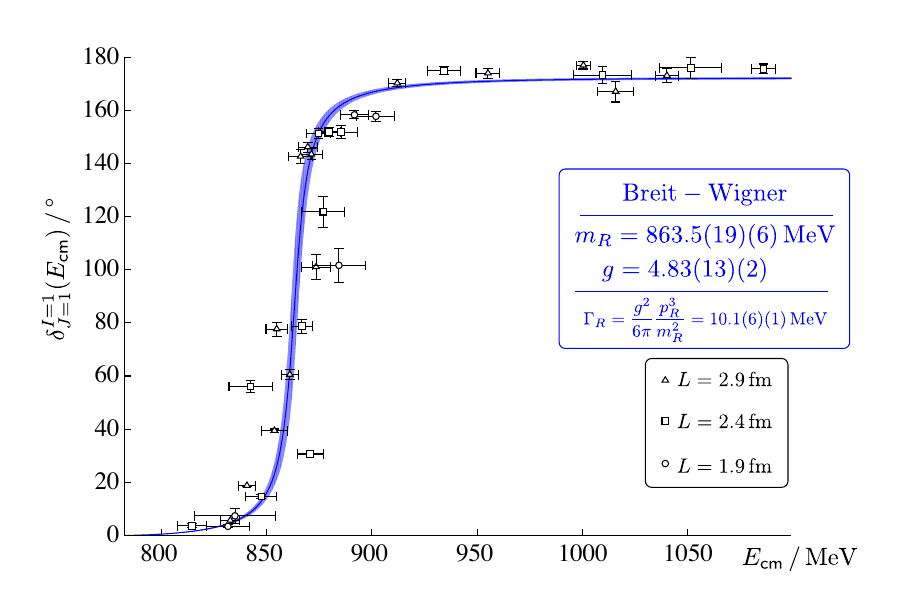}
    \caption[Elastic scattering phase-shift from lattice QCD for $\pi\pi$, $I=1$, $P$-wave scattering]{The 
        elastic scattering phase-shift from lattice QCD for $\pi\pi$, $I=1$, $P$-wave scattering as a 
        function of center-of-mass frame scattering energy and a description by a single Breit-Wigner 
        resonance.
        In the legend, the first error shown on the $\rho$ mass and width is statistical, while the second 
        error is due to the uncertainties in the lattice pion mass and anisotropy parameter.
        The small width of the $\rho$ stems from the lack of phase-space for decay into two 392~MeV pions. 
        From Ref.~\cite{Dudek:2012xn}}
    \label{lqcd:fig:pipiI1}
\end{figure}

These calculations are incomplete.  Most notably, the spectrum is
characterized by states that are resonances unstable under the strong
interactions, and are thereby encapsulated within momentum-dependent
phase shifts which may then be parametrized in terms of a mass and
decay width. In lattice calculations, shifts in the energy spectrum at
finite volume can be related to infinite-volume phase
shifts\cite{Luscher:1986pf,Luscher:1990ux}.  Recently, the energy
dependence of the $\rho$ resonance in $\pi \pi$ elastic scattering has
been mapped in unprecedented detail using this
method\cite{Feng:2010es,Lang:2011mn,Aoki:2011yj,Dudek:2012xn}, and the
mass and width extracted to high precision albeit at unphysically
large quark masses, as illustrated in Figure~\ref{lqcd:fig:pipiI1}.

The calculations cited above lay much of the theoretical and
computational groundwork for the future program of lattice
spectroscopy, and the next few years present an exciting opportunity
for lattice QCD even to \textit{predict} the underlying features of the
spectrum in advance of experiment.  With the first experiments at the
12 GeV upgrade of Jefferson Laboratory anticipated in 2015, an
on-going program at COMPASS at CERN, and the potential \PX\ experiment outlined in 
Chapter~\ref{chapt:hadron-s},
a vibrant program of computational spectroscopy is a key component of
the worldwide lattice effort.

Recent advances in calculating the excited-state spectrum of QCD in
the U.S. have exploited anisotropic lattices with the Wilson-clover fermion action,
which have a fine temporal lattice spacing enabling the resolution of
many levels in the spectrum, but a coarser spatial lattice spacing to
alleviate the cost on currently available computers. This has enabled
the delineation of many energy eigenstates at the sub-percent level
needed to resolve the spectrum, and to identify their continuum
quantum numbers; these calculations have been at pion masses of around
400 MeV and above.  The availability over the next five years of the
emerging generation of leadership-class capability computing, and
dedicated capacity computing, will enable calculations to be performed
at the physical light- and strange-quark masses, at sufficiently fine
lattice spacings to render the use of an anisotropic lattice
redundant, and with sufficient precision to delineate the energy
spectrum. Paramount to the success of such calculations will be a
theoretical effort at developing methods of treating coupled-channel
effects and multi-hadron states that appear above the inelastic
threshold\cite{Liu:2005,Doring:2011,Aoki:2011,Briceno:2012yi,Hansen:2012tf,Guo:2012hv}.

An integral program of first-principles lattice calculations of the
spectrum, a meson spectroscopy effort with different beams, and a
worldwide amplitude-analysis effort will provide an unrivaled
opportunity to understand the low-energy degrees of freedom that
govern the spectrum, and to inform the expected decay channels so as
to guide experiment.

\subsubsection{Nucleon Structure}
\label{lqcd:subsec:NucStruct}

As discussed in Chapter~\ref{chapt:hadron-dy}, the quark orbital angular momentum is accessed in the Ji
decomposition of the proton spin Eq.~(\ref{hadron-dy:eq:Ji}) as moments of generalized parton distribution
functions, which are being aggressively studied by various
groups~\cite{Bratt:2010jn,Sternbeck:2012rw,Liu:2012nz,Alexandrou:2013joa}.
In particular, the quark OAM contribution has been computed by the LHP collaboration in 2+1 flavor
QCD~\cite{Bratt:2010jn}.
In this calculation, the lightest pion mass was about 300 MeV and the largest box was about 3.5 fm on a side.
It was found that the OAM of the up and down quarks is opposite in sign to model expectations, and it is
thought that this may be due to the omission of quark-disconnected contributions, which are computationally
expensive.
The connected contributions from up and down quarks cancel each other out, while a recent study in quenched
QCD by the $\chi$QCD collaboration showed that the disconnected contributions do not cancel each other
out~\cite{Liu:2012nz}.
Whether this continues to hold for the 2+1 flavor theory and at physical quark masses is an important
question.

Important structure quantities like quark momentum fractions and helicities, form factors, axial and tensor
charges, and transverse momentum distributions (TMDs) are also being actively pursued (see the recent review
by Lin~\cite{Lin:2012ev}).
The QCDSF collaboration has recently reported agreement with experiment for the axial charge $g_{A}$ computed
in two-flavor QCD at the physical point with a quoted total error of a bit more than
3\%~\cite{Horsley:2013ayv}.
They observe large finite volume effects in their data, however, which are removed by fitting to SSE chiral
perturbation theory.
Further, they obtain values of $g_A$ below the experimental value for most of their simulated pion masses,
and there is no hint of increasing behavior with decreasing quark mass in their results until one is directly
at the physical point.
Other groups also find low values of
$g_{A}$~\cite{Yamazaki:2008py,Yamazaki:2009zq,Capitani:2012gj,Owen:2012ts,Alexandrou:2013joa}, even down to
$m_{\pi}=170$ MeV~\cite{Lin:2012nv} and 150 MeV~\cite{Green:2012ud}, so the QCDSF result, while interesting,
needs further study.
The LHP collaboration has done a detailed study using several source-sink separations for three-point
correlation functions down to $m_\pi=150$ MeV which indicates large excited state effects for the momentum
fraction, form factors, and charge radii~\cite{Green:2012ud}.
Likewise, the ETM collaboration recently reported results for nucleon structure using 2+1+1 flavor
twisted-mass QCD with $m_{\pi}$ down to 210 MeV.
These results suggest that precise nucleon structure calculations are feasible but will require concerted
effort and significant resources.

The RBC/UKQCD and LHP collaborations are starting computations of hadron structure observables using the
domain-wall fermion action directly at the physical pion mass on a large lattice (5.5 fm) with spacing
$a=0.114$ fm.
Measurements on a finer $a=0.086$ fm lattice are expected to start in about a year's time, and will provide a
continuum limit.
A~Japanese group using the K computer at Kobe has generated an ensemble using improved Wilson fermions on a
10 fm box, also with the physical pion mass.
Planned studies using this very large box will be important in addressing finite size effects.

Nucleon structure calculations are not at the level of precision of their lighter meson cousins, owing to
larger attendant statistical fluctuations and even more rapid exponential decay of correlators.
Even so, experience, new techniques, and growing computer resources are allowing improved calculations.
Recent results provide clear indications that longstanding discrepancies in $g_{A}$, nucleon charge radii,
and structure functions are due to chiral, excited state, and finite-size systematic errors.
The new generation of calculations at the physical pion mass on large lattices should largely eliminate
these, beginning an era of precision nucleon matrix element calculations.
These calculations will be difficult since current ones already struggle to achieve sub-five percent
statistical errors in the best cases when $m_{\pi}$ takes nearly physical values.
It is hoped that new error-reduction methods like all-mode averaging~\cite{Blum:2012uh} and sustained effort
on the new ensembles will allow for smaller errors than have been possible in the past (preliminary results
for $g_{A}$ using domain-wall fermions with $m_{\pi}=170$ are encouraging).
The computation of quark-disconnected diagrams needed for isoscalar quantities, while even more difficult,
is also being actively addressed~\cite{Bali:2011zzc,Liu:2012nz,Alexandrou:2012py}, and will become more
computationally feasible over the next five to ten years with the anticipated increase in computing
capabilities and resources.

%%%%%%%%%%%%%%%%%%%%%%%%%%%%%%%%%%%%%%%%%%%%%%%%%%%%%%%%%%%%
\section{Computational Resources}
\label{lqcd:sec:resources}
%%%%%%%%%%%%%%%%%%%%%%%%%%%%%%%%%%%%%%%%%%%%%%%%%%%%%%%%%%%%

In this section we discuss the computational and software infrastructure resources needed to reach the
scientific goals set out above.
We focus on the efforts and plans in the US, but comparable efforts are ongoing in Europe and Japan.

The lattice gauge theory research community in the United States coordinates much of its effort to obtain
computational hardware and develop software infrastructure through the USQCD Collaboration.
Support for USQCD has been obtained from the high-energy physics and nuclear physics offices of DOE in the
form of (i) funds for hardware and support staff, (ii) computational resources on leadership-class machines
through INCITE awards, and (iii) SciDAC awards for software and algorithm development.
The first has consisted of two 4--5 year grants, the second of which extends until 2014.
Since its inception, the INCITE program has awarded computing resources to USQCD every year.
SciDAC has funded three software projects for lattice QCD, the most recent beginning in 2012.
All three components have been critical for progress in lattice QCD in the past decade.
The primary purpose of USQCD is to support the high-energy and nuclear physics experimental programs in the
US and worldwide.
To this end, USQCD establishes scientific priorities, which are documented in white papers.
USQCD's internal and INCITE computing resources are then allocated, in a proposal driven process, to
self-directed smaller groups within USQCD to accomplish these goals.

At present, members of USQCD are making use of dedicated hardware funded by the DOE through the LQCD-ext
Infrastructure Project, as well as a Cray XE/XK computer, and IBM Blue Gene/Q and Blue Gene/P computers, made
available by the DOE's INCITE Program.
During 2013, USQCD, as a whole, expects to sustain approximately 300 teraflop/s on these machines.
USQCD has a PRAC grant for the development of code for the NSF's petascale computing facility, Blue Waters,
and expects to obtain a significant allocation on this computer during 2013.
Subgroups within USQCD also make use of computing facilities at the DOE's National Energy Research Scientific
Computing Center (NERSC), the Lawrence Livermore National Laboratory (LLNL), and centers supported by the
NSF's XSEDE Program.
For some time, the resources USQCD has obtained have grown with a doubling time of approximately 1.5~years,
consistent with Moore's law, and this growth rate will need to continue to meet the scientific objectives
described previously.

In addition, some components of USQCD have international connections.
HPQCD consists of scientists in the US and the United Kingdom and uses resources funded by the UK Science
and Technology Facilities Council (STFC), as well as USQCD resources.
The RBC Collaboration has access to dedicated Blue Gene/Q computers at the RIKEN BNL Research Center at
Brookhaven National Laboratory, which receives funding from Japan, and, via their collaborators in the UKQCD
Collaboration, at the STFC DiRAC facility at the University of Edinburgh.

Gauge-field configurations must be generated in series, generally requiring high-capability machines such as
the Blue Gene/Q and the Cray XE/XK.
The advent of petascale supercomputers is for the first time enabling widespread simulations with physical up
and down quark masses at small lattice spacings and large volumes.
This development will enable major advances on a range of important calculations.
For the next five years, the US lattice-QCD effort in high-energy physics will generate large sets of
gauge-field ensembles with the domain-wall fermion (DWF)~\cite{Kaplan:1992bt,Furman:1994ky,Vranas:2006zk} and
highly improved staggered quark (HISQ)~\cite{Follana:2006rc} lattice actions.
Each of these formulations has its own advantages.
Further, for the most important calculations, it is helpful to employ more than one lattice formulation in
order to ensure that systematic errors are truly under control.
Computations of operator expectation values on these gauge-field ensembles can be run in parallel and are
well-suited for high-capacity PC and GPU clusters such as the dedicated lattice-QCD facilities at Fermilab
and Jefferson Lab.
Therefore continued support of both the national supercomputing centers and of dedicated USQCD hardware will
be needed to meet the US lattice-QCD community's scientific goals.
 
The software developed by USQCD under a SciDAC grant enables US lattice gauge theorists to use a wide variety
of architectures with very high efficiency, and it is critical that USQCD software efforts continue at their
current pace.
Historically, the advance preparation of USQCD for new hardware has enabled members to take full advantage of
early science time that is often available while new machines are coming online and being tested.
Over time, the development of new algorithms has had at least as important an impact on the field of lattice
QCD as advances in hardware, and this trend is expected to continue, although the rate of algorithmic
advances is not as smooth or easy to predict as that of hardware.

%%%%%%%%%%%%%%%%%%%%%%%%%%%%%%%%%%%%%%%%%%%%%%%%%%%%%%%%%%%%
\section{Summary}
\label{lqcd:sec:summary}
%%%%%%%%%%%%%%%%%%%%%%%%%%%%%%%%%%%%%%%%%%%%%%%%%%%%%%%%%%%%

Lattice-QCD calculations now play an essential role in the search for new physics at the intensity frontier.
They provide accurate results for many of the hadronic matrix elements needed to realize the potential of
present experiments probing the physics of flavor.
The methodology has been validated by comparison 
with a broad array of measured quantities, 
several of which had not been well measured 
in experiment when the first good lattice calculation became available. 
In the US, this effort has been supported in an essential way by hardware and software support provided to
the USQCD Collaboration.

In the next decade, lattice-QCD has the welcome opportunity to play an expanded role in the search for new
physics at the intensity frontier.
This chapter has laid out an ambitious vision for future lattice calculations matched to the experimental
priorities of the planned \PX\ physics program:
\begin{itemize}

\item We will steadily improve the calculations of the hadronic parameters (decay constants, semileptonic
form factors, and mixing matrix elements) needed to obtain the CKM matrix elements and constrain the CKM
unitarity triangle.
We will also continue to improve the determinations of the quark masses and~$\alpha_S$.
We forecast improvements by factors of 2--4 over the next five years, with most quantities having errors at
or below the percent level.
The quark masses and CKM matrix elements enter Standard-Model rates for many rare processes.
Most notably, the anticipated improvement in the $B\to D^* \ell\nu$ form factor (needed for $|V_{cb}|$) will
reduce the uncertainty in the Standard-Model predictions for $K\to\pi\nu\bar\nu$ to the target experimental
precision.

\item We will calculate proton and neutron matrix elements relevant for determining neutrino-nucleon
scattering cross sections, interpreting muon-to-electron conversion measurements as constraints on
new-physics models, constraining TeV- and GUT-scale physics via measurements of EDMs and neutron
$\beta$-decay, and searches for proton decay and neutron-antineutron oscillations.
These calculations are in earlier stages than the precision quark-flavor computations described above.
Further, baryon correlation functions suffer from larger statistical uncertainties than for mesons.
For discovery modes, $\sim$10--20\% precision should be useful and straightforward.
Even for precision matrix elements, such as the axial-vector form factor arising in neutrino scattering,
reducing the errors to $\sim5\%$ will be feasible, certainly over the course of \PX.

\item We will calculate more computationally demanding matrix elements that are needed for the interpretation
of planned (and in some cases old) kaon and charged lepton experiments.
These include the hadronic contributions to the muon $g-2$, long-distance contributions to kaon mixing and to
$K\to\pi \nu\bar\nu$ decays, and the SM prediction for \CP~violation in $K\to\pi\pi$ decays ($\epsilon'$).
Here we require new methods, but the methodology is at a fairly advanced stage of development.
Many of these calculations are in early stages, so future errors are difficult to anticipate.
However, we will devote substantial theoretical and computational effort to these calculations commensurate
with their high experimental priority, so prospects are good for obtaining the errors needed by the
experiments.

\item Key to achieving these goals will be the use of physical light-quark masses.
At the accuracy we propose to obtain for many quantities, we will need to include the effects of isospin
breaking, electromagnetism, and dynamical charm quarks.

\item Implementation of the program outlined here will require dedicated lattice-QCD computing hardware,
leadership-class computing, and efficient lattice-QCD software.
Therefore continued support of USQCD computing infrastructure and personnel is essential to fully capitalize
on the enormous investments in the high-energy physics and nuclear-physics experimental programs.

\end{itemize}
The future success of the \PX\ physics program hinges on reliable Standard-Model predictions on the same time
scale as the experiments and with commensurate uncertainties.
Many of these predictions require nonperturbative hadronic matrix elements that can only be computed
numerically with lattice-QCD.
The lattice-QCD community is well-versed in the plans and needs of the experimental intensity-physics program
over the next decade, and will continue to pursue the necessary supporting theoretical calculations.
Indeed, lattice-QCD calculations for the intensity frontier may play a key role in definitively establishing
the presence of physics beyond the Standard Model and in determining its underlying structure.

%%%%%%%%%%%%%%%%%%%%%%%%%%%%%%%%%%%%%%%%%%%%%%%%%%%%%%%%%%%%

\bibliographystyle{apsrev4-1}
\bibliography{lqcd/refs}
 % Tom and Ruth

\backmatter

% \pagestyle{empty}
% \cleardoublepage
% 
% {\thispagestyle{empty}
% \centering\sffamily
% \Huge
% \color{blue}
% \rule{\textwidth}{3pt}
% \vfill
% INSERT \vfill
% BLUE PAGE \vfill
% HERE
% \vfill
% \rule{\textwidth}{3pt}
% }

% \addcontentsline{toc}{chapter}{Index}
% \printindex

\end{document}